%% file: sBelleSR.tex
\documentclass[floatfix,amsmath,amssymb,12pt,nofootinbib,unsortaddress,superscriptaddress,amsfont,a4paper,preprint,tightenlines,twoside]{revtex4}
\usepackage{graphicx,color} 
\usepackage{dcolumn}  
\usepackage{amsmath}  
\usepackage{times}
\usepackage{enumerate}
\usepackage{array}

\usepackage{pennames}

\makeindex

\begin{document}
\begin{titlepage}
\thispagestyle{empty}
 \begin{flushright}
  \vspace*{1cm}
 \end{flushright}
\vspace*{2cm}
\begin{flushleft}
\Huge
{\bf sBelle Design Study Report}
\end{flushleft}
\vspace*{-3mm}
\rule{\textwidth}{3pt}
\begin{flushright}
\large
{\sl Year 2008 Supplemental Update to the Detector LoI}\\
\end{flushright}
\vspace*{2cm}
\begin{center}
\vspace*{4cm}
\vspace*{2cm}
\LARGE
 sBelle Design Group
\vspace*{4cm}
\end{center}
\end{titlepage}
\begin{titlepage}
\thispagestyle{empty}
\newpage
\mbox{}
\end{titlepage}
\newpage
\title{sBelle Design Study Report}
  \setcounter{page}{1}
  \pagenumbering{roman}
\input{author}

  \begin{abstract}
   In this note, we compile results of various simulation studies for
   the upgrade of the Belle detector. Based on these studies, we propose
   a set of optimum or appropriate parameters for the detector.
  \end{abstract}

  \preprint{KEK Report 2008-7}
  \maketitle

  \newpage
  \tableofcontents

  \clearpage
  \input{intro}

  \clearpage
  \input{beambg}

  \clearpage
  \input{baseline}


  \clearpage
  \input{svd}

  \clearpage
  \input{cdc}

  \clearpage
\input{pidstudy}

  \clearpage
  \input{ecl}

  \clearpage
  \input{kl_klm}

  \clearpage
  \input{material}

  \clearpage
  \input{boost}




\input{closing.tex}

\end{document}

%% file: author.tex
  \author{I.~Adachi}\affiliation{High Energy Accelerator Research Organization (KEK), Tsukuba} 
  \author{M.~Danilov}\affiliation{Institute for Theoretical and Experimental Physics, Moscow} 
  \author{K.~Hara}\affiliation{Nagoya University, Nagoya} 
  \author{T.~Hara}\affiliation{Osaka University, Osaka} 
  \author{T.~Higuchi}\affiliation{High Energy Accelerator Research Organization (KEK), Tsukuba} 

  \author{K.~Inami}\affiliation{Nagoya University, Nagoya} 
  \author{H.~Ishino}\affiliation{Tokyo Institute of Technology, Tokyo} 

  \author{N.~Katayama}\affiliation{High Energy Accelerator Research Organization (KEK), Tsukuba} 
  \author{T.~Kawasaki}\affiliation{Niigata University, Niigata} 
  \author{A.~Kibayashi}\affiliation{High Energy Accelerator Research Organization (KEK), Tsukuba} 
  \author{P.~Kri\v zan}\affiliation{University of Ljubljana, Ljubljana}\affiliation{J. Stefan Institute, Ljubljana} 
  \author{P.~Krokovny}\affiliation{High Energy Accelerator Research Organization (KEK), Tsukuba} 
  \author{Y.~Kuroki}\affiliation{Osaka University, Osaka} 
  \author{A.~Kuzmin}\affiliation{Budker Institute of Nuclear Physics, Novosibirsk} 

  \author{I.~Nakamura}\affiliation{High Energy Accelerator Research Organization (KEK), Tsukuba} 
  \author{E.~Nakano}\affiliation{Osaka City University, Osaka} 
  \author{M.~Nakao}\affiliation{High Energy Accelerator Research Organization (KEK), Tsukuba} 
  \author{S.~Nishida}\affiliation{High Energy Accelerator Research Organization (KEK), Tsukuba} 

  \author{H.~Ozaki}\affiliation{High Energy Accelerator Research Organization (KEK), Tsukuba} 
  \author{P.~Pakhlov}\affiliation{Institute for Theoretical and Experimental Physics, Moscow} 
  \author{R.~Pestotnik}\affiliation{J. Stefan Institute, Ljubljana} 

  \author{Y.~Sakai}\affiliation{High Energy Accelerator Research Organization (KEK), Tsukuba} 
  \author{C.~Schwanda}\affiliation{Institute of High Energy Physics, Vienna} 
  \author{S.~Shinomiya}\affiliation{Osaka University, Osaka} 
  \author{K.~Sumisawa}\affiliation{High Energy Accelerator Research Organization (KEK), Tsukuba} 

  \author{O.~Tajima}\affiliation{High Energy Accelerator Research Organization (KEK), Tsukuba} 
  \author{K.~Trabelsi}\affiliation{High Energy Accelerator Research Organization (KEK), Tsukuba} 
  \author{T.~Tsuboyama}\affiliation{High Energy Accelerator Research Organization (KEK), Tsukuba} 

  \author{S.~Uehara}\affiliation{High Energy Accelerator Research Organization (KEK), Tsukuba} 
  \author{S.~Uno}\affiliation{High Energy Accelerator Research Organization (KEK), Tsukuba} 
  \author{Y.~Ushiroda}\affiliation{High Energy Accelerator Research Organization (KEK), Tsukuba} 


%% file: intro.tex
 \section{Introduction}
  \setcounter{page}{1}
  \pagenumbering{arabic}
 The long awaited upgrade of Belle, which is refered to as
 sBelle~\footnote{The name of the new collaboration/detector has not
 been determined as we write this document. 'sBelle' is not an
 authorized name, but will be used for simplicity in this report.}
 in this note, has become increasingly realistic and is very
 likely to happen in the near future.
 We have proposed a set of detectors in the Letter of Intent (LoI)
 ~\cite{bib:LoI} in 2003, which should work in the harsh beam background
 environment of the upgraded KEKB collider. Since the LoI, we have
 continued the
 R\&D studies mainly in hardware in order to realize such a detector.
 As the design becomes more realistic and more concrete, we found 
 several open questions concerning the parameters of the detector design.

 Indeed, there will be
 several possible parameters such as geometry, structure materials,
 fundamental performance of sensors; and there is room to
 improve the overall performance by finding the best set of parameters.
 When the best set is not feasible either technologically or financially,
 we may need to compromise to some extent.
 In such a case, we must understand what will result. Finally, we
 require that the minimal performance of the new detector be at least as
 good as that of Belle, which is not very conservative because of the
 extremely harsh beam background.

 In May 2007, a task force was formed to lead such discussions in a scientific way,
 and make proposals for the detector parameters from the viewpoint of
 physics analysis. The task force was asked to prepare a report of its studies
 in one year, which is the present report.
 Since we do not have an integrated simulation software tool for
 sBelle (one is under development based on Geant 4 and will require
 some more time to be ready), we began by preparing the simulation
 environments.
 One powerful tool is the Belle Geant 3 based full simulator (gsim) in
 which the parameters such as sensor configurations, material
 densities, and background immunities are modified to represent the sBelle detector.
 Another tool is a fast simulator called fsim6~\cite{bib:fsim6},
 which gives detector responses based on a priori probability density
 functions; where the PDFs are retuned for sBelle.
 In addition, there are standalone Geant 4 simulators for certain
 sub-detectors,
 and a simple track simulator named trackerr.
 Using those tools  we simulated events of various complexity, from  
 simple single tracks  to full $B\bar{B}$ and
 $\tau^+\tau^-$ events in which the real background data obtained
 from Belle can be overlaid to simulate the level of expected conditions
 at sBelle.
 
 Benchmark physics modes in this report are selected mainly because of their
 importance in the sBelle era and 
 sensitivity to the detector parameters to be examined, but also 
 because analysis code and the person in charge were available.
 In spite of very limited manpower, we included the following
 important modes in this report:
 $\PBz\to\phi\PKz$,
 $\PBz\to\PKzS\Pgpz\gamma$,
 $\tau\to\mu\gamma$,
 $B\to\tau\nu$,
 and $B\to\rho\gamma$.
 Studies of other decay modes that are underway will be compiled
 in a special report.

 In the following sections, we first discuss the beam background (section~\ref{sec:beambg}),
 then we recapitulate the design of the sBelle detector that we
 consider as the baseline (section~\ref{sec:baseline}), and present the 
 studies done for each sub-detector~(\ref{sec:svd} to \ref{sec:klm}).
 Specially added are sections about the
 effect of the detector material~\ref{sec:material} and the beam energy
 asymmetry~\ref{sec:boost} that are the key parameters in the overall
 consideration of the detector.

%% file: beambg.tex

\section{Expected beam background}
\label{sec:beambg}

In the forthcoming KEKB upgrade, the luminosity will be  improved 
by increasing the number of bunches and the beam currents,
and by optimizing the beam optics in the interaction region (IR)
as will be discussed in section \ref{subsec:beambg_para}.
Some of these changes will also increase the level of the
beam-induced background for the sBelle detector.
%
A realistic estimation of the beam-induced background level compared
to the present level is important for the new detector design.
By estimating the occupancy, the radiation damage and the dead time for
each sub-detector, we will be able to deduce the impact of the
beam-induced background on physics analyses.
Note that all these analyses have to be 
updated once more when we fix the design of the interaction region.

\subsection{Origins of beam-induced backgrounds}
We can divide the beam-induced background into five types according
to its origin:
 scattering of the beam on residual gas\footnote{
In what follows, we will use the expression 'beam-gas scattering' 
or just 'beam-gas' for this kind of scattering.}, 
Touschek scattering, synchrotron radiation (SR) from
the high energy ring (HER) upstream direction, 
backscattering of SR from HER downstream
and those from electron-positron interactions at the interaction point (IP).
A brief explanation for each of them follows
(details can be found in the LoI~\cite{bib:LoI}).
\paragraph{Beam-Gas}
Beam-gas scattering (bremsstrahlung and Coulomb scattering)
changes the momenta of beam particles,
which then hit the walls of vacuum chambers and magnets.
The shower particles then produced are
the main source of the beam-induced background.
This component is proportional to the product  $I \times P$, 
where $I$ is the beam current 
and   $P$ is the pressure in the ring.

\paragraph{Touschek}
Touschek scattering is intra-bunch scattering.
This type of background is proportional to the bunch current and the
number of bunches, and inversely proportional to the beam size.
The contribution from the HER can be ignored, because
the rate of Touschek scattering is proportional to $E^{-3}$,
where $E$ is the beam energy,
and also to the bunch current density,
which is less than in the LER because
the HER current is smaller than the low energy ring (LER) current.

\paragraph{SR from upstream (SR Upstream)}
Since the energy of radiation is in the keV range, the contribution from
SR is negligible except for the innermost detector, the vertex detector.
The level of this type of background is proportional to the HER current.
It also depends on various parameters,
magnet positions, bending radii and beam orbits. However,
as a rough estimate, we can assume that the intensity is proportional
to the HER current.
The SR power is proportional to $E^2B^2$, where $E$ is the
beam energy and $B$ is the magnetic field strength~\cite{bib:PDG}.
The power is lower for the LER as $E$ is smaller and $B$ is
designed to be smaller in our case. The net
contribution from the LER is negligible in the present IR configuration.
Careful simulation studies both for the HER and the LER will be
performed for the upgraded IR.


\paragraph{Backscattering of SR from downstream (SR Backscattering)}
The final focusing magnet for the HER downstream side is called 'QCS-R',
where R stands for the right hand side in the schematic drawing.
The magnet QCS-R provides the final focusing of the LER beam.
It also works as a bending magnet to separate the outgoing HER beam from
the LER beam.
Strong SR is emitted because of this bending.
This SR hits the downstream chamber, which is $\sim$9\,m away from the
IP, then it backscatters in the detector again
through the IP region.
Careful simulation studies both for the HER and the LER will also be
performed for the backscattering component.

\begin{figure}[h]
\begin{center}
\includegraphics[width=0.75\textwidth]{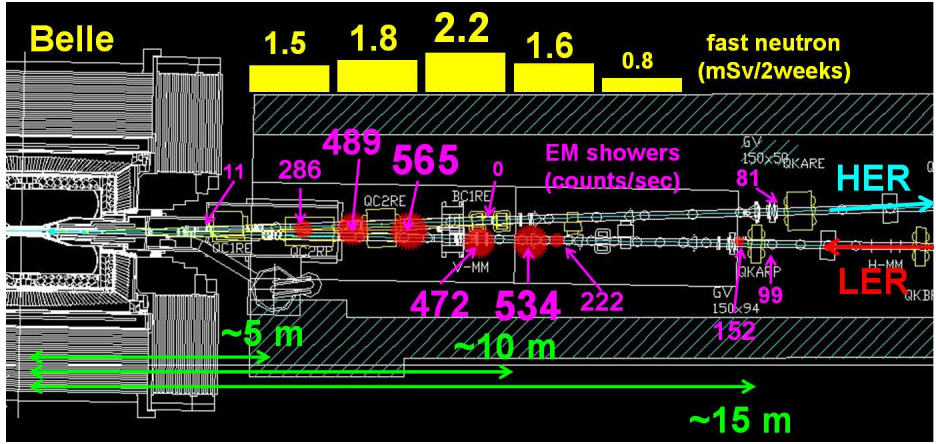}
\caption{
 Measured radiation levels around the beam lines in the HER downstream
 of the Belle detector.
 Neutron dose rates were measured outside of the concrete shield in 2003.
 The electromagnetic (EM) shower rates were measured with a
 scintillation counter in the same year.
 The position resolution of a movable EM shower counter
 is a 150\,mm diameter circle along the beam lines; 
 the counter is surrounded by a 200\,mm thick lead shield and has a
 window diameter of 20\,mm.
 }
 \label{fig:Bhabha_neutron}
\end{center}
\end{figure}

\paragraph{Radiative Bhabha scattering}
The rate of the radiative Bhabha events is proportional to the
luminosity. 
Photons from the radiative Bhabha scattering propagate along the beam axis
direction and interact with the iron of the magnets. In these interactions,
neutrons are copiously produced 
via the giant photo-nuclear resonance mechanism~\cite{bib:PDG}.
These neutrons are the main background source for the outer-most detector, 
the $K_L$ and muon detector (KLM) in the instrumented return yoke of the
spectrometer.
Figure~\ref{fig:Bhabha_neutron} shows the measured radiation levels along
the beam lines in the HER downstream of the Belle detector.
It indicates that 
neutrons are generated along the beam lines within 11\,m from the IP.

In addition, in the radiative Bhabha events
both electron and positron energies decrease.
The scattered particles are therefore over-bent by the QCS magnets and hit  
 the wall of magnets where electromagnetic showers are generated.
In the present Belle spectrometer, this type of background has not been
observed yet.
However, it is expected that the level of this kind of background will 
increase in the 
upgraded KEKB collider because of the considerably different interaction
region design. 
A simulation study for this background is described in 
section~\ref{section:BeamBG_radiativeBhabha}.

\subsection{Background composition for the present KEKB~/~Belle}
The relative contributions of background components are summarized
in Table~\ref{table:bkg_fractions_200X} for the case of the present configuration.
Here, SVD-A (B) means inner (outer) side of the SVD sensors with respect to
the KEKB rings.
This table is based on various background studies described in
the LoI~\cite{bib:LoI}.
We have observed a background contribution originating from 
radiative Bhabhas
only in the KLM; for other sub-detectors, the contributions are estimated
by Monte Carlo (MC)
studies as will be described in subsection~\ref{section:BeamBG_radiativeBhabha}.

\begin{table}[hbt]
\begin{center}
\caption{Relative contributions for individual detector components in the present
 KEKB~/~Belle}
\label{table:bkg_fractions_200X}
\begin{tabular}
{
 @{\hspace{0.2cm}}l@{\hspace{0.2cm}}
 @{\hspace{0.2cm}}c@{\hspace{0.2cm}}   
 @{\hspace{0.2cm}}c@{\hspace{0.2cm}}   
 @{\hspace{0.2cm}}c@{\hspace{0.2cm}}   
 @{\hspace{0.2cm}}c@{\hspace{0.2cm}}   
 @{\hspace{0.2cm}}c@{\hspace{0.2cm}}   
}
\hline \hline
 & SVD~{\it {\footnotesize (A~,~B)}} & CDC & PID &ECL & KLM~{\it {\footnotesize (Barrel~,~E.C.-Fwd~,~E.C.-Bwd)}}
\\
\hline
\\
Beam-Gas (HER) & 0.56~,~0.42 & 0.25 & 0.45 & 0.40 & 0.15~,~0.00~,~0.25
\\
~\hspace{48pt}~(LER)
& 0.20~,~0.13 & 0.40 & 0.40 & 0.40 & 0.12~,~0.20~,~0.00
\\
Touschek
& 0.04~,~0.03 & 0.10 & 0.10 & 0.10 & 0.03~,~0.05~,~0.00
\\
SR Backscattering & 0.18~,~0.40 & 0.24 & -- & -- & -- 
\\
SR Upstream & 0.02~,~0.02 & -- & -- & -- & -- 
\\
Radiative Bhabha 
& -- & 0.01 & 0.05 & 0.10 & 0.70~,~0.75~,~0.75
\\
\hline \hline 
\end{tabular}
\end{center}
\end{table}

\clearpage
\subsection{Machine parameters for each stage of the upgrade}\label{subsec:beambg_para}

The machine parameters 
that are used for the estimation of the beam-induced background 
are summarized in Table~\ref{table:bkg_machine_parameters}
for each stage of the upgrade:
present collider (labeled as '200X'),
initial stage of the upgrade (labeled as '201X')
and ultimate stage of the upgrade (labeled as '202X').
They are based on the previous LoI~\cite{bib:LoI} design 
and recent discussions with the KEKB upgrade design group.
In the '201X' version, the luminosity improvement will be achieved 
 by reducing the value of the $\beta$ function at the IP,
by reducing the electron cloud density in the beam pipe, 
 and by making use of a crab crossing.
In this first stage, the beam currents will not be increased significantly, 
but we will increase the number of bunches in the storage ring 
aiming to increase the beam currents at a later stage.
To reduce the  electron cloud density and  residual gas pressure in the beam pipe, 
all vacuum chambers will be replaced and equipped with ante-chambers\footnote{
Ante-chambers will also help to dispose of high power SR  as heat.}.
 Note that the increase 
of the pressure in the ring is related to the increase in SR 
power; in the initial stage of operation, this effect is enhanced because
the vacuum chambers will not yet be degassed well.

In order to reduce the  value of the $\beta$ function, 
the position of the focusing magnets will be changed.
A smaller horizontal beam size at the IP requires a larger beam size in 
its immediate vicinity near the IR, and this, in turn, requires larger 
focusing magnets (QC1 and QC2 in Fig.~\ref{fig:Bhabha_IR_drawings}). 
Since, in addition, the two beams 
must be well separated to avoid the influence of the magnets of one of the
beams on the other beam,  the beam crossing angle has to be increased
from 22\,mrad to 30\, mrad.
Due to these modifications, the critical energy of SR in QCS-R will almost
double ($\sim$40\,keV); we need to consider a better
IR configuration so that the SR fan does not hit the components near the IR.

\begin{table}[hbt]
\begin{center}
\caption{Assumed machine parameters for beam-induced background estimation.}
\label{table:bkg_machine_parameters}
\begin{tabular}
{
 @{\hspace{0.5cm}}c@{\hspace{0.cm}}
 @{\hspace{0.cm}}l@{\hspace{0.5cm}}
 @{\hspace{0.5cm}}r@{\hspace{0.5cm}}
 @{\hspace{0.5cm}}r@{\hspace{0.5cm}}
 @{\hspace{0.5cm}}r@{\hspace{0.5cm}}
}
\hline \hline
&  & 200X & 201X & 202X \\
\hline
$E$ & : Beam Energy ~(GeV; HER~/~LER) & 8.0~/~3.5 & 8.0~/~3.5 & 8.0~/~3.5 \\
$I$ & : Beam Current ~(A; HER~/~LER) & 1.2~/~1.6 & 1.4~/~2.6 & 4.1~/~9.4 \\
$N$ & : Number of Bunches & 1300 &  5000 & 5000 \\
$i$ & : Bunch Current for LER (mA) & 1.2 & 0.5 & 1.9 \\
$l$ & : Bunch length~(mm)      &    7    &    3    &   3 \\
$\sigma$ & : Beam Size in the Ring ~(scale factor) &    1    &    1    &   1 \\
$\theta$    & : Beam crossing angle~(mrad) & 22 & 30 & 30 \\
$W$ & : Power of SR from QCS-R ~(kW) & 30 & 61 & 179 \\
$P$ & : Pressure in the Ring ($\times 10^{-7}$~Pa) & 1.25 & 3.0 & 5.0 \\
$L$ & : Peak Luminosity (/nb/sec) & 17 & 100 & 500 \\
\hline \hline
\end{tabular}
\end{center}
\end{table}

We estimate how much more background the sBelle detector will receive for
each component during each run period based on the machine parameters
summarized in Table~\ref{table:bkg_machine_parameters} and following the
formulae listed in Table~\ref{table:bkg_scaling_factor}. The results are
summarized in Table~\ref{table:bkg_scaling_factor}. We observe that the
background level increases predominantly due to the increase of the 
beam currents.

\begin{widetext}
 \begin{table}[hbt]
  \begin{center}
   \caption{Background scale factors normalized to '200X'.}
   \label{table:bkg_scaling_factor}
   \begin{tabular}
    {
    @{\hspace{0.5cm}}l@{\hspace{0.5cm}}
    @{\hspace{0.2cm}}l@{\hspace{0.2cm}}
    @{\hspace{0.2cm}}r@{\hspace{0.2cm}}
    @{\hspace{0.2cm}}r@{\hspace{0.5cm}}
    }
    \hline \hline
    & Scaling Formula & 201X & 202X \\
    \hline
    Beam-Gas (HER)& $I \times P$ & 3 &  14  \\
    ~\hspace{48pt}~(LER)     & $I \times P$ & 4 &  24  \\
    Touschek & $N \times i \times i / (\sigma^2 \times l)$ &
    2 -- 3 &  23 -- 46 \\
    &{\footnotesize
    uncertainty in dynamic aperture: 1 -- 2
    }
    \\
    SR Backscattering & $W$ & 2 & 6 \\
    SR Upstream ($r_{min}$ = 2.0\,cm)& $I$ & 1.2 & 3.4 \\
    ~\hspace{62pt}~($r_{min}$ = 1.5\,cm)& {\it simulation} & 14 & 40 \\
    Radiative Bhabha & $L$ 
    & 6 & 30 \\
    \hline \hline 
   \end{tabular}
  \end{center}
 \end{table}

\end{widetext}

\subsection{Simulation study of the radiative Bhabha background contributions}
\label{section:BeamBG_radiativeBhabha}

\begin{figure}[h]
\begin{center}
\includegraphics[width=0.77\textwidth]{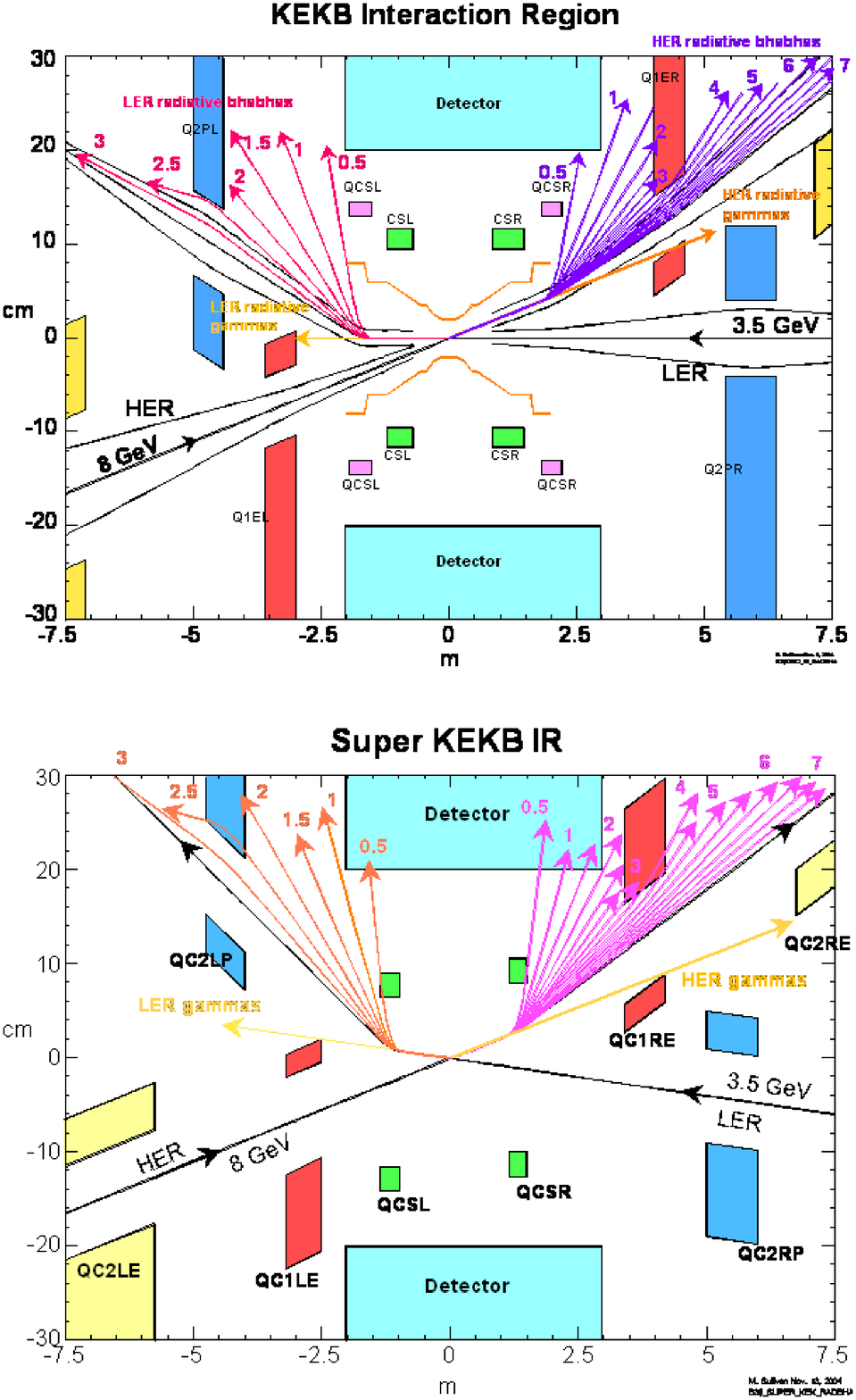}
\caption{
 Trajectories of electrons and positrons of radiative Bhabha events in
 the $x-z$ view of the IR.
 Since their momenta are lower than those of the beams, those particles are
 over-bent by QCS magnets and enter into the detector volume.
 With the new IR design that has a larger beam crossing angle,
 high momentum particles from the HER, namely electrons by default,
 over 1\,GeV can hit the walls of the magnets or the detector.
 This results in a background term proportional to luminosity.
}
 \label{fig:Bhabha_IR_drawings}
\end{center}
\end{figure}

\begin{figure}[h]
\begin{center}
\includegraphics[width=0.77\textwidth]{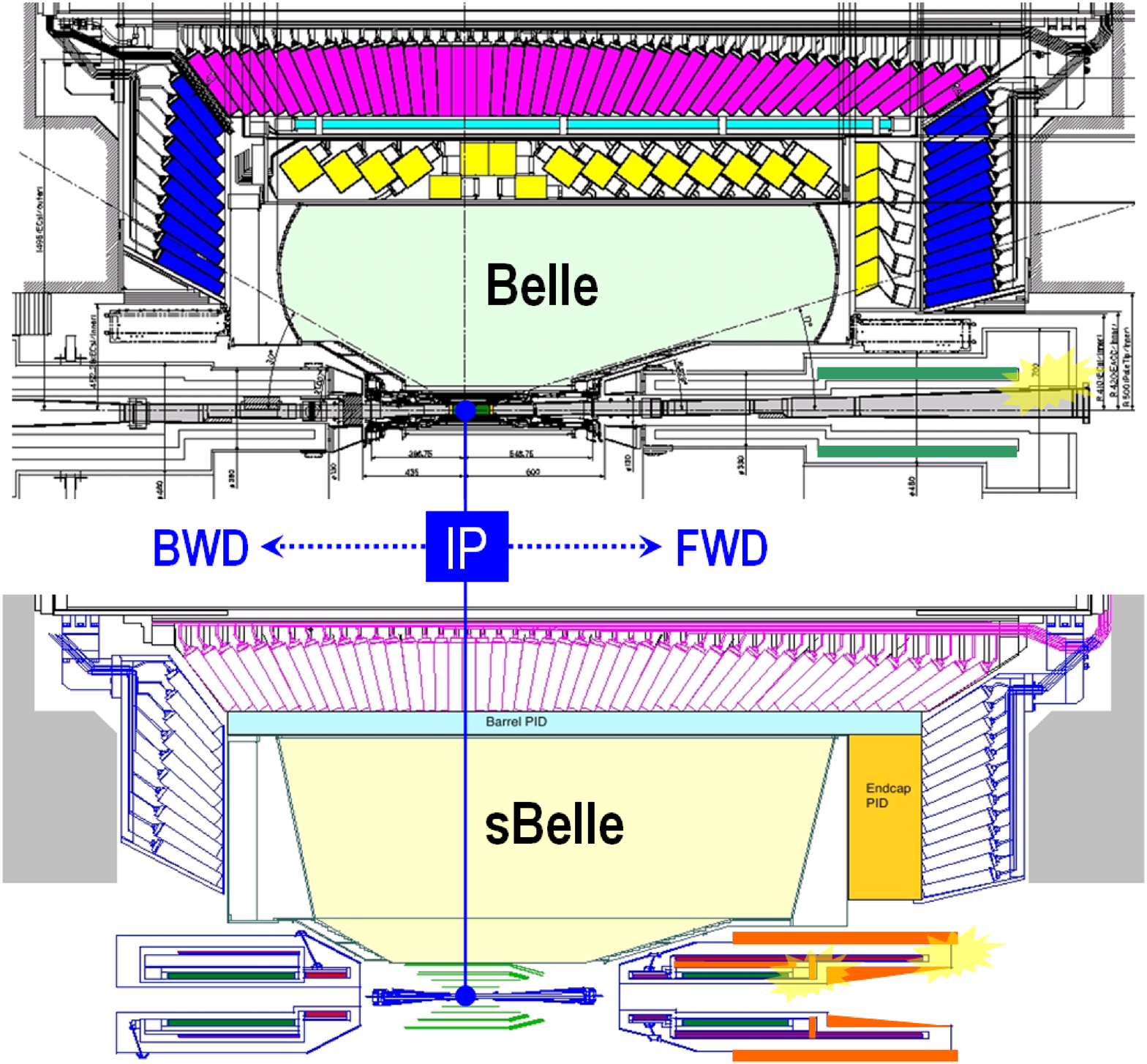}
\caption{
 Top-half views of the Belle with the present QCS configuration (upper)
 and the sBelle with the new QCS configuration (lower).
 The star marks indicate where the over-bent particles hit.
 In the upper configuration,
 most of the over-bent HER particles due to radiative Bhabha scattering
 hit the edge of the QCS magnet. On the other hand, 
 most of the over-bent HER particles strike the inner edge of the QCS
 magnet due to the larger crossing angle in the lower configuration (sBelle).
 }
 \label{fig:Bhabha_QCS_drawings}
\end{center}
\end{figure}

In Fig.~\ref{fig:Bhabha_IR_drawings}, the trajectories of electrons and
positrons
that have lower momenta than the momentum at the nominal beam orbit,
and are hence over-bent by magnets, are shown for the cases of the present
and upgraded designs.
In the upgraded KEKB, both the rate and the energy of the showers from
the over-bent particles are expected to increase, because of the larger
crossing angle.
The top-half view of the detector and QCS magnets are shown in
Fig.~\ref{fig:Bhabha_QCS_drawings}.
Over-bent particles hit the edge of the QCS magnets.
The QCS magnets are located outside of the endcaps in the
configuration of the present Belle and KEKB; most of the 
shower particles from QCS do not hit the detector.
The QCS magnets will be located closer to the IP in the upgraded
detector; hence the shower particles will tend to enter the detector volume.

\begin{figure}[h]
\begin{center}
\includegraphics[width=1.00\textwidth]{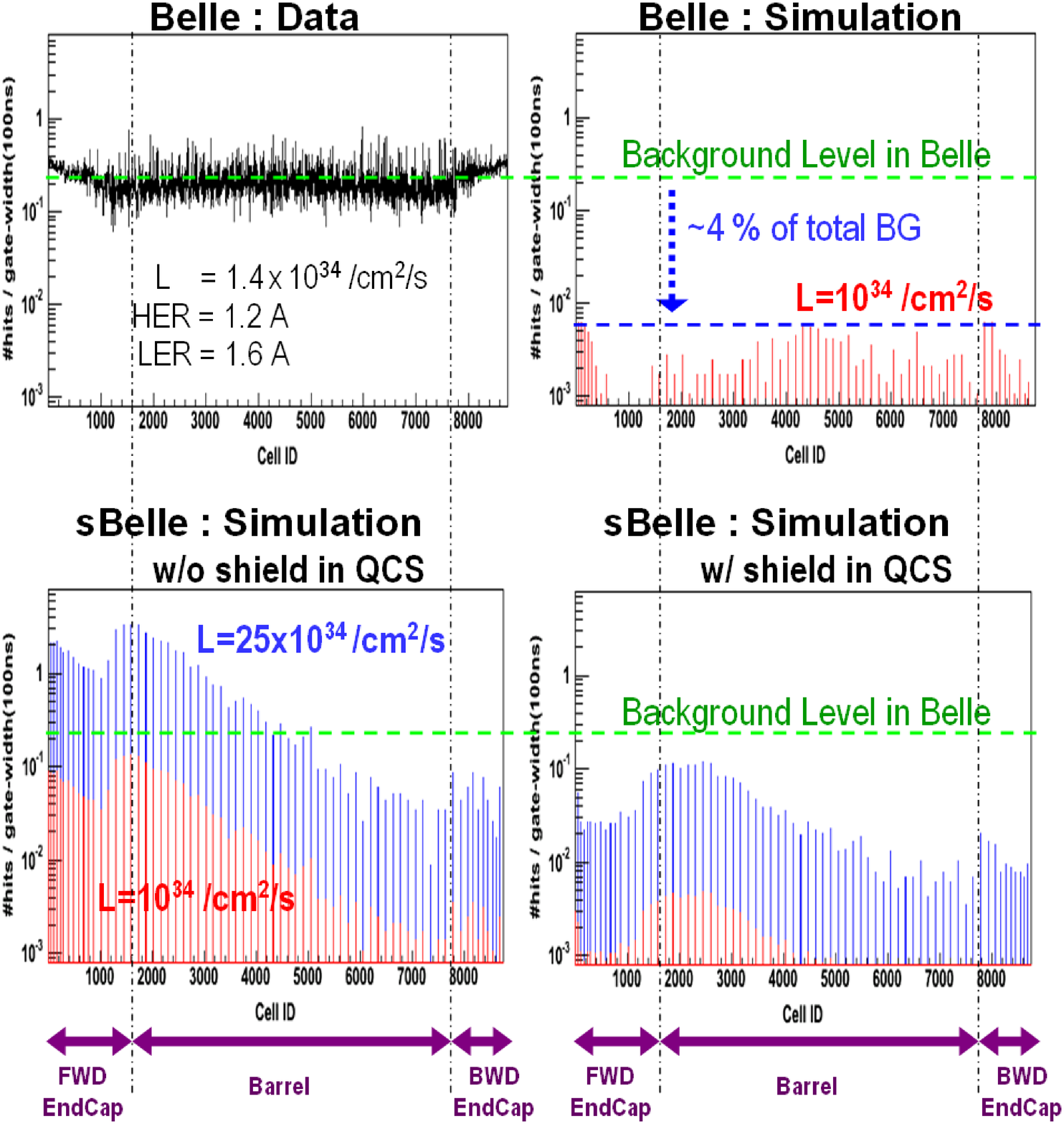}
\caption{
 Total background hit rate per CsI(Tl) calorimeter cell
 as a function of its ID number
 for the real data taken in experiment number 47, Dec. 2005 (upper left).
 The contribution from radiative Bhabha events is
 simulated based on the present IR configuration,
 assuming a luminosity of $1\times 10^{34}$\,/cm$^2$/s (upper right).
 Lower two figures show simulation results for radiative Bhabha events
 based on the upgrade designs: without any shield material around the QCS
 magnets (lower left) and with heavy metal shields (lower right).
 We evaluate two luminosities $1\times 10^{34}$\,/cm$^2$/s
 (red, or light-gray in grayscale) 
 and $25 \times 10^{34}$\,/cm$^2$/s (blue, or gray in grayscale) in the
 lower plots.
 }
 \label{fig:Bhabha_simulation_ecl}
\end{center}
\end{figure}

This background source was studied by MC simulation.
The over-bent electrons and positrons are generated by the 
MC simulator for radiative Bhabha scattering, BBBREM~\cite{bib:BBBREM}.
The development of electromagnetic showers of over-bent particles and the
detector responses are simulated with Geant 4~\cite{bib:geant4}.
In Fig.~\ref{fig:Bhabha_simulation_ecl}, the hit rate of CsI(Tl)
calorimeter cells is plotted as a function of cell ID for the Belle
and sBelle configurations.
In the present configuration, 
the contribution from radiative Bhabhas
is only around 4\,\% of the total background.
It is difficult to observe such a small contribution in the real data.
Indeed we have confirmed that there is no background component in the
calorimeter that
is proportional to the luminosity.
On the other hand, radiative Bhabhas will be a serious background source in the
new IR configuration.
A solution to suppress its contribution is to place heavy metal
shields inside and outside of the QCS magnets;
the contribution can be suppressed to the same level as in the present configuration
for the case of the same luminosity.
Thus, we can simply extrapolate this background as a component proportional to the
luminosity.
The background rates on the backward side are an order of magnitude smaller
than on the forward side as seen from Fig.~\ref{fig:Bhabha_IR_drawings}; 
nevertheless, for safety reasons QCS-L will be shielded as well.

\subsection{Trends in the background levels}
\begin{figure}[h]
\begin{center}
\includegraphics[width=1.00\textwidth]{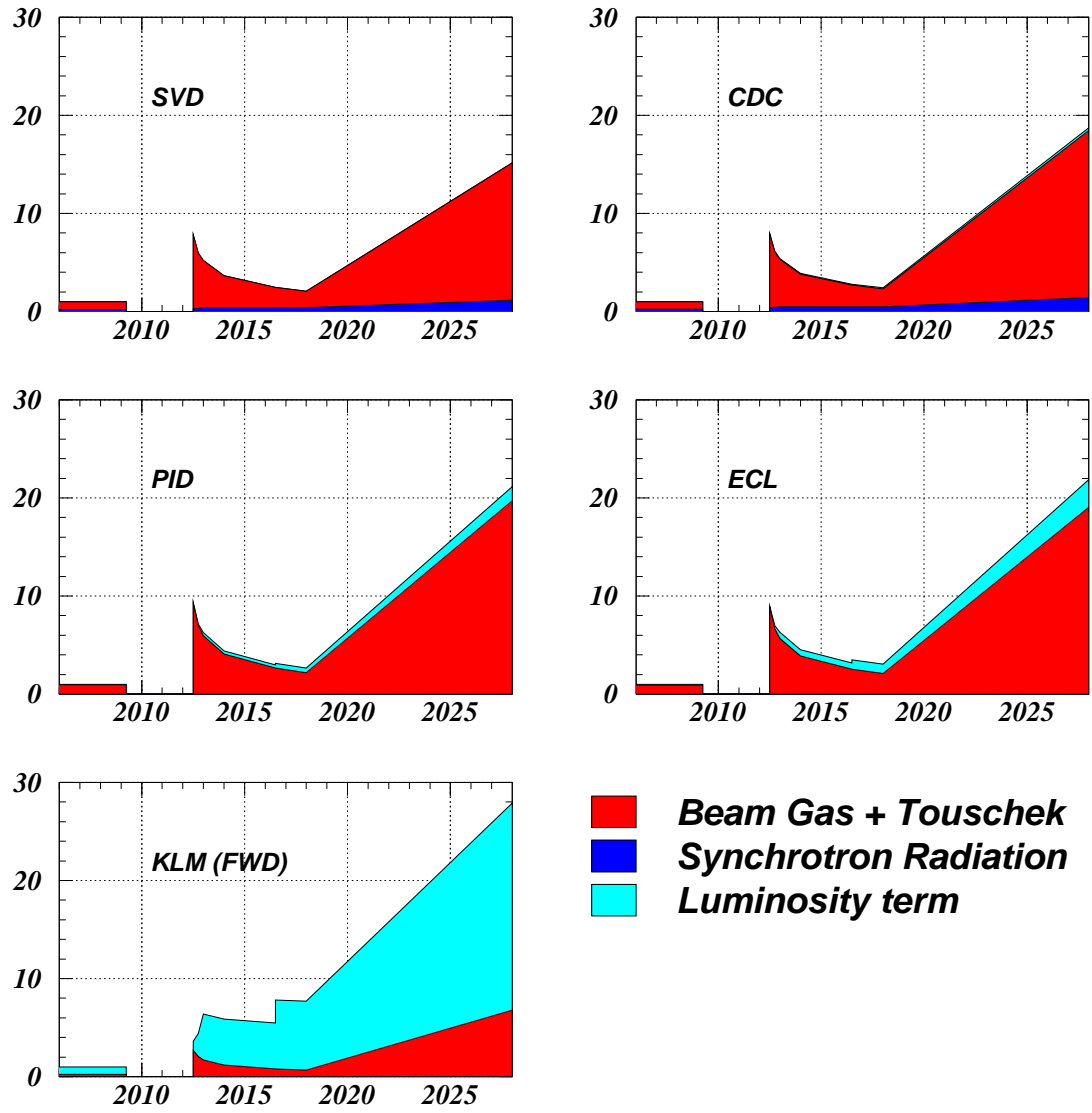}
\caption{
 Extrapolation of the beam-induced background levels for each detector
 normalized to the present level. Time scale is based on a modest
 assumption on how rapidly the beam currents increase. As indicated in the legend,
 beam-gas and Touschek scattering components are
 shown in red, radiative Bhabhas in light blue (brightest in grayscale),
 and SR in blue (darkest in grayscale).
}
 \label{fig:beambkg_trend}
\end{center}
\end{figure}

The expected beam-induced background level for each detector is
estimated from the scale factors (Table~\ref{table:bkg_scaling_factor}),
the relative contributions of components (Table~\ref{table:bkg_fractions_200X}),
and the assumed machine parameters
(Table~\ref{table:bkg_machine_parameters}).
The estimated trend is shown in Fig.~\ref{fig:beambkg_trend}, which is
based on the most conservative constant funding scenario with a
very gradual increase in beam currents.
%
Relative to the present values, we expect several to ten times higher background levels 
in the initial stage of the upgrade, and larger factors as
the beam currents increase.
Beam-gas scattering is the dominant background source for most
detectors, while radiative Bhabhas are the dominant one for the KLM.
General improvements of the beam-gas background from 2013 to 2018 are
based on the additional assumption that the vacuum will improve with the degassing
of the beam chambers.

%% file: baseline.tex
 \section{Baseline Design}
 \label{sec:baseline}
\begin{figure}
 \includegraphics[width=0.9\columnwidth]{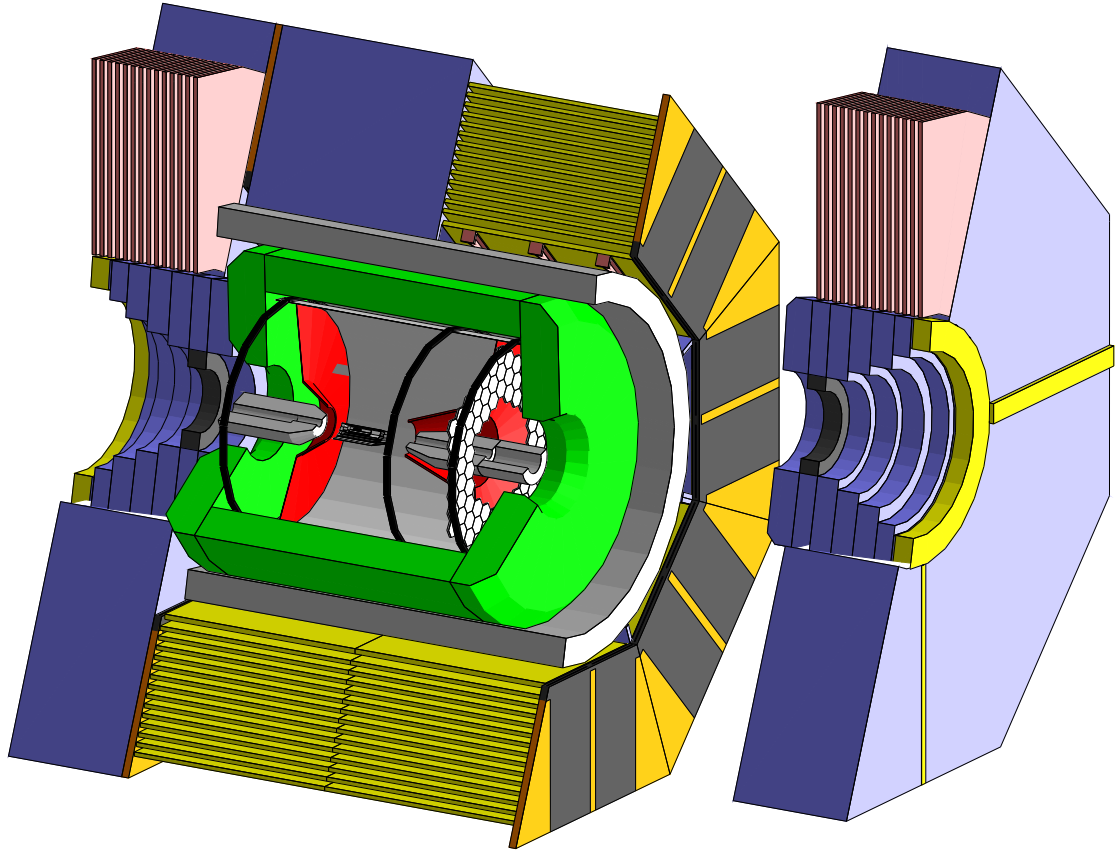}
 \includegraphics[width=0.9\columnwidth]{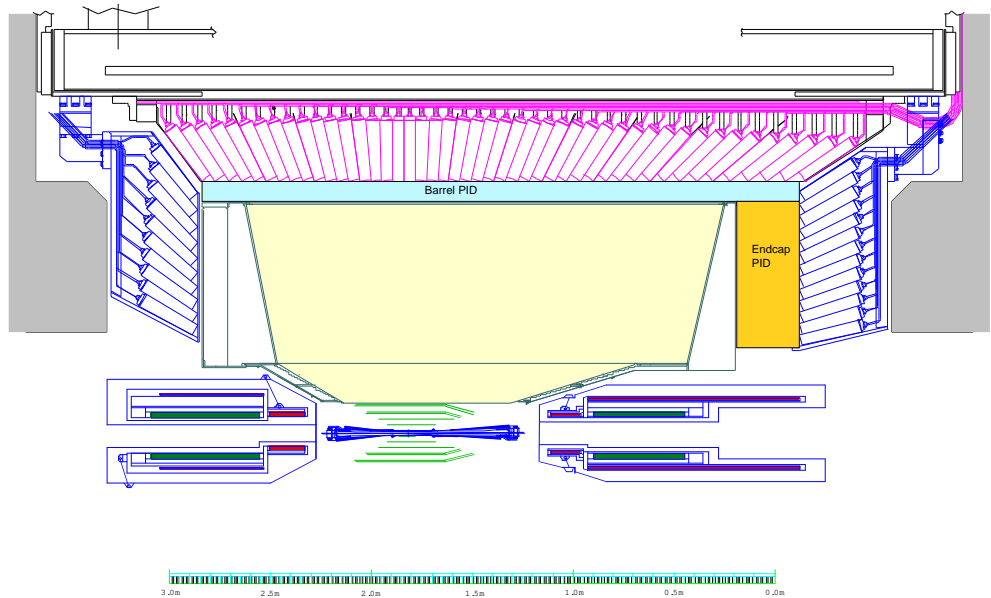}
 \caption{Conceptual design of the sBelle detector in the LoI.
 The cross-sectional view of inner sub-detectors is shown in the bottom
 panel.}
 \label{fig:sBelle}
\end{figure}

 The target detector of this study is the so-called {\it baseline} design, which
 is originally described in the Letter of Intent~\cite{bib:LoI}, and
 which is shown in Fig.~\ref{fig:sBelle}
 and includes some updates and minor changes made since then.
 We will have a 1.5\,cm beam pipe and 6 layers of silicon vertex
 detectors (SVD) that consist of double-sided silicon strip sensors,
 a central drift chamber with about 15k sense wires (CDC),
 a time-of-propagation (TOP) counter as a particle identification device in the
 barrel, aerogel rich (A-RICH) counters in the forward endcap, thallium
 doped CsI
 crystals as an electromagnetic calorimeter (ECL), which will be
 partially replaced
 with pure CsI crystals in the endcaps,
 and 
 a KLM
 based on resistive-plate-counters in the barrel and scintillators in the endcaps.
 
 There are several other proposals for SVD and PID detectors; for
 example, we will include a pixel detector when it is ready, we may
 choose a DIRC
 or imaging TOP as the barrel PID. Whatever works can be a good
 candidate until we make a technical decision. The issues discussed in
 this report are, however, focused on the baseline design.
 This baseline design of each sub-detector is briefly summarized in the following
 subsections.
 
 \input{baseline_svd} 
 \input{baseline_cdc} 

 \input{baseline_pid} 

 \input{baseline_ecl} 

 \input{baseline_klm} 

%% file: baseline_svd.tex
\subsection{SVD} 
\label{sec:baseline_svd}

In the LoI that was released in 2004~\cite{bib:LoI}, 
we have described a 6-layer SVD with a $1.0$\,cm radius beam pipe 
as a possible detector configuration. 
This design was proposed assuming operation at a luminosity of 10$^{36}$/cm$^{2}$/s.
In order to achieve a better impact parameter resolution, 
a smaller radius of the beam pipe was adopted at that time 
($1.5$\,cm for the current Belle detector). 
At the position of the innermost layer, the extrapolated beam background
level is very large (shown in section~\ref{sec:beambg}); it will reach
$\sim$ 5\,MHz/cm$^2$ for beam-induced backgrounds above 10\,keV at the
full spec.
Therefore we concluded that a pixel type sensor is necessary 
as the innermost layer to reduce the occupancy to an acceptable level ($< 10\%$). 
Furthermore, this sensor should be as thin as a strip type sensor ($\sim300\,\mu$m Si) 
to reduce the material budget. 
In terms of the readout time, this should be short to reduce the
detector dead time. 
However we do not yet have a pixel sensor that fulfills these
requirements~\footnote{DEPFET was not known as a candidate when we started
this study.}. 

According to the present schedule for the sBelle, 
operation is anticipated to start in the middle of 2012. 
The primary target luminosity in the early stage is 2$\times$10$^{35}$/cm$^{2}$/s
(``201X'' stage defined in section~\ref{sec:beambg})
and we plan to gradually increase this up to 10$^{36}$/cm$^{2}$/s (``202X'' stage) 
within a few years. 
Because a detector upgrade takes three years according to our past experience, 
we should start the construction in 2009, that is, one year from now. 
Since we have to fix the detector configuration and technology choice immediately, 
we have decided our strategy for SVD upgrade as follows. 
For running in 2012, we will install an upgraded SVD 
with strip type sensors and a $1.5$\,cm radius beam pipe. 
A pixel sensor with a smaller radius beam pipe can 
replace the innermost layer as a further upgrade after a few years of operation. 
Following this new strategy, we propose in this report a new "baseline" SVD design 
for operation starting in 2012.

The specification of the baseline design is as follows. 
The radius of the beam pipe is $1.5$\,cm. 
As described in subsection~\ref{sec:cdc}, 
the inner part of the present CDC cannot be operated at a higher luminosity
because of the harsh beam background. 
Therefore we will replace this part with two additional SVD outer layers, 
which correspond to the fifth and sixth layers. 
The increased number of layers enables stand-alone tracking with the SVD
and improves the tracking efficiency for low momentum particles. 
Moreover, this enlarged SVD requires an increase in the efficiency for 
reconstructing  $\PKzS$ decays. 
However, this also leads to a longer sensor, especially for the outer layers. 
As a result, the noise level will increase. 
To cope with this, a development of a special readout scheme, 
for example the "chip-on-sensor" method, is needed to maintain a good S/N performance. 
Note, however, that this would increase the material budget in the acceptance. 
As for the forward and backward parts, the layers could be slanted or 
have a disk-type shape so that the ladder size and the number of readout channels 
can be reduced without loosing acceptance.

The signals from a strip type sensor should be read out by a front-end chip, 
which has a short shaping time ($\sim 50$\,ns) to reduce the high occupancy 
induced by the harsh beam background. 
Since the L1 trigger rate would be very high ($\sim10$\,kHz), 
a pipeline on the front-end chip is required. 
The VA chip, which has been used for the Belle SVD, 
cannot be used even at the outer layers 
since it needs at least $12.8\,\mu$s to be read out 
and introduces a dead time fraction of more than 15\% at the $10$\,kHz L1 trigger rate. 
The APV25 chip that has been developed for the CMS Si tracker 
fulfills the above requirements and is one possible solution.

In this design report, we will discuss these issues and 
make recommendations for the baseline SVD detector.

%% file: baseline_cdc.tex
\subsection{CDC} 
 \label{sec:baseline_cdc}

The present CDC has been working well for 9 years starting from the beginning of
the Belle experiment.
It is used to reconstruct charged tracks with good
momentum resolution due to its low mass, and hence reduced multiple scattering.
It also provides  particle identification  based on
the characteristic energy loss (dE/dx).
In addition, a powerful level 1 track trigger signal is generated 
with a latency of a few $\mu$s. All 
these features can only be achieved
by using a small cell wire chamber filled with a helium based gas.

We want to maintain a similar performance level for the main tracking device in
the sBelle detector. A new CDC detector should be designed to work
under a beam background rate 20 times higher than the present level.
The inner radius is increased to avoid the
severe beam background
environment in the region that will be covered with the new enlarged SVD.
The outer radius is larger because the barrel part of the particle 
identification device will be thinner.
The overall detector shape is similar to the present CDC with 
three parts, the main part, the conical part and the inner part.
The main part is held between two curved 10\,mm thick aluminum endplates, 
separated by about 2\,m.
The outer surface of the main part is  a cylinder made of 5\,mm
thick carbon fiber reinforced plastic (CFRP).
The conical part is necessary to extend the detector acceptance as much
as possible in the forward direction, while allowing the KEKB machine 
components to be close to the IR.
The inner part has 8 layers with small cells and a cylindrical
inner surface made of 0.4\,mm thick CFRP.

\begin{figure}
 \includegraphics[width=0.9\columnwidth]{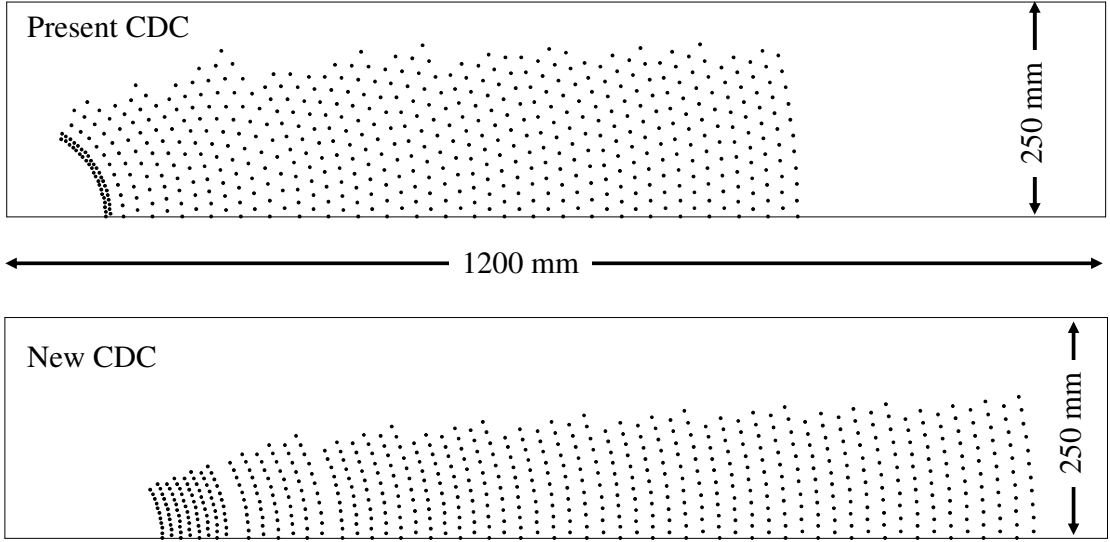}
 \caption{Configurations of CDC sense wires in the present (top)
 and the upgraded (bottom) CDC.}
 \label{fig:cdcwire}
\end{figure}

The wire configuration of the new CDC is shown in
Fig.~\ref{fig:cdcwire}. The cell size will be smaller for the inner
super layers to reduce the occupancy. More stereo layers should allow
for an improved three-dimensional track reconstruction.
For the sense wires as well as for the field wires, the same material and
the same diameter used in the present CDC are selected.
We will also keep the same gas mixture, because we have not 
encountered any problems such as
 radiation damage in the present CDC so far.
Design details are described in the LoI~\cite{bib:LoI}.

New readout electronics should be used to reduce the dead time.
This will be the most effective countermeasure against the expected high
beam background.
ASIC chips with a shorter shaping time will be used for signal
amplification, shaping, and discrimination.
The drift time and the pulse height are measured
separately using pipelined TDCs and slow FADCs. The electronics 
components will be
located near the backward endplate and all information will be
transfered to the electronics hut through optical fibers.

%% file: baseline_pid.tex
%
%

 \subsection{PID} 
 \label{sec:baseline_pid}

To extend our physics reach,
we would like to improve the $K/\pi$ separation capability of the spectrometer by upgrading 
the particle identification (PID) system. An upgrade of the system is also compulsory to cope
with the higher background environment. Another aspect to improve is to reduce the amount of 
 material and make it more uniform since the PID system is located in front of the calorimeter.
 
At this moment, two types of detectors, both of which are based on the 
Cherenkov ring imaging technique, are proposed:
a time-of-propagation (TOP) counter for the barrel region, and 
a proximity focusing Cherenkov ring imaging counter with aerogel radiators (ARICH) for the endcap region.
The baseline design discussed in this report will be outlined below.

\paragraph{TOP counter}

\begin{figure}
\centerline{
\includegraphics[width=8cm]{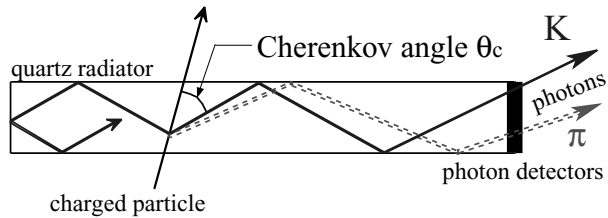}
}
\caption{Schematic side-view of TOP counter and internal reflecting Cherenkov
photons.}
\label{fig:principle}
\end{figure}
In the barrel region of the spectrometer, the present time-of-flight and 
aerogel Cherenkov counters are replaced with
a Time-Of-Propagation (TOP) counter~\cite{bib:TOP1}.  In this counter
the time of propagation of the Cherenkov photons internally reflected inside a quartz radiator
is measured~(Fig.~\ref{fig:principle}).
 The Cherenkov image
is reconstructed from the 2-dimensional information provided by one of the coordinates ($x$) and precise timing,
which is determined by micro-channel plate (MCP) PMTs at the end surfaces of the quartz 
bar.
The array of quartz bars surrounds the outer wall of the CDC; 
they are divided into 18 modules in $\phi$ in the baseline geometry. 
In one module, in order to reduce the possible degradation due to chromatic dispersion, 
each radiator bar is subdivided into two pieces at $z\sim$1070\,mm,
where the ``short'' bar with 750\,mm length is used for the forward side,
 and the ``long'' bar of 1850\,mm length for the backward side.
The short bar is instrumented by PMTs only at the forward side, while the long bar has 
PMTs at both sides.

\begin{figure}
 \centerline{
 \includegraphics[width=0.75\textwidth]{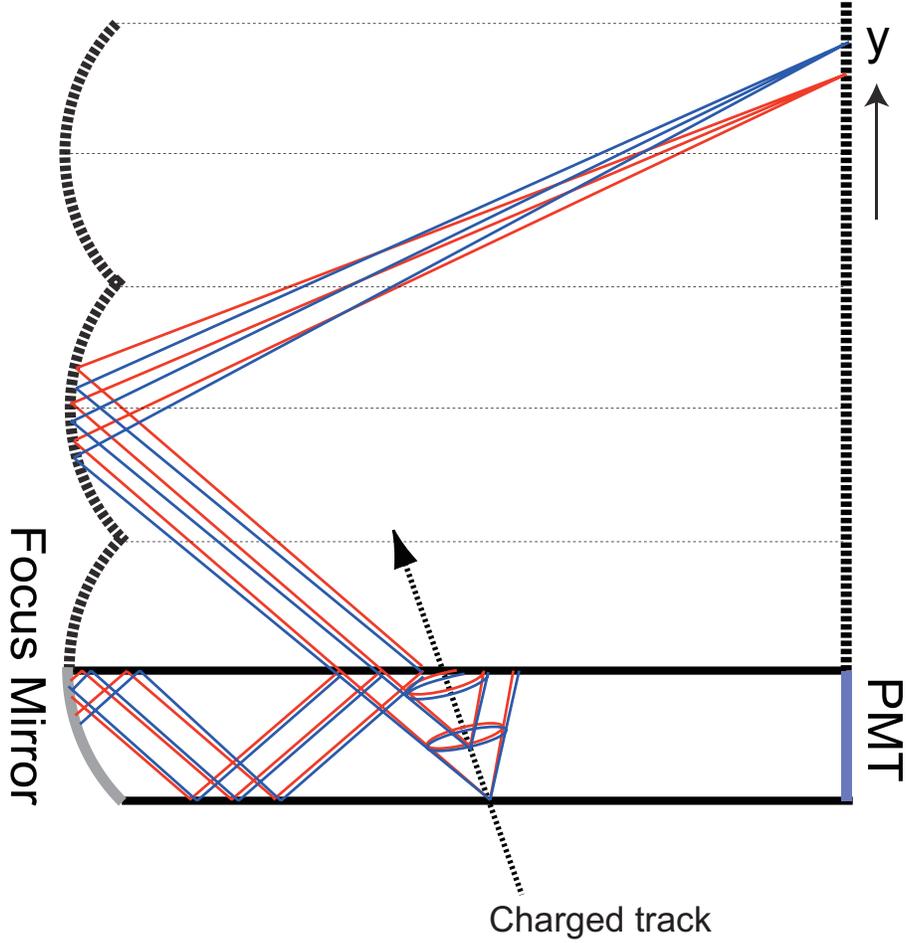}
 }
 \caption{\label{fig:TOPfocus}
 The principle of the focusing scheme in the TOP counter. The virtual extension of the 
 focal surface and of the photon detector plane are shown by the dashed
 curves and dashed lines. }
\end{figure}

\begin{figure}[htbp]
\centerline{
\includegraphics[width=9cm]{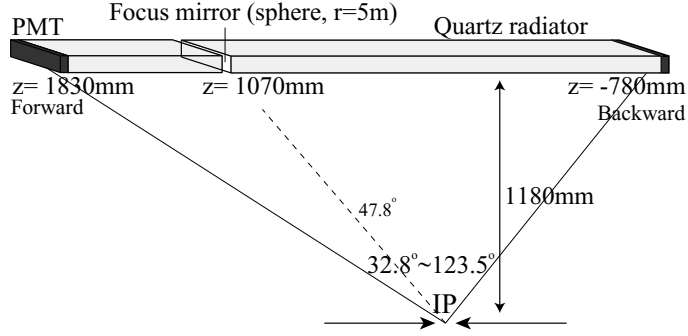}
}
\caption{Baseline design of the focusing type.}
\label{fig:topoverview}
\end{figure}

To improve the performance, a ``focusing scheme'' is being 
considered (Fig.~\ref{fig:TOPfocus}), in which the set 
of MCP-PMTs located at the forward side of the long bar at $z\sim$1070\,mm is replaced by 
a focusing mirror (Fig.~\ref{fig:topoverview}). The  PMTs at the other end are 
rotated to determine the $y$ coordinate in addition to the $x$ position, which enables us to
correct for the chromaticity of Cherenkov light as well as to measure the 3-dimensional information
of emitted Cherenkov photons.

The present support structure for quartz bars occupies a small region in the $\phi$ direction, which
introduces an insensitive region. As will be discussed in Sect.~\ref{subsubsect:top-conf},
this could be recovered by arranging two layers of radiators, where
the inactive area in one radiator layer is covered with the other layer and vice versa.

\paragraph{ARICH counter}

\begin{figure}[h]
\begin{center}
\includegraphics[width=0.75\textwidth]{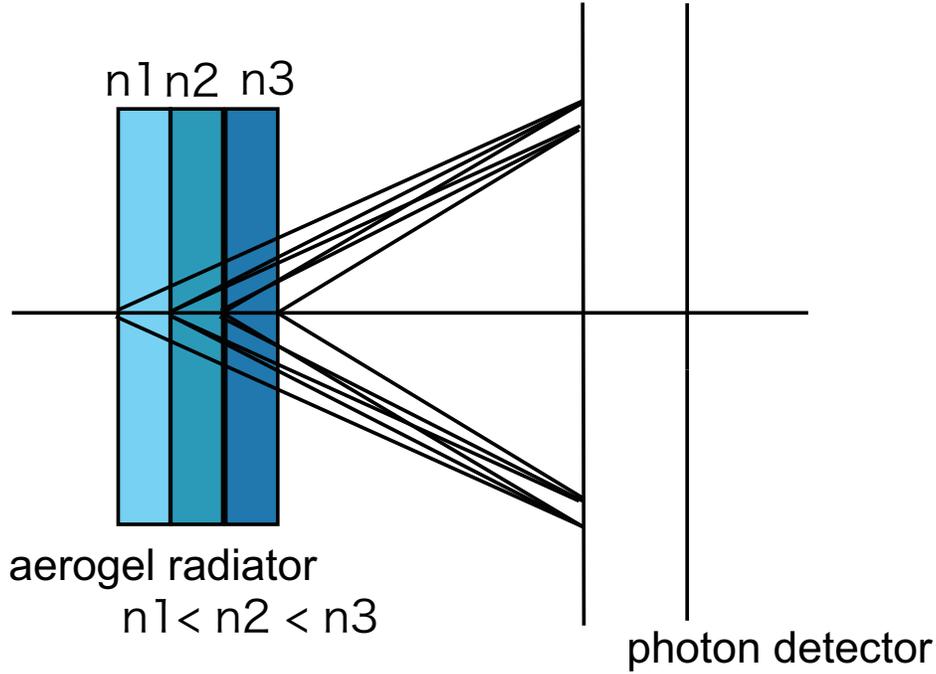}
\caption{\label{fig:ARICH}
A schematic view of ARICH counter with multiple aerogel radiators.
 }
\end{center}
\end{figure}

The ARICH counter is located in front of the forward endcap calorimeter, in the region of the present
forward endcap aerogel counter.
Because of the limited space in the Belle endcap layout, we employ a proximity focusing scheme,
where the Cherenkov expansion distance is 200\,mm.
In the present design, 3 layers of silica aerogels, each 10\,mm thick,  with different refractive indices
from 1.045 to 1.055 are used 
as Cherenkov radiators so that Cherenkov photons produce overlapped images on the photon detector surface, 
as shown in Fig.~\ref{fig:ARICH}.

The photon detector, which has to detect efficiently single photons in the magnetic field of 1.5~T,
has not been selected yet. 
One of the candidates is a hybrid avalanche photon detector (HAPD), where around 600 
HAPDs, each of which has 68\% active area, are aligned in  9 layers in
the radial direction from an inner radius of 435\,mm.
Another choice for the photon detector is the MCP-PMT, which has an excellent timing resolution of 
$\sim50$\,ps for single photons.
This feature allows a time-of-flight measurement by using the Cherenkov photons from the entrance window 
of the MCP-PMT. With this  additional information we could positively identify kaons with momenta 
below the Cherenkov threshold in aerogel ($\approx 1.5$~GeV/$c$).

\begin{figure}[htbp]
\centerline{\includegraphics[width=0.6\columnwidth]{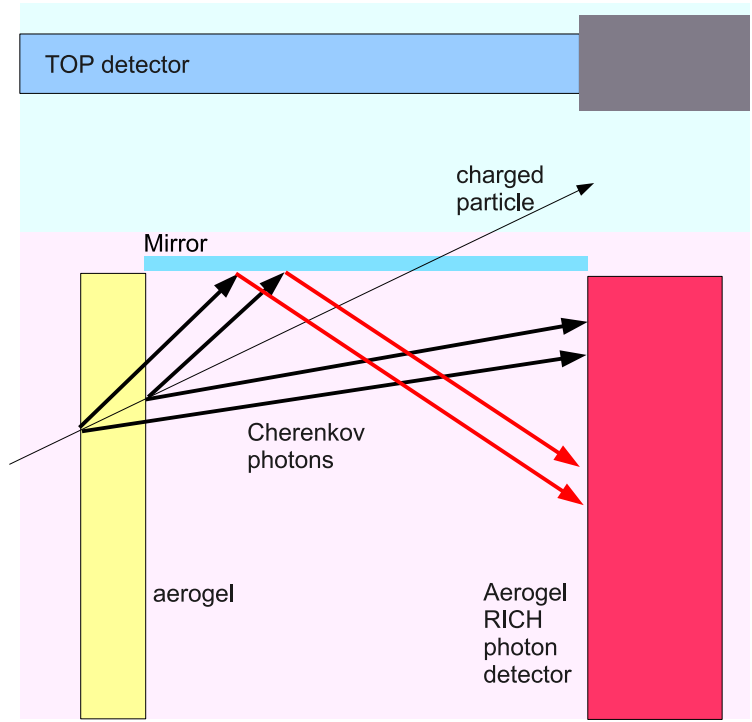}}
  \begin{center}
  \caption[kk]{Propagation of Cherenkov photons of a track hitting the aerogel radiator
 in the vicinity of the boundary to the barrel part,  for the case with an additional planar mirror.}
 \label{fig:arich-mirror}
   \end{center}
\end{figure}

The boundary region between the barrel and the endcap is of concern as well, because 
 this area is used for the MCP-PMT mounting in the TOP counter, and in
the ARICH counter Cherenkov photons emitted from the aerogel radiator tend
to get lost because of detector acceptance~(Fig.~\ref{fig:arich-mirror}).
This deterioration may be alleviated by installing a mirror 
at the outer wall of the ARICH counter in order to reflect photons back to the photon sensor
active area. A potential drawback of this solution is that it may complicate the reconstruction of the ring
images. This option is discussed in Sect.~\ref{subsubsect:arich_perf}.


%% file: baseline_ecl.tex
\subsection{ECL}
\label{sec:baseline_ecl}

The main concern for the electromagnetic calorimeter
(ECL) is the increase of background
due to higher beam currents. The background in ECL will increase
typically as a function of the total stored current multiplied
by the pressure of the residual gas in the beam pipe. 
As the result, the background is expected to be
an order of magnitude larger than at present as discussed in
section~\ref{sec:beambg}.

The basic idea is to replace the present read-out electronics with a 
relatively modern technology.
In the new front-end electronics, waveform sampling will be implemented.
The shaping time will also be shortened from $1\,\mu$sec to $0.5\,\mu$sec.
The waveform is sampled with a $\sim 2\,$MHz sampling frequency.
The sampled waveform is fitted in a FPGA to extract
the amplitude (i.e., energy) and the timing,
which allows us to separate physics signals and beam backgrounds.
With the shorter shaping time and proper waveform analysis,
the background is expected to be reduced by a factor of four to seven
depending on the energy range.

The background rejection by the waveform analysis is limited by
the slow ($\sim 1\,\mu$sec) scintillation light of the CsI(Tl).
In order to further reduce the background, replacement of some CsI(Tl) crystals
with faster ones is desirable. Pure CsI has a fast time constant of
$\sim 30\,$nsec, 30 times shorter than CsI(Tl), while
the light output is 50 to 100 times smaller.
To detect the scintillation light from pure CsI, the photo-detector has 
to have sensitivities to UV light with a gain of 10 to 100 in
a 1.5\,T magnetic field, and has to be as short as a few cm long to fit into
the current mechanical structure.
Fine-mesh photomultipliers~\cite{Kichimi:1992qj} with a small number of multiplication stages
have a gain of more than 20 in 1.5\,T field, and a length of 
about 5\,cm. By replacing the crystals with pure CsI and with a proper
waveform analysis, the background can be reduced by more than a factor
of 100.

Initially, we will upgrade the read-out to do
waveform fitting of the signals from
the existing CsI(Tl) crystals, which is necessary from the very beginning of sBelle
operation.
We will also prepare pure CsI crystals and the corresponding
read-out system for the endcaps by the time we reach a background level
so high that we cannot handle it with waveform fitting alone.
The performance of the ECL and the impact on physics analyses will be discussed
in section~\ref{sec:ecl}. More details of the hardware issues are
discussed in the LoI~\cite{bib:LoI}.

%% file: baseline_klm.tex
 \subsection{KLM}
 \label{sec:baseline_klm}


The detection efficiency of Belle's resistive plate counters (RPCs) depends
strongly on the ambient background rate.  This is due to the long recovery time
of the glass electrodes (resistivity $\sim$10$^{13}~\Omega\cdot$cm)
after the depletion of the charge in a spot near the streamer that forms in
the gas volume upon its ionization by a primary particle.
Neutrons make the dominant background here; the effect of $\gamma$ rays is much smaller.
The expected background rates at SuperKEKB situation are 0.5 -- 4 Hz/cm$^2$
for the barrel and 2 -- 5 Hz/cm$^2$ for the endcap.
Because of the higher background rates in the endcap and the greater
sensitivity of the endcap glass material,
the estimated efficiency is 0\% for the endcap in streamer-mode operation,
and 90\% or more ($\sim$80\%) in the outer (innermost) barrel layers.


In the barrel's inner-layer RPCs, a recovery to $\sim$90\% efficiency
may be achieved through various techniques.
A shorter dead time can be achieved by a modified gas mixture.
From our study using endcap-glass RPCs,
the best gas mixture (Ar/C$_4$H$_{10}$/HFC134a/SF$_6$ = 50/8/37/5)
reduces the dead time by about a factor of three (to 0.15 sec$\cdot$cm$^2$).
However, there is a caveat: the current dead time of the barrel-glass RPCs (0.08 sec$\cdot$cm$^2$)
is already shorter than this improved value, so
the use of this new gas mixture in the inner-layer barrel RPCs may not result
in as marked an improvement in efficiency.
A more drastic measure to suppress the neutron background would be to replace the
innermost RPC layer with a passive polyethylene absorber with a thickness of
4 cm. (This would increase slightly the threshold on transverse momentum of muons
that reach the first sensitive barrel RPC layer.)
By combining these options, the barrel-RPC efficiency degradation is expected to be a few \%.
If the actual background rate turns out to be much higher than the above estimates,
the barrel RPCs will be operated in avalanche mode\cite{Gustavino:2004zh} rather than
streamer mode.  This will require a retrofit of
an on-detector amplifier because of the small signal charge ($\sim$1\,pC).

\begin{figure}[htb]
\includegraphics[width=0.6\textwidth]{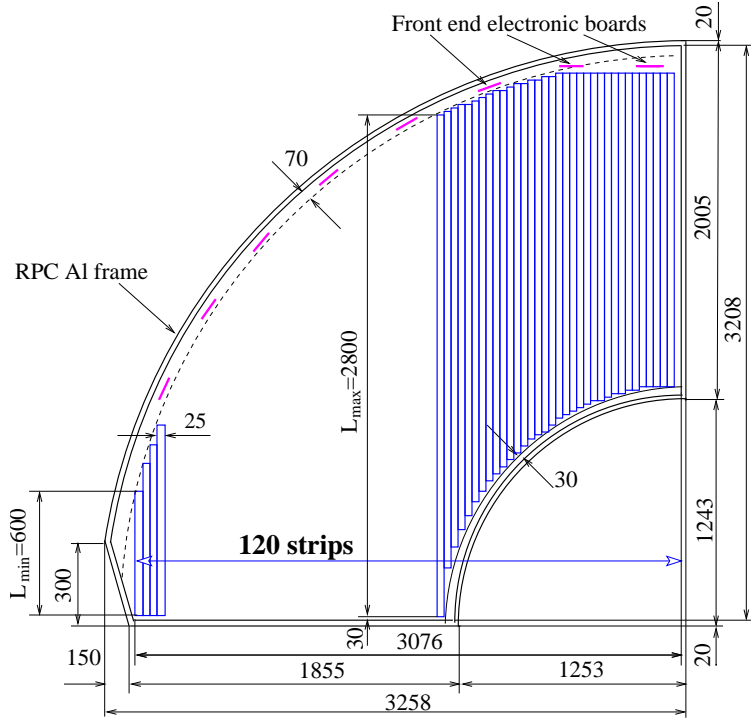} 
\caption{Sketch of the scintillator-strip superlayer.}
\label {klm:sect}
\end{figure}

In the sBelle endcaps, where the beam background
effects are most deleterious, the RPCs will be replaced with a new and
faster detector. In the proposed
system~\cite{bib:LoI} scintillator counters with wave-length-shifting
(WLS) fiber readout are installed in the gaps between the
iron absorber plates of the solenoid yoke.
Each superlayer, contained within an existing RPC module's aluminum frame,
is formed by two independent and orthogonal planes of
scintillator strips.
One quadrant of a scintillator superlayer is shown in Fig.\,\ref{klm:sect}.

Each strip has a cross section of $1.0 \! \times \! 2.5\,$cm$^2$ and
a length of up to $280\,$cm. The strip is extruded from
granulated polystyrene with two dyes (PTP and POPOP) and covered by a
diffraction reflector. The Kuraray multicladded WLS fiber Y11 (200) is
installed in a groove at the strip's surface. To improve the light
collection efficiency, the WLS fiber is glued into the groove with
optical gel and the fiber and groove are covered by a
Superradiant VN2000 foil. The Geiger Photo Diode (GPD) photodetector,
produced by CPTA (Moscow)~\cite{golovin}, is mounted at one
end of the fiber; a mirror is placed at the other end.
The GPD is a matrix of $560$ tiny ($40\!  \times \! 40 \, \mu m^2$)
silicon photodiode pixels operating in Geiger mode. The signals from
all pixels are internally summed to produce the GPD
response. Signals with a distinct number of photoelectrons are well
separated. The typical GPD amplification is about $10^6$. The product of
quantum and geometrical efficiency is $25\!-\!30\,\%$ in the region near
the WLS-fiber spectrum maximum. The radiation hardness of the GPD was
measured directly in the KEKB tunnel and was found to be sufficient for the proposed
system.

The GPD has a quite high single photoelectron noise rate of about
$1\!-\!2\,$MHz at a room temperature. In order to reduce it to an
acceptable level, it is necessary to impose a threshold of a few fired
pixels. The dependence of the strip detection efficiency for a
minimum-ionizing particle (MIP) on this GPD threshold has been
studied using a cosmic ray trigger. The mean
number of fired pixels per MIP is $\sim\!20$, so a
threshold of $\sim\!6$ fired pixels results in a MIP detection efficiency
of as high as 99\% with a suppression of the intrinsic GPD noise rate
to less than $1\,$kHz. This GPD noise rate is much smaller
than the rate due to the accelerator-induced neutron background.
The latter is extrapolated to the SuperKEKB luminosity
from our direct measurements with the test scintillator
KLM module in the KEKB tunnel to be $\simeq 100$\,Hz/cm$^2$. For the
longest strip, the expected rate is $\sim\!70$\,kHz; this does not reduce
the MIP efficiency because of the very short GPD dead time.


%% file: svd.tex
  \section{SVD performance}
  \label{sec:svd}
  \input{general_intro_svd}

  \input{inner_radius_svd}

  \input{APV25_performance}

  \input{SVD5thLayer}

  \input{signal-to-noise}

  \input{SVDPitch}

  \input{slant}

  \newpage
  \input{SVDSummary}

%% file: general_intro_svd.tex
As described in subsection~\ref{sec:baseline_svd}, 
we have decided on the concept of the baseline SVD design, 
that is, a $1.5$\,cm radius beam pipe and 
a 6-layer SVD where all layers are constructed from strip type sensors.
However, we still have many things to determine before finalizing the design.

One of the important issues is the front-end readout chip.
At the target luminosity, the SVD will be exposed to harsh beam background.
According to the estimation shown in section~\ref{sec:beambg},
the beam background in the SVD could be about 15 times higher than the current Belle level.
In this situation, the VA chip that has been used for the current SVD would not
work at all because the rate at which beam background pulses
overlap with the on-time signal pulses would increase drastically.
To overcome this difficulty, we require a front-end chip whose shaping time
is much shorter than that of the VA chip.
Furthermore, the VA chip takes at least $12.8\,\mu$s to be read out.
The expected dead time fraction could then be more than 15\% under the
$10$\,kHz trigger rate that is anticipated in sBelle.
Therefore, we need also a front-end chip that has a pipelined readout scheme.

Other issues are related to the configuration.
As will be described in subsection~\ref{sec:cdc},
the inner radius of the CDC will be enlarged to avoid the severe beam background.
We will install two extra SVD layers in this space instead.
Thanks to this, SVD stand-alone tracking will be available.
and the reconstruction efficiency for Ks events will improve.
However this requires longer sensors increasing the noise level.
We should also be aware of further deterioration with a shorter
shaping time since the intrinsic noise level is proportional to 
$\sqrt{1/t}$, where $t$ is the shaping time.
To maintain good noise performance, a special readout scheme must be developed,
for example a ``chip-on-sensor'' method, is needed.
However, this would increase the material budget in the acceptance
and also the number of readout channels.
One possibility to reduce the number of readout channels is to widen
the pitch size of the readout channel.

Eventually, the design of the enlarged SVD induces many issues to be
optimized; the effect of the poorer noise performance,
the effect caused by the increased material, the optimization of the
readout pitch and the best position of the
fifth layer to improve the reconstruction efficiency for $\PKzS$'s.
Furthermore, the arch structure for the outer layers is also an issue
to be studied.

In this subsection, we will discuss the front-end chip and 
optimization of the configuration \footnote{the effect of the increased
material will be discussed in section~\ref{sec:material}}. 
We will then make recommendations on the baseline SVD design.

%% file: inner_radius_svd.tex
\subsection{Inner radius and resolution}
\label{sec:inrad}
  
The radius of the beam pipe strongly affects the performance of SVD.
In particular, the impact parameter resolutions are determined by the radius 
and the material budget of the beam pipe, 
the position and the material of the innermost layer and 
the position of the second layer.

On the other hand, 
the closer the sensor is to the interaction region,
the higher the expected occupancy.
Since the beam background level in sBelle is 
expected to be 15 times higher 
than in the current Belle configuration, the occupancy 
at the innermost layer will be a crucial problem.
Since the number of fake clusters will increase dramatically and 
even signal clusters would be deformed by beam backgrounds,
the trajectories of tracks reconstructed with fake and/or deformed clusters will be 
shifted from their true trajectories.
Consequently, the hit resolution will deteriorate.
This resolution is evaluated from the residual distribution between the true hit point
in a certain DSSD and the intersection point of the trajectory with the DSSD.
Finer segmentation of the readout channels is one solution to reduce occupancy.
Further segmentation ($< 50\,\mu$m ), however, is technically difficult 
and would increase the number of
the total readout channels. A pixel type sensor, which has a small
amount of material and a fast readout speed, 
has not been developed yet.
Another approach is to suppress the overlaid beam background in the signal
time-window by reducing the shaping time.
In the Belle SVD, the shaping time of the readout chip, VA chip, is about $800$\,ns.
Therefore, assuming a readout chip whose shaping time is $50$\,ns, 
we can reduce the occupancy by a factor of 1/16.
Here, we compare the performance of the following three SVD design candidates. 
We propose the second one as a baseline design for the sBelle SVD. 
The third one is what was proposed in LoI 2004.

\begin{enumerate}

\item ``Modified'':a 4-layer SVD with a $1.5$\,cm radius beam pipe
as for the Belle SVD. 
The readout chip for the two inner layers is replaced 
with one having a shorter shaping time ($50$\,ns) and a pipelined readout.
The readout chips for the remaining outer layers are the same as the
current one. (i.e., the shaping time is $\sim800$\,ns.)

\item ``Baseline'':a 6-layer SVD with a $1.5$\,cm radius beam pipe.
Two outer layers at the radial positions of 13 and 14\,cm are added
to the current Belle SVD configuration.
All layers are read out by chips having a shorter shaping time ($50$\,ns) and
a pipelined readout.

\item ``LoI04'': a 6-layer SVD with a 1.0\,cm radius beam pipe. 
Two outer layers at the radial positions of 13 and 14\,cm are added
to the current Belle SVD configuration.
The two inner layers are replaced with pixel sensors.
All layers are read out by chips having a shorter shaping time ($50$\,ns) and
a pipelined readout.

\end{enumerate}

First, we estimate the occupancy level in each layer for the above SVD candidates. 
Table~\ref{tab:occ} shows a rough estimation of the occupancy 
at a luminosity of 2$\times 10^{35}$/cm$^{2}$/s. 
For simplicity, in this calculation we assume that the occupancy is proportional 
to the shaping time, the area of each 
readout channel and $1/r^{2}$, where $r$ is the radial position 
of the layer from the interaction point. 

If we assume the limit on the occupancy is 15\%, the occupancy of
the third and fourth layers in option 1 nearly reaches this limit.
Because option 1 still employs the VA readout chip for the outer layers, 
which take at least 12.8\,$\mu$s to be read out and does not include any pipeline, 
the dead time fraction could increase up to 15\% or more at the 10\,kHz trigger rate.
For option 2, since we employ a readout chip having a shorter shaping
time, the estimated occupancy for each layer is within an acceptable range.
In option 3, thanks to the pixel sensor, the occupancy is kept at a low level
for the first and second layers.
Moreover, these two options can operate at a 10\,kHz trigger rate
because of the pipelined readout.
Therefore, we need to use a readout chip having a shorter shaping time 
and a pipelined readout for all layers.

\begin{table}  [h]

\caption{\label{tab:occ}
Estimated occupancy at each layer for the different SVD  candidates. The units are in \%.}

\begin{tabular}{l|cccccc} \hline
&1 &2 &3 &4 &5 &6 \\ \hline
Op.1. Modified  &10   & 3  &  15&  15&-&- \\
Op.2. Baseline  &10   & 3  &  1&  1& $<$1 & $<$1 \\
Op.3. LoI04 &$<$1   & $<$1  &  3&  3& $<$1 & $<$1 \\ \hline
%
%
\end{tabular}

\end{table}


We estimated the impact parameter resolution with the TRACKERR program\cite{bib:trackerr}.
The effect of the high occupancy is well-understood from our experience with the Belle SVD.
For example, with 30\% occupancy, 
the intrinsic resolution degrades by 50\% and the hit association is significantly
degraded.

\begin{figure}[h]

 \begin{tabular}{ccc}
 \begin{minipage}{0.5\hsize}
 \begin{center}
  \includegraphics[scale=0.3]{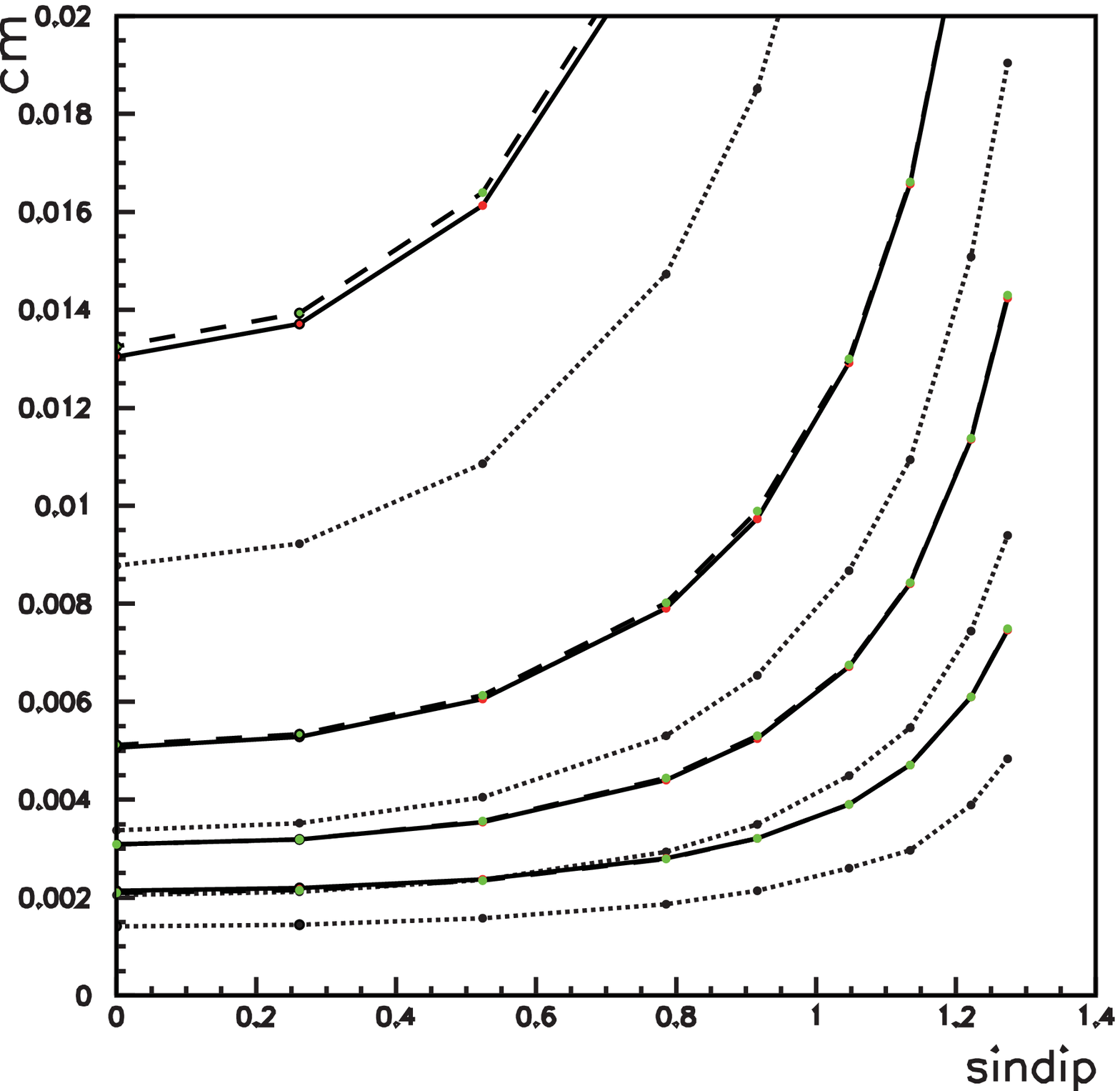}
 \end{center}
  \end{minipage}
 \begin{minipage}{0.5\hsize}
 \begin{center}
  \includegraphics[scale=0.3]{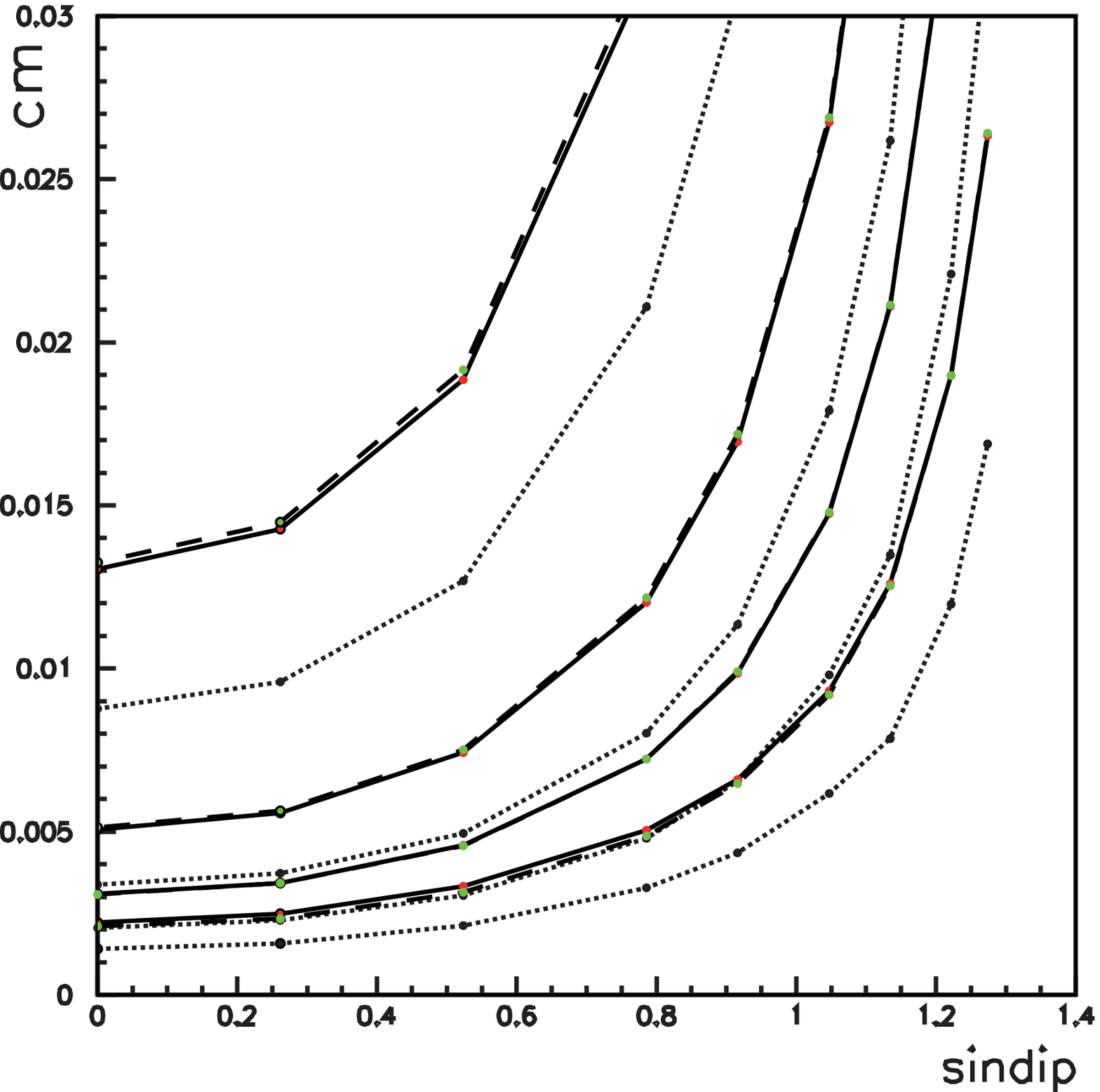}
 \end{center}
 \end{minipage}
 \end{tabular}
  \caption{\label{fig:track_res}
    The impact parameter resolution calculated by TRACKERR for the different SVD configurations.
    Solid, dashed and dotted lines correspond to option 1 ``Modified'', option 2, ``Baseline design''
    and option 3 ``LoI04'', respectively.
    Thanks to the smaller radius beam pipe, option 3 shows better resolution.}
\end{figure}

Figure \ref{fig:track_res} shows the impact parameter resolution 
as a function of the sine of the detector polar angle.
Since the detector configuration for the modified option is the same as that 
of the current Belle SVD except for the readout chip,
the impact parameter resolutions for the modified option
(solid lines in Fig.~\ref{fig:track_res}) are not different from those of the
current Belle SVD.
The baseline design (dashed lines in Fig.~\ref{fig:track_res})
shows performance comparable to
the current Belle SVD for both the $r\phi$ and $z$ directions.
Because of the shorter lever arm for multiple scattering with the $1.0$\,cm radius beam pipe, 
option 3 (dotted lines in Fig.~\ref{fig:track_res})
has better impact parameter resolution.

  

%% file: APV25_performance.tex
\subsection{Beam background effect on the readout chip}
\label{sec:bgeff}

In the previous subsection~\ref{sec:inrad}, 
the requirements for the readout chip
and impact parameter resolutions were obtained.
Actually, option 1 is a very unlikely scenario at this stage.
However, it is a useful to check the effects induced by the beam background
on the readout chip, especially on the innermost and second layers,
partially because we can use the current Belle GEANT3-based full detector simulation
with small modifications and partially because we can also make use of
real beam background data for this check.

To simulate option 1, we replace the VA chip of the innermost and second layers
of the current SVD by another chip whose shaping time is $50$\,ns.
In this simulation, we assume this chip to be ``APV25''.
On the other hand, the same VA chip is used for the remaining outer layers.
The detector dead time caused by the long readout time of the VA chip
is not taken into account throughout this simulation.
We estimate the effect of the beam background on analyses by embedding 
real beam background events taken with a random trigger in the simulated 
$B$ meson decay events.
In order to differentiate the effect of beam background on the readout chip
from other effects, for instance, the effect of the poorly reconstructed tracks
that are smeared with beam background hits in the CDC, we embed beam background events
in the SVD only.
Other details of the simulation procedure are described in \cite{bib:kuroki}.

Here we explain the values used to estimate the performance of the SVD.
In the $B$ factories, a $CP$ violating asymmetry in the time-dependent rates for
$B^0$ and $\bar{B}^0$ decays to a common $CP$ eigenstate can be observed through
the measurement of the distance between the two $B$ decay vertices.
Therefore the precise measurement of the decay vertices is the key point.
For simplicity, the vertex of the $B$ decaying to a $CP$ eigenstate
is described as the ``$CP$'' side vertex and 
that of the accompanying $B$ is the ``tagging'' side vertex.
\begin{figure}[h]
\includegraphics[width=0.8\linewidth]{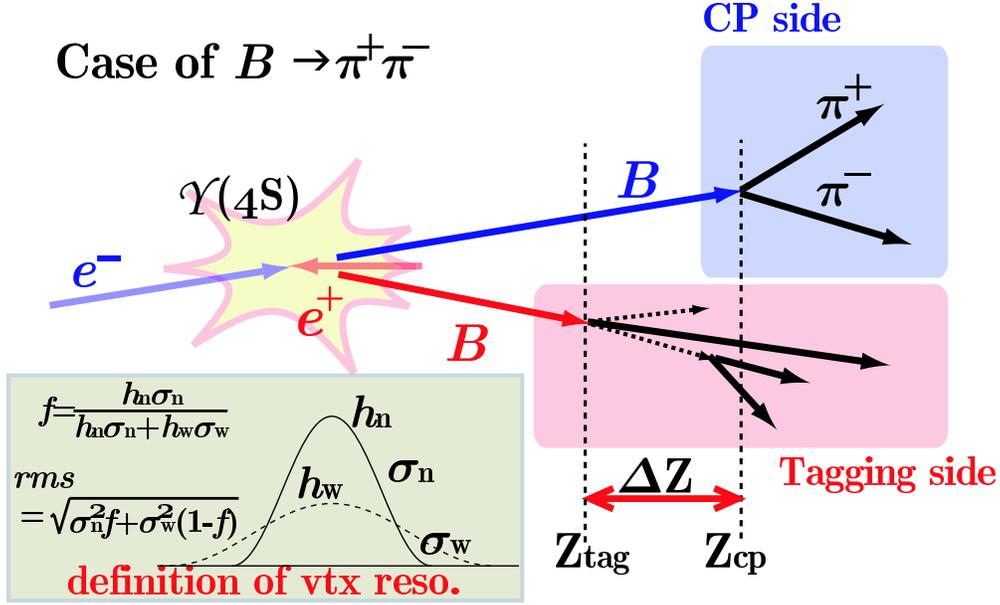}
\caption{\label{fig:vtxprocess} 
Illustration of a typical event vertex topology
}
\end{figure}

A typical event vertex topology is depicted in Fig.~\ref{fig:vtxprocess}, and
the reconstruction procedure is explained as follows.
The neutral $B$ meson is fully reconstructed in the specific
$CP$ eigenstate and then the $CP$ side vertex is determined
by the standard Belle kinematic vertex fitter.
The tagging side vertex is reconstructed from the tracks that are not
used in the reconstruction of the $CP$ side.
Here the types of the particles used in the $CP$ side vertex reconstruction
are identified from generator information.
The distribution of the difference along the beam axis
between the reconstructed $B$ decay vertex position and
the true decay point is fitted with a double gaussian form.
The vertex resolution for the $CP$ and tagging side vertices
is defined as the area-weighted root mean square (for the definition, 
see Fig.~\ref{fig:vtxprocess}) of the extracted $\sigma$ values
of the narrow and wide components.
In the same way, the difference between the distance obtained from
the two reconstructed $B$ decay vertices and the true distance provides 
so-called the $\Delta z$ vertex resolution.

\begin{figure}[htbp]
 \includegraphics[width=0.95\linewidth]{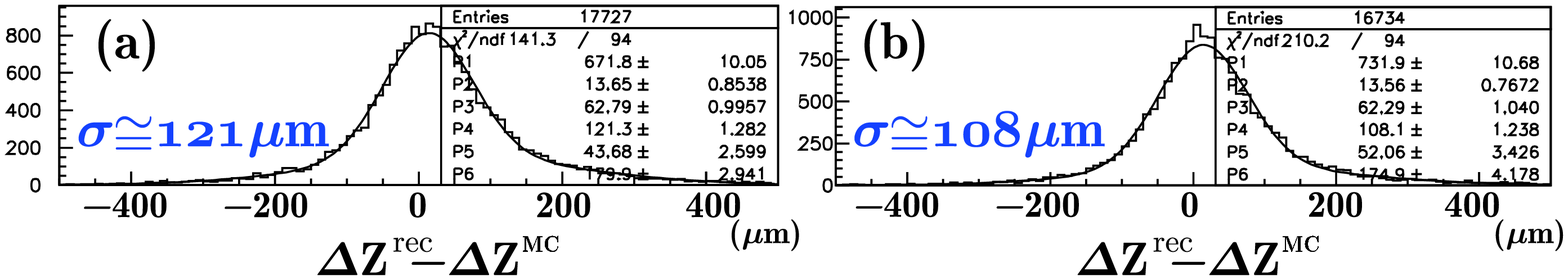}
 \caption{\label{fig:sim_vdif_va1_1}
  (a) Distribution of $\Delta z$ resolution for the SVD with VA
  readout and the nominal beam background level. (b) similar distribution for 
  option 1.}
\end{figure}

\begin{figure}[htbp]
 \includegraphics[width=0.95\linewidth]{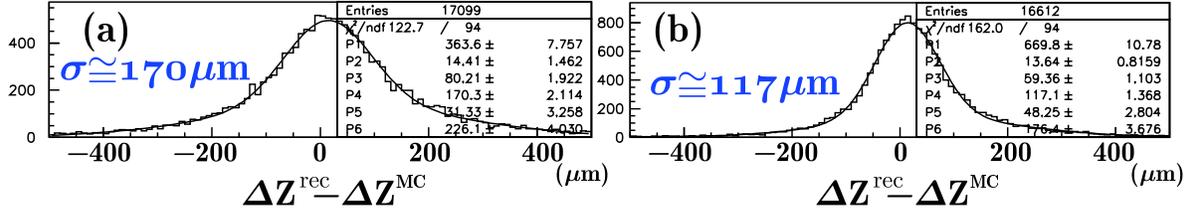}
 \caption{\label{fig:sim_vdif_va1_5}
  (a) Distribution of $\Delta z$ resolution of the SVD with VA
  with a beam background level five times larger.
  (b) similar distribution for option 1.}
\end{figure}

Figures~\ref{fig:sim_vdif_va1_1} (a) and (b) show distributions of $\Delta z$ resolution
for the $B \to J/\Psi Ks$ events with the nominal beam background 
under the assumption of the current SVD and that of option 1, respectively.
Figure~\ref{fig:sim_vdif_va1_5} shows similar distributions 
with a beam background level five times larger.
By comparing these figures, we found that for the current
SVD the resolution deteriorates by a factor of 1.4 times
when the beam background level is five times larger, whereas for option 1
the resolution is about the same as for the current SVD with
the nominal beam background.


\begin{figure}[htbp]
  \begin{center}
  \includegraphics[width=0.8\linewidth]{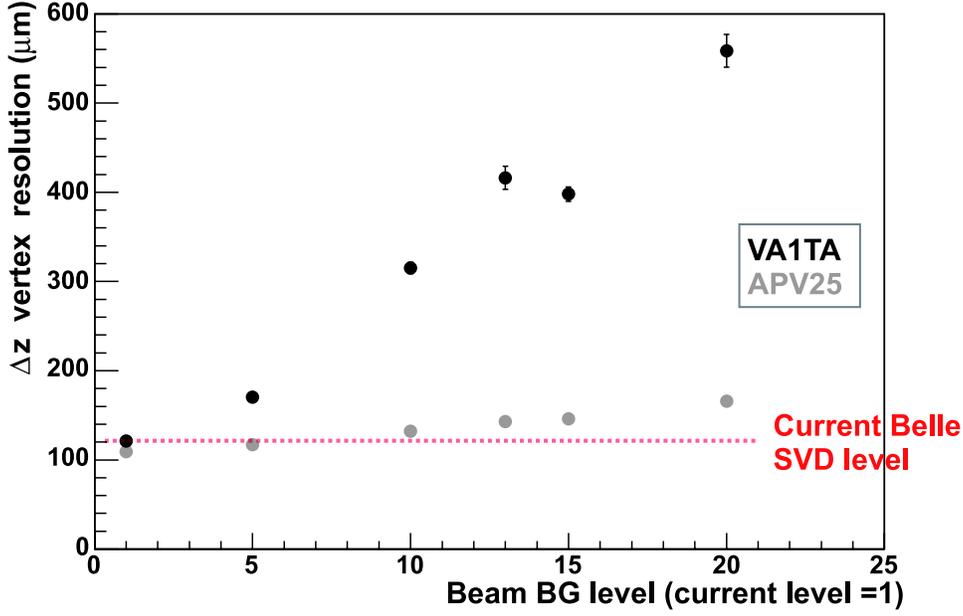}
  \caption{\label{fig:sim_vdifbgl}
    $\Delta z$ vertex resolution as a function of the beam background level 
    relative to the current Belle background level:
    APV25 (black) and VA1TA (gray)}
  \end{center}
\end{figure}

Figure~\ref{fig:sim_vdifbgl} shows the $\Delta z$ vertex resolution 
as a function of the beam background level.
As shown in Fig.~\ref{fig:beambkg_trend}, we expect roughly a beam
background six times larger than the current level in the SVD
for the early stages of sBelle.
By changing the readout chips of the innermost and second layers from the
VA chip to the APV25, we can maintain the $\Delta z$ vertex resolution at the
level that has been achieved by the current SVD with six times
higher beam background.
In other words, the key to maintaining the current level of the
vertex resolution under worse beam background conditions is to suppress
the beam background contribution by reducing the shaping time of the readout chip,
especially for the innermost and second layers.
Here we should recall that we have ignored the dead time
induced by the long readout time of the VA chip in this study.
As discussed in the previous subsection~\ref{sec:inrad},
the VA chip can not be used with a $10$\,kHz trigger rate.
From these points of view, the APV25, which has a short shaping time ($50$\,ns) 
and a pipelined readout 
and which is also a commercially available chip at this stage, is one
possible solution.

%% file: SVD5thLayer.tex
\subsection{Outer radius and $\PKzS$ vertex reconstruction efficiency}
\label{sec:svd5th}

As discussed in Section~\ref{sec:baseline_svd},
to improve the reconstruction efficiency for the decay $B \to K^{*0} \gamma$,
we have decided to extend the SVD volume in the radial direction and to enlarge the
current 4-layer configuration to a 6-layer one.
Although there are no strict limits on the positions of the additional layers,
a realistic outer radius of the SVD is currently assumed to be $16$\,cm.
Practical considerations about the support structure
suggest that the outermost ($=$ sixth) layer
should be located $14$\,cm from the interaction point.

\begin{figure}[h]
\includegraphics[scale=0.7]{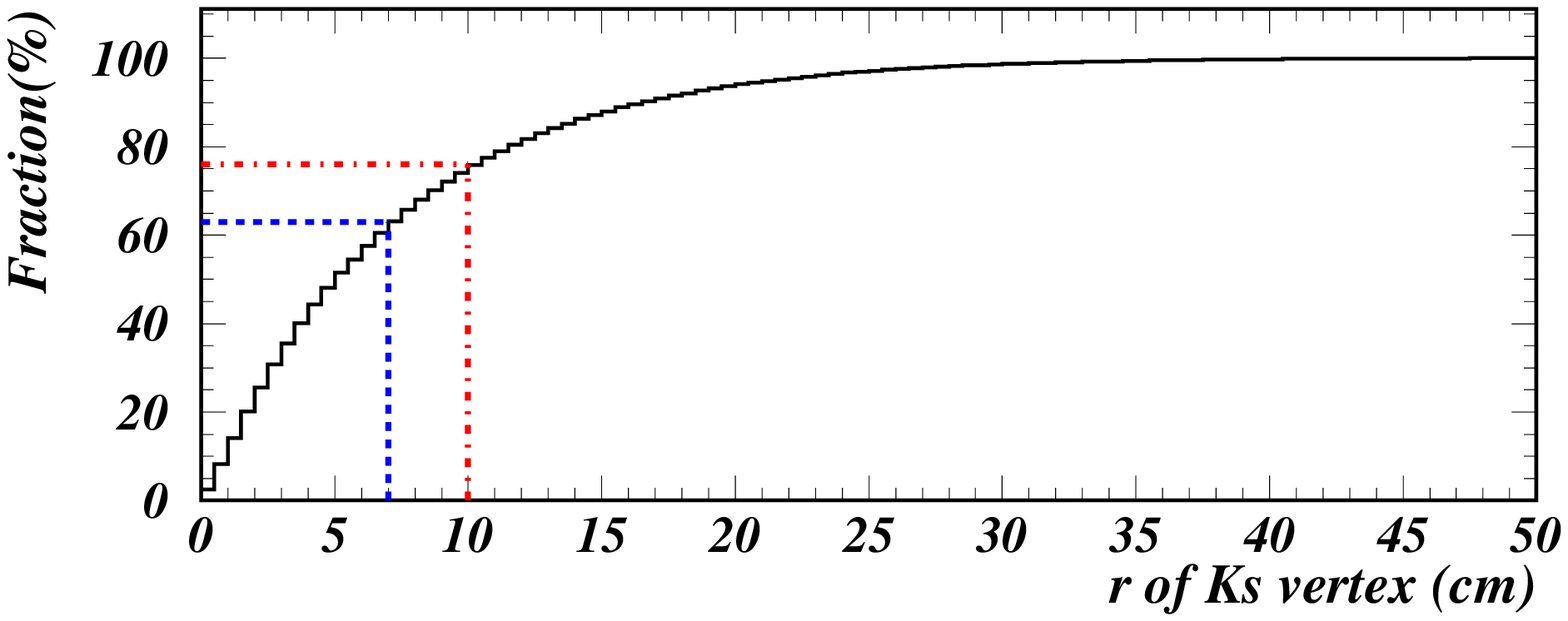}
\caption{\label{fig:ksdecpos} The fraction of $\PKzS$'s that decay within
 radius $r$ for the $B \to K^{*0} \gamma$ mode.
 The dashed lines show the location of the second outermost layer of the current SVD
 ($r \simeq 7$\,cm) and the corresponding fraction, and the
 dot-dashed lines the case in which 
 the second outermost layer in the upgraded SVD is placed at $r=10$\,cm.
}
\end{figure}

In reconstructing this decay mode, only two charged pions from 
the $\PKzS$ decay  can be used for the $B$ decay vertex
determination. Since for each track   at least two SVD hits are required,
the location of the fifth layer is important.
Figure~\ref{fig:ksdecpos} shows the fraction of $\PKzS$ mesons that decay within
a radius $r$  for this decay  mode.
The second outermost layer of the current SVD (i.e., third layer) is placed at
roughly $7$\,cm. In this case
 only $\sim 60\%$ of the $\PKzS$ decays from $B \to K^{*0} \gamma$ can be detected.
In the new SVD, we can increase  
the number of reconstructed $\PKzS$'s considerably by changing the position of 
the second outermost layer ($=$ fifth layer).
For instance, if the fifth layer is placed at $r=10$\,cm, $75\%$ of $\PKzS$'s can be 
detected in this decay mode.

However, even if the number of reconstructed $\PKzS$ were be increased,
they would be useless without a sufficiently precise measurement
of the $\PKzS$ decay vertex; in addition, a good resolution in the corresponding 
$B$ decay vertex position, important to measure the $CP$ asymmetry,  
would be degraded.
In this section, we present the results of a simulation study to optimize the location
of the fifth layer in terms of both the reconstruction efficiency and the
vertex resolution.

For this simulation study we make use of the GEANT3-based full detector 
simulator that is modified for the 6-layer configuration as described 
in Section~\ref{sec:newlayer}.
The position of the fifth layer is varied from 
$r=10$\,cm to $14$\,cm while the outermost layer is fixed at $r=14$\,cm.
The standard Belle software is used for tracking, $\PKzS$ vertexing,
$\pi^0$ reconstruction and $B$ decay vertex determination.
If there are several $B$ candidates in an event, the best $B$ candidate
is selected by requiring the minimum mass difference between the
reconstructed $B$ masses and the nominal value.
The vertex resolution is estimated in the same way as described in
section~\ref{sec:vertexing}.

\begin{figure}[h]
\includegraphics[width=0.9\linewidth]{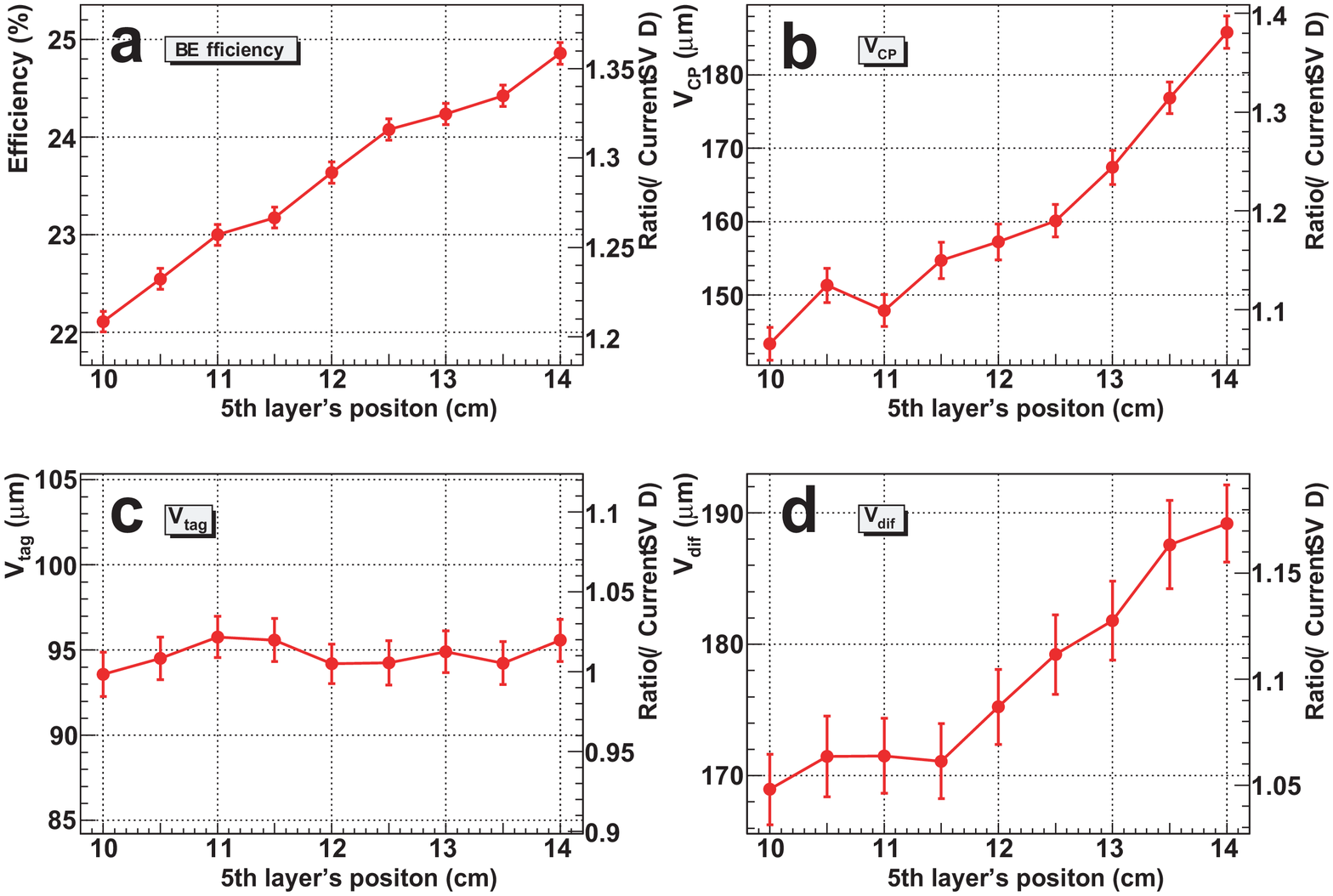}
\caption{\label{fig:ksg_bres} (a) : The reconstruction efficiency
for $B \to K^{*0} \gamma$.
(b) and (c) are the $CP$ side and tagging side vertex resolutions, respectively.
(d) shows the $\Delta z$ vertex resolution.
Here all results are shown as a function of the position of the fifth layer
in steps of $0.5$\,cm.
}
\end{figure}

The detection efficiency for $B \to K^{*0} \gamma$ decays
is shown in Fig.~\ref{fig:ksg_bres}(a) as a function
of the location of the fifth layer.
By increasing the radius of the fifth layer, the efficiency increases
by $29\%$  for $r=12$\,cm relative to the current 4-layer SVD,
and by $\sim 36\%$ for $r=14$\,cm.
On the other hand, the $B$ decay vertex resolution deteriorates for  
larger radii of the fifth layer.
It degrades by $\sim17\%$ relative to the 4-layer SVD at $r=12$\,cm
and $38\%$ at $r=14$\,cm (See Fig.~\ref{fig:ksg_bres}(b)).
For the tagging side vertex resolution, which is not affected 
by long-lived particles such as $\PKzS$'s, the vertex resolution does not
depend on the location of the fifth layer (Fig.~\ref{fig:ksg_bres}(c)).
As expected, the $\Delta z$ vertex resolution shows a tendency similar to the
$B$ decay vertex resolution (Fig.~\ref{fig:ksg_bres}(d)).
Up to $r=11.5$\,cm, the resolution is constant, and then deteriorates
as the radius increases.

To search for the optimal position, we need to consider the improvement of
the efficiency and the degradation of the resolution simultaneously.
Ref~\cite{bn111} shows the correlation between the vertex resolution and the
necessary statistics to extract one of the $CP$ asymmetry parameters 
${\cal S}$.
The minimum number of events ($N_s$) required to measure $\cal S$ with a 
statistical significance $s$ can be expressed as
\begin{equation}
N_s \simeq \left( \frac{s}{d \cdot \cal S} \right)^2
\label{eq:necessary}
\end{equation}
where $d$ is the so called dilution factor; $d$ is  a function of $\cal S$,  the 
$B^0-\bar{B}^0$ mixing parameter $\delta m/\Gamma$ and the detector
response function.
For simplicity, a double Gaussian function which is evaluated from the
$\Delta z$ distribution is assumed as the detector response function.
In this study, the wrong tag probability and beam background effects are not 
taken into consideration.

\begin{figure}[h]
\includegraphics[width=0.9\linewidth]{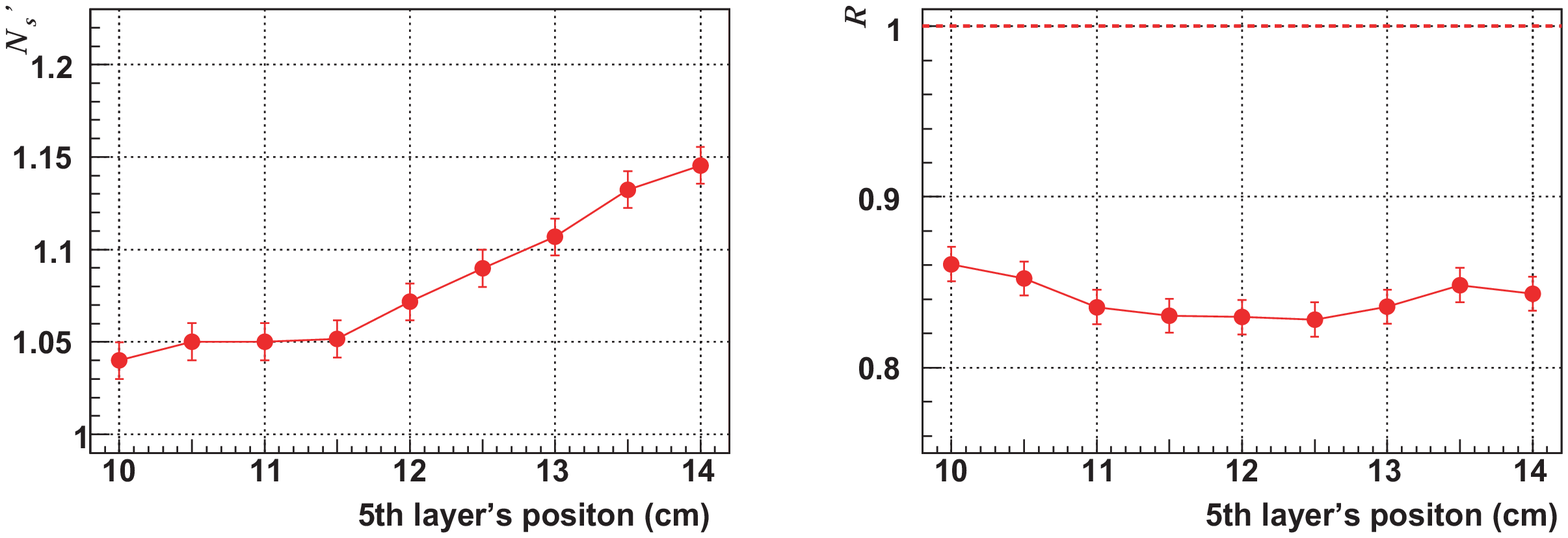}
\caption{\label{fig:necessary2} The necessary number of events 
for the 6-layer SVD $N_s$ normalized to the corresponding number for the  4-layer case,
as a function of the location of the fifth layer (left). 
 The ratio of $N_s/\epsilon$ for the 6-layer and 4-layer SVD configurations (right).
}
\end{figure}

Assuming the value of $\cal S$ as  expected in the Standard Model, ${\cal S}=0.03$,
the necessary number of events for the 6-layer SVD $N_s$ normalized to the
corresponding number for the  4-layer case
is shown in Fig.~\ref{fig:necessary2}  as a function of the position of the fifth layer.
Apparently, we need more statistics to observe $\cal S$ with
the same statistical significance with the 6-layer SVD,
because the average vertex resolution is degraded due to $\PKzS$ 
mesons that decayed between the third and the fifth layers.
However, in this figure of merit the effect of the better detection efficiency 
in a larger SVD volume is not taken into account.
Once this efficiency $\epsilon$ is considered by taking the ratio $N_s/\epsilon$,
the optimal position of the fifth layer can be determined.
The right side of Fig.~\ref{fig:necessary2} shows the  ratio of $N_s/\epsilon$
for the 6-layer and 4-layer SVDs .
The minimum at about $12$\,cm indicates the best position for the fifth layer.

\begin{figure}[h]
\includegraphics[width=0.9\linewidth]{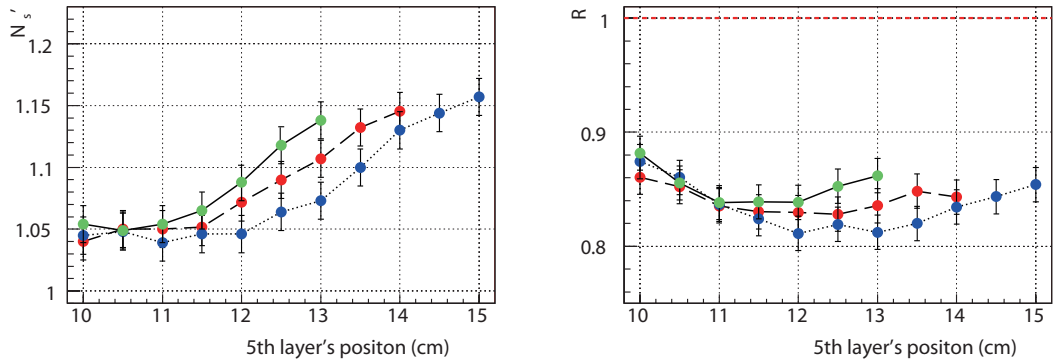}
\caption{\label{fig:vary6th} The necessary number of events 
for the 6-layer  SVD $N_s$ normalized to the corresponding number for the  4-layer case,
as a function of the location of the fifth layer (left). 
 The ratio of $N_s/\epsilon$ for the 6-layer and 4-layer SVD configurations (right). 
The solid, dashed and dotted lines correspond to cases with 
the outermost layer at $r=13$\,cm, $14$\,cm and $15$\,cm, respectively.
}
\end{figure}

So far, the outermost layer has been fixed at $r=14$\,cm, and
 the estimated best position of the fifth layer was found to be around $r=12$\,cm.
To generalize this result, similar simulation studies were performed
with  the outermost layer at $r=13$\,cm and $r=15$\,cm.
As in the previous study, the position of the fifth layer is
varied from $10$\,cm to the allowed maximum radius in steps of $0.5$\,cm.
The left side of Fig.~\ref{fig:vary6th} shows $N_s^\prime$ for the three 
cases. We note that the track parameter resolution for tracks that are associated with
SVD hits only in the fifth and sixth layers depends on
the distance between these two layers.
As this distance decreases, the $B$ vertex resolution degrades,
and a larger number of events is required to observe $\cal S$ with the
same statistical significance. 
The $N_s/\epsilon$ ratio for the 6-layer SVD normalized by that for the 4-layer one is shown 
on the right side of Fig.~\ref{fig:vary6th}.
Again, the minimum indicates the best position of the fifth layer.
From these plots, we conclude that about 2\,cm inside the outermost layer
seems to be the optimum for all cases.

\begin{figure}[h]
\includegraphics[width=0.9\linewidth]{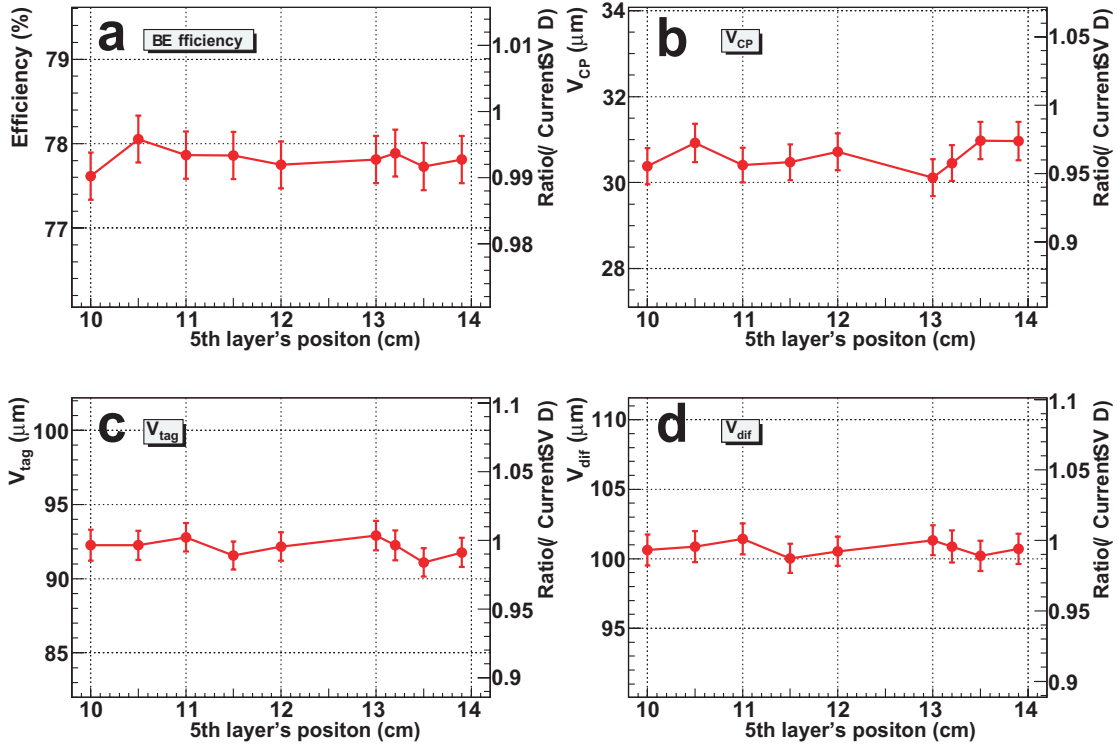}
\caption{\label{fig:pipi_res} (a) : The reconstruction efficiency
for $B \to \pi^+ \pi^-$ decay.
(b) and (c) are the $CP$ side and tagging side vertex resolutions, respectively.
(d) shows the $\Delta z$ vertex resolution.
Here all results are shown as a function of the position of the fifth layer
in steps of $0.5$\,cm.
}
\end{figure}

\begin{figure}[h]
\includegraphics[width=0.9\linewidth]{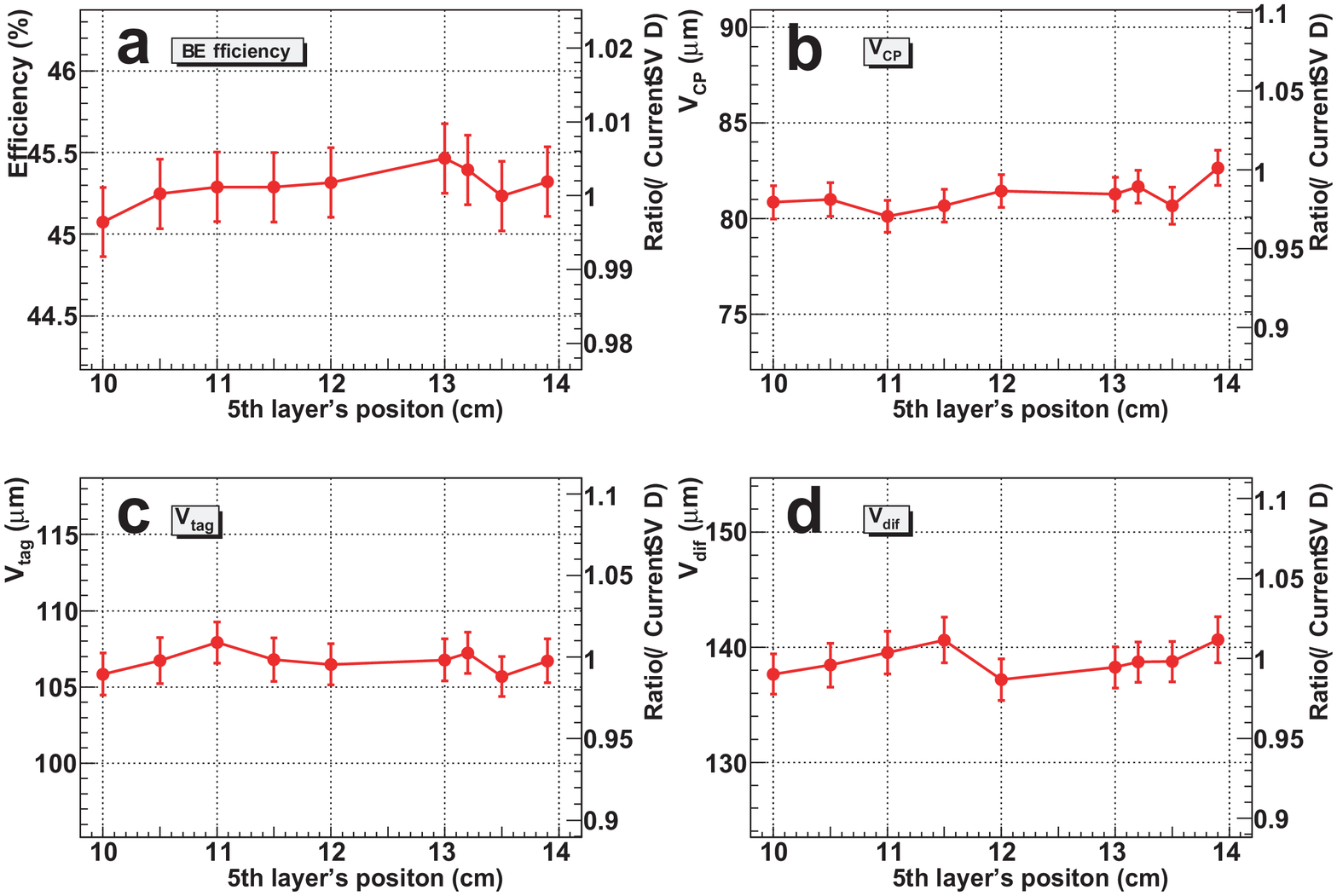}
\caption{\label{fig:phiks_res}  (a) : The reconstruction efficiency
for $B \to \phi\PKzS$ as a function of the position of the fifth layer;
(b) and (c) are the $CP$ side and tagging side vertex resolutions, respectively,
and (d) shows the $\Delta z$ vertex resolution.
}
\end{figure}

The above optimization was performed for a specific decay channel
$B \to K^{*0} \gamma$. It is mandatory  to check whether the position of the fifth layer
affects the vertexing performance of  other important $B$ decay modes.
The detection efficiency and the vertex resolution for the
$B \to \pi^+ \pi^-$ and $B \to \phi\PKzS$ modes were investigated.
The outermost layer is fixed at $r=14$\,cm in this case.
For $B \to \pi^+ \pi^-$, all tracks used in the vertex reconstruction
are produced in the vicinity of the interaction point.
Therefore, it is expected that the position of the fifth layer does
not affect the efficiency nor the vertex resolution.
Figure~\ref{fig:pipi_res} shows the results of this study, and, as expected, 
we find no variation with the  position of the fifth layer.
In the case of the $B \to \phi\PKzS$, although a $\PKzS$ is involved in this
decay mode, the decay point of $B$ is mainly determined by the decay point 
of $\phi$. Hence, it is again expected that
the position of the fifth layer does not affect the performance; this is 
confirmed by Fig.~\ref{fig:phiks_res}.

We conclude that the
second outermost layer be located $2$\,cm inside the outermost layer.


%% file: signal-to-noise.tex
\subsection{S/N ratio and track matching efficiency}
\label{sec:svdsn}
%
In order to provide SVD hit information for  tracks reconstructed in CDC,
the tracking acceptance of SVD should be the same or larger than that of CDC.
Therefore, the SVD layers have to cover a wider range of angles with respect to the beam axis.
For the Belle SVD, the double sided sensors (DSSDs) are assembled into a ``ladder structure''.
In particular for the outer layers, 
several DSSDs are connected and read out by a single front-end chip, which is placed
outside the tracking acceptance to reduce the material within the spectrometer acceptance.
Since the number of DSSDs increases with the  the length of the ladder structure,
the detector capacitance connected to a single front-end chip increases,
and the noise level becomes large.

Furthermore, to reduce the occupancy in the upgraded SVD,
we plan to employ a new front-end chip type APV25, 
which has been developed for the CMS Si tracker, 
with a shorter shaping time ($\sim$50\,ns) than the presently used, VA1TA chip. 
Generally, the shorter shaping time results in a higher intrinsic electronic noise.
In the current Belle SVD, the S/N ratio is around 35 
for the inner layers and around 16 for the outer layers;
the APV25 chip has a noise level four times higher than the VA1TA chip. 
Because a degraded  S/N ratio affects the total performance of the SVD,
it is important to evaluate the effect of the poor S/N ratio at the outer layers. 
In the case of a degraded S/N ratio, we expect that the matching efficiency and 
the tracking resolution will deteriorate.
Here, the matching efficiency is defined as the probability 
that tracks reconstructed in CDC have hit information in at least two SVD layers.
When the noise level becomes very large, 
the required threshold level could exceed the signal level. 
In that case, hit signals could be missed, and the matching efficiency reduced. 
Even if we  can find the correct signal hit, 
the high noise level distorts the SVD hit position information
and degrades the track resolution.

To evaluate these effects, we carried out simulation studies.
For this purpose, the current SVD configuration 
that is implemented in the standard Belle full detector simulator is 
replaced with the 6-layer one as shown in Fig.~\ref{fig:6lyr}.
As the detailed configuration for the SVD upgrade has not been determined yet,
for the additional two layers we assume the same ladder modules as used in the fourth layer 
of the present Belle SVD,  and place them at $r=13$\,cm and $r=14$\,cm.

\begin{figure}[h]
 \begin{tabular}{ccc}
 \begin{minipage}{0.5\hsize}
 \begin{center}
  \includegraphics[scale=0.3]{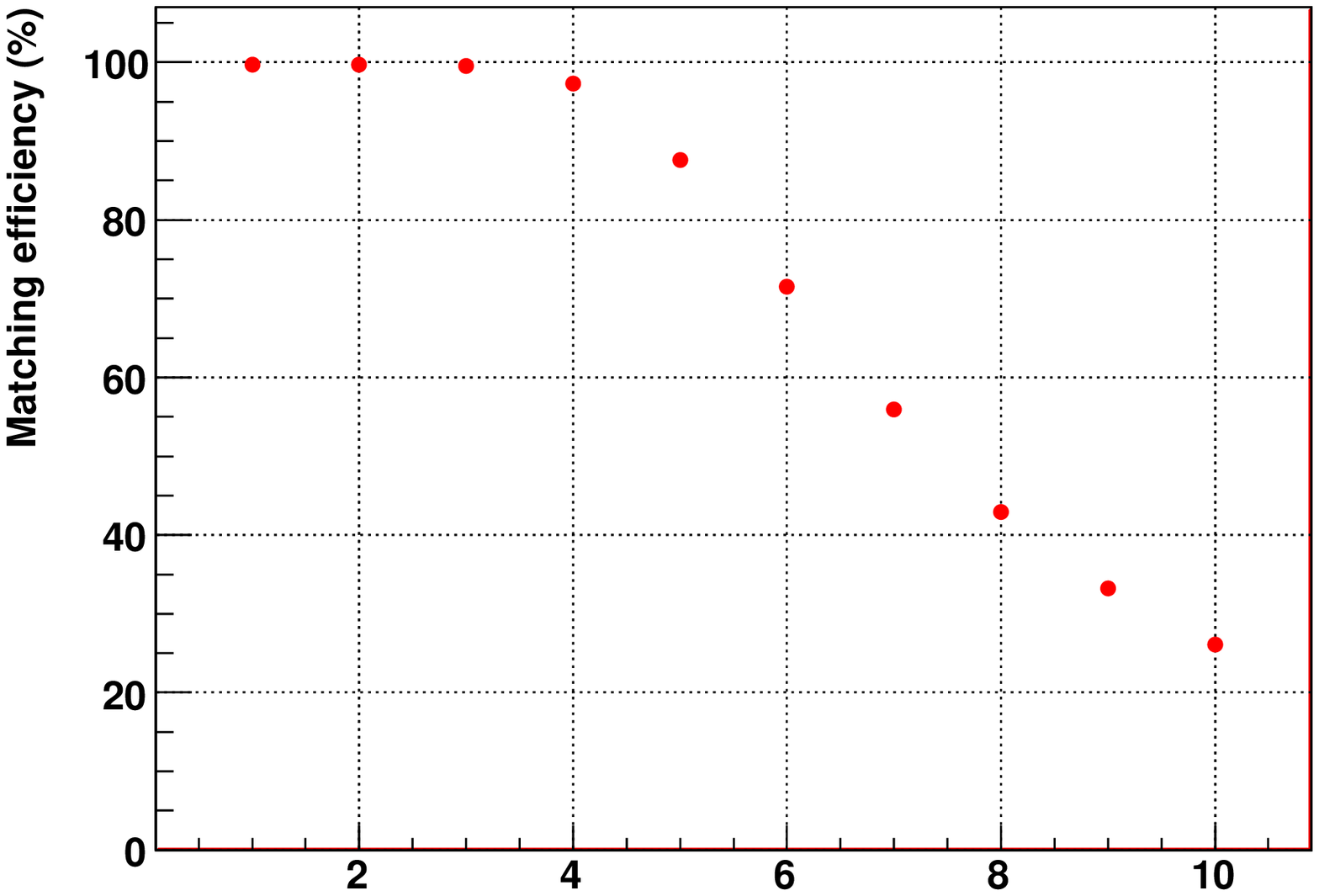}
 \end{center}
  \end{minipage}
 \begin{minipage}{0.5\hsize}
 \begin{center}
  \includegraphics[scale=0.3]{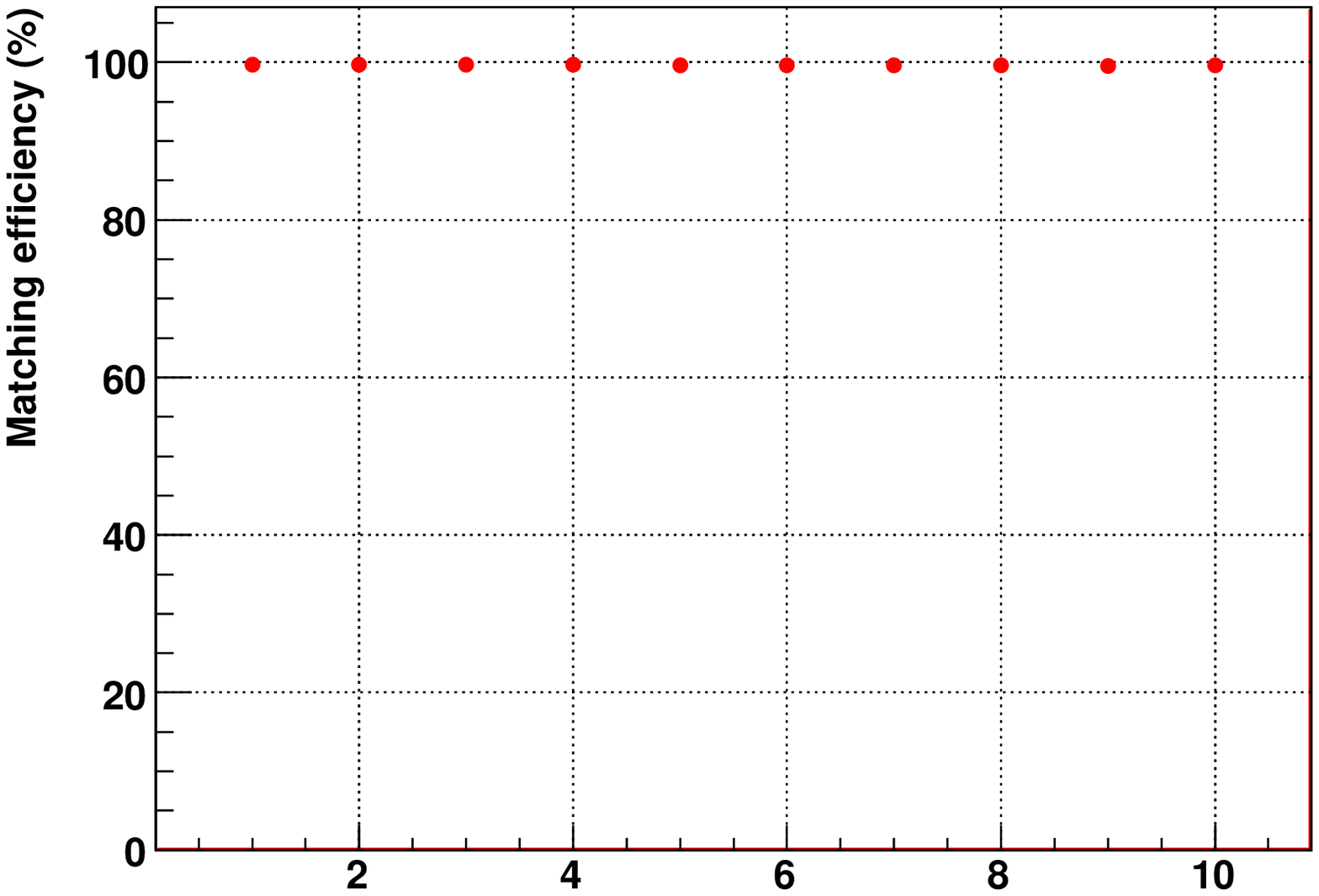}
 \end{center}
 \end{minipage}
 \end{tabular}
  \caption{The matching efficiency as a function of the noise level normalized to the 
  noise level of the layer 4 in the current Belle SVD.  The noise level is varied on all
  layers for the left figure, and on only layers 5 and 6 for the right figure.
}
 \label{fig:SN56ME}
\end{figure}

We have checked the matching efficiency with various noise levels in layers and 5 and 6.
At first, we simulated generic $B$ meson decays.
In Fig.~\ref{fig:SN56ME},  the matching efficiency is shown
normalized to the current Belle SVD performance as a function of the noise level.
We increase the noise level on all layers in the figure on the left, and 
only on layers 5 and 6 for the figure on the right.
For the latter case we conclude that 
the matching efficiency is higher than 99\% for any noise level in layers 5 and 6 
because there are still four other layers where matched hits can be found.


\begin{figure}[htbp]

 \begin{center}
  \includegraphics[scale=0.5]{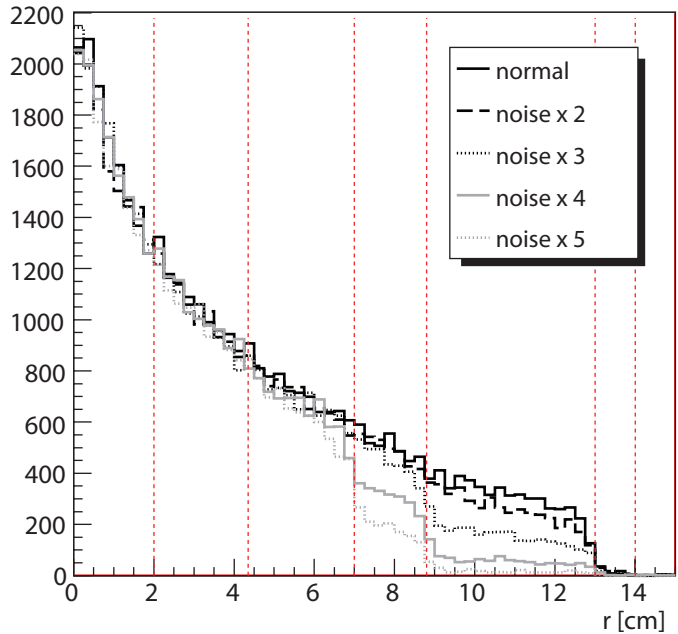}
  \caption{The distribution of the $\PKzS$ vertex, whose
  daughter tracks are matched with two layers of SVD hits,
  in $B^{+}\to K_{S}\pi^{+}$ for different noise levels in layers 5 and 6. The positions of
individual layers are indicated by dashed lines.
  Thanks to layers 5 and 6, 20\% more $\PKzS$'s are reconstructed within the SVD. 
  The benefit is, however, diluted at higher noise levels.}
 \label{fig:56trk}
 \end{center}
\end{figure}

\begin{figure}[htbp]
 \begin{center}
  \includegraphics[scale=0.5]{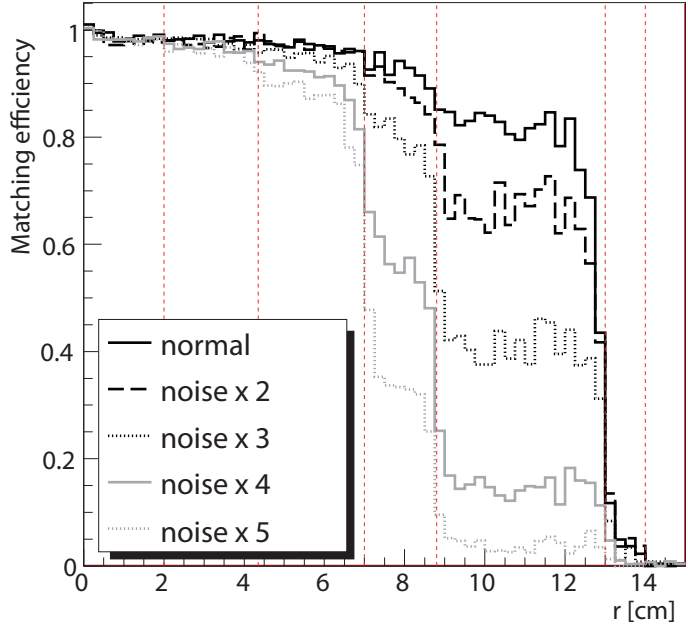}
  \caption{The matching efficiency as a function of the $\PKzS$ vertex position 
  in $B^{+}\to K_{S}\pi^{+}$ 
  for different noise levels. If the noise level is 3 times higher
  than that of the present Belle SVD, the gain from the extra coverage 
  by the outer layers is reduced by 50\%.}
\label{fig:th3-56me}
 \end{center}
\end{figure}

The main motivation for the proposed  6 layer configuration is 
to improve the efficiency for the $\PKzS$ reconstruction. 
Due to the finite lifetime of $\PKzS$, the $\PKzS$ daughter tracks originate
from a vertex far from the IP. In this case, the effect of the poor S/N
performance in the outer layers becomes severe.
Figure \ref{fig:56trk} shows the vertex position of the $\PKzS$ in
$B^{+}\to K_{S}\pi^{+}$ decays for the case where two layers of SVD hits 
could be associated with both daughter tracks.
Thanks to the additional two layers, 
we reconstruct an additional 20\% of $\PKzS$ vertices. 
Figure \ref{fig:th3-56me} shows the matching efficiency with various noise levels.
We conclude that about 60\% of tracks, decaying between layers 4 and 5,
will be lost when the noise levels in layers 5 and 6 are three times higher
than in the current Belle SVD.

\begin{figure}[htbp]
 \begin{center}
  \includegraphics[scale=0.5]{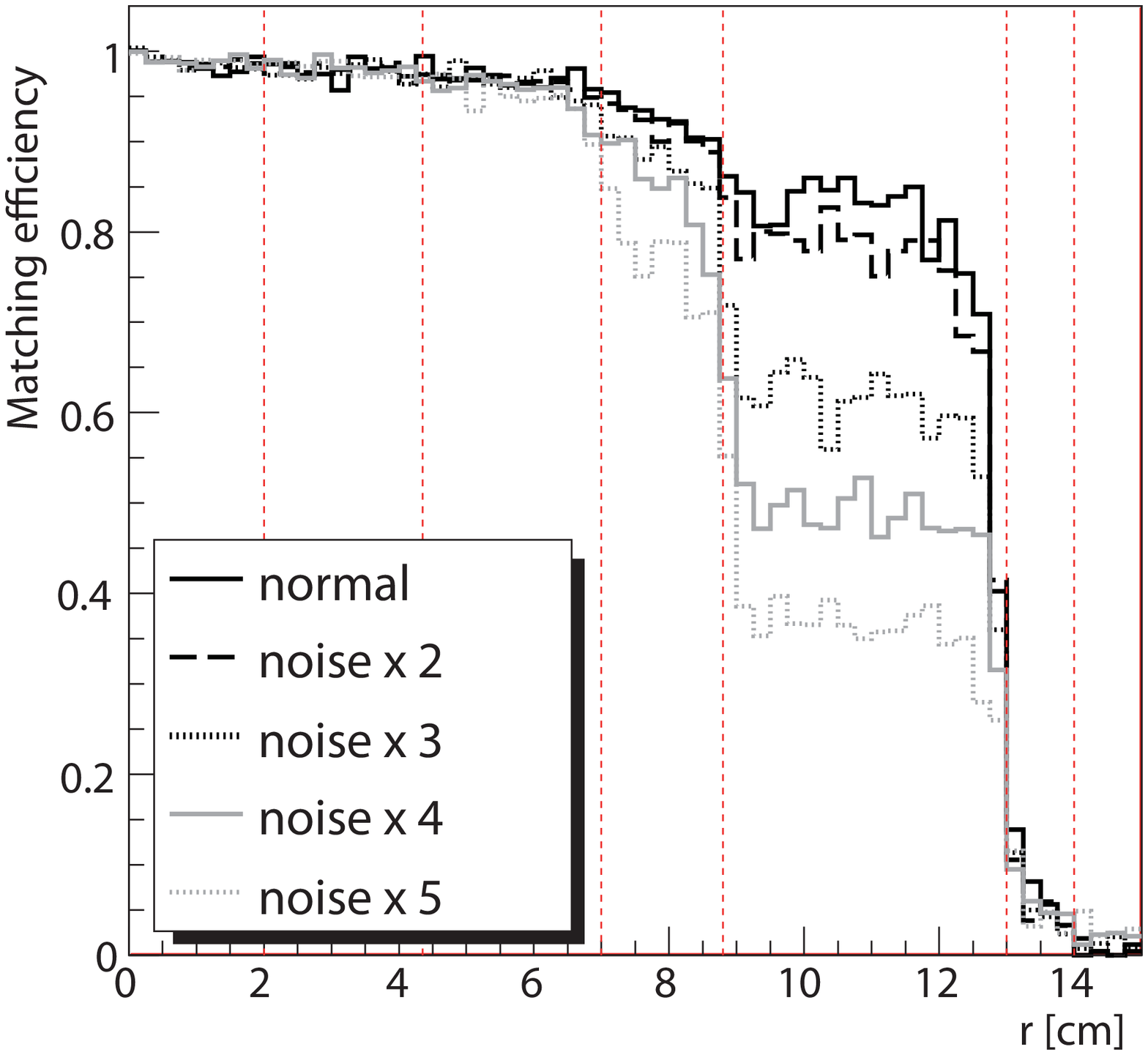}
  \caption{The matching efficiency as a function of the $\PKzS$ vertex position
  in $B^{+}\to K_{S}\pi^{+}$, 
  where the threshold for SVD hit finding is reduced from 3$\sigma$ to 2$\sigma$. 
 }
\label{fig:th2-56me}
 \end{center}
\end{figure}

In the studies discussed above, we search for SVD hits in the DSSDs by applying
a threshold cut on the energy
deposited per strip at 3$\sigma$, where $\sigma$ is the noise level.
We tried to recover the matching efficiency by lowering this threshold.
Figure \ref{fig:th2-56me} shows the matching efficiency obtained 
with a threshold of 2$\sigma$.
The losses in matching efficiency can be reduced by up to $40$\% 
by lowering the threshold. Note, however, that data have to be sparsified
when recorded in order to reduce the stored data size. 
With a lower threshold, we need to store 20 -- 30 times more data, 
which is not trivial from the DAQ point of view.

%
%
We conclude that we have to be careful not to increase the noise level 
when we design the sensor and the readout scheme for the outer layers. 
It must not exceed a  level which is 3 times larger than the one in layer 4
of the  present Belle SVD.
A possible solution is to develop a low noise DSSD and/or front-end chips. 
We have started a project to develop our own front-end chip, 
which has better S/N performance and a fast readout capability. 
Another solution is to avoid ganging of DSSDs.
This can be realized by using a ``chip-on-sensor'' or a FLEX readout. 
However, both options will increase the amount of material and the number of readout channels, 
and will be challenges in the technical design of the SVD upgrade.


%% file: SVDPitch.tex
\subsection{Readout pitch of outer layers}
\label{sec:svdpitch}

\begin{figure}[h]
\includegraphics[width=0.5\linewidth]{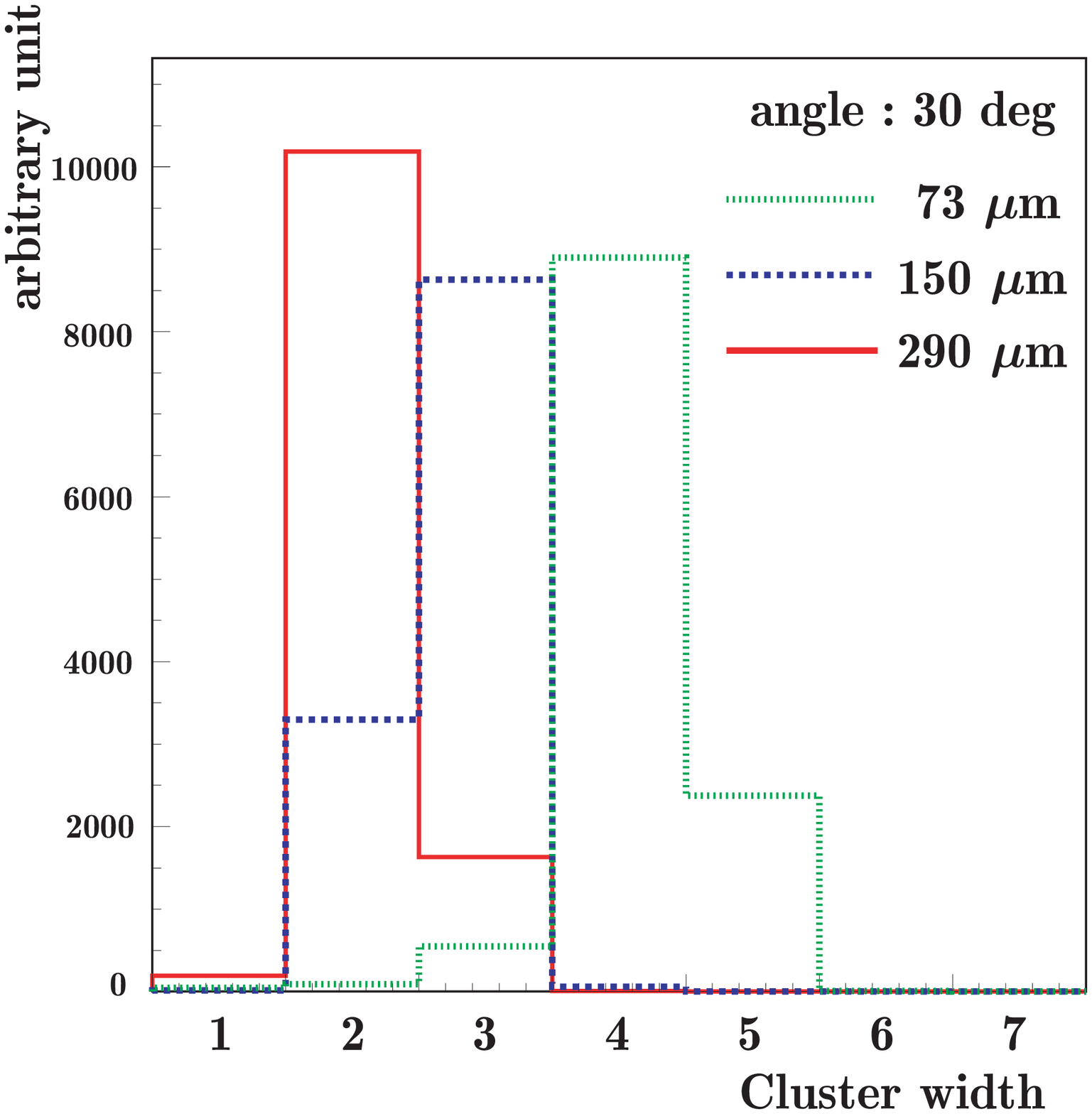}
\caption{\label{fig:clstwidth} Distributions of the z-side cluster width in the
fifth and sixth layers for tracks that are produced at the IP
with a polar angle of 30 degrees.
The transverse momentum of tracks is fixed to $2.0$\,GeV/c.
The dotted line, dashed line and solid line corresponds to
the case where the z-side strip pitch is set to $73\,\mu$m, $150\,\mu$m and
$290\,\mu$m, respectively.
In this simulation, the collected charges on the floating strip are divided
equally and added to the adjacent strips, while those of the readout strip
are left untouched.}
\end{figure}

Due to the enlarged SVD volume in the radial direction,
we can increase the reconstruction efficiency for $\PKzS$ events.
A much larger SVD volume requires an increase in	
the number of readout channels as described in section~\ref{sec:svd}.
It is therefore important to study whether or not the number of 
readout channels can be reduced without degrading the performance, 
especially for the outer layers.
For this purpose, we use the GEANT3-based simulation again
that was modified for the study in subsection~\ref{sec:svd5th}.
This time, the strip pitch of the outermost (=sixth) and fifth layers
is artificially increased.
Note that this modification is applied only to the z-side strips, i.e., strips along the  
$z$ direction, since it is expected 
that a corresponding change in the other direction could degrade the $B$ vertex resolution.
We compare the performance of the detector with the initial z-side strip 
pitch at $73\,\mu$m, and with three alternatives, 
 $110\,\mu$m, $150\,\mu$m and $290\,\mu$m. We only vary the strip pitch for the 
outer two layers (5 and 6).
The positions of the outer two layers are fixed to $r=12$\,cm and $14$\,cm.
Figure~\ref{fig:clstwidth} shows the distributions of the z-side cluster width
in the fifth and sixth layers for tracks that are produced at the coordinate origin
with a polar angle of 30 degrees.
For tracks with a shallow incidence angle, 
the wider the strip pitch is, the smaller the cluster width becomes.

\begin{figure}[h]
\includegraphics[width=0.9\linewidth]{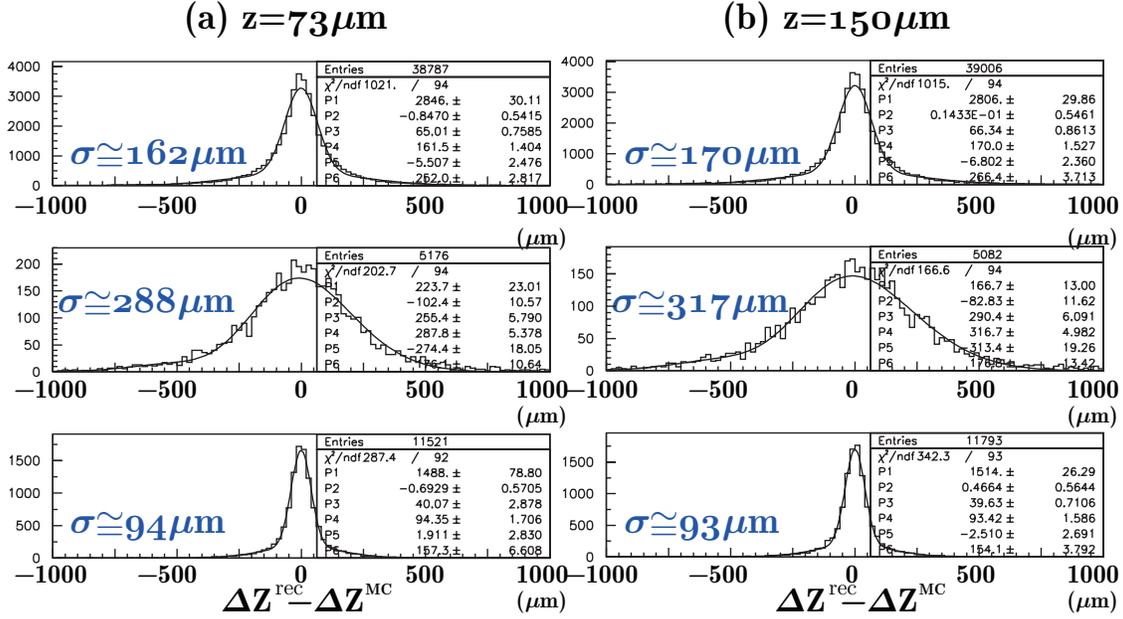}
\caption{\label{fig:z73vsz150} Decay vertex resolution for $B\to K^{*0} \gamma$ decays  
with z-side strip pitches of $73\,\mu$m (column a)  
and $150\,\mu$m (column b).
For the top of each (a) and (b), we require at least two SVD hits
to reconstruct each charged pion. 
The middle figures show the distributions for the case 
where at least one of the two charged pions is required to have
an SVD hit in the outermost two layers.
The bottom figures are the distributions for the case 
where hits are required in all SVD layers 
for at least one of the two charged pions.
}
\end{figure}

In general, a strip-type sensor with a wider pitch has
worse spatial resolution than that with a narrow pitch.
If we increase the strip pitch for the outer layers,
the vertex resolution for the $\PKzS$'s that decay
between the fourth and fifth layers could degrade.
On the other hand, the vertex resolution for the $\PKzS$'s
that decay in the vicinity of the interaction point should not change much,
since the better spatial resolution of the inner layers has a greater
impact on the vertex resolution than the poorer resolution of the outer layers.
In order to check these expectations, the vertex resolution for the
$B \to K^{*0} \gamma$ decay is investigated.
The analysis procedure is the same as in subsection~\ref{sec:svd5th}.
The  vertex resolutions for the case 
where the z-side strip pitches are set to $73\,\mu$m and $150\,\mu$m
are shown in Fig.~\ref{fig:z73vsz150}.
Shown at the top are the resolutions 
for the case where at least two SVD hits are
required to reconstruct each charged pion, which is a decay product of the $\PKzS$
from the $K^{*0}$ decay chain.
The middle figures in Fig.~\ref{fig:z73vsz150} show the same distributions
for the case where at least one of the two charged pions is required to have
SVD hits only in the outer two layers.
From these plots, a clear degradation caused by the wider strip pitch
is observed in the vertex resolution.
In contrast, when we require hits in all SVD layers for at least one of 
the two charged pions, no degradation is observed as shown in similar
distributions at the bottom of Fig.~\ref{fig:z73vsz150}.
The results including cases with other pitch values
 are summarized in Table~\ref{tab:ksres+pitch}.

Table~\ref{tab:ksres+pitch} shows that 
there is no sizable degradation in the vertex resolution,
if we require SVD hits in all layers for the charged pions used to reconstruct
the $\PKzS$ (see ``hits in all lyrs''in Table~\ref{tab:ksres+pitch}).
Furthermore, when a charged track has SVD hits in the
outermost and fifth layers only, the vertex resolution becomes worse
(see ``hits in 5th \& 6th lyrs only'' in Table~\ref{tab:ksres+pitch}).
In particular, in the case where the strip pitch is set to $290\,\mu$m
the vertex resolution degrades by roughly $20$\% 
(for the more specific case where at least one charged pion is 
required to have SVD hits in the outermost and fifth layers only,
which is denoted as ``at least 1 track'' in Table~\ref{tab:ksres+pitch}).
In this case, the merit of the enlarged SVD volume, that is, 
the better reconstruction efficiency for $\PKzS$'s, could be reduced
by the poorer resolution.
For the case of $110\,\mu$m strip pitch, the vertex resolutions are 
comparable to the case of $73\,\mu$m strip pitch, except when
both charged pions have SVD hits in the outermost
and fifth layers only.
With the requirement that a charged track must have
at least two SVD hits without any constraint on which layers were hit  
(the standard procedure in $\PKzS$ reconstruction in Belle,
see ``$\geq 2$ lyr hits'' in Table~\ref{tab:ksres+pitch}),
we observe a degradation by about  $4$\% even for the case of $110\,\mu$m strip pitch
compared to the case of $73\,\mu$m strip pitch.
If only these results are taken into account,
strip pitches in the range up to $110$ -- $150\,\mu$m seem to be acceptable.
However, as will be discussed in section~\ref{sec:material}, we also have to
consider the effect of the increased material in order to maintain
a S/N ratio of $\sim15$ even in the outer layers.
Because the effect of the increased material cannot be avoided, a
further degradation caused by the wider strip pitch is not acceptable.

To conclude this subsection we note that outer layers with a wider strip pitch
are not an optimal solution for the $\PKzS$ reconstruction in
the analysis of $B \to K^{*0} \gamma$ decays.
One possibility is to introduce sensors with a wider strip pitch in the 
forward and backward region of the outer layers, 
while keeping the current strip pitch ($\sim73\,\mu$m)
in the main cylindrical part.
As reported in \cite{bib:widepitch}, the intrinsic resolution of the
sensors with a wider strip pitch could be better than that with narrow strip
pitch for particles with a shallow incidence angle (larger than $50$ degrees).
Thanks to this, the vertex resolution  in the vicinity of the
interaction point should not be affected.
However, even this modification would degrade the performance of
$\PKzS$'s that decay between the fourth and fifth layers.

As an alternative solution, the introduction of an arc-shaped detector
or a slanted detector would reduce the number of readout
channels in the outer layers. Note, however, that in this case the SVD detector 
would occupy a smaller overall volume; as a consequence, the 
 $\PKzS$ reconstruction  efficiency  would be reduced.

\begin{table}[htbp]
\caption{\label{tab:ksres+pitch} Decay vertex resolution for $B\to K^{*0} \gamma$ decays.
Charged pions from the   
$\PKzS$ decay are required to have at least two SVD hits (denoted as ``default'') or 
more (denoted as ``hits in all lyrs'' and ``hits in 5th + 6th lyrs only'').
}
%
%
\begin{tabular}{|l|c|c|c|c|c|}
\hline
(z-side strip pitch)  & $\geq 2$ lyr hits. &  \multicolumn{2}{c|}{hits in all lyrs} &  \multicolumn{2}{c|}{hits in 5th \& 6th lyrs only} \\
\cline{2-6}
 & (default) & (at least 1 track) & (for both tracks) & (at least 1 track) & (for both tracks) \\
\hline
\hline
$73\,\mu$m & 162$\pm$1 ($\mu$m) & 94$\pm$2 ($\mu$m) & 66$\pm$1 ($\mu$m) & 288$\pm$5 ($\mu$m) & 234$\pm$4 ($\mu$m)\\
\hline
$110\,\mu$m & 168$\pm$1 ($\mu$m) & 91$\pm$1 ($\mu$m) & 61$\pm$1 ($\mu$m) & 283$\pm$5 ($\mu$m) & 300$\pm$5 ($\mu$m) \\
\hline
$150\,\mu$m & 170$\pm$2 ($\mu$m) & 93$\pm$2 ($\mu$m) & 62$\pm$1 ($\mu$m) & 317$\pm$5 ($\mu$m) & 304$\pm$4 ($\mu$m) \\
\hline
$290\,\mu$m & 176$\pm$1 ($\mu$m) & 97$\pm$2 ($\mu$m) & 63$\pm$2 ($\mu$m) & 342$\pm$4 ($\mu$m) & 352$\pm$5 ($\mu$m) \\
\hline
\end{tabular}
\end{table}


%% file: slant.tex
\subsection{Slant angle}
\label{sec:svdslant}
In the silicon vertex detector (SVD) of the Belle experiment, all the DSSD sensors are placed 
parallel to the beam axis.  The 2004 Super KEKB LoI introduced slanted sensors 
to the SVD configuration in the forward region. 
These slanted sensors reduce the length of the SVD ladders in the outer layers. 
Note that even in present SVD2 with four layers, the length of the layer 4 ladders 
is about 50\,cm.
  
In the design, we tried to minimize the number of DSSD types in order to reduce 
the cost of masks used in the silicon process.
However, the ladders in the outer layers are still long, and a reduction of ladder 
length might be beneficial. 
In addition, it was pointed out that the dip angles of 
particles entering the DSSDs become very shallow in the forward region.
The effective material thickness and cluster size increase by a factor of 3 -- 4 
compared to 
those in the central region, and vertex reconstruction and position
resolution degrade.

To address this problem, an optimization of the sensor configuration is carried out.
Slanted sensors are introduced both in the forward and backward regions,
 and the slant angles are varied to find the best solution.

\begin{figure}
\begin{center}
\includegraphics[width=120mm]{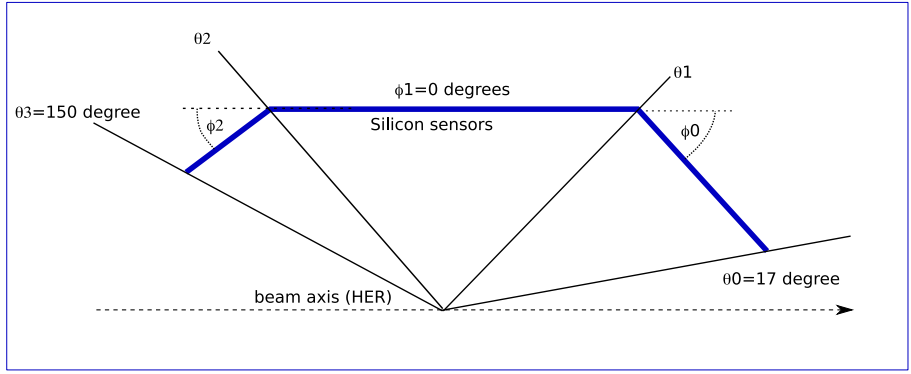}
\end{center}
\caption{\label{fig:slantdesign} The schematic of a SVD ladder, with slanted sensors in 
the forward and backward regions with the definition of angles used for the optimization.}
\end{figure}

The optimization is based on the minimization of the 
 average incident angle of tracks originating in the collision point. 
Parameters used in this optimization are shown in Figure~\ref{fig:slantdesign}.
The angles  $\theta_1$ and $\theta_2$ are  polar border angles of the planar sensor section,
    $\phi_0$,  $\phi_1$ and  $\phi_2$  are the slant angles; 
  $\theta_1$, $\theta_2$, $\phi_0$, and  $\phi_2$  are the free parameters, and
  $\phi_1=0$ is fixed as the middle sensor is kept parallel to the beam axis.

The optimization is carried out by minimizing either the average incident angle  or 
its square, 
\begin{eqnarray}
V_1&= \overline{ {\rm Incident \, angle} } &= \int {\rm(Incident \, angle)} \cdot d \cos \theta  \\
V_2&= \overline{ {\rm Incident \, angle} ^2} &= \int{\rm(Incident \, angle)}^2\cdot d \cos \theta
\end{eqnarray}
for tracks within the acceptance ($17^\circ < \theta < 150 ^\circ$).

In addition to the LoI design (slanted sensors only in the forward region) 
and the optimized design, these variables are also 
calculated for a cylindrical detector with no slanted sensors,
 as well as for 
a configuration with disks in the forward and backward regions ($\phi_2=-\phi_0=90 ^\circ$).
The total DSSD area $S$ for layer 6 ($R$=14\,cm) is calculated to estimate the cost of the detector.

The optimization using $V_1$ and $V_2$ gives almost the same answer
within 1$^\circ$ -- 2$^\circ$.
After the optimization, the average incident angle is close to 1/3 of the LoI design
value, and 
the sensor area $S$ is reduced by a factor of more than two. The values of 
$V_1$ and $V_2$ for the non-slanted design and LoI design are almost the same. This is due to the fact that
in the LoI design the slanted region covers only a very small solid angle. 
Note also that in the LoI design we only save 20\% of the sensor area compared to the cylindrical
non-slanted solution.

\begin{table}
\begin{center}
\caption{Table for the optimization}
\begin{tabular}{ccccccccccc}
\hline
       &$\theta_0$ & $\theta_1$  &$\theta_2$ & $\theta_3$ & $\phi_0$ & $\phi_1$ & $\phi_2$ & $V_1$ & $V_2$ & $S$  (cm$^2$) \\
\hline
No Slant & 17  & - & - & 150 & -  & 0 & - & 29 & 34 & 6161 \\
\hline
LoI  &17 & 34 & -  &150  & -15  & 0 &   & 28        & 33 & 5018\\
\hline
Disk & 17 &  45 & 135 & 150 & -90 & 0 & 90 & 24 & 28& 3431 \\
\hline
Optimized &17 & 70 & 108  &150 & -41  & 0 & 36 & 11      & 13 & 2419\\
\hline
\end{tabular}
\end{center}
\end{table}

%% file: SVDSummary.tex
\subsection{Summary of recommendations for SVD}

\noindent
{\bf Front-end chip}

In terms of the impact parameter resolution, 
the innermost layer should be located as close to the beam pipe as possible.
This means that the innermost layer should be operable
under a harsh beam background that is estimated to be 15 times 
higher than in the current Belle detector.
According to subsection~\ref{sec:inrad}, this problem can be solved by adopting 
a front-end chip that has a shorter shaping time, such as $\sim50$\,nsec. 
Furthermore, from the MC study described in subsection~\ref{sec:bgeff}, 
the vertex resolution for $B \to J/\Psi\PKzS$ can be maintained at
the same level that has been achieved by the current SVD 
even if the beam background level is six times higher. 
Here in this simulation, the front-end chips of the innermost and 
second layers are assumed to be "APV25's". 
Therefore, increased occupancy from the beam background can be reduced to 
a manageable level by adopting a front-end chip that has a shorter shaping time, 
especially for the innermost and second layers.
For the readout speed, the current front-end chip "VA1" takes
$13\,\mu$sec to be read out with the maximum possible clock rate.
This is not acceptable with a high trigger rate, such as $10$\,kHz. 
In order to cope, we need a front-end chip that has a pipelined readout 
scheme even for the outer layers.
Therefore, {\bf a front-end chip that has both a shorter shaping time ($\sim50$\,nsec) 
and a pipelined readout scheme is required for all layers
to overcome the higher beam background without 
degrading the performance}.
\\

\noindent
{\bf Geometrical design}

To increase the vertexing efficiency for the $\PKzS$
from the $B \to K^{0*} \gamma$ decay chain, 
the SVD volume is enlarged in the radial direction. 
Because at least two SVD hits for each daughter pion
are required to reconstruct the $\PKzS$,
the position of the second to outermost layer is important. 
As demonstrated by the simulation in subsection~\ref{sec:svd5th}, 
{\bf the best position of the second to outermost layer is found to be $2$\,cm 
inside the outermost layer}.

Due to the expansion of the SVD volume, the detector, 
especially in the outermost and fifth layers, becomes longer.
This leads to an increase of the detector capacitance and leakage current
in the front-end chip.
Consequently, the noise level becomes larger.
Furthermore, the shorter shaping time also results in higher intrinsic electronic noise.
The simulation study performed in subsection~\ref{sec:svdsn} shows that
the S/N ratio for all the layers should be better than 
a third of that in the current Belle SVD,
which is roughly 35 for the inner layers and about 16 for the outer layers.
Otherwise, the efficiency for matching tracks between the SVD and the CDC will degrade.
In particular in $\PKzS$ vertexing,
we would lose $60$\% of tracks that decay between the fourth and fifth layers
if the noise level is three times higher than in the current Belle SVD.
{\bf A possible solution to overcome the poorer S/N ratio is to develop a system with
low-noise sensors (DSSD) and/or front-end chips}.
However, this has not been acomplished yet.
{\bf Another solution is to make use of a relatively established technology,
e.g., ``chip-on-sensor'', which can avoid the ganging of sensors}.
{\bf However, to realize this we need to clarify the material budget of the chips and
the cooling systems (the choice of the coolant, the flow rate of the coolant circulation, 
the material of the cooling pipe and must test the temperature/pressure tolerance,
the leackge of the coolant, corrosion, clogging and 
the connection of the cooling pipe to the outside),
the number of readout channels, the support structure 
and the space for the cabling and so on}.

The effect of the additional material is studied in subsection~\ref{sec:matsvd}.
For physics that aims at $CP$ violation,
for instance, $B \to J/\Psi\PKzS$, $\pi^+ \pi^-$ and $D^+ D^-$,
increasing the amount of material
in the innermost and second layers degrades the
$\Delta z$ vertex resolution by roughly $10$\%.
Therefore {\bf we recommend that the chip-on-sensor should not be adopted for the
innermost and second layers}.
On the other hand, {\bf for the outer layers (the outermost and second
to outermost layers),
we have to compromise because the degradation from S/N 
is more serious than that from the increase in material budget}.

Another side effect of the expansion of the SVD volume is
an increase of the number of readout channels. 
Moreover, as described in subsection~\ref{sec:svdsn}, 
the chip-on-sensor technology will be applied to maintain 
the current level of S/N, $\sim15$, especially for the outer layers. 
In this case, the number of readout channels would increase drastically. 
In order to explore the possibility that the number of readout channels 
in the outer layers can be reduced without degrading the performance, 
a MC study was carried out with $B \to K^{0*} \gamma$ decay, 
because this mode is sensitive to the performance of the outer layers. 
In this simulation, the z-side strip pitch, which is nominally $73\,\mu$m, 
was increased by a factor of two or four
and then the $CP$ side vertex resolution was examined. 
From the results obtained in subsection~\ref{sec:svdpitch} we conclude that
the wider strip pitch seems to be feasible. 
However, considering the degradation caused by the increased material 
to maintain the current level of S/N, 
{\bf further degradation from a wider strip pitch is no longer acceptable}. 
As an alternative solution, {\bf the introduction of an arch-shaped detector or 
a slanted detector is recommended}.

When the arch-shaped detector or the slanted detector configuration is considered, 
one can optimize by minimizing the average incidence angle of tracks from the 
interaction point to the sensor. 
Thus the "optimized" design shown in subsection~\ref{sec:svdslant} is determined. 
Thanks to this optimization, 
we find a better resolution than the case without an arch-shaped or 
slanted detector. 
Furthermore, the total area covered by sensors also can be reduced, 
which directly reduces the cost. 
However, if the fifth layer is located at $r=12$\,cm in the optimized design, 
roughly $8$\% of $\PKzS$'s would be lost. 
Consequently, among the four options shown in subsection~\ref{sec:svdslant}, 
the LoI design is the best in terms of $\PKzS$ vertexing efficiency. 
However, {\bf there is still much room to optimize the arch-shaped or slanted part 
to optimize the tracking performance, the track matching efficiency 
between CDC and SVD, the detection efficiency for the $\PKzS$ and the effect of the
beam background.
This should be clarified before the final decision on the SVD design}.

Although we did not discuss the middle layers (the third and fourth layers),
these are also important for the SVD stand-alone tracking.
At this moment, we have not decided the position of these layers and
the choice of technology for the readout system.
The decision will depend on several issues such as a possible mechanical
structure for these layers,
how many readout channels are acceptable,
and what level of the S/N can be achieved;
these issues have not been studied in detail yet.
We will fix these in parallel with the hardware development.

%% file: cdc.tex
 \section{CDC performance}
 \label{sec:cdc}

\newcommand \ie{{\it i.e.}}


 There are three requirements to be able to operate the new CDC with
 higher beam background.
 First, the dead time for the readout electronics must be reduced.
 At present, we are using charge to time conversion chips
 and multi hit TDCs for measurements of the drift time and the energy
 loss.  The conversion takes some time (800 -- 2200\,nsec) and the drift time
 cannot be measured during that time. This contributes to the dead time. The new
 system records the timing with a pipelined TDC and the charge with an FADC
 separately. The dead time for the new system depends on the shaping time for
 the new ASIC chip and will be around 200\,nsec. This is significantly
 better than in the present system. Second, the cell size must be reduced.
 This means that the number of cells in one layer increases and the maximum
 drift time decreases. The number of cells for the first super layer
 (where the most serious background is expected) increases from 64 to 160 (A
 small cell chamber was installed instead of the cathode chamber in
 2003. It has 128 cells for just two layers in order to test the small
 cell principle and has been working well for 5 years.) Finally, we will
 increase the number of stereo layers in each super layer to obtain higher
 efficiency for three dimensional tracking. A detailed wire configuration is
 described in the LoI~\cite{bib:LoI}.  We studied
 the effects of background on tracking performance including these
 improvements included in the upgrade; the results are described in the
 subsections that follow.

\subsection{Tracking efficiency and mass resolution}
We approximately estimate the tracking performance in sBelle using the 
Geant3-based detector simulator for Belle. The same CDC and SVD as used in Belle 
are assumed except for the dead time of the readout electronics for the CDC, 
which is set to 200\,nsec.
The high beam background environment expected at sBelle
is simulated by superimposing real random 
triggered events (taken with the Belle detector) on each simulated 
signal event.

We assume three beam background levels in the simulation study:
\begin{itemize}
\item nominal ($1 \times$ bkg): 1 times bkg both in the CDC and SVD,
where bkg is the nominal level of beam background observed in the Belle 
detector;
\item moderately-high ($5 \times$ bkg): 
3 times bkg in the CDC and 1 times bkg in the SVD; 
\item highest ($20 \times$ bkd): 
13 times bkg in the CDC and 1 times bkg in the SVD.
\end{itemize}
\noindent
Here note that 
the background level in the highest (moderately-high) case is 20 
(5) times nominal, which approximately corresponds to 13 (3) times bkg in the 
sBelle CDC, whose cell sizes are smaller than those of the Belle CDC.
The beam background in the SVD is assumed to be nominal in all cases, because 
readout electronics with much shorter ($\sim 1/16$) 
shaping time than in the Belle SVD will be used for the sBelle SVD.

The track-finder in the CDC used in the study is one that has been
recently updated. It has higher tracking efficiency than that used to
process data so far in Belle, especially for events with high charged multiplicity. 
In addition, the stand-alone track-finder for the SVD, which 
became available recently is used. This finder aims at efficient
tracking of low momentum particles such as the slow pion from $D^\ast\to
D^0\pi$ decay.
 
The physics channels studied are,
$ B^0~\to~J/\psi K^0_S$ ($J/\psi~\to~\mu^+\mu^-;~K^0_S~\to~\pi^+\pi^-$),
which is an example of a typical channel, and $ B^0 \to D^{*+} D^{*-}$ 
($~D^{*\pm} \to D^0\pi^\pm_s;~D^0 \to K3\pi$), 
which is an example of the most difficult channels in terms of tracking.

\subsubsection{Result for $B^0 \to J/\psi K^0_S$ }
Summarized in Table~\ref{cdc_eff1} are the reconstruction efficiencies and 
relative efficiencies for the three background cases.~\footnote{The event selection is performed with the following criteria: 
$-60~<~(M(\mu^+\mu^-)-M_{J/\psi})~<~36~{\rm{MeV}}/c^2$, 
$ |M(\pi^+\pi^-)~-~M_{K^0_S}|~<~16~{\rm{MeV}}/c^2 $,
$ 5.270~<~M_{\rm{bc}}~<~5.290~{\rm{GeV}}/c^2$ and $|\Delta E|~<~40$~MeV. 
These are somewhat simpler criteria than those applied in the data analysis in 
Belle~\cite{bib:JpsiKs_analysis}.}

\begin{table}[hbtp]
\caption{\label{cdc_eff1}
Reconstruction efficiencies and relative efficiencies
for $B^0\to J/\psi K^0_S$ for the three background cases.
The efficiency in Belle is also shown in the last row for 
comparison.}
\begin{center}
\begin{tabular}{|c|c|c|c|} \hline
bkg level  & eff. (\%)   & eff. ratio - 1 & eff. ratio - 1 \\ 
               &             & wrt $1\times$ bkg (\%) 
                             & wrt Belle (\%)                \\ \hline\hline
$ 1\times$ bkg &  58.7       &  $\equiv 0$   & $+11.3$       \\ \hline
$ 5\times$ bkg &  57.7       &    $-1.7$     & $+~9.4$       \\ \hline
$20\times$ bkg &  53.6       &    $-8.8$     & $+~1.5$       \\ \hline\hline
$ 1\times$ bkg (Belle)&  52.7       &    -          & $\equiv 0$ \\ \hline
\end{tabular}
\end{center}
\end{table}

As shown in column 3 of the table, 
the reconstruction efficiency in the highest (moderately-high) background case
drops only by $\sim 9~(2)\%$ with respect to the nominal background case.
Furthermore as shown in column 4, 
nearly the same efficiency as in Belle can be maintained even 
in the highest background case, thanks to the hardware (reduced dead time and 
smaller cell size) and software improvements.
Note that the efficiencies in the table are those derived from the CDC 
track-finder, with the SVD stand-alone track-finder turned off. 
The difference in efficiency between the 
SVD track-finder off and on is very small, $\sim 0.1 - 0.2\%$ level, 
because the final state does not include slow particles.

Summarized in Table~\ref{cdc_resol1} are the mass and $\Delta E$ resolutions.
The degradations of the resolutions due to background are not large, 
at the few \% level.

\begin{table}[hbtp]
\caption{\label{cdc_resol1}
Mass and $\Delta E$ resolutions for $B^0\to J/\psi K^0_S$ for the three
background cases. Here the resolution is simply defined as $\sigma$ of a 
single-Gaussian fit to the distribution in question.
The resolutions in Belle are also shown in the last row for comparison.}
\begin{center}
\begin{tabular}{|c|c|c|c|} \hline
bkg level  & $M(\mu^+\mu^-)$ & $M(\pi^+\pi^-)$  & $\Delta E$  \\ 
               & (MeV/$c^2$)     & (MeV/$c^2$)  & (MeV)       \\ \hline\hline
$ 1\times$ bkg &  8.9            &    2.2       &  7.2     \\ \hline
$ 5\times$ bkg &  8.9            &    2.2       &  7.4     \\ \hline
$20\times$ bkg &  9.1            &    2.3       &  7.5     \\ \hline\hline
$ 1\times$ bkg (Belle)&  8.8            &    2.1       &  7.1     \\ \hline 
\end{tabular}
\end{center}
\end{table}

\subsubsection{Result for $B^0 \to D^{*+}D^{*-} $}
The final state includes ten charged particles, 
two of which are low momentum pions from 
$D^\ast$'s, hence this is regarded as one of the most difficult channels 
in term of charged particle tracking.

Summarized in Table~\ref{cdc_eff2} are the reconstruction efficiencies and 
relative efficiencies for the three background cases.~\footnote{
The event selection is performed with the following criteria: 
$ |M(K3\pi) - M_{D^0}|~<~16~{\rm{MeV}}/c^2 $,
$ |\delta M - \delta M_{nominal}| < 3~{\rm{MeV}}/c^2 $ 
($\delta M \equiv  M(K3\pi\pi_s) - M(K3\pi)$),
$ |M_{\rm{bc}} - M_{B^0}| < 10.5~{\rm{MeV}}/c^2$ and 
$|\Delta E|~<~40$~MeV. 
These are somewhat simpler criteria than those applied in the Belle data 
analysis~\cite{bib:D*D*_analysis}.}

\begin{table}[hbtp]
\caption{\label{cdc_eff2}
Same as Table~\ref{cdc_eff1} but for $B^0\to D^{*+}D^{*-}$.}
\begin{center}
\begin{tabular}{|c|c|c|c|} \hline
bkg level  & eff. (\%)   & eff. ratio - 1 & eff. ratio - 1 \\ 
               &             & wrt $1\times$ bkg (\%) 
                             & wrt Belle (\%)            \\ \hline\hline
$ 1\times$ bkg &  7.3       &  $\equiv 0$ & $+119$       \\ \hline
$ 5\times$ bkg &  6.7       &    $ -9$    & $+~99$       \\ \hline
$20\times$ bkg &  4.4       &    $-41$    & $+~30$       \\ \hline\hline
$ 1\times$ bkg (Belle)    &  3.3       &    -        & $\equiv 0$   \\ \hline
\end{tabular}
\end{center}
\end{table}

As shown in column 4 of the table, the reconstruction efficiency for 
the nominal (and 5 times) background case is  
approximately twice that of Belle, thanks to the hardware and software 
improvements. The efficiency in the highest 
background case, however, significantly drops (column 3) in contrast to the
$J/\psi K^0_S$ channel. This is because many (2.5 times more) 
low momentum charged particle tracks must be reconstructed 
in this channel compared to $J/\psi K^0_S$. 
Nevertheless, an efficiency higher than in Belle can be maintained (column 4).

\begin{figure}[h]
\begin{center}
\includegraphics[width=0.56\linewidth]{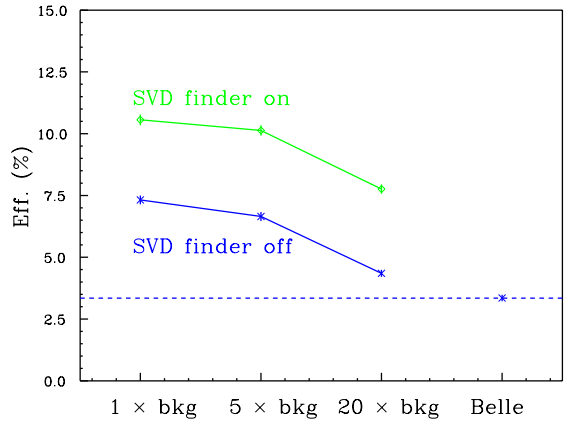}
\caption{\label{cdc_eff2fig}
Reconstruction efficiencies for $B^0\to D^{*+}D^{*-}$ with the SVD 
stand-alone track-finder on and off.}
\end{center}
\end{figure}
\noindent 
Shown in Fig.~\ref{cdc_eff2fig} are the reconstruction efficiencies with the 
SVD stand-alone track-finder on,  
along with the efficiencies with the SVD finder off (the same 
efficiencies as in Table~\ref{cdc_eff2}). 
As seen in the figure, 
the SVD finder improves the efficiency significantly, because this finder 
is very efficient for the low momentum pions from $D^\ast$'s.

Summarized in Table~\ref{cdc_resol2} are the mass and $\Delta E$ resolutions.
No degradation is found (within fitting errors) 
either in the $D$ mass or in the $D^\ast - D$ mass 
difference resolution, whereas there is a $\sim 10\%$ degradation
in the $\Delta E$ resolution. 

\begin{table}[hbtp]
 \caption{\label{cdc_resol2}
 Same as Table~\ref{cdc_resol1} but for $B^0\to D^{*+}D^{*-}$. 
 These are results obtained without the SVD track-finder.
$\delta M \equiv  M(K3\pi\pi_s) - M(K3\pi)$.
}
 \begin{center}
  \begin{tabular}{|c|c|c|c|} \hline
   bkg level  & $M(K3\pi)$          & $\delta M $   & $\Delta E$  \\ 
   & (MeV/$c^2$)     & (MeV/$c^2$)  & (MeV)    \\ \hline\hline
   $ 1\times$ bkg &  5.1            &    0.72       &  7.2     \\ \hline
   $ 5\times$ bkg &  5.0            &    0.72       &  7.5     \\ \hline
   $20\times$ bkg &  5.1            &    0.70       &  7.9     \\ \hline\hline
   $ 1\times$ bkg (Belle) &  4.9            &    0.73       &  7.1 \\ \hline 
  \end{tabular}
 \end{center}
\end{table}

\subsection{Summary of CDC performance}
The CDC will be upgraded to handle much higher beam background.
Quick simulation studies including the effects of the upgraded components, such as the
smaller cell size, optimized time windows, as well as the new tracking
algorithm and the SVD standalone tracker, have been carried out.
We found that the new CDC will maintain a track reconstruction
performance that is similar to the present CDC.

%% file: pidstudy.tex
\newcommand{\GeV}{\mathrm{GeV}}
\newcommand{\abi}{\mathrm{ab}^{-1}}
\newcommand{\Mbc}{M_{\mathrm{bc}}}
\newcommand{\DE}{\Delta E}
\newcommand{\qq}{q\bar{q}}


\section{PID performance}

\input{top}

\input{arich_perf}


\input{pid_rhogam}


%% file: top.tex




\subsection{TOP counter}
\label{subsubsect:top-conf}

\subsubsection{Overview}

In a TOP counter, Cherenkov photons are guided to the photon detector by
total internal reflections (Fig.~\ref{fig:principle}).
The detector measures the precise arrival time and position of the photons.
The determination of the velocity, $\beta$, of the particle is derived from
a combination of two different contributions:
\begin{list}{}{}
\item[(i)] The propagation time for a photon emitted at Cherenkov
angle, $\theta_c$, is a function of $\theta_c$.
\item[(ii)] The time of flight for the charged particle from the
interaction point to the counter is a function of $\beta$ of the
particle and will therefore influence the arrival time of the Cherenkov
photon at the photon detector.
\end{list}
The time-of-arrival difference between Cherenkov photons from a 3\,GeV/$c$ $K$ and $\pi$
hitting a TOP counter at 1\,m from the photo detector, is about $75$\,ps.
The time of flight difference is about $50$\,ps for a 1\,m flight path.
Compared to a DIRC counter~\cite{bib:DIRC}, the TOP counter reconstructs the
ring image using all timing information, including the TOF.
This improves the separation power.

The baseline design is given in Fig.~\ref{fig:topoverview}.
The quartz radiators are $75 \times 40 \times 2$\,cm$^3$ and
$185 \times 40 \times 2$\,cm$^3$.
The quartz is cut at $\theta = 47.8$~degrees,
where $\theta$ is the polar angle in the Belle detector. The end
surface of the longer
part of the bar (covering $\theta > 47.8$~degrees) is a 
spherical focusing mirror with a 5\,m radius of curvature.
The mirror's position is optimized to improve the PID power,
because the separation between ring images is large when the propagation length
is long, however, the timing resolution becomes worse
due to 
dispersion as  described later.
Cherenkov photons are detected by a multi-anode micro-channel-plate
photo-multiplier-tube (MCP-PMT). The time resolution is excellent,
$<40$\,ps \cite{bib:MCPtiming}.
The anodes are arranged in a linear array with 5\,mm pitch.
The MCP-PMTs are attached to the ends of the quartz bars as shown in
Fig.~\ref{fig:topoverview}. A multi-alkali photocathode is used in the 
standard version of such a MCP-PMT.

\subsubsection{GaAsP photocathode}

\begin{figure}
\centerline{
\includegraphics[width=8cm]{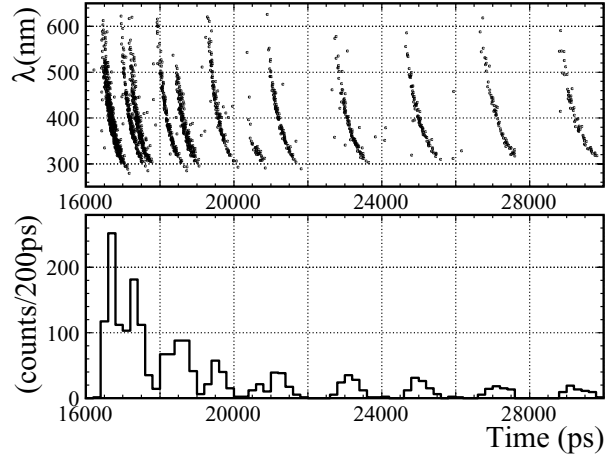}
}
\caption{Wavelength ($\lambda$) dependence of the detection time (top) and
the projected time distribution (bottom).}
\label{fig:lam-dep}
\end{figure}

Fig.~\ref{fig:lam-dep} shows a simulation of the time distribution
for a channel around the center of the TOP counter.
It is clear that the improvement of the PID power is limited by the
broadening of the time resolution due to the chromaticity of
Cherenkov photons.
From Fig.~\ref{fig:lam-dep} we observe that the detection time is shifted
by about 500\,ps if wavelength changes from 300\,nm to 600\,nm.

\begin{figure}
\centerline{
\includegraphics[height=5cm]{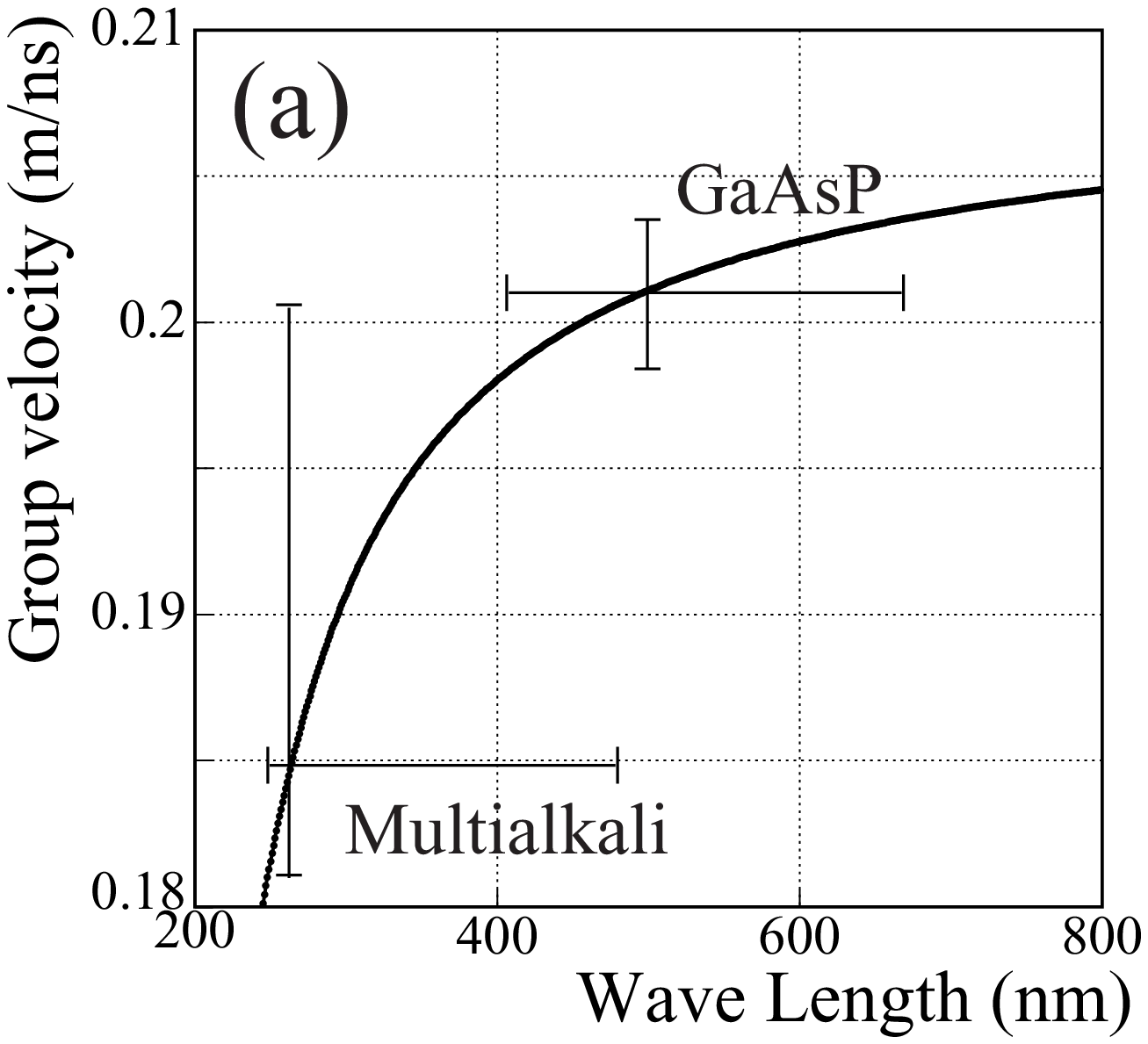}
\includegraphics[height=5cm]{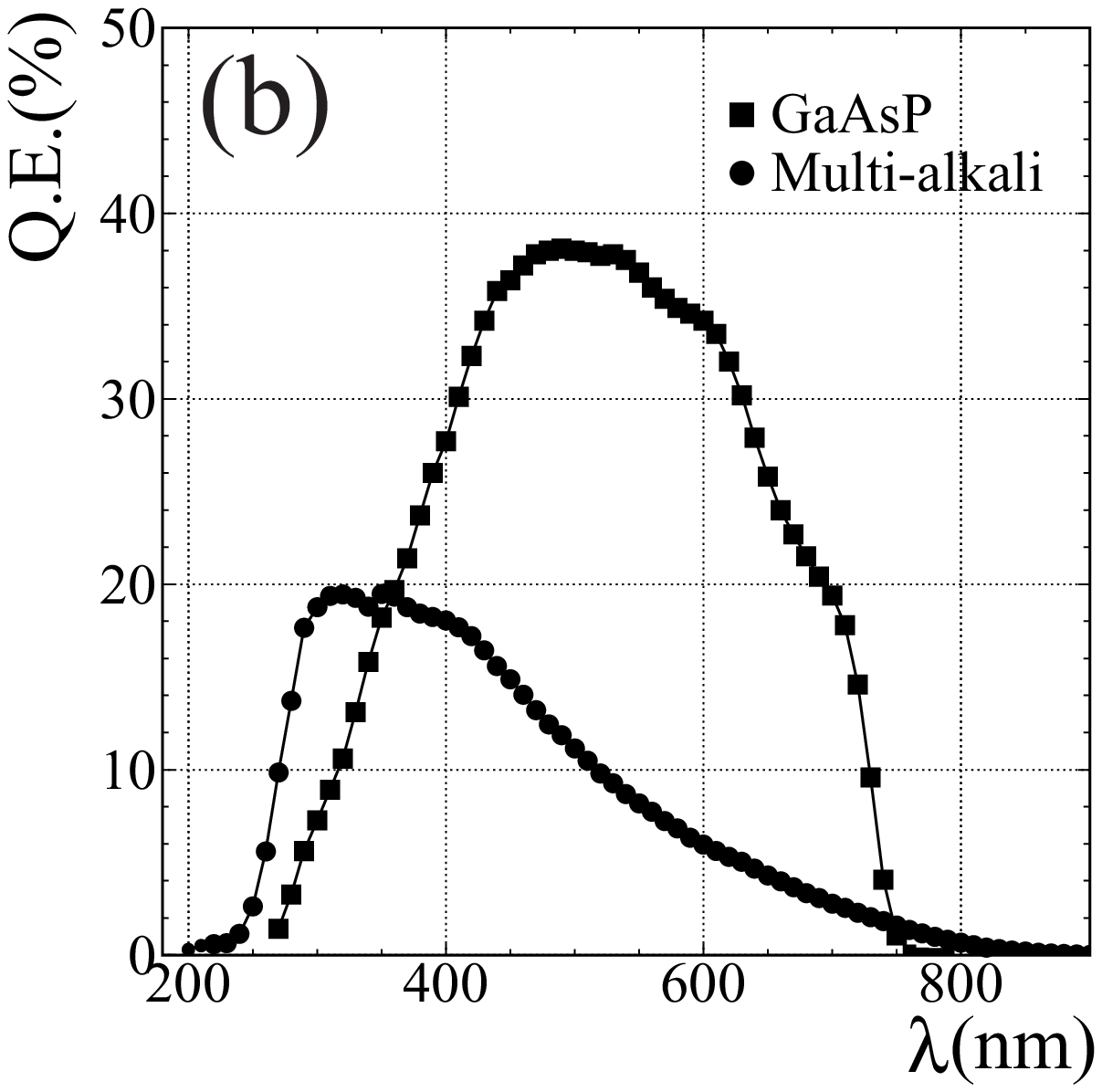}
}
\caption{Light propagation velocity
inside quartz as a function of the wavelength $\lambda$ (a), 
and quantum efficiency (b).}
\label{fig:qe-velo}
\end{figure}

One way to minimize this effect is to restrict
the sensitivity window of the photon detector as
shown in Fig.~\ref{fig:qe-velo}(a), and make use of the reduced
 velocity variation at longer wavelengths.
Fig.~\ref{fig:qe-velo}(b) shows the QE of multi-alkali and GaAsP
photocathodes as a function of wavelength.
A GaAsP photocathode has a higher QE at longer wavelengths
compared to the alkali photocathodes. 
The light propagation velocity spread, corresponding to the FWHM 
interval of the QE, is
$0.181$-$0.201$ m/ns for multi-alkali and
$0.198$-$0.204$ m/ns for GaAsP photocathodes, 
as shown in Fig.~\ref{fig:qe-velo}(a).
The use of a GaAsP photocathode will therefore improve the time resolution
for propagated photons.

\begin{figure}
\centerline{
\includegraphics[width=5cm]{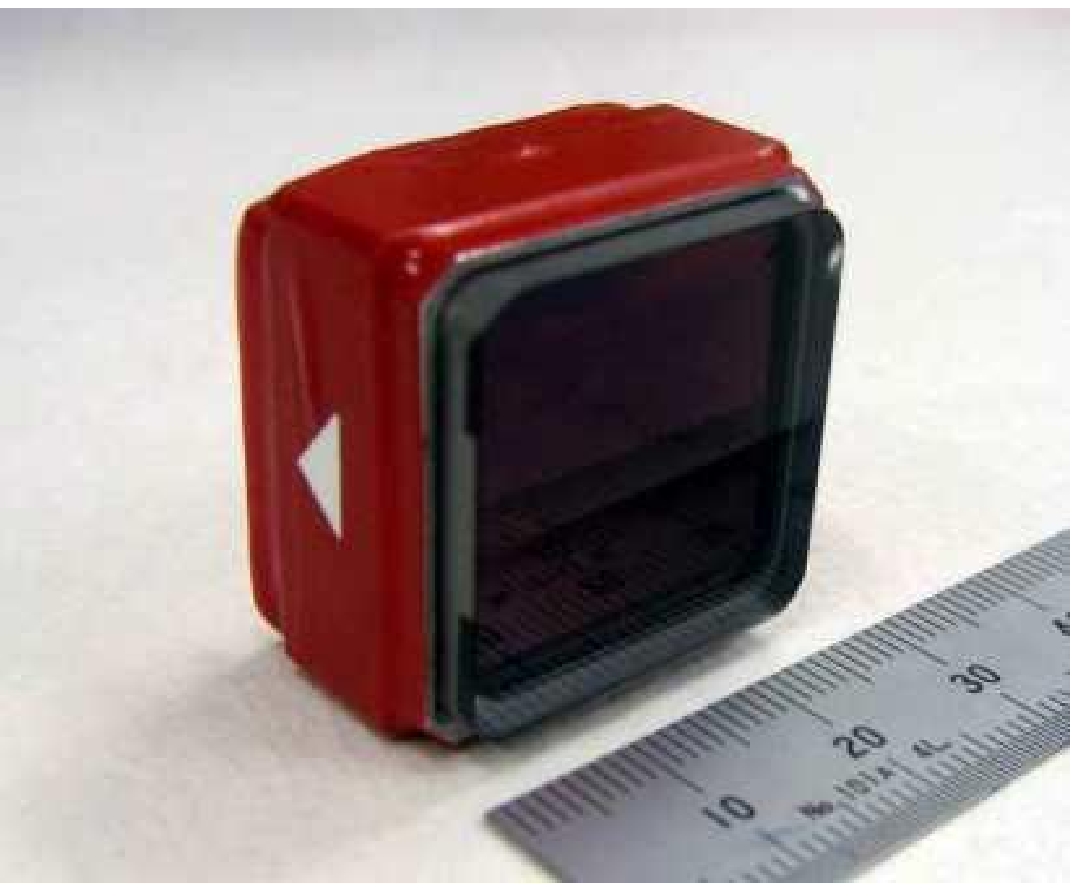}~~~~
\includegraphics[width=3cm]{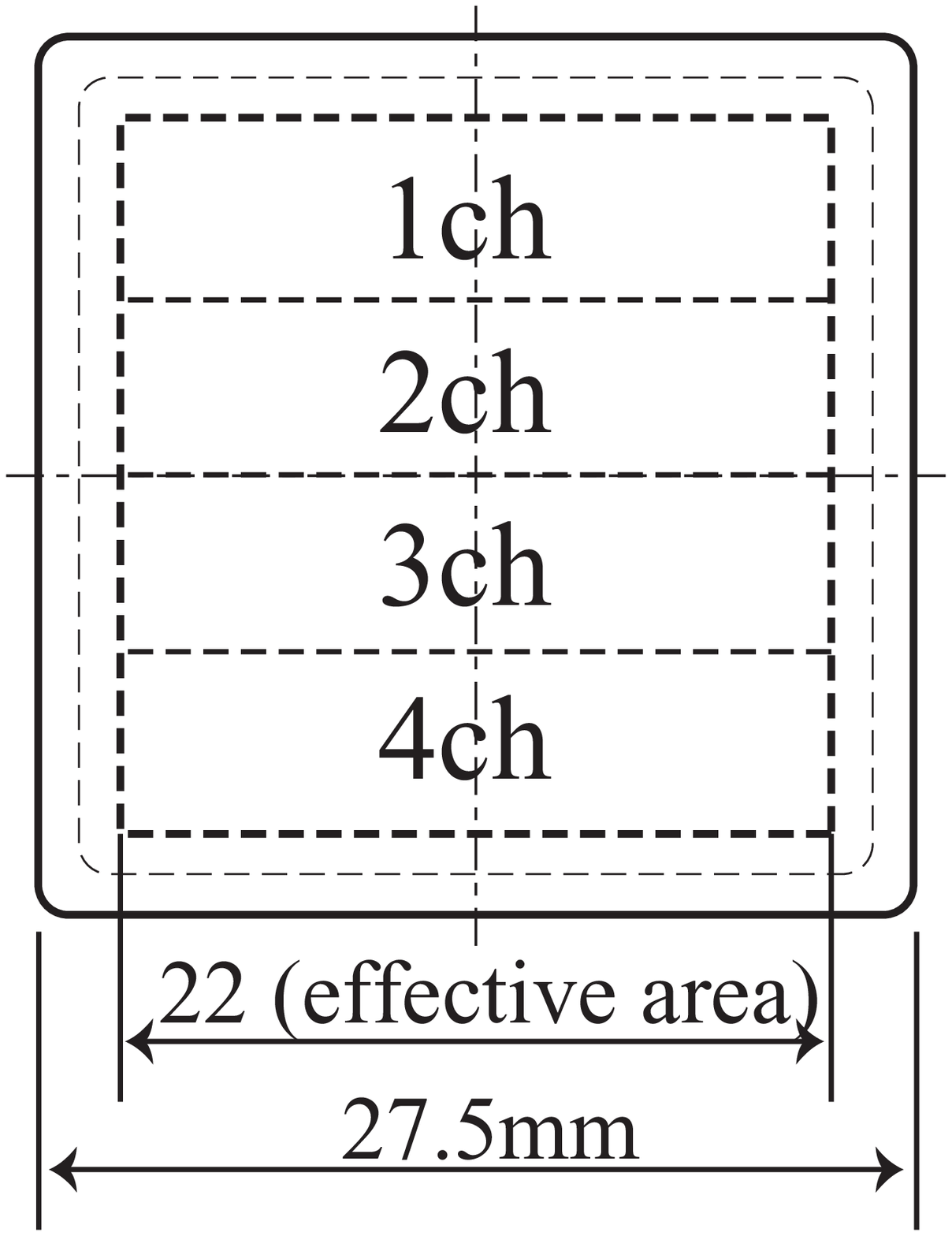}
}
\caption{Picture of a prototype MCP-PMT with the anode channel layout.}
\label{fig:picture}
\end{figure}

We have been developing a square-shaped MCP-PMT with a GaAsP photocathode
with Hamamatsu photonics K.K. (HPK).
Fig.~\ref{fig:picture} shows a picture of the prototype.
It has four anode strips, a GaAsP photocathode and an aluminum protection
layer on the first MCP to protect the photocathode from damage by
feedback ions.
The anode layout is also shown in Fig.~\ref{fig:picture}.

We checked the time response for single-photon detection,
using a picosecond laser  (HPK; PLP-02-SLDH-041).
The wavelength and pulse width are, respectively, $405\pm10$\,nm and 34\,ps.
The light intensity is reduced by a diffuser and filters
down to the single-photon level.
The photons are led to the PMT window by an optical fiber.
The output of the MCP-PMT is fed into an
attenuator (Agilent; frequency $<18$ GHz)
and an amplifier (HPK; C5594; gain = 36 dB; frequency,
50 kHz $- 1.5$ GHz), to avoid noise at the discriminator.
The charge and timing are measured 
by an ADC, 0.25\,pC/count, and a TDC, 25\,ps/count.
A discriminator (Phillips Scientific, model 708)
is used with an threshold of 20\,mV.

\begin{figure}
\centerline{
\includegraphics[width=9cm]{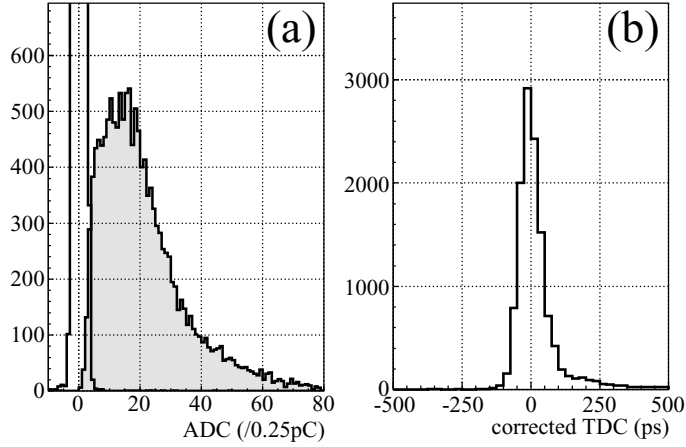}
}
\caption{ADC (a) and TDC (b) distributions of the
GaAsP photocathode MCP-PMT for single-photon detection.
In (a), the shaded histogram is the output charge for single photons
within the corresponding time window, and the open histogram is the pedestal.}
\label{fig:adctdc}
\end{figure}

Fig.~\ref{fig:adctdc} gives the charge and timing distribution
for single photons. The signal output is fast
with a rise time of $\sim500$\,ps. The gain is in the order of
$\sim 0.6 \times 10^6$.
We also obtained a good time resolution of $\sim$35\,ps around the main peak.

\subsubsection{Focusing system}

Even when using the GaAsP photocathode MCP-PMT, the remaining chromatic
effect contributes a 100\,ps spread (rms) to the time of propogation. 
This is of the same
order of magnitude as the time difference between $K$ and $\pi$ detection.
To minimize the chromatic effect, we can introduce a focusing system.
The refractive index, $n$, of quartz is well described by the following
approximation within the range $200<\lambda<900$\,nm:
\[
n(\lambda) = 1.44 + \frac{8.2}{\lambda-126}
\]
for $\lambda$ in nm.
The Cherenkov angle, $\theta_c$, depends on the photon wavelength
according to
\[
\cos \theta_c = \frac{1}{n(\lambda) \beta}.
\]
We can therefore correct the chromaticity directly by using
the $\lambda$ dependence of $\theta_c$.

\begin{figure*}[htb]
\centerline{
\includegraphics[width=12cm]{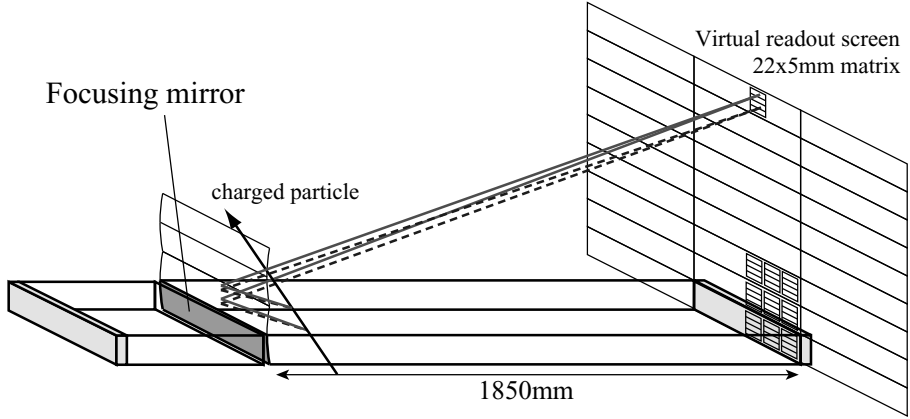}
}
\caption{Schematic view of the focusing system.}
\label{fig:focus}
\end{figure*}

Fig.~\ref{fig:focus} shows the schematic set-up.
A focusing mirror is introduced at the end of the longer quartz bar,
and replaces PMTs in the previous  TOP counter (3-readout type) verion.
The PMTs at the other end of the bar are
 rotated to measure the position of the photon impact in $x$ and $y$.
The cross section of the quartz bar is a rectangle. We can therefore expand
the light trajectory into the mirror-image region and
create a virtual readout screen, a matrix of $22 \times 5$\,mm$^2$
readout channels  as shown in Fig.~\ref{fig:picture}.
 Cherenkov photons with different $\theta_c$ will focus onto
different PMT channels, and we thereby obtain $\lambda$ information
from the $y$ detection position through $\theta_c$.
We can reconstruct the ring image from 3-dimensional information
on time, $x$ and $y$.

Across the restricted $\lambda$ range, $\theta_c$ varies by $\sim12$\,mrad. 
The bar length from the mirror to the MCP-PMT is $1850$\,mm, as shown in
 Fig.~\ref{fig:focus}. The corresponding difference in the $y$ photon
detection position is  $\sim20$\,mm, 
about four times the pad size, and comparable to the quartz thickness.
We can therefore improve our estimate of 
$\lambda$ and light propagation velocity, and consequently  
obtain improved
performance, even with this narrow mirror and readout plane.
The focusing TOP counter does not require a large focusing mirror and
readout channels with high granularity.

\begin{figure}
\centerline{
\includegraphics[width=11cm]{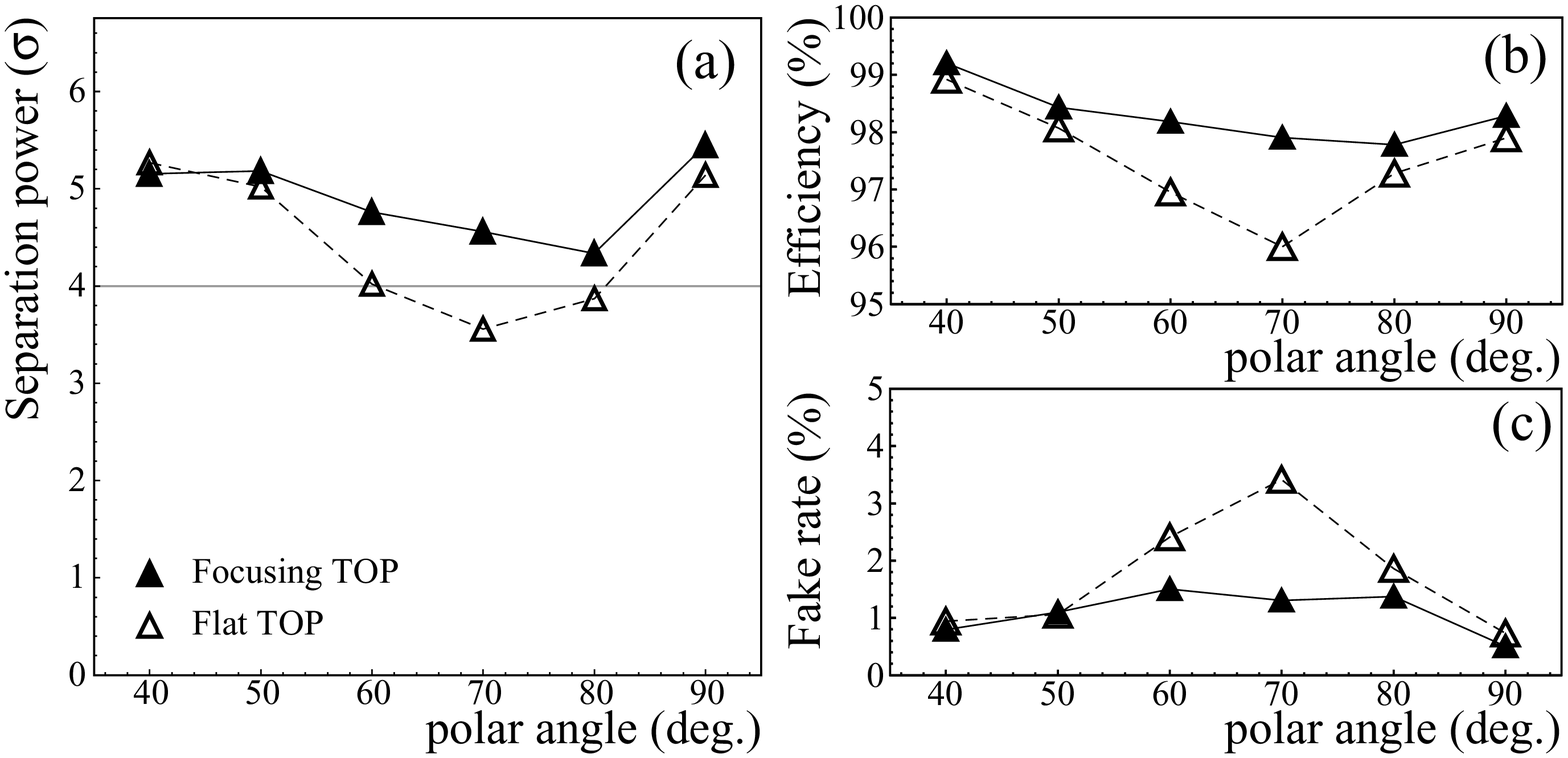}
}
\caption{(a) Separation power, (b) kaon efficiency and
(c) pion fake rate for the 3-readout type ($\vartriangle$)
and the focusing type ($\blacktriangle$).}
\label{fig:sep}
\end{figure}

The estimated performance of the focusing TOP is shown in Fig.~\ref{fig:sep}.
We evaluated the performance of the TOP counter
with the GaAsP photocathode by simulating the Cherenkov photon generation,
light propagation inside the quartz and the time fluctuation of the MCP-PMT.
We have included in the simulation all chromatic effects.
We have also applied a cut-off at $400$\,nm to the QE distribution
and a collection efficiency of 36\% for the MCP-PMT.
The fake rate is improved by a factor
of two and we obtain $4.3\sigma$ separation for 4\,GeV/$c$ tracks.


\subsubsection{TOP configurations}

In the baseline design, we plan to use the MCP-PMT with a GaAsP photocathode.
The key issue for the Belle upgrade is the lifetime of the photocathode.
Because of the high beam background rate (we expect an increase by about
a factor of 20 with respect to the present rate in the TOF counter), 
many Cherenkov photons from gamma-converted electrons will be detected.
The current prototype MCP-PMT has an inadequate lifetime for the high
intensity run.
Although we are trying to improve the lifetime of the GaAsP photocathode,
we are also developing a MCP-PMT with a multi-alkali photocathode
as a backup option.
Our study~\cite{bib:mcp-lifetime} shows a satisfactory lifetime, which 
was achieved when the multi-alkali photocathode has a thin
aluminum layer on the first MCP surface. This layer protects the 
photocathode from damage by feedback ions.
However, such a layer also 
 reduces the collection efficiency (CE) by 40\%. 
To recover the collection efficiency and maintain a long lifetime,
we are developing a new type of MCP-PMT by changing the internal configuration,
either with a protection layer on the second MCP surface, or with an additional MCP layer.

There are two possible radiator configurations,
the 3-readout type and the focusing type.
We plan to use the focusing type as the baseline design.
Although the 3-readout type has simpler ring images compared to
the focusing type,
the ring-image separation of the focusing type is better than that of
the 3-readout type.

The possible TOP configurations are summarized in Table~\ref{tbl:top-config}.
The focusing TOP with a GaAsP photocathode MCP-PMT has the best performance
with $4.2\,\sigma$ separation for $K/\pi$ tracks at 4\,GeV/$c$ and $\theta=70$~degrees.
The focusing type using a multi-alkali photocathode MCP-PMT with improved CE
has a similar performance.
\begin{table}
\caption{TOP configurations and their performance in terms of
separation  of
$K$ and $\pi$ at 4\,GeV/$c$ and $\theta=70^\circ$.}
\label{tbl:top-config}
\begin{center}
\begin{tabular}{l|c|l}\hline
Configuration & Separation & Status \\ \hline
A) 3-readout and multi-alkali p.c. & $2.8 \sigma$ & Ready \\
B) 3-readout and GaAsP p.c.        & $3.5 \sigma$ & PMT production and lifetime test \\
C) Focusing and multi-alkali p.c.  & $2.5 \sigma$ & Focusing system test \\
~~C') C with CE=60\%                & $4.0 \sigma$ & Lifetime test \\
D) Focusing and GaAsP p.c.         & $4.2 \sigma$ & PMT production and lifetime test \\
\hline
\end{tabular}
\end{center}
\end{table}

\subsubsection{$\phi$ coverage}

\begin{figure}
\includegraphics[width=0.4\linewidth]{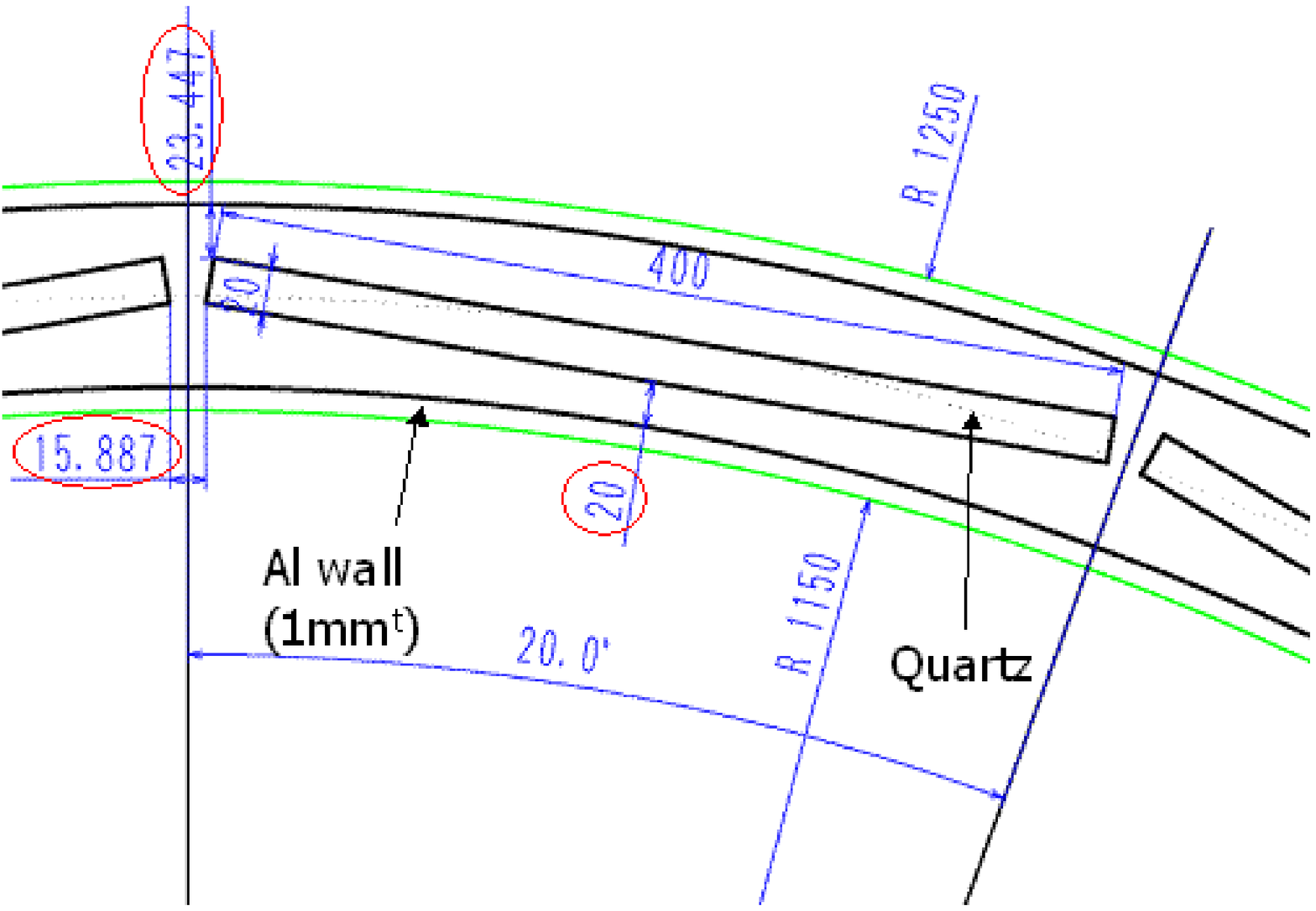}~~~~
\includegraphics[width=0.4\linewidth]{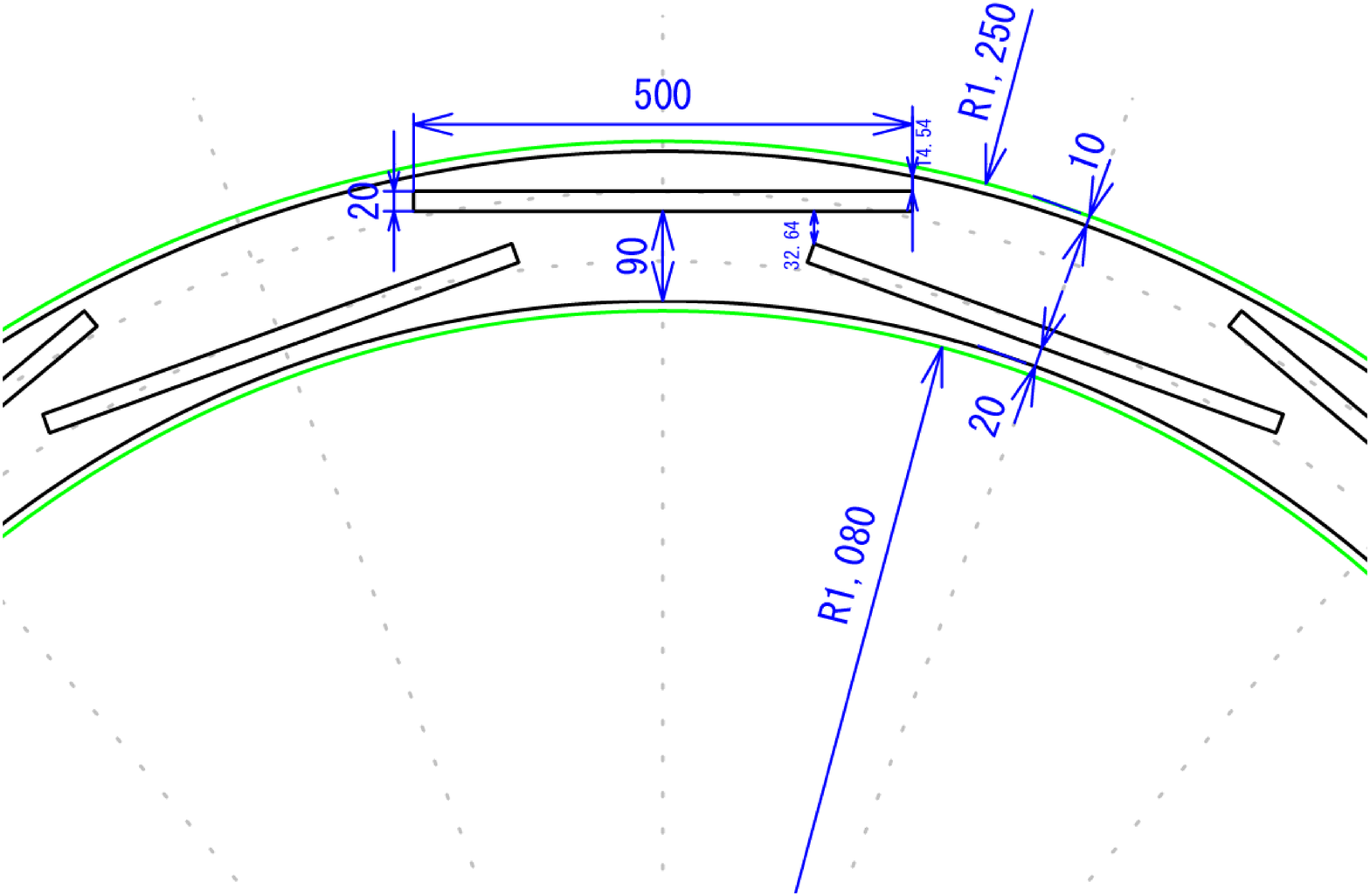}
\caption{Quartz radiator configurations in the $r-\phi$ plane
for a basic TOP (left) and for a layout with overlaps (right).}
\label{fig:top-rphiview}
\end{figure}

The layout of the proposed TOP quartz radiators is shown
in Fig.~\ref{fig:top-rphiview}(left), and is similar to the structure of Babar's DIRC~\cite{bib:DIRC}.
In this case, there is a gap between radiators. In addition,
the TOP counter has an insensitive region of about 1\,cm from the quartz edge
 due to the short path length of the curled tracks in the radiator bar.
In total, we have about 10\% (or less) dead space in the $\phi$ direction
in the barrel region.

To reduce this dead space, we consider an overlapping layout as shown in
Fig.~\ref{fig:top-rphiview}(right).
We adopt this staggered layout to avoid charge asymmetry.
In this case, we need wide quartz bars (50\,cm width) and
a complicated support structure.
We also have to reduce the inner radius from 1150\,mm to 1080\,mm.








%% file: arich_perf.tex
\subsection{Aerogel RICH Performance }
\label{subsubsect:arich_perf}

\begin{figure}[htb]
  \begin{center}
    \includegraphics[width=.3\columnwidth]{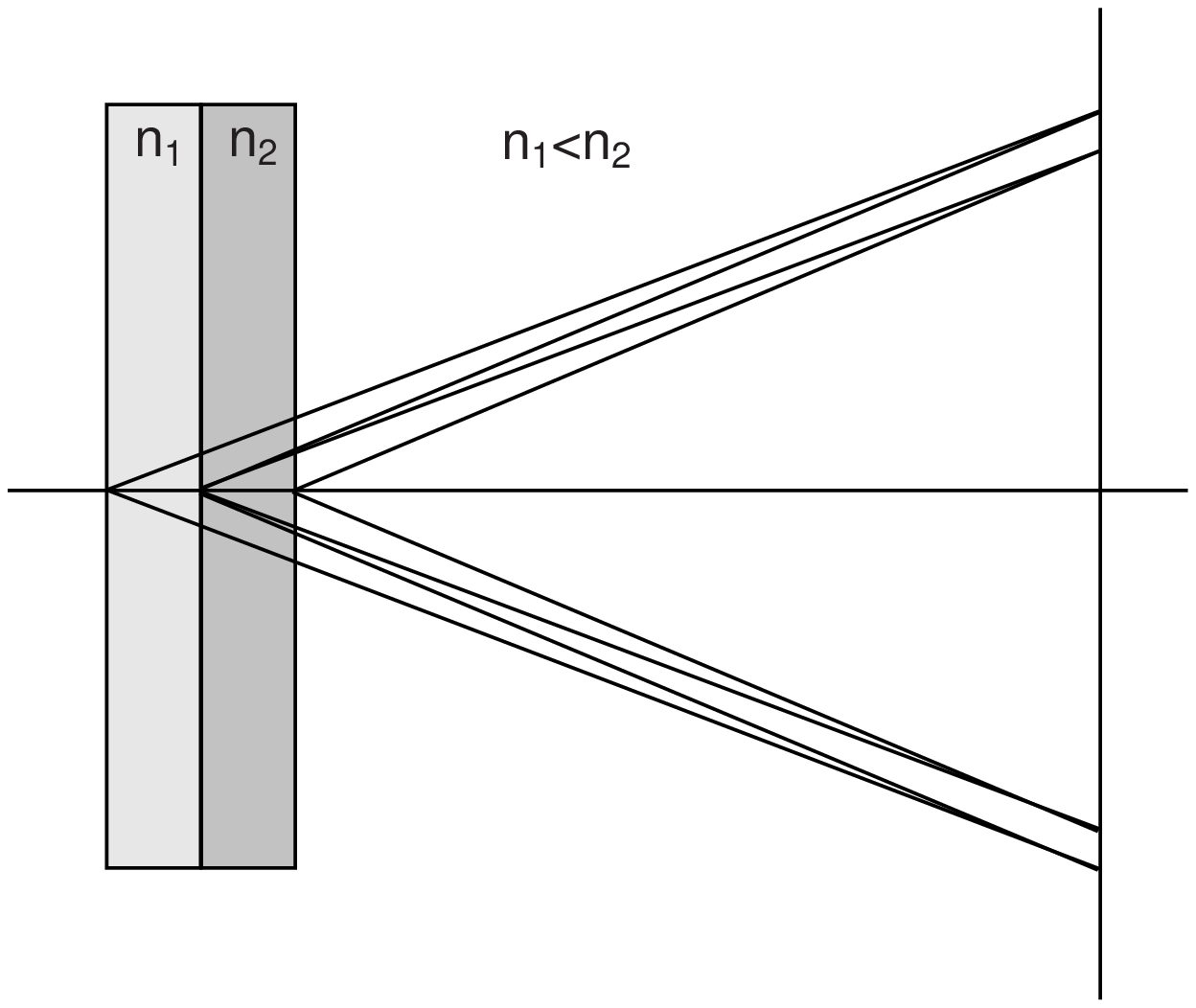}
    \includegraphics[width=.65\columnwidth]{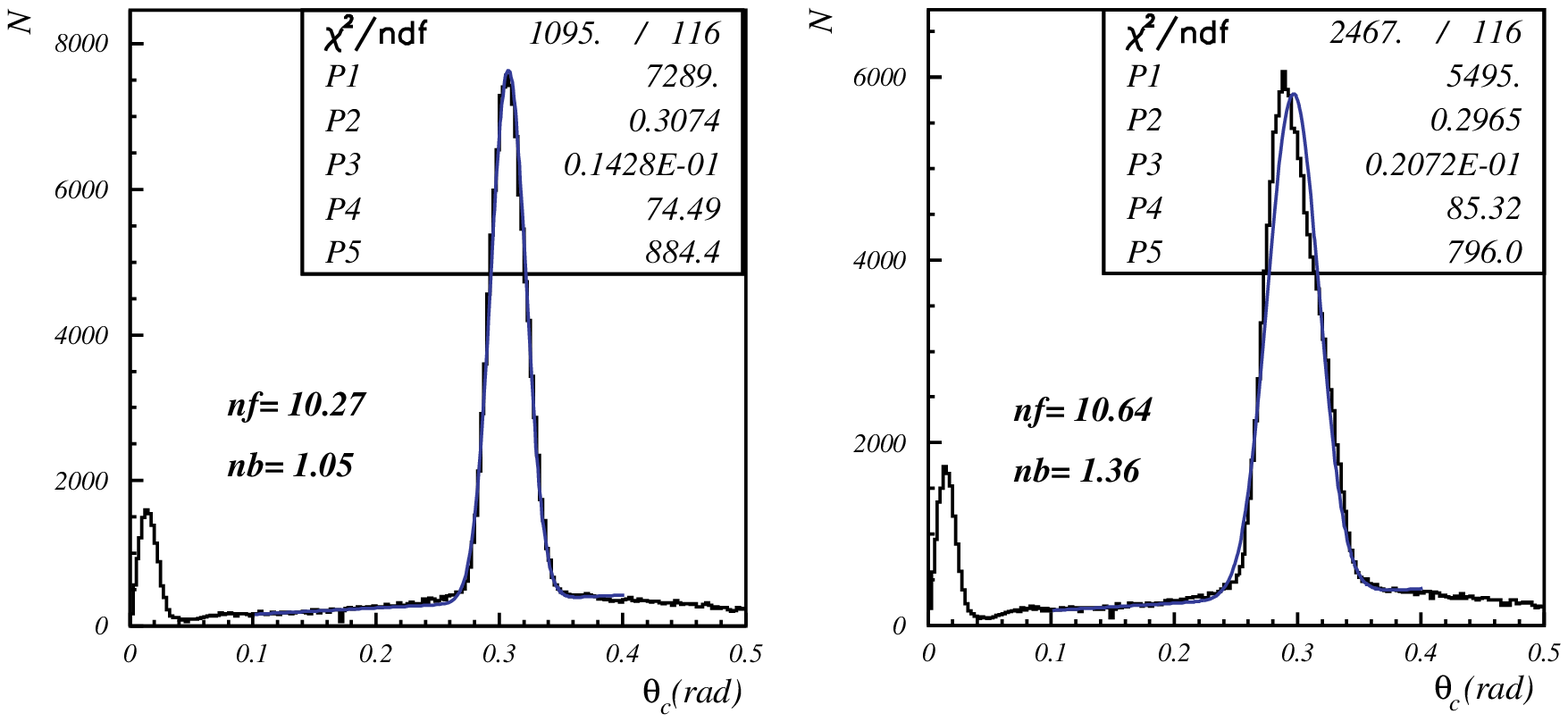}
    \caption{\sl Proximity focusing RICH with  a nonhomogeneous aerogel radiator 
          in the  focusing configuration(left); 
        the accumulated distribution of Cherenkov photon hits depending
          on the corresponding Cherenkov angle  for a 4cm homogeneous radiator
          (right) and a focusing configuration with $n_{1}$~=~1.046,
          $n_{2}$~=~1.056 (center).  \label{foc} }
  \end{center}
\end{figure}

The key issue in the performance of a RICH counter is
the Cherenkov angle resolution per track $\sigma_{\rm track} = \sigma_{\theta}/\sqrt{N}$.
With a thicker radiator, the number of detected photons increases, but in a
proximity focusing RICH the single photon resolution degrades because of the 
emission point uncertainty. 
As it turns out for a given geometry, the optimal thickness is around $20~\mathrm{mm}$ 
\cite{2ndBeamTest,nim-multi}.
However, this limitation can be overcome in a proximity focusing RICH 
with a non-homogeneous radiator~\cite{nim-multi,pk-hawaii,danilyuk,multi-rad-ana}.  
By appropriately  choosing the
refractive indices of consecutive aerogel radiator layers, one may achieve overlapping
of the corresponding Cherenkov rings on the photon detector (Fig.~\ref{foc}) 
\cite{multi-rad-ana}. 
This represents a sort of focusing
of the photons within the radiator, and eliminates or at least considerably reduces the spread
due to emission point uncertainty.  Note that such a tuning of  
 refractive indices for individual layers is only possible with aerogel, 
which may be produced with any desired refractive index in the range 1.01-1.07~\cite{aerogel-new}. 
The dual radiator combination can readily be extended to
more than two aerogel radiators. In this case, the indices of aerogel layers should
gradually increase from the upstream to the downstream layer.
In Fig.~\ref{foc}, we compare the data for two 4~cm thick radiators;
one with
aerogel tiles of equal refractive index (n = 1.046), the other with the
focusing arrangement ($n_{1} = 1.046 , n_{2} = 1.056$). The
improvement is clearly visible.
The single photon resolution $\sigma_\theta=14.3$~mrad for the dual radiator 
is considerably smaller than the corresponding value for the single refractive
index radiator 
($\sigma_\theta=20.7$~mrad), while the number of detected photons is the same in 
both cases.


The particle identification capabilities of the counter were evaluated by using 
simulated data. The backgrounds that are not included in the simulation (Rayleigh 
scattered \v Cerenkov photons from the same track, 
\v Cerenkov photons emitted  by the same track in the photon detector window,
beam related background hits and electronic noise) 
were added  according to observed or expected rates. 
\begin{figure}[htbp]
  \centerline{\includegraphics[width=0.7\columnwidth]{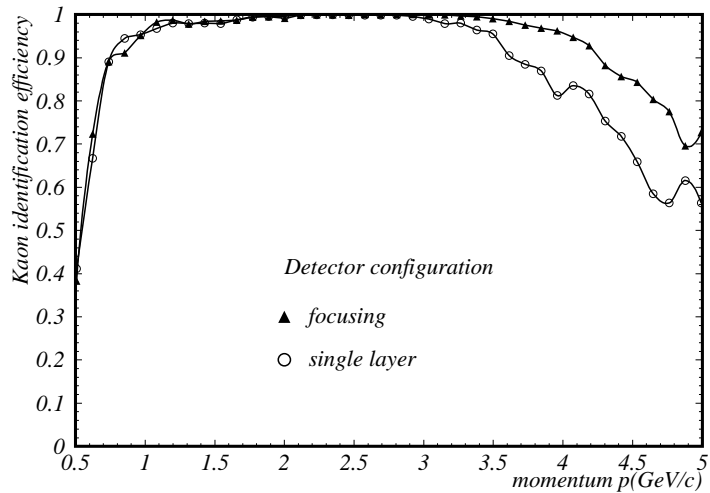}}
  \begin{center}
  \caption[kk]{\em Particle identification capability of the aerogel RICH counter:
 kaon efficiency at 1\% pion misidentification probability for the focusing radiator 
configuration (triangles) compared to the homogenous radiator (open circles).}
   \label{fig:effi-arich-1}
   \end{center}
\end{figure}
In the analysis of simulated events, the likelihood for the observed hit pattern is calculated 
for each hypothesis, and momentum dependent selection criteria are chosen for 
a given fake hypothesis probability~\cite{arich-lklh}. The resulting identification efficiency for kaons
is shown in Fig.~\ref{fig:effi-arich-1}; as expected we observe a clear advantage of 
using a focusing radiator compared to   the homogenous one.

\begin{figure}[htbp]
  \centerline{\includegraphics[width=0.6\columnwidth]{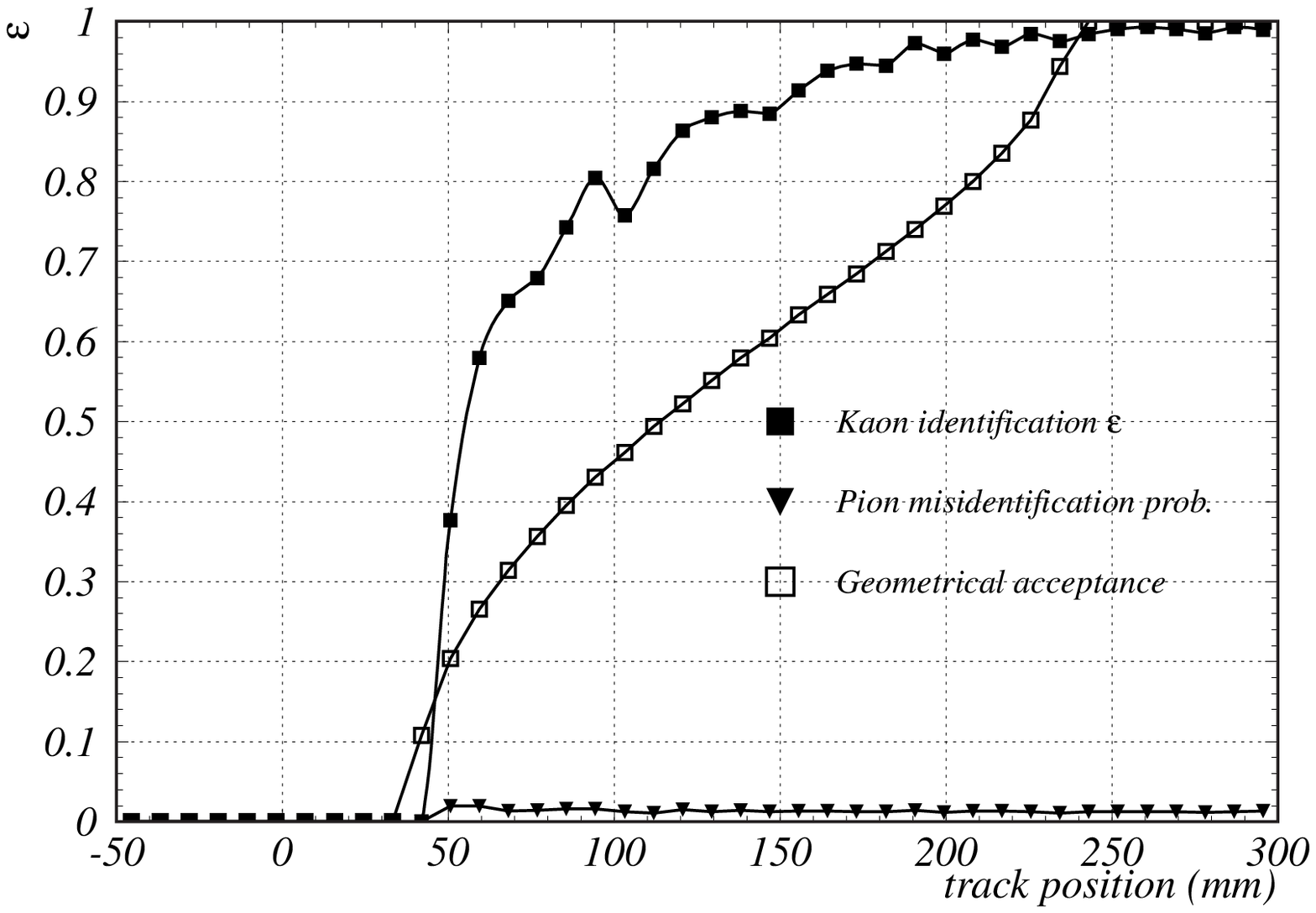}
              \includegraphics[width=0.6\columnwidth]{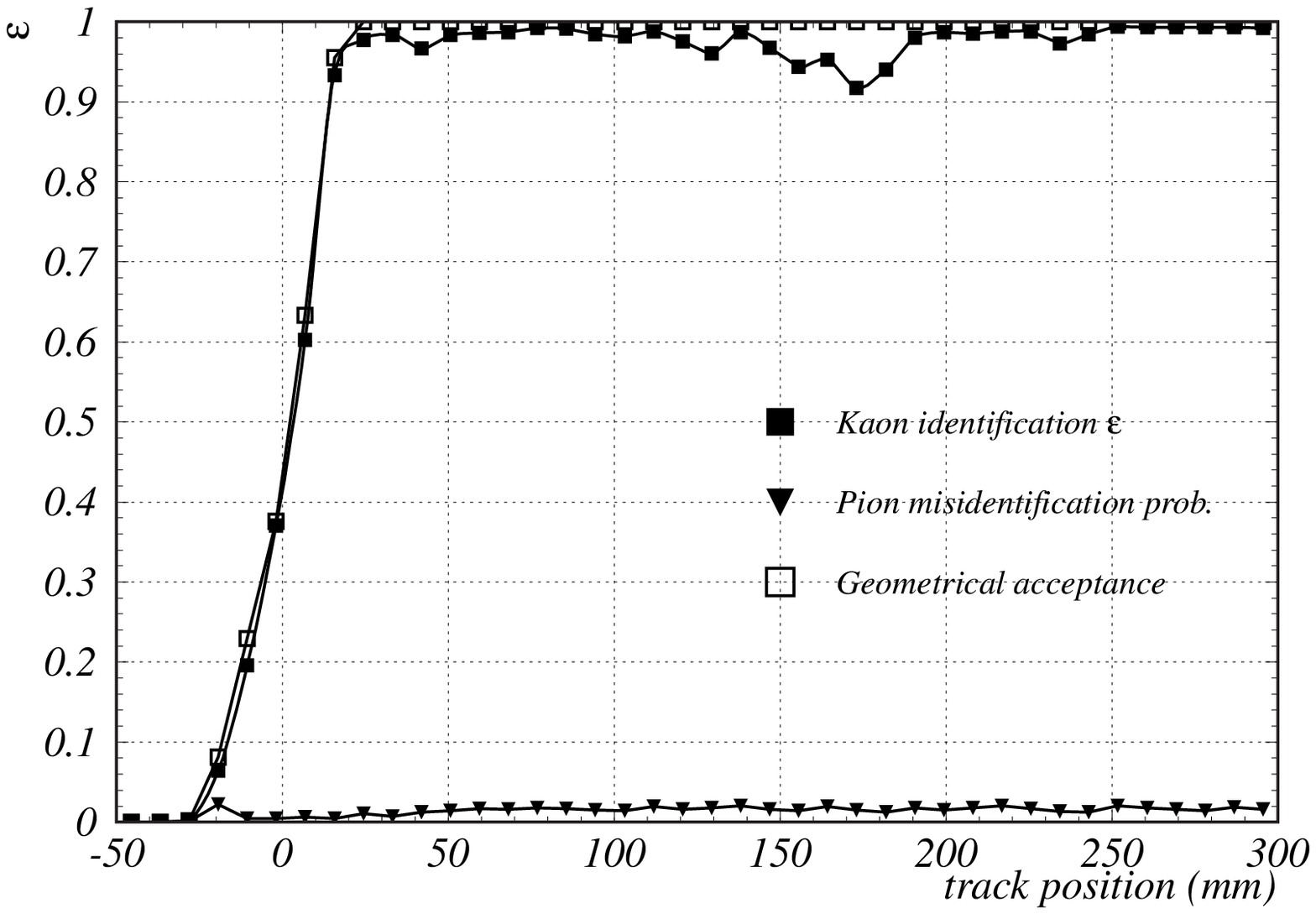}}
  \begin{center}
  \caption[kk]{\em Particle identification capability in the vicinity of the boundary to the barrel:
 kaon efficiency at 1\% pion misidentification probability for the case with an additional planar mirror
(right) and without it (left), shown as function of the distance from the boundary.}
 \label{fig:effi-arich-2}
   \end{center}
\end{figure}
We have also investigated the peformance of the counter in the region close to the 
boundary to the barrel PID device. As can be seen from  
Fig.~\ref{fig:arich-mirror}, a sizable  fraction of 
Cherenkov photons emitted by tracks close to this boundary do not hit the photon detector. 
This problem could in principle be overcome by inserting a planar mirror as illustrated in 
Fig.~\ref{fig:effi-arich-2}.

%% file: pid_rhogam.tex
  \subsection{Impact on $B \to \rho\gamma$ analysis}

  One of the important modes in which PID can contribute is
  $B \to \rho\gamma$.
  This mode suffers from a severe background from $B \to K^\ast\gamma$
  decays when the $K$ from the $K^\ast$ is misidentified as a $\pi$.
  We have therefore estimated the effect of an upgrade of the PID 
  system on the analysis of this decay mode.

  \subsubsection{TOP and ARICH performance with the simulator}

  In order to estimate the impact on the $B \to \rho\gamma$ analysis,
  we use gsim and fsim6 
  simulators to describe the present Belle detector (ACC+TOF+$dE/dx$)
  and the future sBelle detector (TOP+ARICH+$dE/dx$), respectively.
  

  In fsim6 with the nominal TOP configuration, the effective $\phi$ coverage
  is assumed to be $90\%$, where the $10\%$ insensitive region
  comes from the gap between the quartz bars and
  from the edge of the quartz bars where the PID efficiency is expected
  to be very low.
  In order to  precisely estimate the actual $\phi$ coverage,
  we have  to make a real design of the supporting structure and study 
  carefully the TOP performance for tracks that pass at the edges of bars,
  but we think that the assumption of $90\%$ coverage is conservative. 
  For the staggered quartz bar configuration, we assume the
  effective $\phi$ coverage to be $100\%$.
  This assumption may be too optimistic, especially for low momentum tracks,
  but it is practical to study the impact of $\phi$ coverage on 
  the performance.

\begin{figure}
   \begin{center}
    \begin{tabular}{cc}
     \multicolumn{1}{l}{(a)} & \multicolumn{1}{l}{(b)} \\
     \includegraphics[scale=0.78]{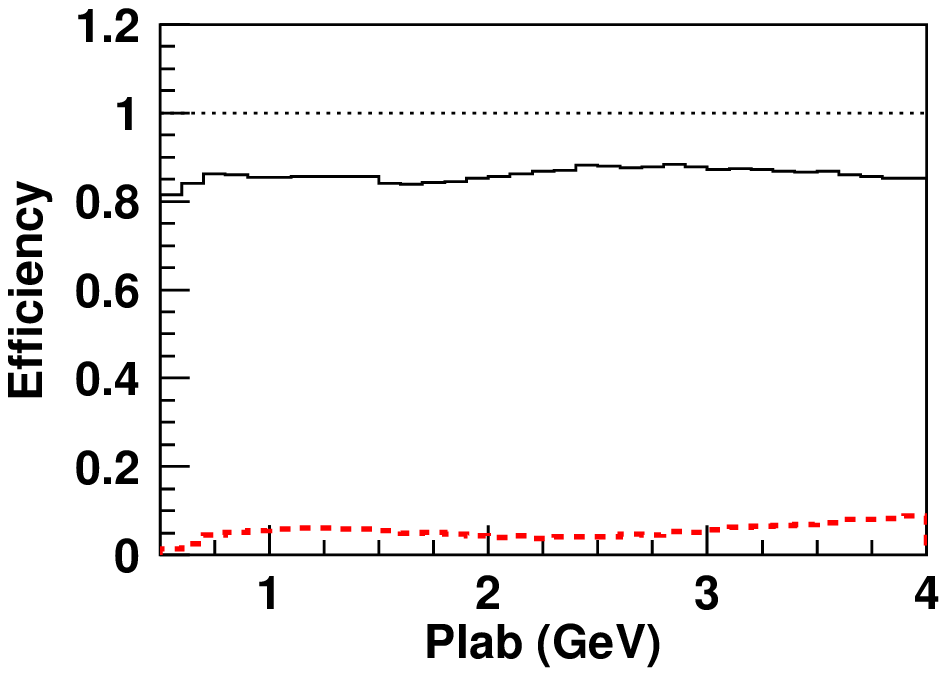}
     & \includegraphics[scale=0.78]{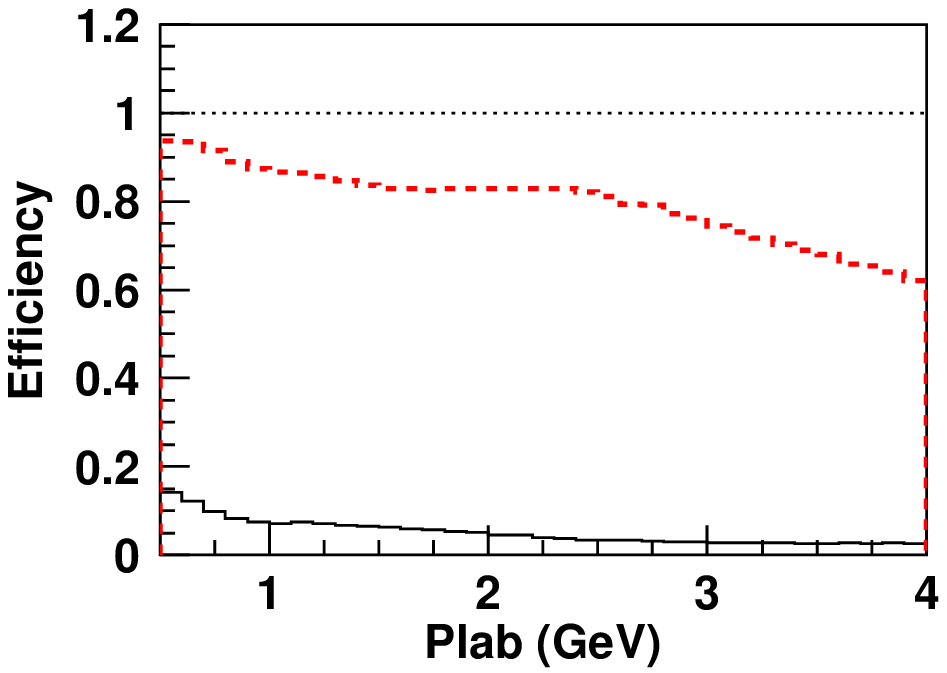} \\
     \multicolumn{1}{l}{(c)} & \multicolumn{1}{l}{(d)} \\
     \includegraphics[scale=0.78]{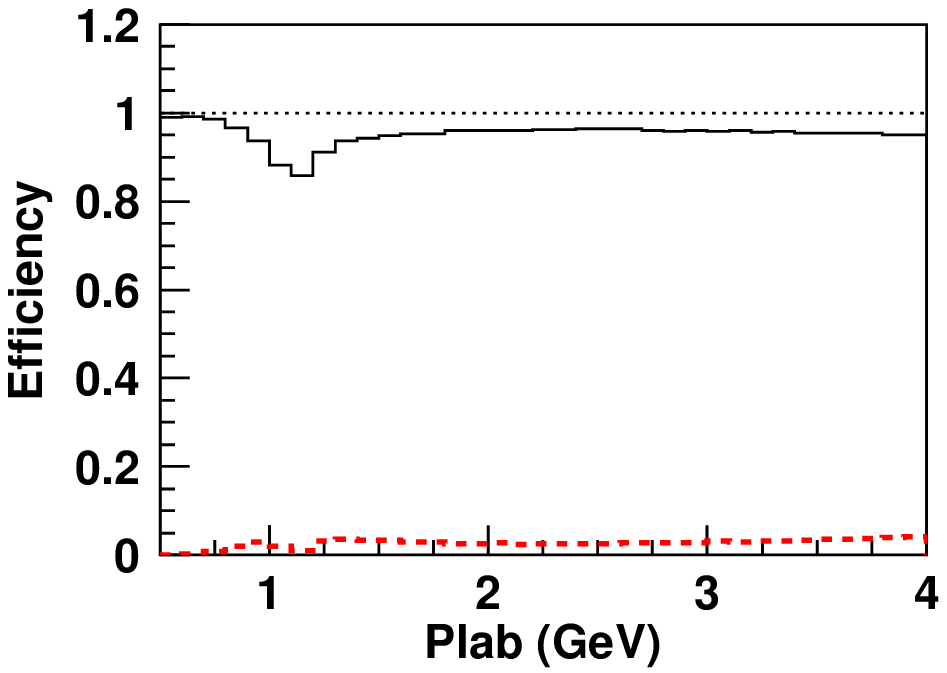}
     & \includegraphics[scale=0.78]{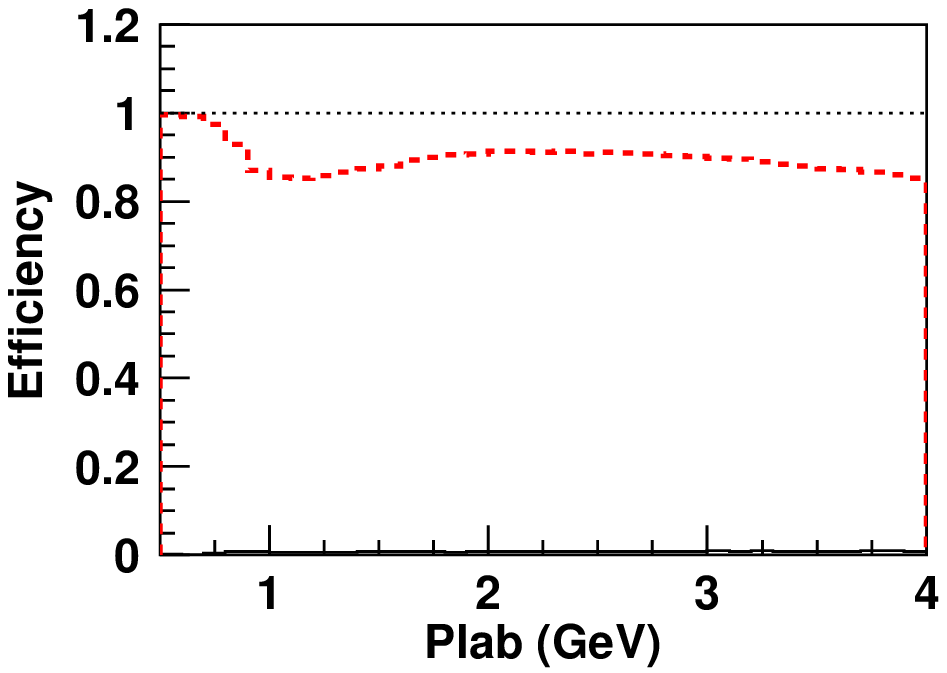} \\
     \multicolumn{1}{l}{(e)} & \multicolumn{1}{l}{(f)} \\
     \includegraphics[scale=0.78]{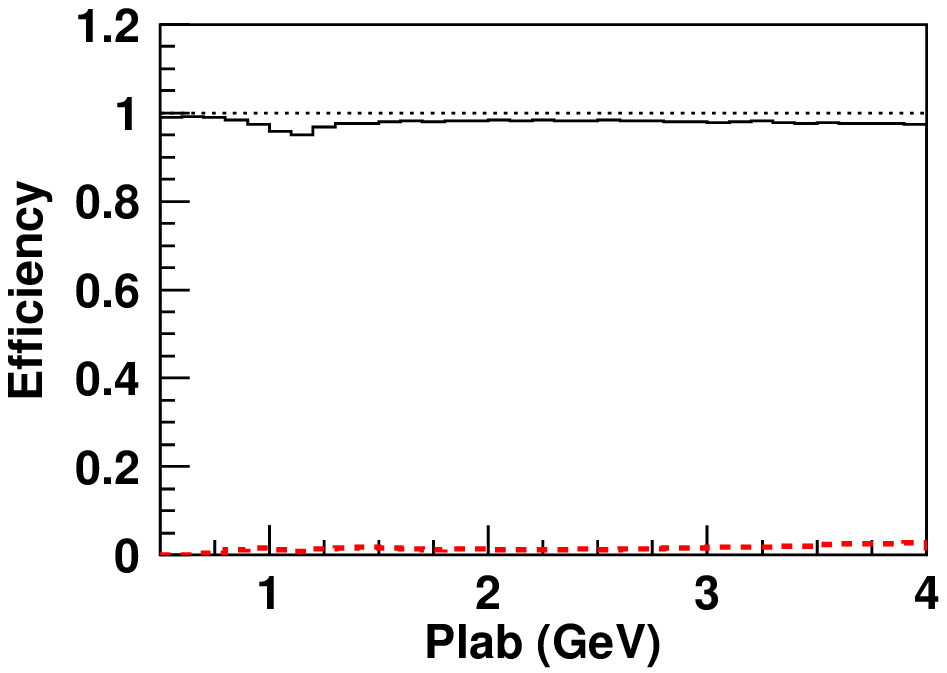}
     & \includegraphics[scale=0.78]{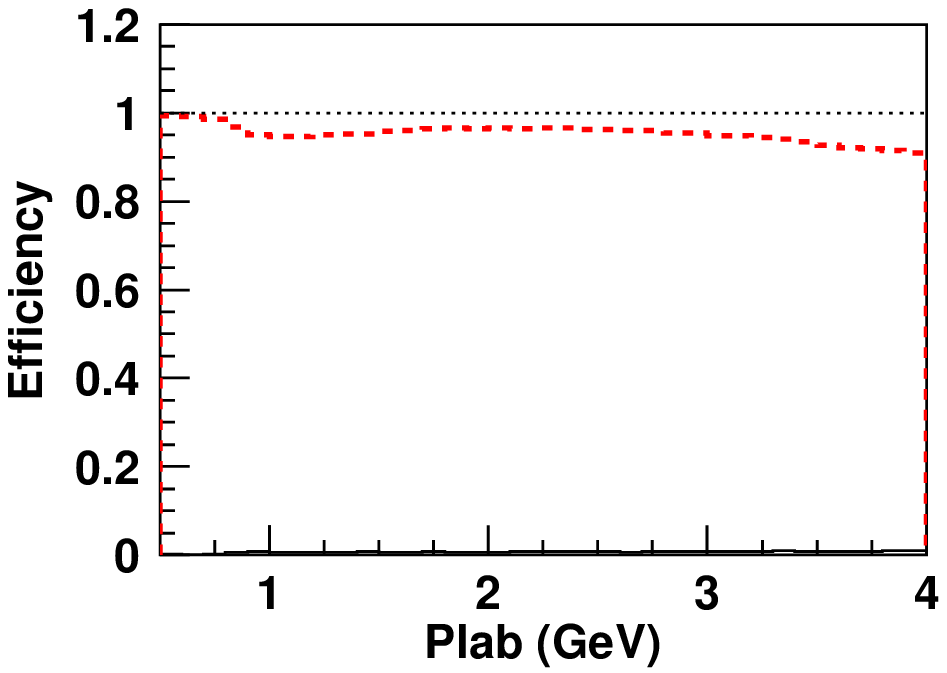} \\
    \end{tabular}
    \caption{\label{fig:acc_top-eff}%
    Momentum dependence of the efficiencies for $K$'s (solid) and $\pi$'s (dashed)
    (a) for $\mathit{LR}(K,\pi) > 0.6$ with the present barrel PID,
    (b) for $\mathit{LR}(K,\pi) < 0.1$ with the present barrel PID,
    (c) for $\mathit{LR}(K,\pi) > 0.6$ with TOP + $dE/dx$,
    (d) for $\mathit{LR}(K,\pi) < 0.1$ with TOP + $dE/dx$,
    (e) for $\mathit{LR}(K,\pi) > 0.6$ with TOP + $dE/dx$ without a $\phi$ gap,
    (f) for $\mathit{LR}(K,\pi) < 0.1$ with TOP + $dE/dx$ without a $\phi$ gap.
    }
   \end{center}
  \end{figure}

  We first study  the   PID system performance for single tracks,
  generated uniformly over the  laboratory solid angle ($\cos\theta$).
  Figure~\ref{fig:acc_top-eff} shows the efficiencies and mis-identification
  probabilities 
  of $K$ or $\pi$ for two kinds of $K$-$\pi$ selection criteria
  using TOP and $dE/dx$
  for two TOP configurations (Fig.~\ref{fig:acc_top-eff} (a) $\sim$ (d)).
  Here, the performance of $dE/dx$ at sBelle is assumed to be identical
  to the performance in the present Belle detector.
  The performance of the present PID (ACC, TOF and $dE/dx$)
  estimated using gsim is also shown for comparison
  (Fig.~\ref{fig:acc_top-eff} (e), (f)).
  We can observe a significant improvement of the performance
  with the TOP detector.
  We can also see the effect of the $\phi$ gap.

\begin{figure}
   \begin{center}
    \begin{tabular}{cc}
     \multicolumn{1}{l}{(a)} & \multicolumn{1}{l}{(b)} \\
     \includegraphics[scale=0.78]{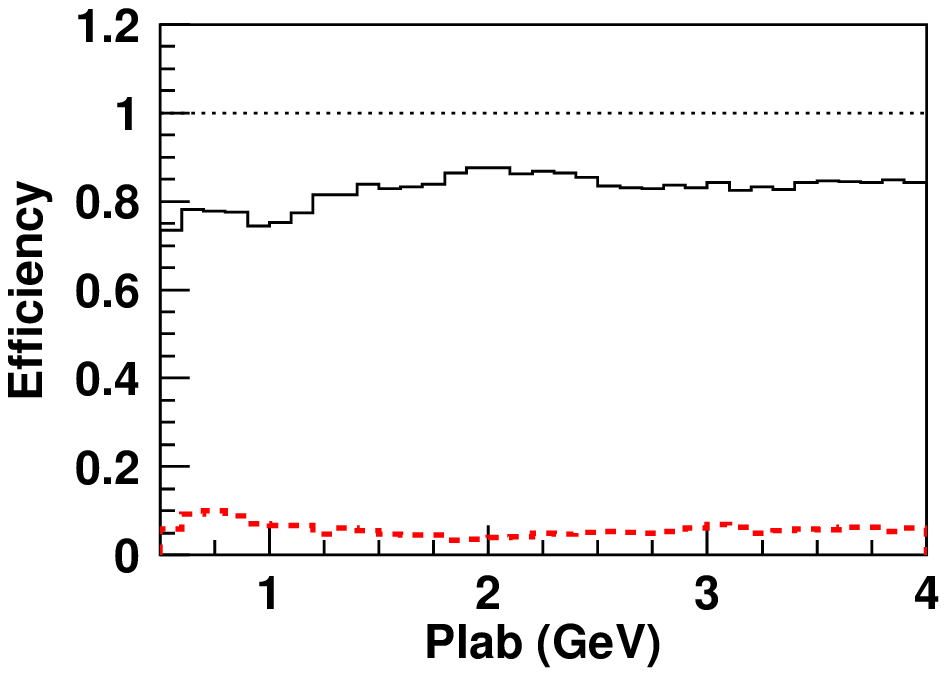}
     & \includegraphics[scale=0.78]{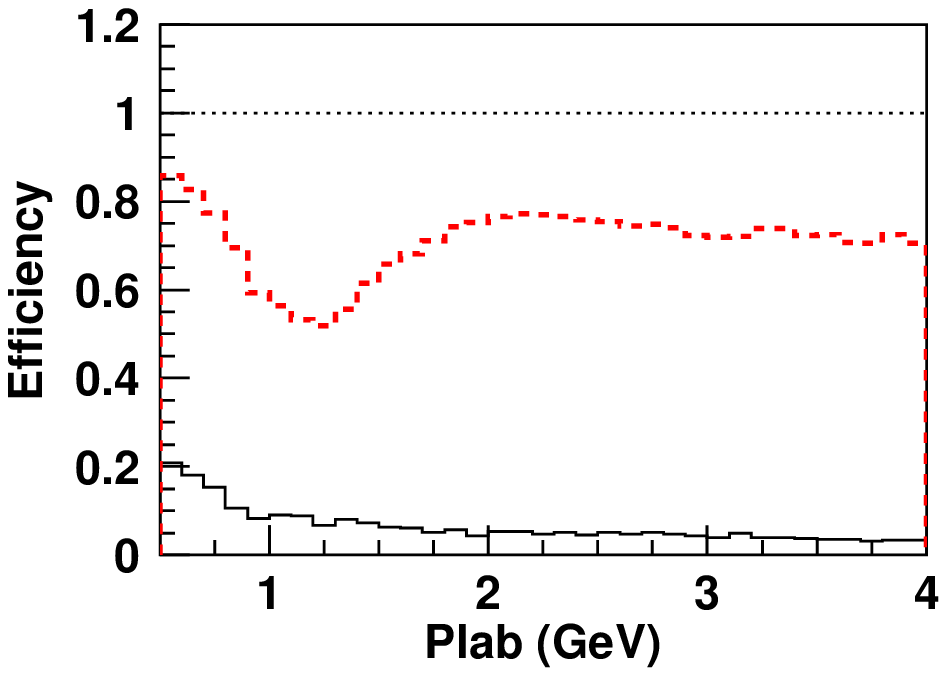} \\
     \multicolumn{1}{l}{(c)} & \multicolumn{1}{l}{(d)} \\
     \includegraphics[scale=0.78]{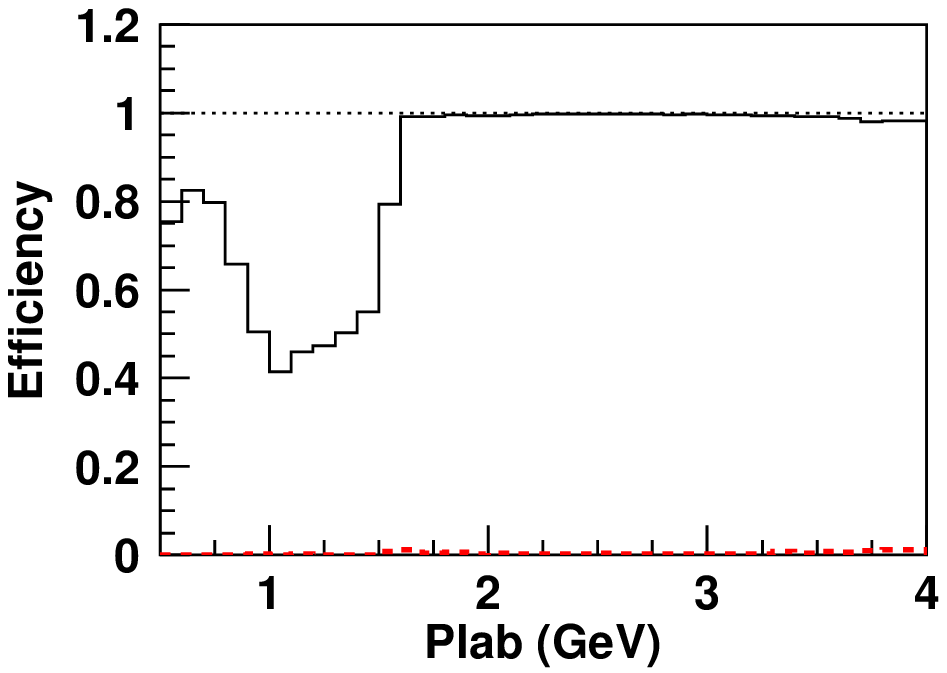}
     & \includegraphics[scale=0.78]{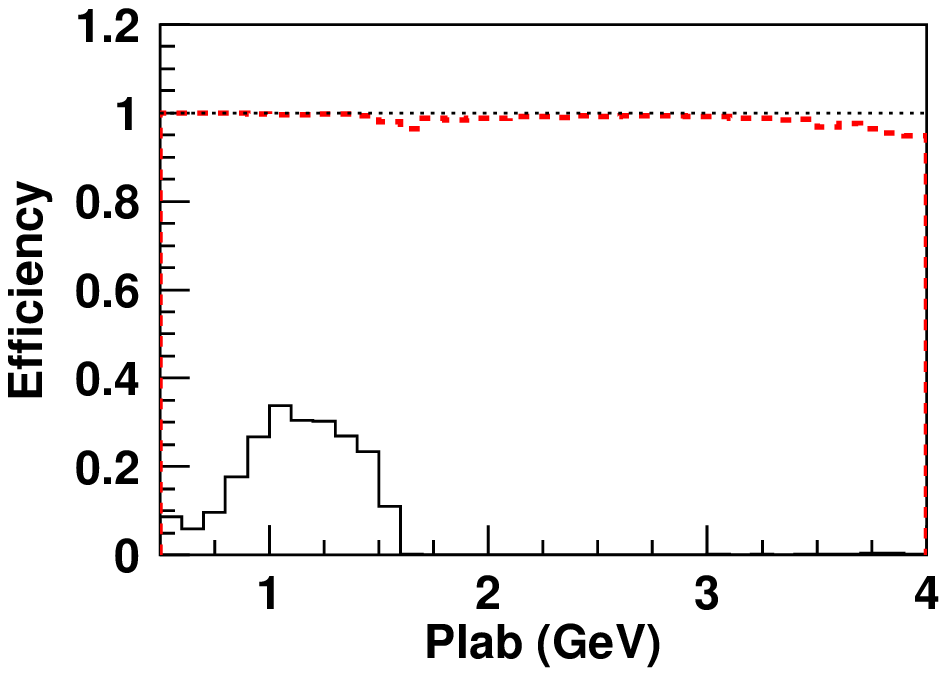} \\
    \end{tabular}
    \caption{\label{fig:acc_arich-eff}%
    Momentum dependence of the efficiencies for $K$'s (solid) and $\pi$'s (dashed)
    (a) for $\mathit{LR}(K,\pi) > 0.6$ with the present endcap PID
    (b) for $\mathit{LR}(K,\pi) < 0.1$ with the present endcap PID
    (c) for $\mathit{LR}(K,\pi) > 0.6$ with ARICH + $dE/dx$
    (d) for $\mathit{LR}(K,\pi) < 0.1$ with ARICH + $dE/dx$.
    }
   \end{center}
  \end{figure}

  Similar distributions can be obtained for the endcap region.
  Figure~\ref{fig:acc_arich-eff} shows the efficiencies and mis-identification
  proabilities for $K$'s or $\pi$'s with the ACC and an ARICH.
  The performance can be expressed in terms of the separation,
  which can be calculated from the efficiency and mis-identification
  rate for certain selection criteria.
  Table~\ref{tab:acc_arich_k_06} (\ref{tab:acc_arich_pi_01})
  lists the $K$ ($\pi$) efficiencies and
  $\pi$ ($K$) mis-identification rates and the $K$-$\pi$ separations
  for a particular choice of selection criteria based on the likelihood
  ratio for the kaon and pion hypotheses,
  $\mathit{LR}(K,\pi) > 0.6$ ($\mathit{LR}(K,\pi) < 0.1$),
  for three momentum regions.
  The momentum regions denoted as $1\,\GeV$, $2\,\GeV$ and $4\,\GeV$ correspond
  to the intervals $0.8 < p < 1.2\,\GeV$, $1.8 < p < 2.2\,\GeV$ and
  $3.5 < p < 4.0\,\GeV$, respectively.

  \begin{table}
   \begin{center}
    \caption{\label{tab:acc_arich_k_06}%
    $K$ efficiency (Eff.), $\pi$ mis-identification rate (Fake)
    and separation (Sep.) for the selection $\mathit{LR}(K,\pi) > 0.6$.
    }
    \catcode`;=\active \def;{\phantom{0}}
    \begin{tabular}{|c|ccc|ccc|ccc|}
     \hline
     & \multicolumn{3}{c|}{$1\,\GeV/c$} & \multicolumn{3}{c|}{$2\,\GeV/c$} &
     \multicolumn{3}{c|}{$4\,\GeV/c$} \\ \hline
     & Eff.($\%$) & Fake($\%$) & Sep. & Eff.($\%$) & Fake($\%$) & Sep.
     & Eff.($\%$) & Fake($\%$) & Sep. \\ \hline
     ACC & $71.2$ & $6.4;$ & $2.1$
		 & $83.6$ & $3.2$ & $2.8$ & $38.8$ & $14.4$ & $0.8$ \\
     ACC + $dE/dx$ & $76.2$ & $7.3;$ & $2.2$
		 & $86.9$ & $3.8$ & $2.9$ & $84.4$ & $;6.0$ & $2.6$ \\
     ARICH & $36.0$ & $0.08$ & $2.8$
		 & $95.8$ & $0.6$ & $4.2$ & $93.1$ & $;1.4$ & $3.7$ \\
     ARICH + $dE/dx$ & $48.2$ & $0.06$ & $3.2$
		 & $99.5$ & $0.4$ & $5.2$ & $98.6$ & $;1.0$ & $4.5$
     \\ \hline
    \end{tabular}
   \end{center}
  \end{table}

  \begin{table}
   \begin{center}
    \caption{\label{tab:acc_arich_pi_01}%
    $\pi$ efficiency (Eff.), $K$ mis-identification rate (Fake)
    and separation (Sep.)
    for the selection $\mathit{LR}(K,\pi) < 0.1$.
    }
    \catcode`;=\active \def;{\phantom{0}}
    \begin{tabular}{|c|ccc|ccc|ccc|}
     \hline
     & \multicolumn{3}{c|}{$1\,\GeV/c$} & \multicolumn{3}{c|}{$2\,\GeV/c$} &
     \multicolumn{3}{c|}{$4\,\GeV/c$} \\ \hline
     & Eff.($\%$) & Fake($\%$) & Sep. & Eff.($\%$) & Fake($\%$) & Sep.
     & Eff.($\%$) & Fake($\%$) & Sep. \\ \hline
     ACC & $59.6$ & $;9.1$ & $1.6$
		 & $64.1$ & $6.5;;$ & $1.9$ & $;3.5$ & $2.6;$ & $0.1$ \\
     ACC + $dE/dx$ & $59.9$ & $;9.2$ & $1.8$
		 & $75.7$ & $5.2;;$ & $2.3$ & $71.1$ & $3.5;$ & $2.4$ \\
     ARICH & $99.5$ & $37.4$ & $2.9$
		 & $94.8$ & $0.12;$ & $4.7$ & $89.6$ & $0.38$ & $3.9$ \\
     ARICH + $dE/dx$ & $99.8$ & $28.8$ & $3.4$
		 & $98.7$ & $0.086$ & $5.4$ & $96.3$ & $0.31$ & $4.5$
     \\ \hline
    \end{tabular}
   \end{center}
  \end{table}

  \subsubsection{Luminosity gain in $B \to \rho\gamma$}
  
  We consider the following PID configurations for the barrel region:
  \begin{itemize}
   \item[(B1)] Present ACC only. This corresponds to the case
	       when the present PID device is used at Super KEKB
	       and not only TOF but also $dE/dx$ are conservatively assumed to
	       be unusable.
   \item[(B2)] Present ACC and $dE/dx$ only. This corresponds to
	       the case when the present PID device is used at Super KEKB
	       and $dE/dx$ is assumed to work as at present.
   \item[(B3)] Present Belle PID (ACC, TOF, $dE/dx$).
	       This configuration is for reference,
	       and does not correspond to any PID configuration at Super KEKB.
   \item[(B4)] TOP detector with the 3-readout scheme and a multi-alkali MCP,
	       i.e., lower performance TOP.
	       $dE/dx$ is also assumed to work as at present.
   \item[(B5)] TOP detector with the focusing scheme and a GaAsP MCP.
	       $dE/dx$ is also assumed to work as at present.
   \item[(B6)] TOP detector with the focusing scheme and a GaAsP MCP
	       without a $\phi$ gap.
	       $dE/dx$ is also assumed to work as at present.
  \end{itemize}
  The simulations for (B1), (B2) and (B3) are done using the Geant 3
  simulator gsim for the present Belle detector.
  The simulation for (B5) and (B6) is done with fsim.
  We do not have a simulator for configuration (B4), and
  we assume that the mis-identification rate is 5 times higher
  at $4\,\GeV$ and two times higher below $3\,\GeV$ compared to (B5).

  Similarly, the following configuration is assumed
  for the forward endcap region:
  \begin{itemize}
   \item[(F1)] Present ACC only. This corresponds to the case
	       when the present PID device is used at Super KEKB
	       and $dE/dx$ is consertively assumed to be unusable.
   \item[(F2)] Present Belle PID (ACC and $dE/dx$).
	       This corresponds to the case when the present PID device
	       is used at Super KEKB
	       and $dE/dx$ is assumed to work as present.
	       This configuration is also a reference for the present Belle PID.
   \item[(F3)] ARICH detector.
   \item[(F4)] ARICH detector with TOF information.
  \end{itemize}
  The simulations for (F1), (F2) are done using gsim for the present Belle.
  The simulation for (F3) is done with fsim.
  We do not have a simulator for configuration (F4), and
  we simply assume that the mis-identification rate below $1.5\,\GeV$
  is half of that in (F3).

  We reconstruct $B^0 \to \rho^0\gamma$ and $B^+ \to \rho^+\gamma$
  for $B \to \rho\gamma$ and $B \to K^\ast\gamma$ MC samples.
  The selection criteria are similar to those in the recent Belle analysis.
  For simplicity,
  the contribution from the $\qq$ background is assumed to
  be independent of the PID configuration.
  and is fixed to $2000$ events for $7.5\,\abi$.
  We estimate the figure of merit (FOM) from the number of
  signal ($B \to \rho\gamma$) events $N_S$
  and background ($B \to K^\ast\gamma$ and $\qq$ background) events $N_B$
  in the signal region.
  Here, the signal region is defined by $\Mbc > 5.27\,\GeV$ and
  $\DE_{\mathrm{min}} < \DE < 0.08\,\GeV$,
  where $\DE_{\mathrm{min}}$ is chosen for each PID configuration
  to maximize the FOM.

  \begin{table} 
   \begin{center}
    \begin{minipage}{0.43\textwidth}
     \begin{center}
      \caption{\label{tab:FOM_rho0gamma}%
      Number of signal ($N_S$) and background ($N_B$) events,
      figure of merit (FOM), and the $\DE$ lower limit
      $\DE_{\mathrm{min}}$ chosen
      for each barrel and forward PID configuration
      for $B^0 \to \rho^0\gamma$ analysis at $7.5\,\abi$.
      See the text for an explanation of the configuration label.}
      \catcode`;=\active \def;{\phantom{0}}
      \begin{tabular}{cccccc}
       \hline\hline
       Barrel & Forward & $N_S$ & $N_B$ & FOM & $\DE_{\mathrm{min}}$ [$\GeV$]
       \\ \hline
       B1 & F1 & $;427$ & $3998$ & $;6.4$ & $-0.10$ \\
       B1 & F2 & $;474$ & $4021$ & $;7.1$ & $-0.10$ \\
       B1 & F3 & $;510$ & $3846$ & $;7.7$ & $-0.10$ \\
       B1 & F4 & $;510$ & $3836$ & $;7.7$ & $-0.10$ \\
       B2 & F1 & $;714$ & $4791$ & $;9.6$ & $-0.20$ \\
       B2 & F2 & $;800$ & $5254$ & $10.3$ & $-0.30$ \\
       B2 & F3 & $;813$ & $4609$ & $11.0$ & $-0.20$ \\
       B2 & F4 & $;813$ & $4590$ & $11.1$ & $-0.20$ \\
       B3 & F1 & $;930$ & $5207$ & $11.9$ & $-0.25$ \\
       B3 & F2 & $;987$ & $5242$ & $12.5$ & $-0.25$ \\
       B3 & F3 & $1032$ & $5026$ & $13.3$ & $-0.25$ \\
       B3 & F4 & $1032$ & $5002$ & $13.3$ & $-0.25$ \\
       B4 & F1 & $;924$ & $3389$ & $14.1$ & $-0.30$ \\
       B4 & F2 & $;982$ & $3422$ & $14.8$ & $-0.30$ \\
       B4 & F3 & $1027$ & $3208$ & $15.8$ & $-0.30$ \\
       B4 & F4 & $1027$ & $3180$ & $15.8$ & $-0.30$ \\
       B5 & F1 & $;924$ & $2832$ & $15.1$ & $-0.30$ \\
       B5 & F2 & $;982$ & $2865$ & $15.8$ & $-0.30$ \\
       B5 & F3 & $1027$ & $2651$ & $16.9$ & $-0.30$ \\
       B5 & F4 & $1027$ & $2623$ & $17.0$ & $-0.30$ \\
       B6 & F1 & $1060$ & $2857$ & $16.9$ & $-0.30$ \\
       B6 & F2 & $1118$ & $2890$ & $17.7$ & $-0.30$ \\
       B6 & F3 & $1163$ & $2677$ & $18.8$ & $-0.30$ \\
       B6 & F4 & $1163$ & $2648$ & $18.8$ & $-0.30$ \\ \hline\hline
      \end{tabular}
     \end{center}
    \end{minipage}
    \hfil
    \begin{minipage}{0.43\textwidth}
     \begin{center}
      \catcode`;=\active \def;{\phantom{0}}
      \caption{\label{tab:FOM_rhopgamma}%
      Number of signal ($N_S$) and background ($N_B$) events,
      figure of merit (FOM), and the $\DE$ lower limit
      $\DE_{\mathrm{min}}$ chosen
      for each barrel and forward PID configuration
      for $B^+ \to \rho^+\gamma$ analysis at $7.5\,\abi$.
      See the text for an explanation of the configuration label.}
      \begin{tabular}{cccccc}
       \hline\hline
       Barrel & Forward & $N_S$ & $N_B$ & FOM & $\DE_{\mathrm{min}}$ [$\GeV$]
       \\ \hline
       B1 & F1 & $1014$ & $3775$ & $14.7$ & $-0.30$ \\
       B1 & F2 & $1034$ & $3681$ & $15.1$ & $-0.30$ \\
       B1 & F3 & $1076$ & $3656$ & $15.6$ & $-0.30$ \\
       B1 & F4 & $1076$ & $3626$ & $15.7$ & $-0.30$ \\
       B2 & F1 & $;997$ & $3197$ & $15.4$ & $-0.30$ \\
       B2 & F2 & $1017$ & $3103$ & $15.8$ & $-0.30$ \\
       B2 & F3 & $1059$ & $3078$ & $16.5$ & $-0.30$ \\
       B2 & F4 & $1059$ & $3048$ & $16.5$ & $-0.30$ \\
       B3 & F1 & $1132$ & $3100$ & $17.4$ & $-0.30$ \\
       B3 & F2 & $1152$ & $3006$ & $17.9$ & $-0.30$ \\
       B3 & F3 & $1194$ & $2981$ & $18.5$ & $-0.30$ \\
       B3 & F4 & $1194$ & $2951$ & $18.5$ & $-0.30$ \\
       B4 & F1 & $1121$ & $2950$ & $17.6$ & $-0.30$ \\
       B4 & F2 & $1141$ & $2856$ & $18.0$ & $-0.30$ \\
       B4 & F3 & $1183$ & $2831$ & $18.7$ & $-0.30$ \\
       B4 & F4 & $1183$ & $2802$ & $18.7$ & $-0.30$ \\
       B5 & F1 & $1121$ & $2584$ & $18.4$ & $-0.30$ \\
       B5 & F2 & $1141$ & $2490$ & $18.9$ & $-0.30$ \\
       B5 & F3 & $1183$ & $2465$ & $19.6$ & $-0.30$ \\
       B5 & F4 & $1183$ & $2436$ & $19.7$ & $-0.30$ \\
       B6 & F1 & $1174$ & $2597$ & $19.1$ & $-0.30$ \\
       B6 & F2 & $1194$ & $2503$ & $19.6$ & $-0.30$ \\
       B6 & F3 & $1236$ & $2478$ & $20.3$ & $-0.30$ \\
       B6 & F4 & $1236$ & $2448$ & $20.4$ & $-0.30$ \\ \hline\hline
      \end{tabular}
     \end{center}
    \end{minipage}
   \end{center}
  \end{table}

  \begin{figure}
   \begin{center}
    \begin{tabular}{cc}
     \multicolumn{1}{l}{(a)} & \multicolumn{1}{l}{(b)} \\
     \includegraphics[scale=0.78]{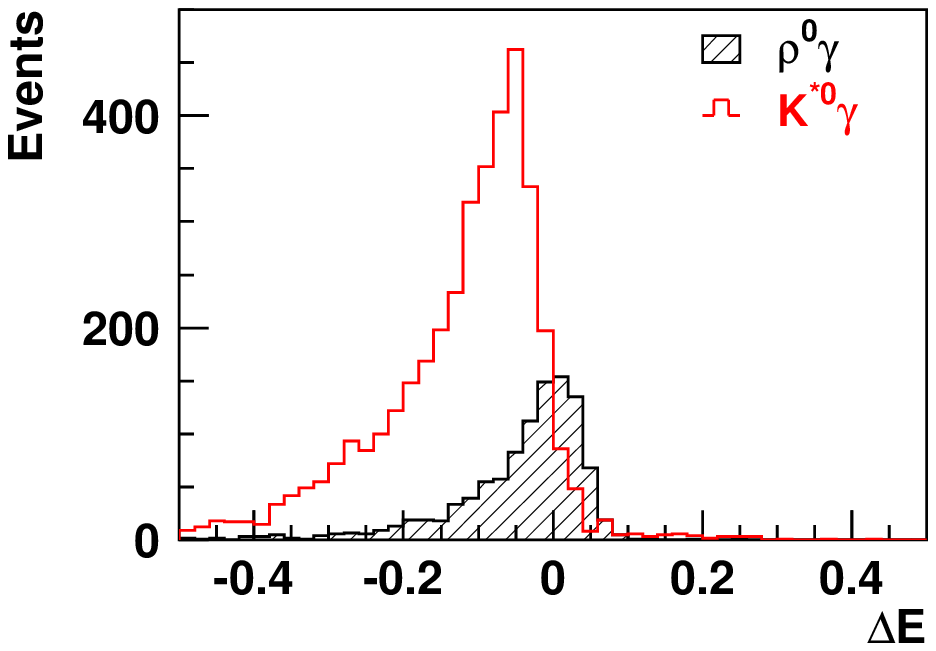}
     & \includegraphics[scale=0.78]{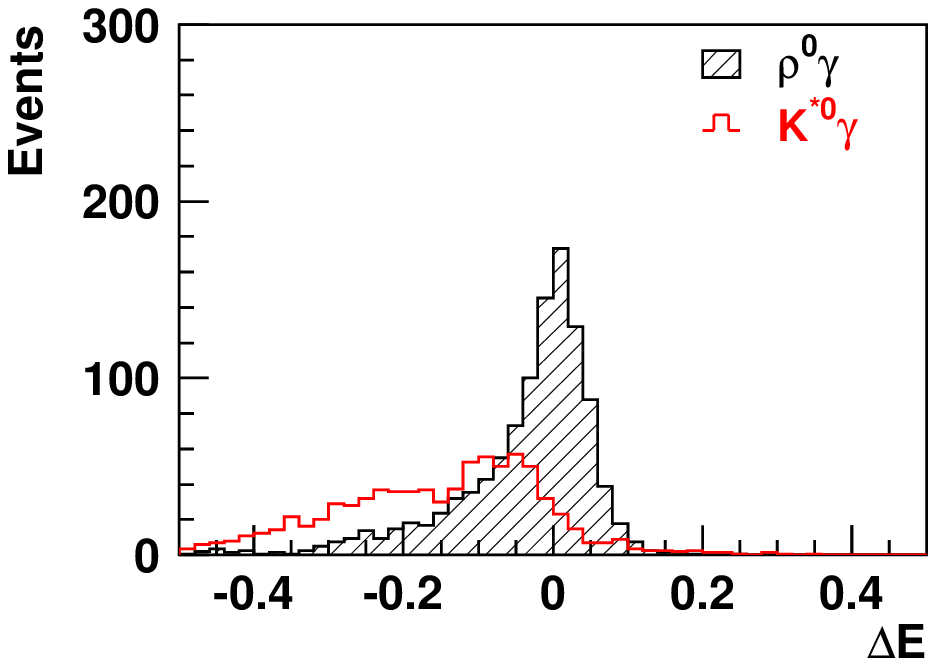} \\
     \multicolumn{1}{l}{(c)} & \multicolumn{1}{l}{(d)} \\
     \includegraphics[scale=0.78]{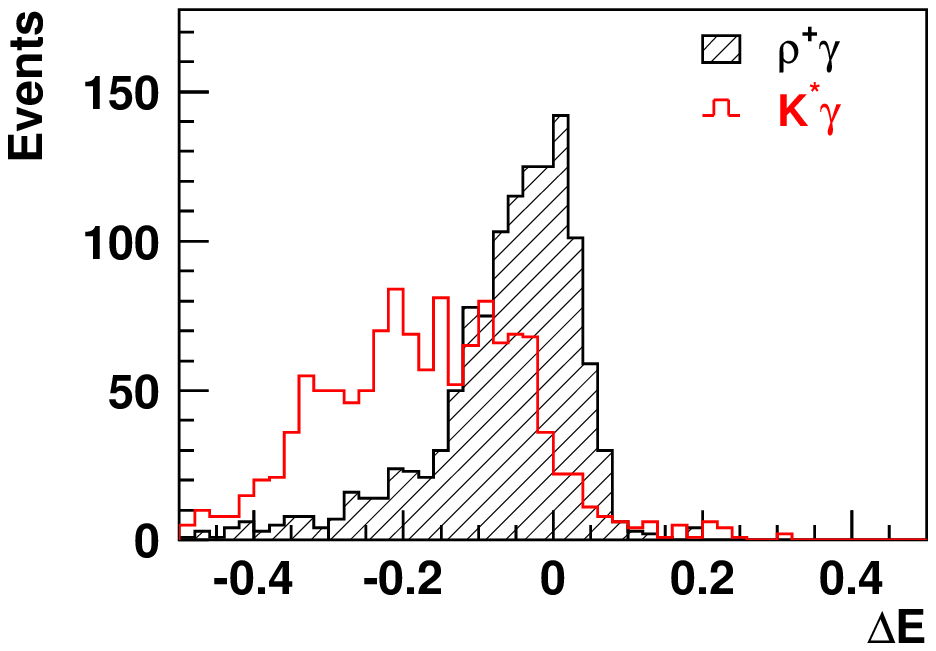}
     & \includegraphics[scale=0.78]{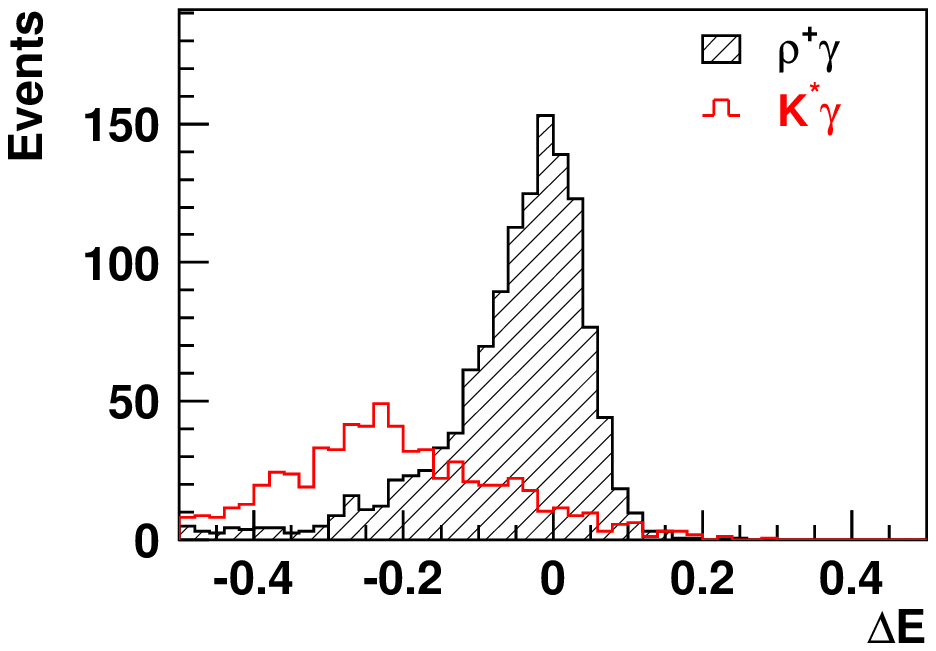} \\
    \end{tabular}
    \caption{\label{fig:rzrp_de}%
    Expected $\DE$ distribution at $7.5\,\abi$ for
    (a) $B^0 \to \rho^0\gamma$ with present PID performance (B3 and F2),
    (b) $B^0 \to \rho^0\gamma$ with TOP and ARICH (B5 and F3),
    (c) $B^+ \to \rho^+\gamma$ with present PID performance (B3 and F2),
    (d) $B^+ \to \rho^+\gamma$ with TOP and ARICH (B5 and F3).
    }
   \end{center}
  \end{figure}

  Tables~\ref{tab:FOM_rho0gamma} and \ref{tab:FOM_rhopgamma}
  list the FOM for various detector configurations.
  The $\DE$ distributions for two typical configurations,
  PID configuration B3 and F2 that correspond to the present PID performance
  and configuration B5 and F3 corresponding to TOP and ARICH,
  are shown in Fig.~\ref{fig:rzrp_de}.
  As seen from Fig.~(a) and (b), the suppression of the $K^\ast\gamma$ component
  is quite significant with the TOP and the ARICH.
  The contribution of the $K^\ast\gamma$ component below $-0.2\,\GeV$ contains
  events that are not due to a simple mis-identification between $K$ and $\pi$
  but come from mis-reconstruction.
  In (b), because the mis-identification is highly
  suppressed by good PID devices, such mis-reconstructed events become visible.
  For the $B^+ \to \rho^+\gamma$ case ((c) and (d)),
  we can also observe a significant improvement in the analysis,
  although the effect is smaller if compared to the $\rho^0\gamma$ channel
  because mis-reconstructed events already represent a significant fraction 
  of the background in the present detector.

  Tables~\ref{tab:lumgain_rho0gamma} and \ref{tab:lumgain_rhopgamma}
  show the ``luminosity gain''
  compared to the present PID performance (configuration B3 and F2).
  The luminosity gain is defined as a square of the ratio of the FOM,
  and is the luminosity ratio that would be necessary
  to obtain the same sensitivity with the nominal PID configuration.
  The rows corresponding to B1 and B2 of the two tables show that the
  present Belle detector cannot be used at sBelle because
  the TOF counter cannot operate reliably in such an environment.
  The rows corresponding to B4 and B5 of Table~\ref{tab:lumgain_rho0gamma}
  show that the introduction of the TOP detector is essential for
  this mode.
  The F3 and F4 columns show that the ARICH further improves the performance.

  The reduction of the $\phi$ gap in the TOP counter also seems to be of some
  benefit as can be seen from B6, where
  the coverage of the TOP counter in the $\phi$ direction
  is assumed to be $100\%$.  However, this assumption might be too simple for 
  the staggered-bar option, and more studies are needed to resolve this question.

  \begin{table}
   \begin{center}
    \begin{minipage}{0.43\textwidth}
     \begin{center}
      \caption{\label{tab:lumgain_rho0gamma}%
      Luminosity gain for $B^0 \to \rho^0\gamma$
      for various detector configurations.
      }
      \begin{tabular}{|c|c|c|c|c|}
       \hline
       & F1 & F2 & F3 & F4 \\ \hline
       B1 & $0.26$ & $0.32$ & $0.38$ & $0.38$ \\ \hline
       B2 & $0.59$ & $0.68$ & $0.78$ & $0.78$ \\ \hline
       B3 & $0.90$ & $1.00$ & $1.12$ & $1.13$ \\ \hline
       B4 & $1.27$ & $1.40$ & $1.59$ & $1.60$ \\ \hline
       B5 & $1.45$ & $1.60$ & $1.83$ & $1.85$ \\ \hline
       B6 & $1.83$ & $1.99$ & $2.25$ & $2.27$ \\ \hline
      \end{tabular}
     \end{center}
    \end{minipage}
    \hfil
    \begin{minipage}{0.43\textwidth}
     \begin{center}
      \caption{\label{tab:lumgain_rhopgamma}%
      Luminosity gain for $B^+ \to \rho^+\gamma$
      for various detector configurations.
      }
      \begin{tabular}{|c|c|c|c|c|}
       \hline
       & F1 & F2 & F3 & F4 \\ \hline
       B1 & $0.67$ & $0.71$ & $0.77$ & $0.77$ \\ \hline
       B2 & $0.74$ & $0.79$ & $0.85$ & $0.86$ \\ \hline
       B3 & $0.95$ & $1.00$ & $1.07$ & $1.08$ \\ \hline
       B4 & $0.97$ & $1.02$ & $1.09$ & $1.10$ \\ \hline
       B5 & $1.06$ & $1.12$ & $1.20$ & $1.21$ \\ \hline
       B6 & $1.14$ & $1.21$ & $1.29$ & $1.30$ \\ \hline
      \end{tabular}
     \end{center}
    \end{minipage}
   \end{center}
  \end{table}


%% file: ecl.tex
 \section{ECL performance}
 \label{sec:ecl}
 For the electromagnetic calorimeter, ECL, overlapping beam-background
 events will degrade the performance. The mean shift of energy distribution due to this
 background source can be calibrated in principle, but the energy
 resolution will be worse and cannot be recovered unless we resolve the
 contribution from the background.  Therefore, we will upgrade the
 readout electronics so that we can sample and fit the signal
 waveform. With waveform fitting two overlapping waveforms can be
 resolved if they are separated by more than about one
 peaking time. From a toy MC study, in which we assume a flat background
 distribution in time, the background reduction factor is estimated to
 be seven in the energy range above 20\,MeV where the background hits
 create fake clusters.

 On the other hand, in the energy range below
 20\,MeV, the effect of the beam-background is similar to electric
 noise; we call this low energy background ``pileup noise''.  The
 waveform fitting does not have great suppression power for the pileup
 noise, compared to its suppression of fake clusters; the
 suppression factor is two for the pileup noise.
 The magnitude of the pileup noise, $\sigma_p$, is proportional to the
 fluctuation of the background, which is,
 \[
  \sigma_p\propto\sqrt{\nu\tau (\overline{\epsilon^2}-\overline{\epsilon}^2)},
 \]
 where $\nu$ is the rate of the photons, $\tau$ is effective duration
 of the signal, $\overline{\epsilon}$ and $\overline{\epsilon^{2}}$ are
 the mean and the square mean of the photon energies.
 Therefore, the background suppression factor for the pileup noise by
 waveform fitting is four ($\sqrt{N}/2 = \sqrt{N/4}$).
 A concern is only in the
 low energy photon reconstruction, which should be examined in the
 analysis of $B\to D^\ast K, D^\ast\to D\gamma$ or $D\pi^0$.

 For higher background levels when the suppression by waveform fitting
 is not sufficient, CsI(Tl) has to be replaced
 with pure CsI crystals. This is planned for both endcaps. 
 Because of the shorter time constant for pure CsI ($\sim 30$\,ns)
 compared to CsI(Tl) ($\sim 1\,\mu$s), a background reduction of about 30 is expected.
 Since the light output will be lower and the wavelength of the 
 scintillation light will be in the UV region, photomultiplier tubes 
 have to be used instead of PIN photodiodes.
 The biggest issue for pure CsI crystals is the cost. The 
 number of crystals that we can replace will depend on the available budget.
 In the following subsections, effects of a partial upgrade with pure
 CsI are discussed.

 \subsection{Partition of endcaps}
 If only a part of the endcaps is upgraded with pure CsI crystals, a 
 choice on the location of the pure CsI part has to be made.
 There are several possible partitions of the two regions.  Two ways of
 partitioning are under consideration, 
 \begin{itemize}
  \item {\bf (A)}: 
	Forward innermost 4 rings (224), middle 5 rings (448) and
	outermost 4 rings (480)\\
	Barrel (6624)\\
	Backward innermost 6 rings (480), and outermost 4 rings (480),
  \item {\bf (B)}: 
	Forward innermost 5 rings (288), next 2 rings (192), next 2
	rings (192) and outermost 4 rings (480)\\
	Barrel (6624)\\
	Backward innermost 3 rings (192), next 1 ring (96), next 2 rings
	(192) and outermost 4 rings (480),
 \end{itemize}
 where the numbers in parentheses are the numbers of crystals in each partition.
 The difference between the two turn out to be minor, 
 and are not crucial  for the conclusion of the studies in this
 section. 
 
 \subsection{Study of $B\to\PKzS\Pgpz\gamma$ decays}
 Assuming that the readout electronics will be upgraded and that
 waveform fitting will give us a background suppression by a factor of 7 as
 expected, 20 times more background corresponds to 3 times more
 background for the case with CsI(Tl), while we will be essentially free from
 background for the case with pure CsI. 
 The effect of a partial upgrade with pure CsI on $B\to\PKzS\Pgpz\gamma$ is studied
 with partitioning scheme A.
 Using exp 55 MC, we overlaid 3 times more background to the part with
 CsI(Tl), while no background is overlaid on the part with pure CsI.
 For detectors other than ECL, 3 times more background is overlaid in
 all cases for technical reasons.

 Several combinations of partitions are tested. Each configuration is
 identified with a 6 digit identifier like 033000, where each digit
 represents the background multiplication factor; the leftmost
 digit refers to the most backward (largest in theta) partition,
 and the rightmost digit to the most forward (smallest in theta)
 partition. Hence in the example given above, 033000, corresponds to 
 no background in the backward innermost 6 rings,
 3 times more background in the backgward outermost 4 rings, 3 times
 more background in the barrel, and no background in the forward endcap.
 Note that the present conditions correspond to 111111.

\begin{figure}
  \resizebox{0.46\columnwidth}{!}{\includegraphics{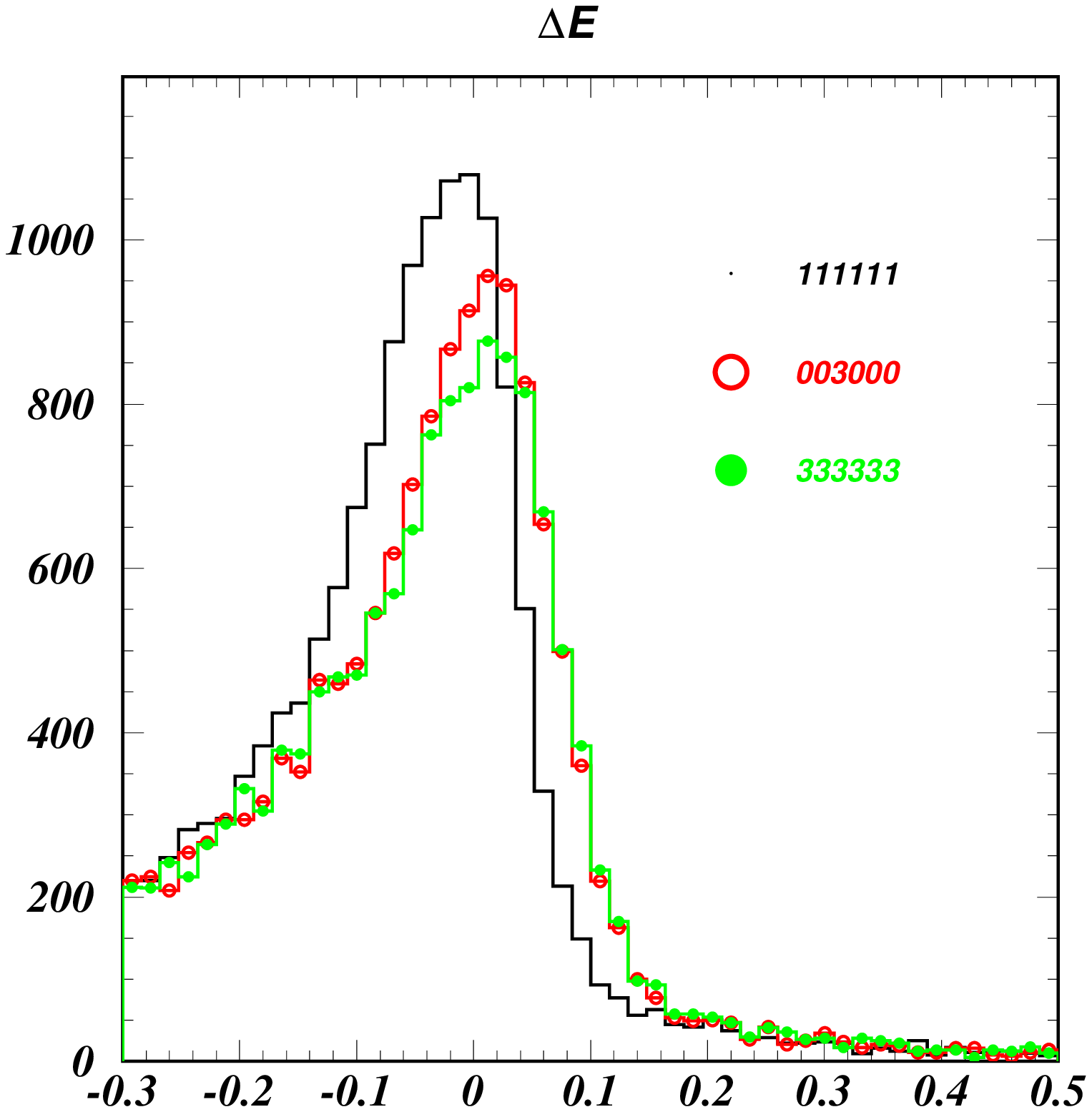}}
  \caption{ $\Delta E$ distributions for $B\to\PKzS\Pgpz\gamma$ decays for several
  background conditions.
  }
  \label{fig:de_kspi0gam}
 \end{figure}

 As shown in Fig.~\ref{fig:de_kspi0gam}, in the case of 333333 (green,
 with filled circles),
 the mean of the $\Delta E$ distribution is shifted towards the positive
 side, the width becomes larger, and the signal yield is lower compared
 to 111111 (black solid histogram); these are consequences of the
 background superposition.  We observe a similar deterioration also in
 the case of 003000 (red, with open circles), namely
 the case with both forward and backward endcaps fully upgraded to pure
 CsI. Other configurations (003300, 003333, 033000, 033300, 033330,
 033333, 333000, 333300 and 333330) are also tested; it is found that the 
 corresponding performance is between the performances for 003000 and 333333.
 This deterioration corresponds to a 10 to 15 percent loss of
 luminosity, depending on the number of pure CsI crystals.

\begin{figure}
  \resizebox{0.46\columnwidth}{!}{\includegraphics{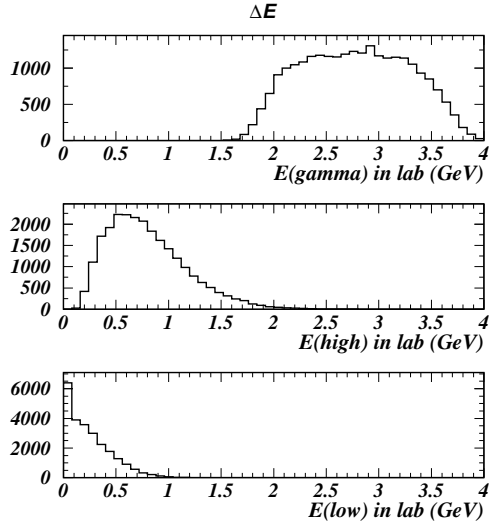}}
  \caption{ Laboratory energy distribution of three photons in the
  $B\to\PKzS\Pgpz\gamma$ decay sequence: the primary photon from the $B$
  decay (top) and the two photons from the $\Pgpz$ decay ordered by energy (middle and bottom).}
  \label{fig:energies_kspi0gam}
 \end{figure}

 Note that in this decay photons with relatively high energy
 are produced as  shown in Fig.~\ref{fig:energies_kspi0gam}.
 The analysis can be performed using the barrel ECL.
 Therefore, the impact of having pure CsI in the endcaps on this mode
 is not very large, although it is preferable to have a larger number of 
 pure CsI crystals. 

 \input{ecl_frec}


%% file: ecl_frec.tex
\subsection{Full reconstruction of $B$}

\begin{figure}[htbp]
   \resizebox{0.4\columnwidth}{!}{\includegraphics{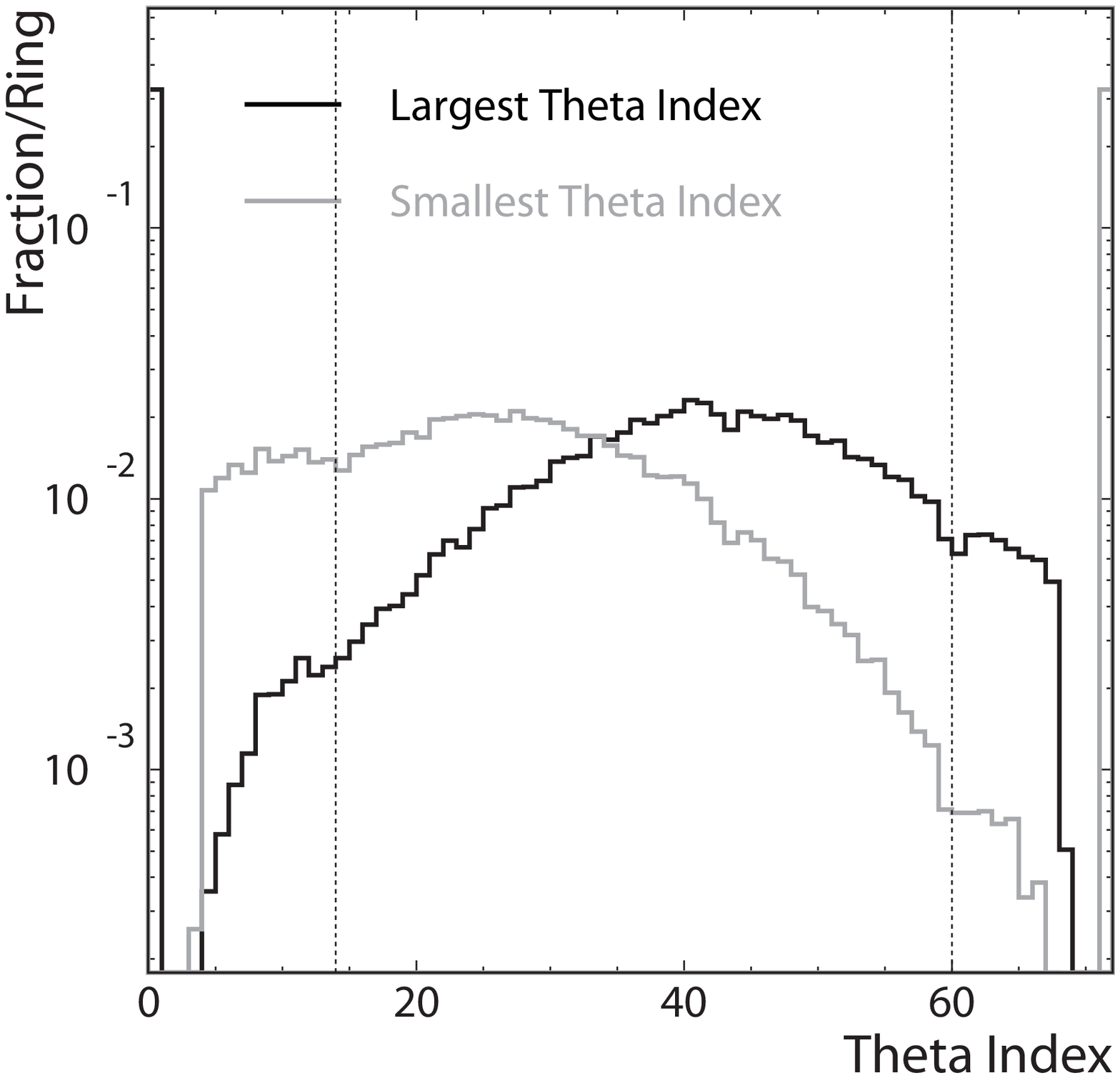}}
   \resizebox{0.4\columnwidth}{!}{\includegraphics{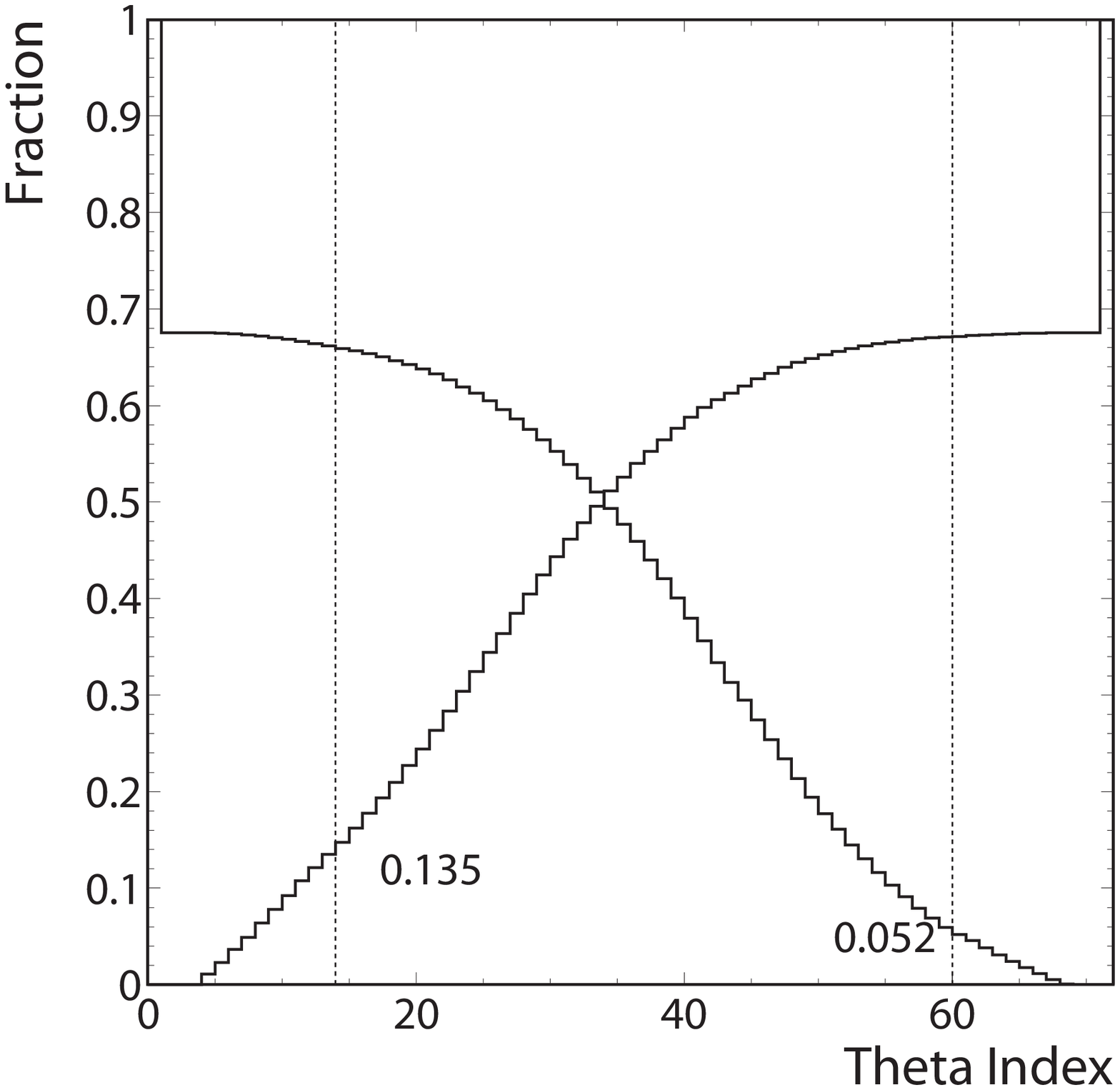}}
    \caption{Left: Minimum/Maximum $\theta$ of photons in fully
      reconstructed B events with nominal background.
      Right: Fraction of events integrated over angle.}
    \label{fig:frac_gam}
\end{figure}

Events where one of the $B$ meson decays is fully reconstructed, are of particular 
importance in studies of rare decays such as $B\to\tau\nu$ and $B\to K\nu\nu$.
The main role of ECL in full reconstruction is to identify photons and \Pgpz's.
It is therefore useful to understand the contribution of photons in the fully
reconstructed events.
Figure\,\ref{fig:frac_gam}(left) shows the smallest and largest
$\theta$ indices of the photons in each of the correctly reconstructed
MC  events without overlaid background. Peaks at the theta indices = 0 and 70
correspond to events without  photons in the final state.
Those events, which account for about 30\% of total events currently used
for the full reconstruction, are not relevant for the ECL performance study.
Figure\,\ref{fig:frac_gam}(right) shows the integrals of the two distributions.
From the figure the maximum contributions for the forward (index 1--13) and
for the backward side (index 59--69) are estimated to be at most 14\% and 5\%,
respectively.

Also note that the introduction of waveform sampling reduces the pile-up
by a factor of seven.
This means that the performance will be better than at present until 
the beam background
becomes seven times higher. This is expected to occur when
the beam current is about three times higher.

The effect of partial replacement was studied as described in the previous
subsection, except that the partitioning was according to scheme B.
In this case the nominal background level  was chosen as
two times the background level of experiment 49. The background level in the
region with the crystals replaced with pure CsI is again assumed to be zero.
The reconstruction efficiency is estimated using MC events
where one of the $B$ mesons decays to the final states included in the full-reconstruction
algorithm. The other $B$ is forced to decay to a neutrino. 
The combinatorial background from real $B$ decays is estimated using
generic $B$ decay MC.
To obtain the contribution from combinatorial background, the $m_\mathrm{bc}$
distribution is fitted with a Gaussian signal and with Argus background shapes.
The integral of the fitted Argus background is taken as the combinatorial
background. Note that the  peaking background in the $m_\mathrm{bc}$
distribution is not considered here. From the MC study,
about one third of the Gaussian signals in the generic $B$ MC
events actually originates from the combinatorial background.
The background from light flavor production is estimated
using udsc-mixed MC events.

\begin{figure}[htb]
  \begin{center}
   \resizebox{0.8\columnwidth}{!}{\includegraphics{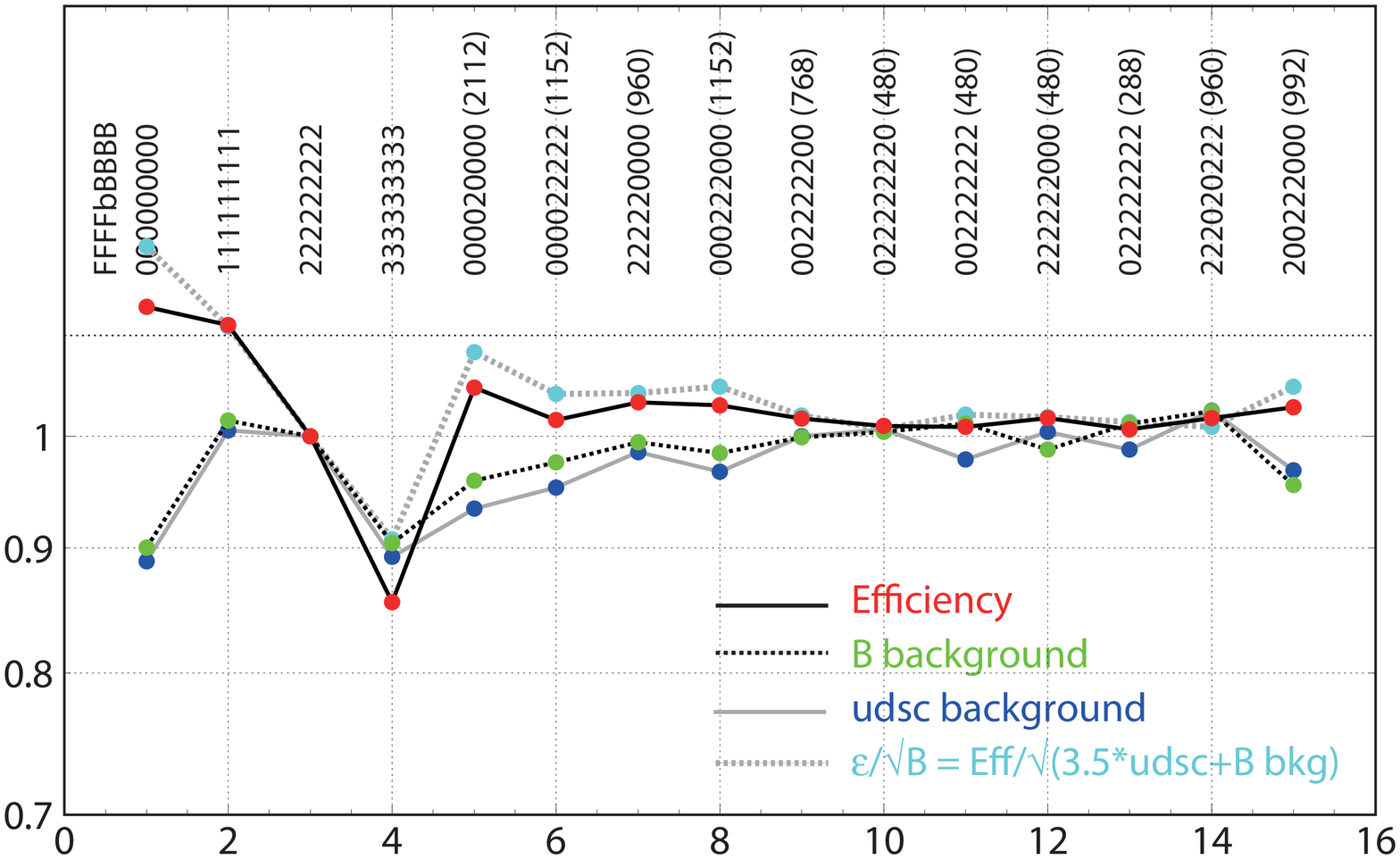}}
    \caption{Effect of a partial upgrade of the ECL. Black solid line,
   black dashed line and gray solid line show signal efficiency, $B$
   combinatorial background, and
   background from light flavors normalized to the case with
   two times background overlaid, respectively.
   Also shown by a gray dashed line is the efficiency divided by
   the square root of background. Numbers in parentheses are the numbers
   of crystals replaced with pure CsI. The nine digit numbers indicate 
   the background level factor for each of the nine detector regions.}
    \label{fig:result}
  \end{center}
\end{figure}

Figure\,\ref{fig:result} summarizes the effect of background in
various partial upgrade conditions. The configuration type is indicated
with a convention similar to the previous subsection. In this case there 
are four subdivisions in each of the endcaps. So the number of digits is
nine, FFFFbBBBB, where F/b/B are the forward/barrel/backward, respectively.
In the figure the variables are normalized to the case with
two times overlaid background. The numbers in the parentheses are the
numbers of crystals replaced by pure CsI.
The efficiency increase and background reduction by a full replacement, i.e.,
with 2112 pure CsI crystals, is about 5\%. The effect is very small when the number
of replaced CsI crystals is below 960.

%% file: kl_klm.tex



\section{$\PKzL$ reconstruction}
 \label{sec:klm}
In the design studies for the existing Belle detector, the KLM subsystem's segmented geometry
was chosen to enhance the reconstruction of $\PKzL$ mesons arising from
$B \to J/\psi \PKzL$ decays for time-dependent $CP$ asymmetry measurements.
Indeed, $\PKzL$ reconstruction in Belle has led to high quality $CP$
asymmetry measurements in $B \to J/\psi \PKzL$ as well as in other rare $B$ decay modes such as
$\phi \PKzL$ and $\eta' \PKzL$.

In the sBelle era, $\PKzL$ reconstruction will be driven not only by these 
time-dependent $CP$ asymmetry measurements but also as a veto of backgrounds
in rare decay modes with neutrinos such as $B\to\tau\nu$.
In this section, we study the extent to which we can improve the sensitivity of
$B\to\tau\nu$ via enhancement of $\PKzL$ reconstruction compared to the
current Belle detector and discuss 
the possibility of implementing such enhancements.

\subsection{Impact of $\PKzL$ reconstruction improvement to $B\to\tau\nu$}
 A large Monte Carlo simulation sample shows that
 about $78\pm1$\% of the background events remaining in the signal region of $B\to\tau\nu$ have
 at least one $\PKzL$ among the final state particles.
 In particular, the dominant background component that peaks in the signal region arises from
 $b\to c$ semileptonic decays where the charm-meson daughter decays to a
 $\PKzL$.
 Therefore, an effective $\PKzL$ veto would be very useful to reduce this background; in particular,
 an enhancement of $\PKzL$ detection efficiency is very desirable.

 Based on the GEANT simulation of the existing KLM and the measured detection efficiency (from 
 $J/\psi\PKzL$ data), the $\PKzL$ veto efficiency is estimated to be 35\%.
 We use these data to estimate the improvement in the significance of the $B\to\tau\nu$ signal
 with a higher $\PKzL$ veto efficiency.
 For the signal and background ratio, we use our published measurement of $B\to\tau\nu$ obtained
 from 447 million $B\bar{B}$ events.
 The loss in signal efficiency due to accidental $\PKzL$ vetoes, which is estimated to be about 15\%
 from MC simulation, is taken into account in this study.

\begin{figure}
 \begin{center}
  \includegraphics[width=0.85\textwidth]{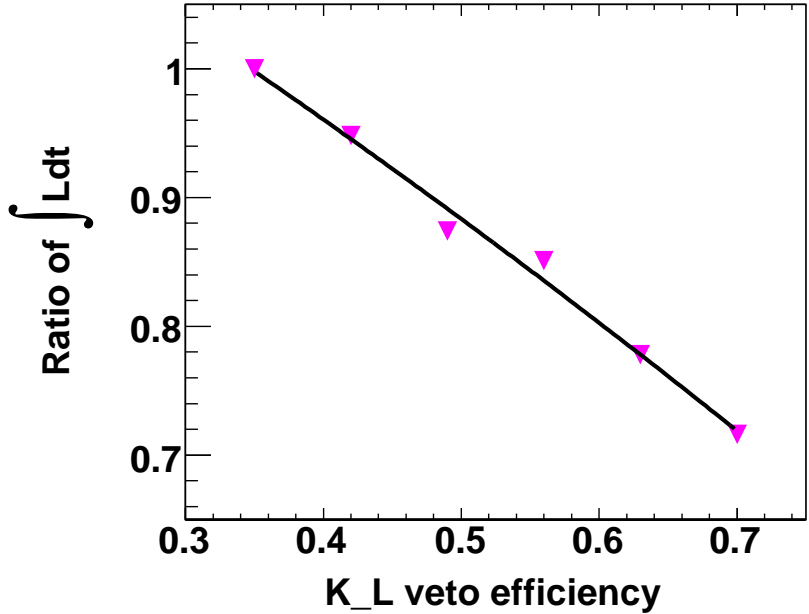}
  \caption{
  \label{fig:kl-taunu}
  The required-luminosity dependence on the $K_L$ veto efficiency in
  $B\to\tau\nu$.
  The vertical axis is the relative luminosity needed to achieve a given
  signal sensitivity; smaller values are better.
  }
 \end{center}
\end{figure}

 Figure~\ref{fig:kl-taunu} shows the $B\to\tau\nu$ sensitivity in terms of the required luminosity ratio
 to achieve a particular sensitivity for several $\PKzL$ veto efficiency cases.
 Smaller values on the vertical axis (luminosity ratio) are better.
 From the figure, a $PKzL$ veto efficiency of 70\% (i.e., double the present value) corresponds
 to a remarkable 30\% reduction in the required luminosity.
  
\subsection{Possibility of $\PKzL$ reconstruction improvement}

The present algorithm for $K_L$ identification is based on requiring
nearby hits in at least two RPC superlayers (``KLM-only'')
or an ECL cluster matching geometrically with an RPC hit (``ECL-KLM'').
(All RPC hits are three-dimensional: the orthogonal cathode strips give two
coordinates, while the RPC plane's position gives the third.)
One possible way to increase the efficiency without changing the
present iron-plate configuration is to relax the requirement of the KLM-only
class to at least one hit in a detector superlayer.
This increases the efficiency of $K_L$ reconstruction by a factor of
$\sim 1.3$, according to the GEANT3 simulation.
A more robust estimation, using better hadronic-shower simulation, will
be developed with the new GEANT4 Monte Carlo simulation.

Relaxing the $K_L$ identification requirement to a single superlayer
increases the fraction of neutrons misidentified as $K_L$ mesons.  This is
exacerbated with increased luminosity, since all of the neutrons arise
from beam backgrounds.

In the scintillator endcaps, the independent operation of the two orthogonal
scintillator planes within a superlayer
considerably reduces this neutron background, since one neutron typically is
absorbed in a single scintillator strip.  This is a distinct advantage over
the RPCs, where a neutron interaction within the active volume produces
hits on both orthogonal cathode planes.
Such backgrounds can be further suppressed in the scintillator endcaps by
using fast electronics and taking advantage of the sub-ns time resolution 
of the scintillator detectors. Using a signal gate of $10\,$ns, 
the neutron background drops to $\sim\!0.5$ per event in the
entire endcap KLM. 
This $K_L$ misidentification rate gives too high an accidental signal veto
of $B\to\tau\nu$, but this estimate may be overestimated because
\begin{itemize}
\item the background neutron rate may actually be smaller, while it
enters as a square.
\item we ignored the fact that some fraction of the strips in two
  layers do not overlap
\item the outermost endcap superlayers (with the highest neutron rate) are
  not useful for the $K_L$ veto and can be excluded.
\end{itemize}
There are additional possibilities to reduce the background rate:
\begin{itemize}
\item use a higher threshold (at $\sim 0.6$~MIP) for $K_L$
  identification in case of a single superlayer cluster. This reduces
  the final coincidence rate by a factor of $\sim$4. However this
  option requires a modification of the electronics (an additional
  discriminator).
\item require an additional nearby hit within the same
  superlayer. The final random coincidence rate will be reduced by a
  roughly a factor of 250. However, the efficiency of such requirement
  cannot be determined without a full GEANT4 simulation.
\item use three scintillator planes per superlayer.
\end{itemize}

\subsection{Summary}
An enhancement of $\PKzL$ identification will have a large and
beneficial impact on the measurement of the decay modes
with neutrinos such as $B\to\tau\nu$.
One way to enhance the $\PKzL$ identification is to relax the $K_L$
definition in the scintillator endcaps to one or more struck superlayers.
Our preliminary MC study shows that a factor of 1.3 improvement in
the $K_L$ identification efficiency is possible.
This corresponds to about a 30\% reduction in the luminosity required to
achieve a given $B\to\tau\nu$ signal sensitivity.
Of concern is the increase of the $K_L$ misidentification rate due to
stray neutrons. A conservative estimate indicates that
$\sim\!0.5$ neutrons will be misidentified as $K_L$ mesons per event
in the sBelle scintillator-based endcap.
However, it may be possible to improve the background rate by optimizing
the electronics and the reconstruction algorithm.


%% file: material.tex
 \section{Material effects}
\label{sec:material}
 In detector design, the material budget is always a big
 issue. This is especially true in a $B$-factory experiment, since the momenta of
 particles that we measure are relatively low and the effect of multiple
 scattering can be critical to determine $B$ decay vertices with good
 precision. We must
 pay special attention to this issue, because some of the layers of the 
upgraded SVD  are expected to have
 readout chips on sensors instead of at the edges of the 
sensor chain as in the present detector. 
 This implies that all the necessary massive structure
 for cooling will also be inside the detector acceptance.
 
 The robustness of the calorimeter performance should be checked.
 Photon conversions
 at detector components in front of the calorimeter can result in detection
 inefficiencies. If a conversion occurs just in front of the calorimeter,
 we can still reconstruct the electromagnetic shower, but components with
  sizable  amount of material placed far from the ECL should be avoided.
 In this respect, replacing the current ACC detector with the TOP counter
 is expected to improve the calorimeter performance.
 
  \input{Track+Vertex}

  \input{eclmat}

%% file: Track+Vertex.tex
\subsection{Track and vertex reconstruction}
\label{sec:matsvd}

As discussed in the beam background section~\ref{sec:beambg},
the SVD will be exposed to harsh
 beam background under the target luminosity.
The beam background in the SVD could be about 15 times higher than in the current
 Belle detector.
To cope with this, we decided to replace the current VA1TA readout chips
with APV25 chips.
Since the APV25 has a pulse width $16$ times shorter than the VA1TA,
the rate at which beam background pulses overlap with the signal will
decrease accordingly.

However, to realize the shorter pulse width we have to compromise on the
signal-to-noise (S/N) ratio.
In particular, for the outer layers several silicon sensors could be
connected and read out by a chip located at one end of the sensor chain.
In that case the S/N ratio would be about $4$ times worse
than for the current SVD.
Recently, an attractive new technique has been proposed, the so-called
``chip-on-sensor'', where readout chips are attached 
to each silicon sensor to reduce the deterioration of the S/N ratio.
However, this leads to an increase in the amount of material
within the tracking acceptance region.
Furthermore, the total number of readout channels would drastically increase.

For sBelle, the $B \to K^{*0}\gamma$ decay (reconstructed in $\PKzS \pi^0 \gamma$)
is one of the most important modes 
used to explore physics beyond the Standard Model.
Only charged decay products of the $\PKzS$ from the $K^{*0} \to \PKzS \pi^0$ decay chain 
are available for the reconstruction of the $B$ decay vertex.
To improve the detection efficiency for $B \to K^{*0} \gamma$, 
we have decided to extend the SVD detector volume in the radial direction 
from a 4-layer to  a 6-layer configuration, while keeping 
the same tracking acceptance for the CDC.
However, this change requires longer silicon sensors, which 
in turn leads to a  deterioration of the  S/N ratio of the readout chips.
The two additional layers also contribute to an increase in the 
amount of material.

In terms of the CDC, to reduce the high occupancy induced by the harsh beam background,
we will make the cell size smaller and
increase the number of sense wires by roughly a factor of two.
This proposal implies that the amount of material in the CDC will increase and
that the tracking performance could deteriorate.

Therefore, to estimate the effects that stem from the required
beam background tolerance and the impact on physics, simulation studies are performed.
The in-depth report related to the SVD itself (optimization of the layer position,
a study of a wider readout pitch size, the effect of a poorer S/N
ratio) was presented in  section~\ref{sec:svd}.
In this section, we discuss the impact of the 
additional material on the charged particle tracking and vertexing, while its 
effect on the calorimetry is described in the next subsection (\ref{sec:eclmat}).

\subsubsection{Tracking Performance}{\label{sec:tracking}}

In order to estimate the effect of the extra material on the tracking performance,
we utilize a GEANT3-based full detector simulation that has also been used in
the current Belle experiment.
In this simulation the density parameters of the silicon sensors in the SVD and 
those of wires in the CDC are intentionally multiplied by a factor of two or three.

Single muons are generated at the coordinate origin in a random direction with
a specific momentum ($0.2\,{\rm GeV/c}$, $0.5\,{\rm GeV/c}$, $1.0\,{\rm GeV/c}$ and
$2.0\,{\rm GeV/c}$) and then are reconstructed by the standard Belle tracking algorithm,
which is modified to correspond to each density case.
Here, to clearly see the dependence on the amount of material, no beam background is
assumed throughout this study. The effect of beam background on the
tracking is discussed in section~\ref{sec:cdc}.

\begin{figure}[h]
\includegraphics[width=0.8\linewidth]{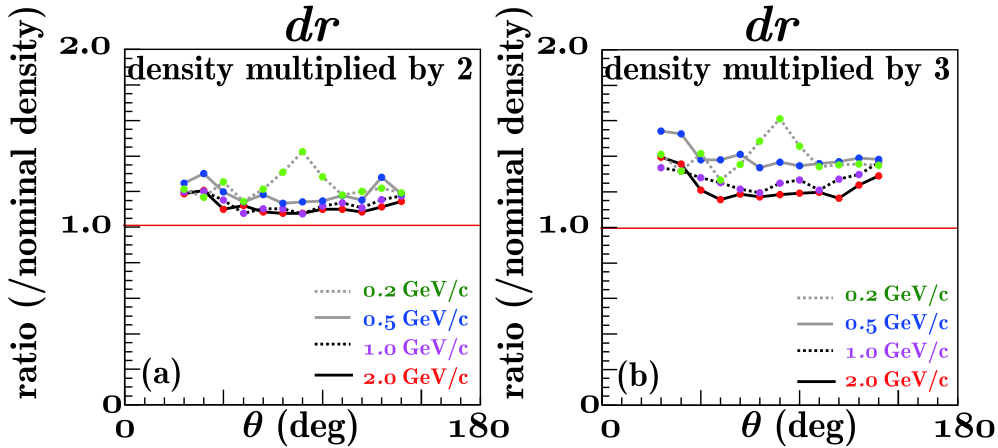}
\caption{\label{fig:dr_4single} $dr$ resolution as a function of the 
polar angle relative to the beam axis, 
 normalized to the corresponding resolution estimated in
the nominal density case. The resolution degradation can be observed as the deviation
from 1.0. Plots (a) and (b) correspond to the cases 
when the densities of the silicon sensors in the SVD 
and the wires in the CDC are increased by a factor of two and three, respectively.}
\end{figure}

\begin{figure}[h]
\includegraphics[width=0.8\linewidth]{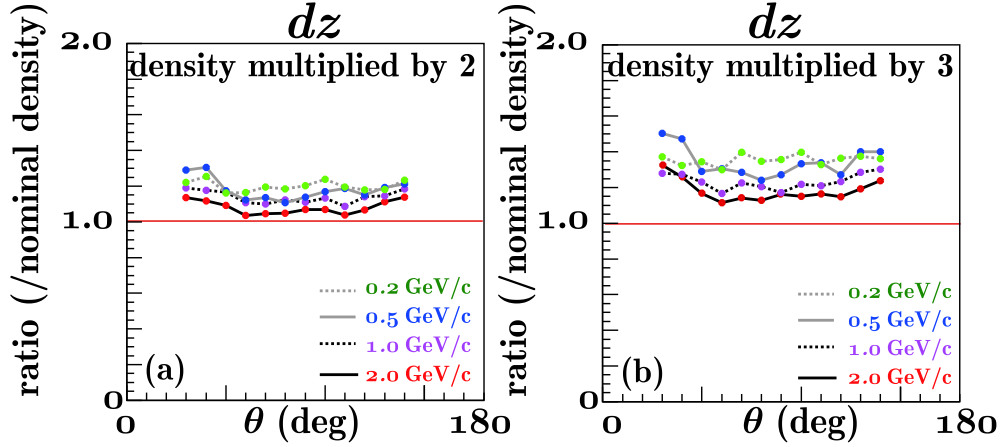}
\caption{\label{fig:dz_4single} $dz$ resolution as a function of the
polar angle relative to the beam axis, 
 normalized to the corresponding  resolution estimated in
the nominal density case. The resolution degradation can be seen as the deviation
from 1.0. Plots (a) and (b) correspond to the cases 
when the densities of the silicon sensors in the SVD 
and the wires in the CDC are increased by a factor of two and three, respectively.}
\end{figure}

Figures~\ref{fig:dr_4single} and \ref{fig:dz_4single} show the relative resolution
for the tracking parameters ``$dr$'' and ``$dz$'' with respect to the nominal density case,
respectively. 
In each of the two figures, plot (a) corresponds to the case when the density of materials
under consideration is multiplied by a factor of two, and plot (b) to the case 
when it is multiplied by a factor of three.
The resolution degrades by $5\%$ -- $10\%$ depending on the momentum
for both ``$dr$'' and ``$dz$'' in case (a);
$8\%$ -- $20\%$ degradations are observed in case (b).

\begin{figure}[h]
\includegraphics[width=0.8\linewidth]{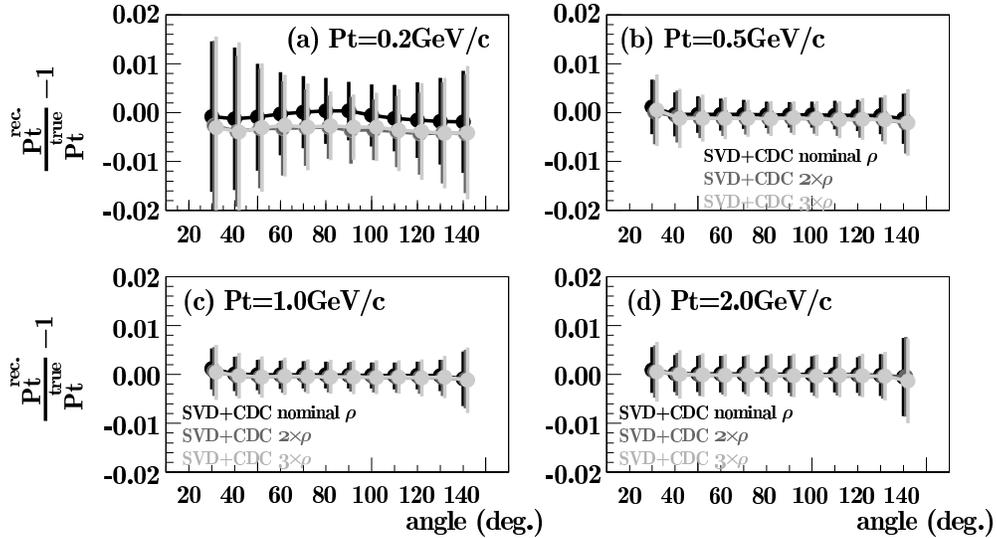}
\caption{\label{fig:pt_4single} $P_t$ relative bias and resolution  
 as a function of the polar angle relative to the beam axis. 
 The difference between the reconstructed $P_t$ and the input is shown as the deviation
 from 0.0. 
 Black points and error bars correspond to the nominal density case.
 Dark gray (light gray) corresponds to the case when the densities of the silicon sensors
 in the SVD and the wires in the CDC are multiplied by a factor of two (three).}
\end{figure}

Another tracking parameter, the transverse momentum component $P_t$, 
is also checked with the same conditions as above,
but the transverse momentum of particle is fixed instead of the full momentum.
The deviation of the reconstructed transverse momentum from the input value is
plotted as a point in the Fig.~\ref{fig:pt_4single} as a function of the polar angle
relative to the beam axis.
The relative resolution of the transverse momentum is also shown 
by the points with error bars.
Although the reconstructed transverse momentum matches well to the input
for relatively high momentum tracks, the resolution in the case when the density
is increased by a factor of two (three) becomes $5\%$ -- $15\%$ ($15\%$ -- $30\%$) 
worse than in the case of nominal density, depending on the polar angle.
For low momentum tracks, due to the energy loss in the SVD and CDC, a small bias
(below $1\%$) is observed in the reconstructed transverse momentum, 
as shown in Fig.~\ref{fig:pt_4single}(a).         
In terms of the resolution, the degradation ranges from $4\%$ -- $20\%$
depending on the polar angle, for the case when the density is 
increased by a factor of two.

\begin{table}[htbp]
 \caption{\label{tab:ptres}The deviation of the reconstructed $P_t$ from the input and
 the resolution for the typical polar angle. Both are nomalized by the input $P_t$.
 }
 \scriptsize
 \begin{tabular}{|l|c|c|c|c|c|c|}
  \hline
  Case & \multicolumn{6}{c|}{$P_t=0.2$GeV/c} \\
  \cline{2-7}
  & \multicolumn{2}{c|}{$40^{\circ}$deg} & \multicolumn{2}{c|}{$80^{\circ}$deg} & \multicolumn{2}{c|}{$130^{\circ}$deg} \\
  \cline{2-7}
  & ${P_t^{rec.}}/{P_t^{true}}-1$ & res./$P_t^{true}$ & ${P_t^{rec.}}/{P_t^{true}}-1$ & res./$P_t^{true}$ &  ${P_t^{rec.}}/{P_t^{true}}-1$ & res./$P_t^{true}$ \\
  \hline
  nominal $\rho$ & -0.00127$\pm$0.00025 & 0.01460$\pm$0.00032 & 0.00039$\pm$0.00010 & 0.00674$\pm$0.00011 & -0.00175$\pm$0.00015 & 0.00884$\pm$0.00016  \\
  $2\times\rho$ & -0.00383$\pm$0.00027 & 0.01543$\pm$0.00035 & -0.00299$\pm$0.00010 & 0.00648$\pm$0.00010 & -0.00409$\pm$0.00018 & 0.01035$\pm$0.00020  \\
  $3\times\rho$ & -0.00397$\pm$0.00024 & 0.01837$\pm$0.00039 & -0.00268$\pm$0.00011 & 0.00703$\pm$0.00013 & -0.00419$\pm$0.00020 & 0.01130$\pm$0.00019  \\

  \hline
  \hline
  Case & \multicolumn{6}{c|}{$P_t=0.5$GeV/c} \\
  \hline
  nominal $\rho$ & -0.00035$\pm$0.00008 & 0.00449$\pm$0.00007 & -0.00045$\pm$0.00004 & 0.00271$\pm$0.00004 & -0.00066$\pm$0.00006 & 0.00374$\pm$0.00005  \\
  $2\times\rho$ & -0.00081$\pm$0.00010 & 0.00531$\pm$0.00008 & -0.00117$\pm$0.00005 & 0.00314$\pm$0.00004 & -0.00138$\pm$0.00008 & 0.00448$\pm$0.00004  \\
  $3\times\rho$ &  -0.00114$\pm$0.00011 & 0.00601$\pm$0.00009 & -0.00121$\pm$0.00006 & 0.00358$\pm$0.00005 & -0.00142$\pm$0.00009 & 0.00504$\pm$0.00007  \\
  \hline
  \hline
  Case & \multicolumn{6}{c|}{$P_t=1.0$GeV/c} \\
  \hline
  nominal $\rho$ & 0.00010$\pm$0.00006 & 0.00349$\pm$0.00005 & -0.00012$\pm$0.00004 & 0.00276$\pm$0.00003 & -0.00023$\pm$0.00005 & 0.00323$\pm$0.00005  \\
  $2\times\rho$ & -0.00034$\pm$0.00008 & 0.00407$\pm$0.00006 & -0.00055$\pm$0.00005 & 0.00310$\pm$0.00004 & -0.00070$\pm$0.00007 & 0.00371$\pm$0.00005  \\
  $3\times\rho$ &  -0.00026$\pm$0.00009 & 0.00471$\pm$0.00007 & -0.00051$\pm$0.00005 & 0.00348$\pm$0.00004 & -0.00054$\pm$0.00007 & 0.00432$\pm$0.00006  \\
  \hline
  \hline
  Case & \multicolumn{6}{c|}{$P_t=2.0$GeV/c} \\
  \hline
  nominal $\rho$ & 0.00014$\pm$0.00007 & 0.00378$\pm$0.00005 & 0.00008$\pm$0.00006 & 0.00376$\pm$0.00005 & -0.00015$\pm$0.00007 & 0.00429$\pm$0.00006  \\
  $2\times\rho$ & -0.00003$\pm$0.00008 & 0.00438$\pm$0.00005 & -0.00014$\pm$0.00006 & 0.00402$\pm$0.00005 & -0.00044$\pm$0.00008 & 0.00451$\pm$0.00006  \\
  $3\times\rho$ &  -0.00003$\pm$0.00009 & 0.00496$\pm$0.00007 & -0.00018$\pm$0.00007 & 0.00439$\pm$0.00006 & -0.00038$\pm$0.00009 & 0.00497$\pm$0.00009  \\
  \hline

  \hline
 \end{tabular}
\end{table}

To summarize, for single muon tracks in the more realistic case 
when the amount of material increases by a factor of two,
the $dr$ and $dz$ resolutions degrade by $5\%$ -- $10\%$, while the $P_t$ resolution
deteriorates by  $5\%$ -- $15\%$.
However, the more practical consequences of the increased material, such as
the effect on the vertexing performance, should be
checked in a high track-multiplicity environment.
For this purpose, we examine the effect on the $B$ vertex resolution 
with typical physics events we are interested in.
This is discussed in the next subsection (\ref{sec:vertexing}).

\subsubsection{Vertexing Performance}{\label{sec:vertexing}}

First of all, to see the influence of the additional material 
located in the inner part of SVD,
we use the modes $B \to J/\Psi\PKzS$, $B \to \pi^+ \pi^-$
and $B \to D^+D^-$ ($D^{\pm} \to K^{\mp} \pi^{\pm} \pi^{\pm}$),
since the $CP$ side vertex can be reconstructed
from charged particles produced in the vicinity of the interaction point.
For the study of the first two physics modes,  $50,000$ events were simulated,
while for the latter we used $100,000$ simulated events, all without the beam background.
Here the fitting parameters used in this study are the same as those defined in the 
subsection~\ref{sec:bgeff}.  

\begin{figure}[h]
\includegraphics[width=0.7\linewidth]{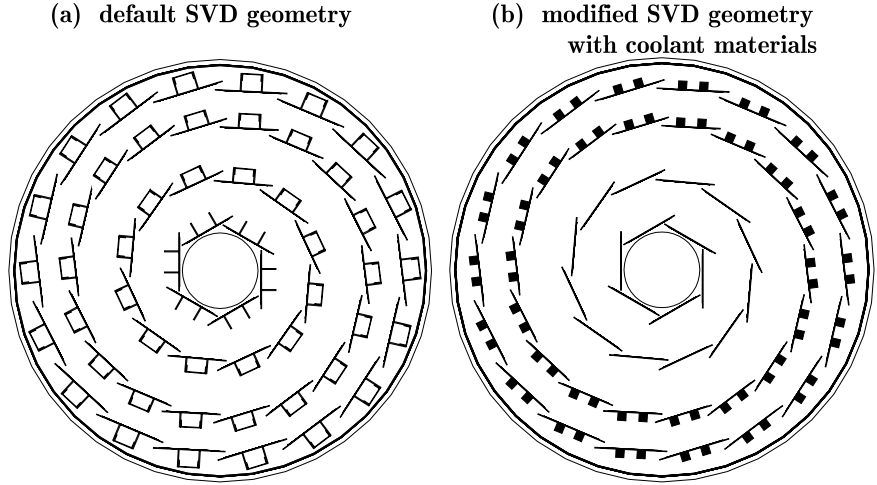}
\caption{\label{fig:SVDgeom} (a) the r-$\phi$ view of the current SVD geometry, 
four layers of units, each of which is composed of several
double-sided silicon strip detectors. The support materials, the so-called ``ribs'' made
from ``Zylon'', are attached on each unit to reinforce the unit structure. The
effect of the material in the ``ribs'' is negligible.
(b) the modified SVD geometry by replacing the ``ribs'' on the third and outermost
layer with two cooling tubes made from CFRP.}
\end{figure}

The following simulation conditions are examined by modifying the GEANT3-based
full detector simulation.
\begin{itemize}
\item[(A)] The densities of the silicon sensors in the SVD and
the wires in the CDC are set to the nominal values.
\item[(B)] The densities of the silicon sensors in the SVD and
the wires in the CDC are multiplied by a factor of three
\item[(C)] The densities of the silicon sensors in the SVD and
the wires in the CDC are multiplied by a factor of two.
\item[(D)] The density of the silicon sensors in the SVD is multiplied by a factor of two,
but that of the wires in the CDC is set to the nominal value.
\item[(E)] The density of the silicon sensors in the first and
second innermost layers in the SVD is multiplied by a factor of two,
while the densities of the other layers in the SVD and 
the wires in the CDC are unmodified.
\item[(F)] The density of the silicon sensors in the two
outermost layers (i.e., fourth and third layers) in the SVD
is multiplied by a factor of two, while the densities of the other layers in the SVD
and the wires in the CDC are unmodified.
\item[(G)] The same as the condition for (F), with the addition of
two 4mm $\times$ 4mm solid CFRP rods, which are
attached on each silicon sensor to model the coolant materials
for the chip-on-sensor as shown in Fig.~\ref{fig:SVDgeom}.
\end{itemize}

\begin{figure}[h]
\includegraphics[width=0.8\linewidth]{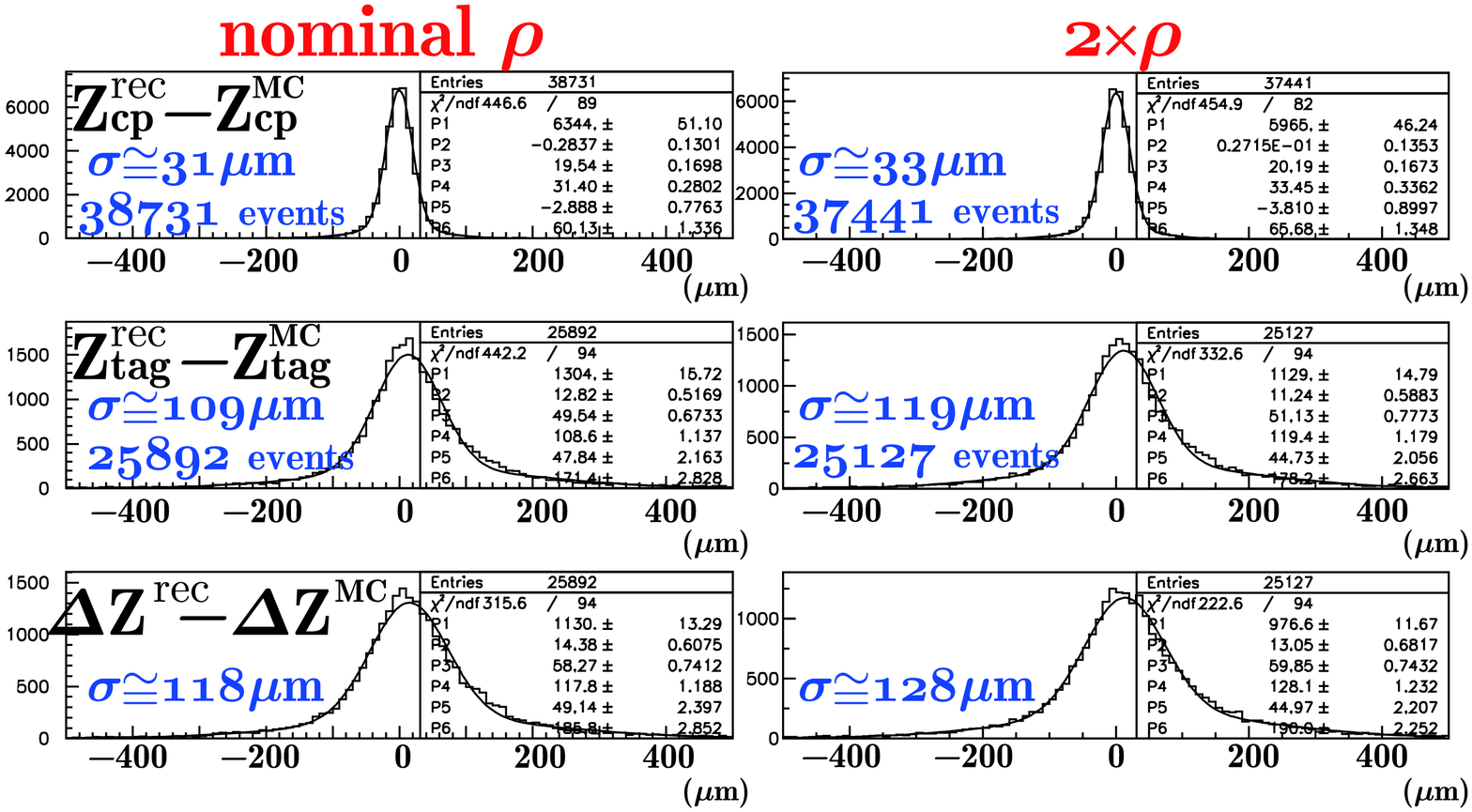}
\caption{\label{fig:vtxres} Distributions of the vertex resolution
for different simulation conditions. 
From top to bottom,
the three histograms show the distributions of 
the difference between the reconstructed $B$ decay vertex and true
decay point for $CP$-side (upper), tagging-side (middle)
and the difference of the displacement for the two $B$ decay vertices 
from the true value (lower).
The results for the nominal density case are shown in the left column and
the case that the density under consideration is increased by a factor two
is shown on the right.
}
\end{figure}

Figure~\ref{fig:vtxres} shows the fitted vertex resolutions
for case(A) (left) and (B) (right).
Tables~\ref{tab:cpres} and \ref{tab:tagres} also show the vertex resolution and
the efficiency for each case listed above.
Moreover the estimated $\Delta z$ vertex resolution for each mode is summarized 
in Table~\ref{tab:deltaz}.

\begin{table}[htbp]
\caption{\label{tab:cpres}$CP$ side vertex resolution. (A)-(G) are explained in the text.
{$\downarrow$} and {$\uparrow$}
represent the degradation and improvement, respectively.
The relative degradation or improvement with respect to the nominal density case is also
shown.
}
\begin{tabular}{|l|c|c|c|c|c|c|}
\hline
Case & \multicolumn{2}{c|}{$B \to \pi^+ \pi^-$} & \multicolumn{2}{c|}{$B \to J/\Psi\PKzS$} & \multicolumn{2}{c|}{$B \to D^+ D^-$} \\
\cline{2-7}
  & res. ($\mu$m) & \# of events & res. ($\mu$m) & \# of events & res. ($\mu$m) & \# of events \\
\hline
\hline
(A) & 31.4$\pm$0.3 & 38,731 & 35.6$\pm$0.4 & 24,859 & 43.0$\pm$0.6 & 13,225 \\
nominal $\rho$ & $-$ & $-$ & $-$ & $-$ & $-$ & $-$\\
\hline
(B) & 35.4$\pm$0.3 & 36,268 & 44.5$\pm$0.5 & 23,853 & 57.6$\pm$1.0 & 11,777\\
$3\times\rho$ : all SVD \& CDC layers & 13\% {$\downarrow$} & 6.4\% {$\downarrow$} & 25\% {$\downarrow$} & 4.0\% {$\downarrow$} & 34\% {$\downarrow$} & 10.9\% {$\downarrow$}\\
\hline
(C) & 33.5$\pm$0.3 & 37,441 & 40.0$\pm$0.4 & 24,369 & 51.2$\pm$0.8 & 12,366\\
$2\times\rho$ : all SVD \& CDC layers & 7\% {$\downarrow$} & 3.3\% {$\downarrow$} & 12\% {$\downarrow$} & 2.0\% {$\downarrow$} & 19\% {$\downarrow$} & 6.5\% {$\downarrow$}\\
\hline
(D) & 33.0$\pm$0.3 & 38,153 & 40.1$\pm$0.4 & 24,316 & 51.6$\pm$0.8 & 12,520\\
$2\times\rho$ : all SVD layers only & 5\% {$\downarrow$} & 1.5\% {$\downarrow$} & 13\% {$\downarrow$} & 2.2\% {$\downarrow$} & 20\% {$\downarrow$} & 5.3\% {$\downarrow$}\\
\hline
(E) & 33.7$\pm$0.3 & 38,289 & 40.0$\pm$0.4 & 24,556 & 50.6$\pm$0.8 & 12,713\\
$2\times\rho$ : SVD 1st \& 2nd layers only & 7\% {$\downarrow$} & 1.1\% {$\downarrow$} & 12\% {$\downarrow$} & 1.2\% {$\downarrow$} & 18\% {$\downarrow$} & 3.9\% {$\downarrow$}\\
\hline
(F) & 31.3$\pm$0.3 & 38,162 & 36.5$\pm$0.4 & 24,534 & 41.6$\pm$0.6 & 12,559\\
$2\times\rho$ : SVD 3rd \& 4th layers only & $-$ & 1.5\% {$\downarrow$} & 3\% {$\downarrow$} & 1.3\% {$\downarrow$} & 3\% {$\uparrow$} & 5.0\% {$\downarrow$}\\
\hline
(G) & 30.3$\pm$0.3 & 37,901 & 35.3$\pm$0.4 & 24,221 & 43.5$\pm$0.7 & 12,388\\
assumption (F) + cooling tubes & 4\% {$\uparrow$} & 2.1\% {$\downarrow$} & $-$ & 2.6\% {$\downarrow$} & 1\% {$\downarrow$} & 6.3\% {$\downarrow$}\\
\hline
\end{tabular}
\end{table}

\begin{table}[htbp]
\caption{\label{tab:tagres}The tagging side vertex resolution. 
(A)-(G) are explained in the text.
{$\downarrow$} and {$\uparrow$} represent the degradation and improvement, respectively.
The relative degradation or improvement with respect to the nominal density case is also
shown.}
\begin{tabular}{|l|c|c|c|c|c|c|}
\hline
Case & \multicolumn{2}{c|}{$B \to \pi^+ \pi^-$} & \multicolumn{2}{c|}{$B \to J/\Psi\PKzS$} & \multicolumn{2}{c|}{$B \to D^+ D^-$} \\
\cline{2-7}
  & res. ($\mu$m) & \# of events & res. ($\mu$m) & \# of events & res. ($\mu$m) & \# of events \\
\hline
\hline
(A) & 108.6$\pm$1.1 & 25,892 & 108.8$\pm$1.4 & 15,833 & 111.0$\pm$1.9 & 8,564 \\
nominal $\rho$ & $-$ & $-$ & $-$ & $-$ & $-$ & $-$\\
\hline
(B) & 124.7$\pm$1.3 & 24,495 & 125.2$\pm$1.6 & 15,344 & 124.4$\pm$2.2 & 7,729\\
$3\times\rho$ : all SVD \& CDC layers & 15\% {$\downarrow$} & 5.4\% {$\downarrow$} & 15\% {$\downarrow$} & 3.1\% {$\downarrow$} & 12\% {$\downarrow$} & 9.8\% {$\downarrow$}\\
\hline
(C) & 119.4$\pm$1.2 & 25,127 & 116.6$\pm$1.5 & 15,715 & 115.0$\pm$1.8 & 7,989\\
$2\times\rho$ : all SVD \& CDC layers & 10\% {$\downarrow$} & 3.0\% {$\downarrow$} & 7\% {$\downarrow$} & 0.7\% {$\downarrow$} & 4\% {$\downarrow$} & 6.7\% {$\downarrow$}\\
\hline
(D) & 120.2$\pm$1.2 & 25,609 & 118.3$\pm$1.6 & 15,487 & 113.3$\pm$1.9 & 8,123\\
$2\times\rho$ : all SVD layers only & 11\% {$\downarrow$} & 1.1\% {$\downarrow$} & 9\% {$\downarrow$} & 2.2\% {$\downarrow$} & 2\% {$\downarrow$} & 5.1\% {$\downarrow$}\\
\hline
(E) & 119.9$\pm$1.2 & 25,585 & 116.5$\pm$1.4 & 15,619 & 119.1$\pm$2.1 & 8,301\\
$2\times\rho$ : SVD 1st \& 2nd layers only & 10\% {$\downarrow$} & 1.2\% {$\downarrow$} & 7\% {$\downarrow$} & 1.4\% {$\downarrow$} & 7\% {$\downarrow$} & 3.1\% {$\downarrow$}\\
\hline
(F) & 109.9$\pm$1.1 & 25,372 & 108.7$\pm$1.4 & 15,590 & 107.6$\pm$1.7 & 8,128\\
$2\times\rho$ : SVD 3rd \& 4th layers only & 1\% {$\downarrow$} & 2.0\% {$\downarrow$} & $-$ & 1.5\% {$\downarrow$} & 3\% {$\uparrow$} & 5.1\% {$\downarrow$}\\
\hline
(G) & 109.0$\pm$0.4 & 25,322 & 106.4$\pm$1.3 & 15,458 & 106.0$\pm$1.8 & 8,044\\
assumption (F) + cooling tubes & $-$ & 2.2\% {$\downarrow$} & 2\% {$\uparrow$} & 2.4\% {$\downarrow$} & 5\% {$\uparrow$} & 6.1\% {$\downarrow$}\\
\hline
\end{tabular}
\end{table}

\begin{table}[htbp]
\caption{\label{tab:deltaz}$\Delta$z vertex resolution.
(A)-(G) are explained in the text.
{$\downarrow$} and {$\uparrow$} represent the degradation and improvement, respectively.
The relative degradation or improvement with respect to the nominal density case is also
shown.}
\begin{tabular}{|l|c|c|c|}
\hline
Case & {$B \to \pi^+ \pi^-$} & {$B \to J/\Psi\PKzS$} & {$B \to D^+ D^-$} \\
\cline{2-4}
  & res. ($\mu$m) & res. ($\mu$m) & res. ($\mu$m) \\
\hline
\hline
(A) & 117.8$\pm$1.2 & 117.8$\pm$1.4 & 123.8$\pm$2.0 \\
nominal $\rho$ & $-$ & $-$ & $-$ \\
\hline
(B) & 133.0$\pm$1.3 & 137.6$\pm$1.6 & 147.5$\pm$2.7 \\
$3\times\rho$ : all SVD \& CDC layers & 13\% {$\downarrow$} & 17\% {$\downarrow$} & 19\% {$\downarrow$} \\
\hline
(C) & 128.1$\pm$1.2 & 127.3$\pm$1.6 & 134.1$\pm$2.8 \\
$2\times\rho$ : all SVD \& CDC layers & 9\% {$\downarrow$} & 8\% {$\downarrow$} & 8\% {$\downarrow$} \\
\hline
(D) & 128.5$\pm$0.5 & 126.5$\pm$1.6 & 129.9$\pm$2.2 \\
$2\times\rho$ : all SVD layers only & 9\% {$\downarrow$} & 7\% {$\downarrow$} & 5\% {$\downarrow$} \\
\hline
(E) & 127.2$\pm$1.2 & 127.1$\pm$3.0 & 135.7$\pm$1.8 \\
$2\times\rho$ : SVD 1st \& 2nd layers only & 8\% {$\downarrow$} & 8\% {$\downarrow$} & 10\% {$\downarrow$} \\
\hline
(F) & 117.3$\pm$1.2 & 118.9$\pm$1.5 & 121.6$\pm$1.8 \\
$2\times\rho$ : SVD 3rd \& 4th layers only & $-$ & 1\% {$\downarrow$} & 2\% {$\uparrow$} \\
\hline
(G) & 116.4$\pm$1.1 & 116.4$\pm$1.4 & 122.5$\pm$2.0 \\
assumption (F) + cooling tubes & 1\% {$\uparrow$} & 1\% {$\uparrow$} & 1\% {$\uparrow$} \\
\hline
\end{tabular}
\end{table}

These results imply that the additional material in the CDC does not affect
the vertex resolution at all.
Another crucial point is that, as expected, the material in the innermost and the second
SVD layers degrades the vertex resolution while
those in the outermost and third layers do not affect it.
Furthermore, the effect of the material attached to the outer layers
such as the coolant is not large.
The reconstruction efficiency including the vertexing efficiency
decreases at most by roughly $6\%$ because of the extra material. 
Note, however, that it is expected that 
this loss can be partially recovered from the SVD stand-alone tracking 
(under development).
This is especially important for the $B \to D^+ D^-$ mode.

Therefore, the chip-on-sensor on the outer layers is a possible technology
for the vertex detector as far as the  physics modes discussed above are concerned.
However, a chip-on-sensor for the inner layers does not seem to be a viable solution.

As a next step, the $B$ vertex resolution of a mode that could be more sensitive to
the material in the SVD outer layers has to be checked.
For this purpose, we use the $B \to K^{*0} \gamma (K^{*0} \to\PKzS\Pgpz)$ decay mode.
One reason is that only two charged pions from the $\PKzS$ decay 
can be used for the $B$ vertex reconstruction, and
in addition we require at least two SVD hits to reconstruct the $\PKzS$.
That is, the $B$ vertex of some events could be reconstructed from two tracks,
each of which has only two SVD hits in the outer layers.
Therefore, the additional material in the outer part of the SVD,
e.g., the chip-on-sensor, could affect the $B$ vertex resolution for this mode.

The simulation studies for cases (A) and (G) are done in the same way.
Table~\ref{tab:ksres} shows the $CP$ side vertex resolutions for each case.
The averaged vertex resolution for the case (G) is degraded by roughly $7\%$ from the 
nominal density case (A).
To clearly see the impact of the material in the SVD outer layers, the 
reconstructed events are categorized according to the SVD hit pattern.
When events are reconstructed by requiring SVD hits in all four layers for each track
the vertex resolution for the case (G) 
is consistent to that for the nominal density case (A) 
(See ``hits in all lyrs'' in Table~\ref{tab:ksres}).
On the other hand, in cases where SVD hits are required only in the third and fourth
layers for each track, the vertex resolution degrades by $\sim 15\%$
(See ``hits in 3 \& 4 lyrs'' in Table~\ref{tab:ksres}).

\begin{table}[htbp]
\caption{\label{tab:ksres}$CP$ side vertex resolution for $B \to K^{*0} \gamma$ decay.
(A) and (G) are explained in the text. The results are still preliminary.}
\begin{tabular}{|l|c|c|c|}
\hline
Case & \multicolumn{3}{c|}{$B \to K{*0} \gamma$} \\
\cline{2-4}
  & $\geq 2$ lyr hits. & hits in all lyrs & hits in 3 \& 4 lyrs \\
\hline
\hline
(A) & 128$\pm$2 ($\mu$m) & 78$\pm$2 ($\mu$m) & 163$\pm$4 ($\mu$m) \\
nominal $\rho$ & $-$ & $-$ & $-$ \\
\hline
(G) & 137$\pm$2 ($\mu$m) & 77$\pm$2 ($\mu$m) & 188$\pm$6 ($\mu$m) \\
$2\times\rho$ : SVD 3rd \& 4th layers only + coolant & 7\% {$\downarrow$} & $-$ & 15\% {$\downarrow$} \\
\hline
\end{tabular}
\end{table}

From this study an increase in the material in the SVD outer layers 
is found to be undesirable.
However, without applying the chip-on-sensor technique on the outer layers,
the S/N ratio could degrade by a factor of four and then
the SVD-CDC track-matching efficiency for the tracks from $\PKzS$ would 
become much worse as discussed in subsection~\ref{sec:svdsn}.
Therefore we have to make a compromise.

In conclusion, the amount of material in the outer layers of SVD has
to be increased in order to mantain a good S/N ratio, and as a 
consequence the vertex resolution for the $B \to K^{\ast0} \gamma$ decay channel
degrades by $7\%$ on average.
For the other physics modes, $B \to J/\Psi\PKzS$, $B \to \pi^+ \pi^-$
and $B \to D^+D^-$, the vertex resolution is mainly degraded by the 
material in the SVD inner layers.
Therefore, it is strongly recommended that the amount of 
material in the innermost and second layers should be minimized.

\subsubsection{Effect of the additional layers}{\label{sec:newlayer}}

To improve the reconstruction efficiency for the $B \to K^{*0} \gamma$ decay,
we decided to extend the SVD volume in radial direction from the
current 4-layer configuration to a 6-layer one.
This implies that the amount of the material corresponding to 
the additional two layers increases.
In this section, another simulation study is carried out by modifying the GEANT3-based
full detector simulator.

\begin{figure}[h]
\includegraphics[width=0.7\linewidth]{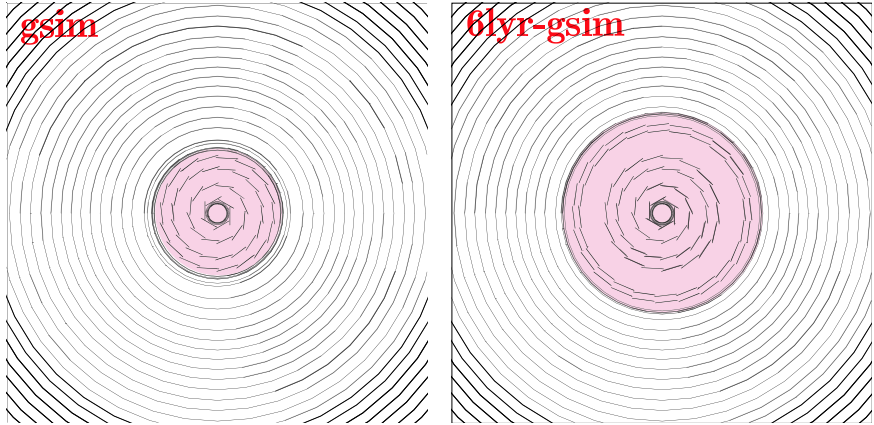}
\caption{\label{fig:6lyr} The figure on the left is the current SVD, which consists of
4 layers. The figure on the right is the 6-layer SVD configuration 
in which the three inner CDC layers
are replaced by two new SVD layers.} 
\end{figure}

As shown in Fig.~\ref{fig:6lyr}, the geometry implemented in the current
detector simulator is modified for this purpose.
The three inner CDC layers are removed and then the new extra two SVD layers are
installed in the extended SVD volume.
The fifth and sixth layer are located $13$\,cm and $14$\,cm
from the interaction point in the radial direction, respectively,
while the first to the fourth layers remain in their current positions.
The $30$ additional silicon sensor units of the same type as used for the fourth layer,
are placed so that they cover the fourth layer without any gaps in the acceptance.
In this study we asume that the density of the silicon sensors is not modified at all,
i.e., they are the nominal density.
The thickness of each silicon sensor is set to $300\,\mu$m, which is the same
as the silicon sensors installed in the current SVD.

The same physics modes are investigated as in the previous section.
The resolutions for the $CP$ and tagging side vertices are evaluated in the same way
and summarized in Table~\ref{tab:6lyr}.
In this case, no degradation is observed in the vertex resolution.
That is, the extra material corresponding to the additional two SVD layers
does not change the vertex resolution.
The $B$ vertex resolution for $B \to K^{*0} \gamma$ decay with the 6-layer SVD
is discussed in the SVD section, because this is related to the optimization of the
fifth layer position.

\begin{table}[htbp]
\caption{\label{tab:6lyr}$CP$ and tagging side vertex resolutions for 
$B \to J/\Psi\PKzS$, $B \to \pi^+ \pi^-$ and $B \to D^+D^-$.}
\begin{tabular}{|l|c|c|c|c|c|c|}
\hline
Case & \multicolumn{2}{c|}{$B \to \pi^+ \pi^-$} & \multicolumn{2}{c|}{$B \to J/\Psi\PKzS$} & \multicolumn{2}{c|}{$B \
to D^+D^-$}
\\
\cline{2-7}
  & $CP$ ($\mu$m) & tagging ($\mu$m) & $CP$ ($\mu$m) & tagging ($\mu$m) & $CP$ ($\mu$m) & tagging ($\mu$m) \\
\hline
\hline
default & 31.4$\pm$0.3 & 108.6$\pm$1.1 & 35.8$\pm$0.4 & 108.8$\pm$1.4 & 43.0$\pm$0.6 & 111.0$\pm$1.9 \\
\hline
6-lyr. & 28.5$\pm$0.2 & 106.2$\pm$1.1 & 33.8$\pm$0.3 & 104.6$\pm$1.4 & 40.0$\pm$0.6 & 99.2$\pm$1.7 \\

\hline
\end{tabular}
\end{table}

%% file: eclmat.tex
\subsection{Calorimetry}
\label{sec:eclmat}
\begin{figure}
 \includegraphics[width=0.3\columnwidth]{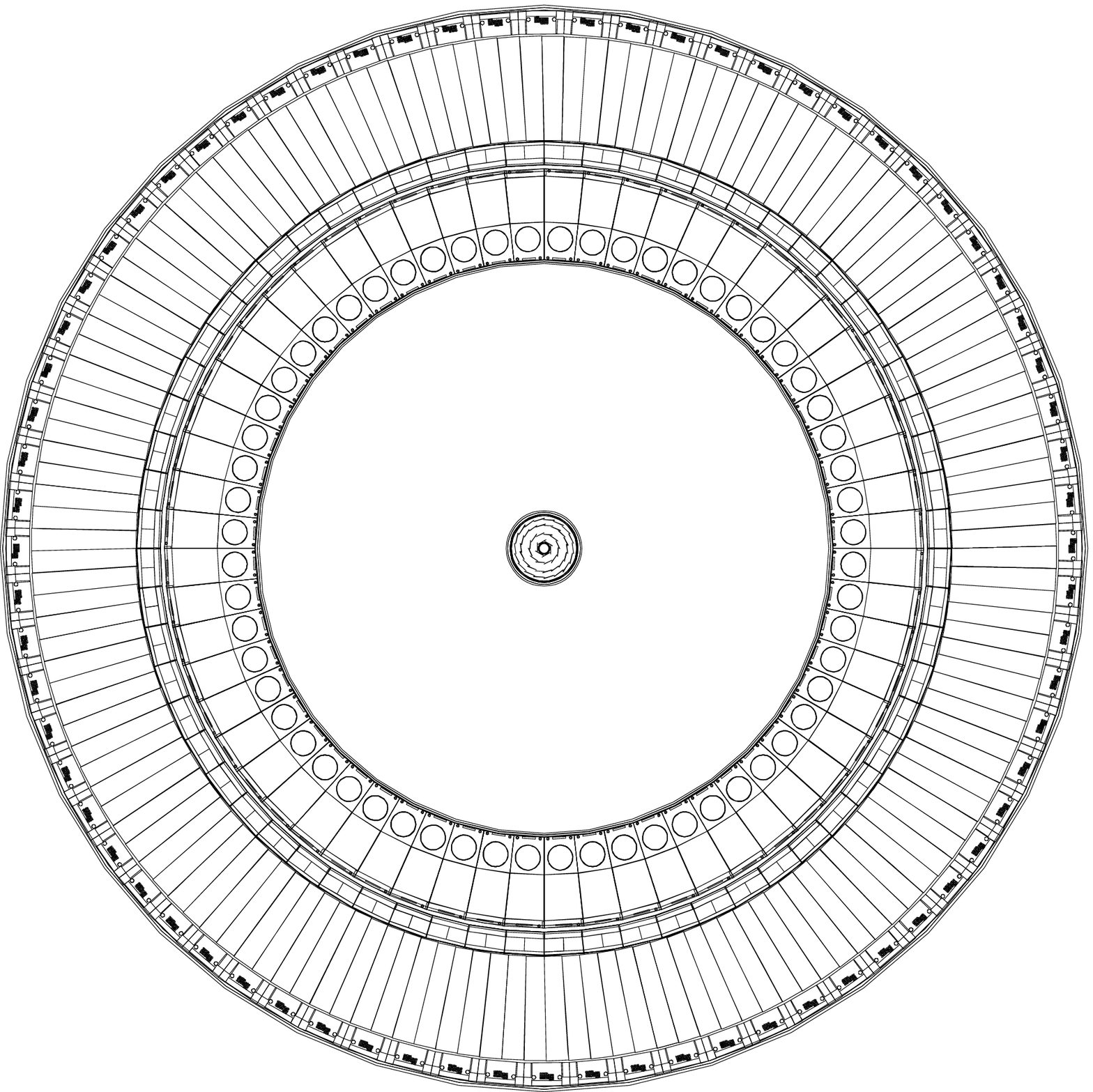}
 \includegraphics[width=0.3\columnwidth]{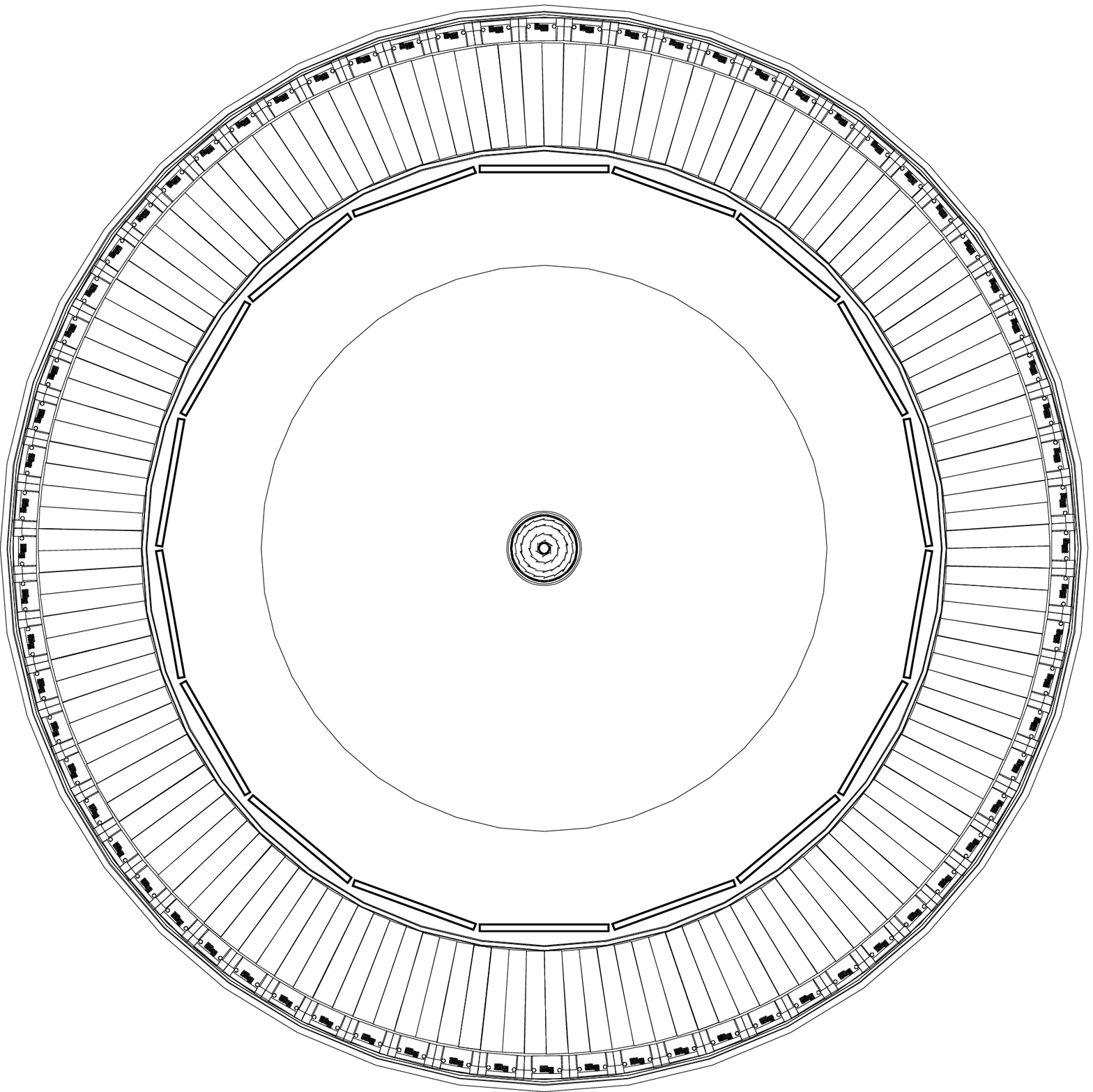}
 \includegraphics[width=0.3\columnwidth]{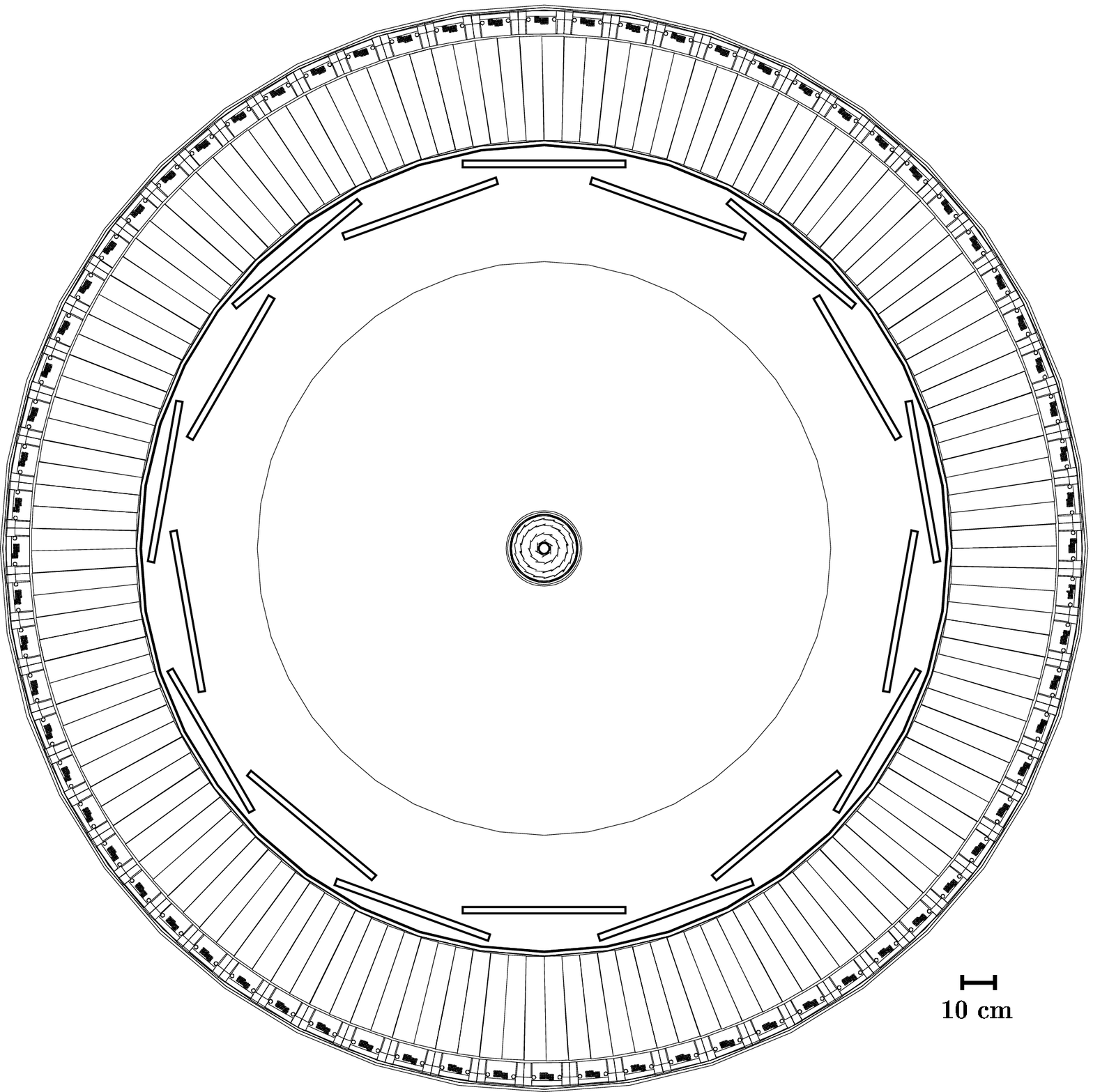}
 \caption{Barrel PID configurations that were tested to determine the
 effect of additional material on calorimetry
 From left to right, the ACC of Belle, single inlined TOP and staggered TOP.}
 \label{fig:topgeom}
\end{figure}

The effect of material on the efficiencies of photons and \Pgpz{}s are
studied usig the full simulation described in the previous section.
Here the densities of materials in the tracking devices (SVD and CDC) are
multiplied by
a factor of three to conservatively account for the increase in material
due to the upgrade.
An A-RICH detector is assumed for the upgraded endcap PID device,
while a TOP counter replaces the current barrel ACC and TOF counters.
Two types of TOP counter geometry were tested. One has no geometrical
overlap between adjacent quartz bars (sBelle geometry) while
the bars overlap in the other geometry (sBelle2); these two geometries are shown
in Fig.~\ref{fig:topgeom}.

Single particle Monte Carlo events are generated for photons and
\Pgpz{}s. Beam-induced background data obtained in experiment 49
are overlaid on the event.
Generated events are processed with the standard photon and \Pgpz\ finders.
\begin{figure}[htb]
 \begin{center}
  \includegraphics[width=.4\textwidth]{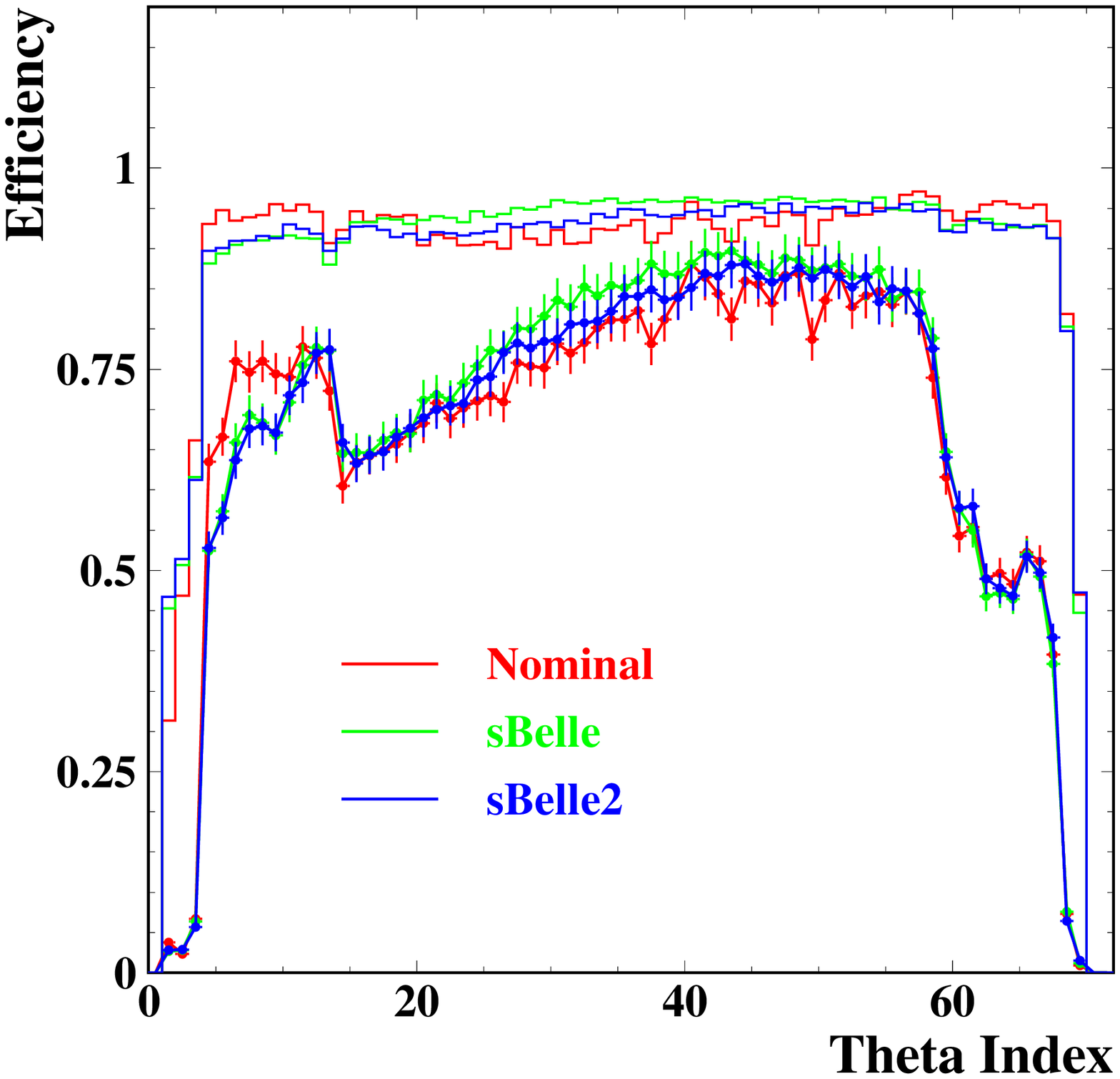}
  \includegraphics[width=.4\textwidth]{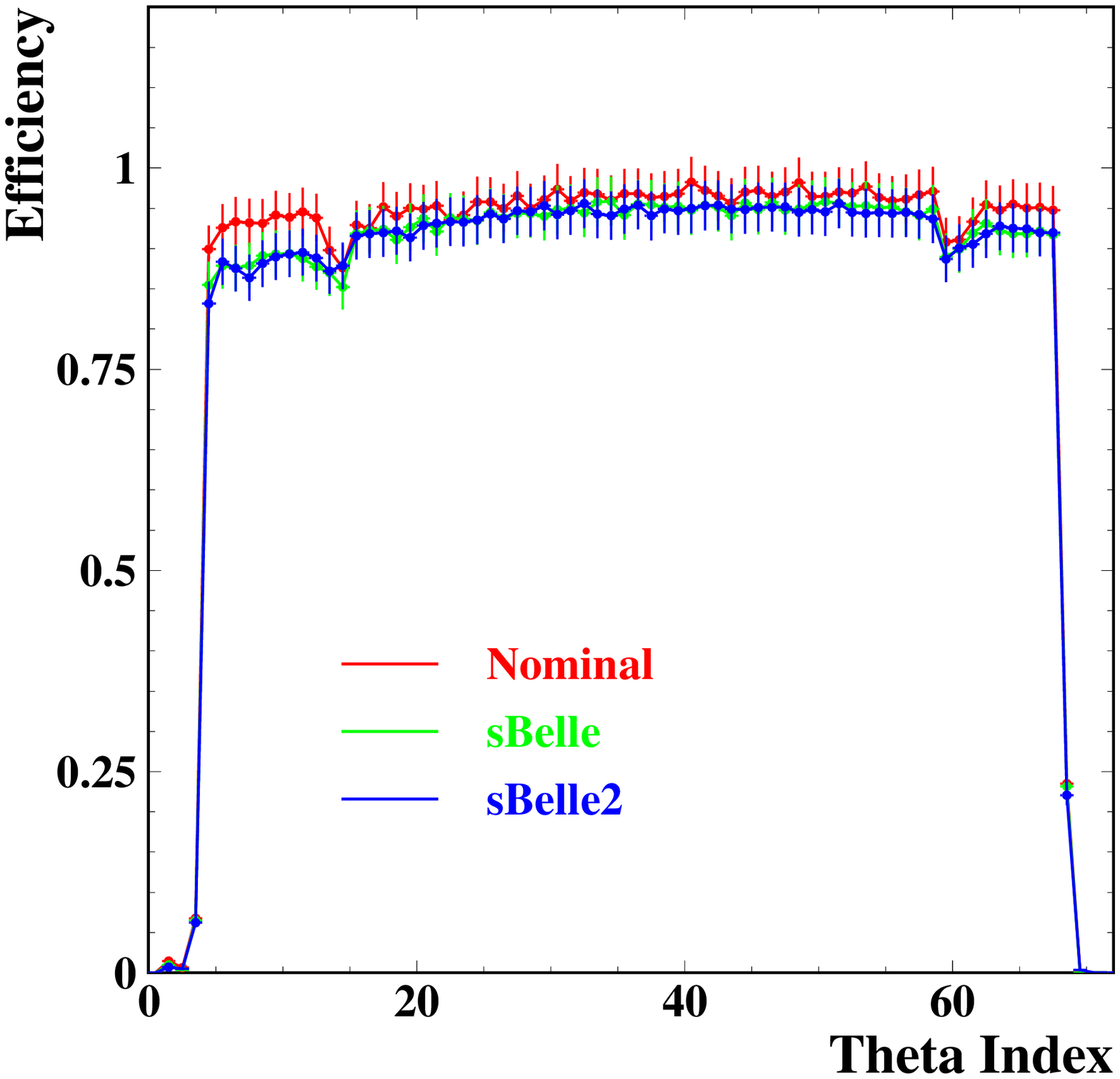}
  \caption{\sl Photon finding efficiencies as functions of
  $\theta$ indices of the calorimeter cell for 100\,MeV (Left) and
  1\,GeV (Right) photons. Open histograms in the left figure
  show the cluster finding efficiency before photon selection. 
  }
  \label{fig:gameff}
 \end{center}
\end{figure}

Figure\,\ref{fig:gameff} shows the photon finding efficiencies
for 100\,MeV and 1\,GeV photons. Three different colors correspond to
the three
detector configurations. As one can imagine, the efficiency depends on
the amount of material. The efficiency for the barrel increases slightly
as the amount of material decreases, while the efficiency for the forward endcap
decreases because of the extra material in the CDC end plate.
Although there are changes due to material, the effect is not very large.

 \begin{figure}[htb]
   \begin{center}
    \includegraphics[width=.4\textwidth]{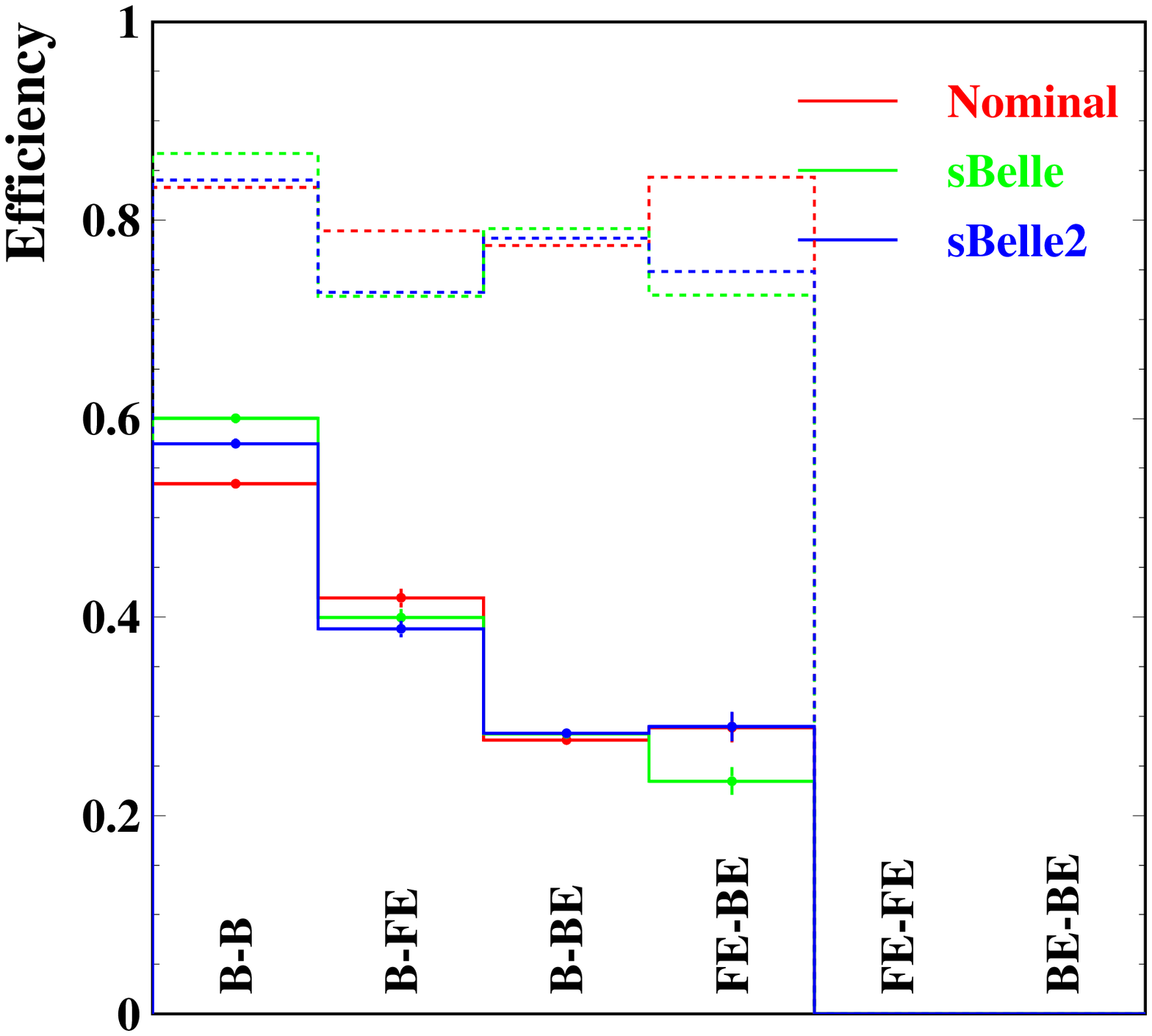}
    \includegraphics[width=.4\textwidth]{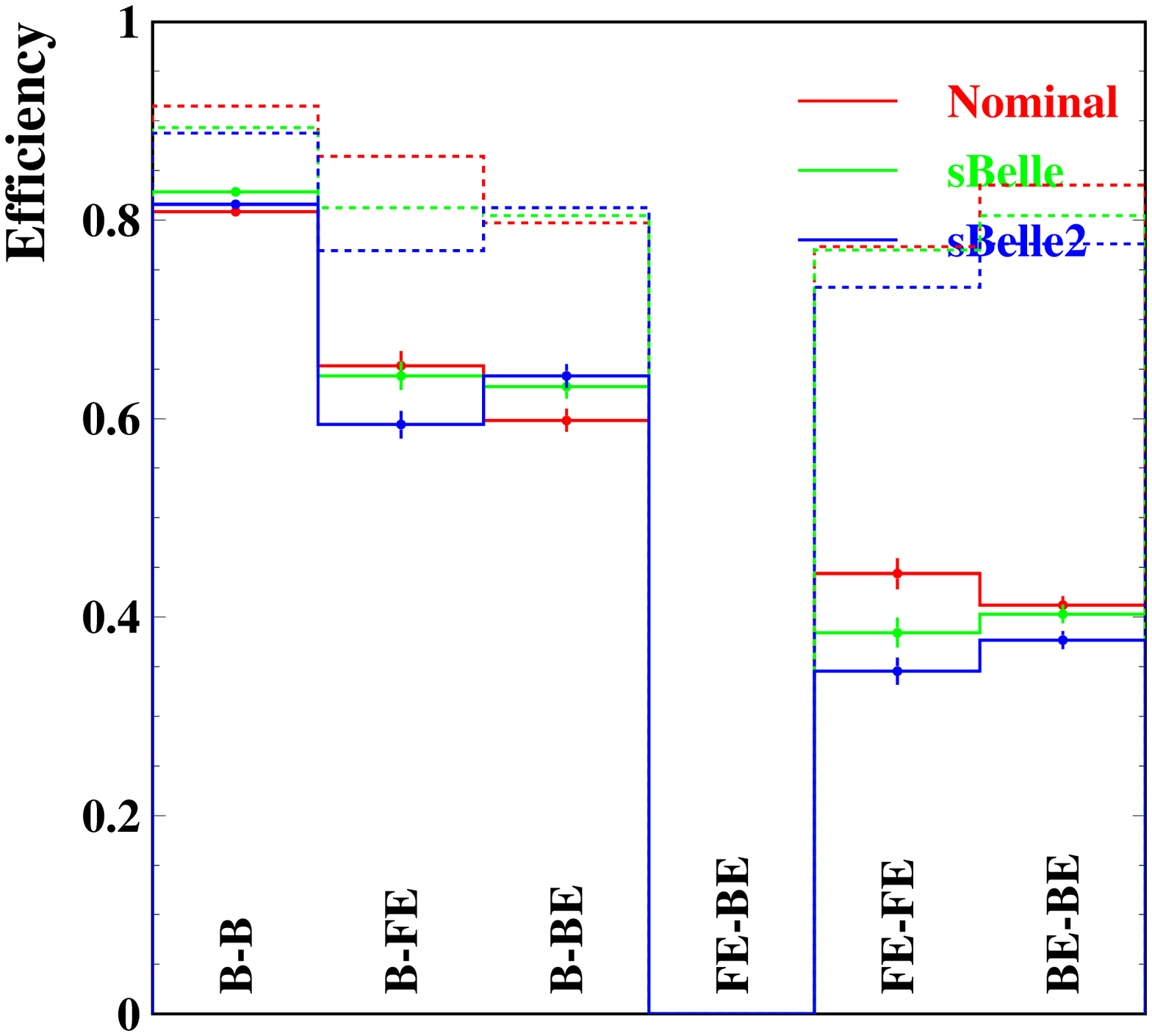}
    \caption{\sl \Pgpz\ finding efficiencies for 100\,MeV(Left) and
    1 GeV(Right) \Pgpz{}s as functions of combinations of detector regions
    B(barrel), FE(forward endcap) and BE(backward endcap).
    The dotted histograms show the efficiency to find two clusters
    for the two photons from a \Pgpz.
    }
    \label{fig:pi0eff}
   \end{center}
 \end{figure}
Figure\,\ref{fig:pi0eff} shows the \Pgpz\ finding efficiencies
for 100\,MeV and 1\,GeV \Pgpz{}s for combinations of detector regions.
One finds that the tendencies are the same as those for photons.

%% file: boost.tex
 \section{Beam Energy Asymmetry}
 \label{sec:boost}

 A larger beam energy asymmetry provides larger boost to the
 $\Upsilon(4S)$ system with respect to the laboratory frame so that the
 produced $B$ meson pair travels further along the beam axis,
 which allows us to measure
 the decay time difference ($\Delta t$) between the two $B$ mesons
 with higher precision. On the other hand, the additional boost distorts the
 solid angle, which is uniform in the center-of-mass frame, when seen in
 the laboratory frame. We place the detectors as close to the beam axis as
 possible to cover more solid angle with sensitive devices,
 however, there must be gaps around the beam axis
 in order to allow the beam and the SR to
 pass through as well as accommodate the readout electronics and cables from
 the sensors.
 Therefore, a larger boost will reduce the total
 detector coverage, which is not beneficial to any of the analyses. 

 When the energy asymmetry of the present KEKB was determined,
 discussions were based on
 the impact in the $\sin 2\phi_1$ measurement with $\PBz\to
 J/\psi\PKzS$, which was the critical measurement
 in the early stages of the Belle experiment.
 Now, the $\sin 2\phi_1$ measurement is already a precision measurement,
 and there are other important analyses that will be the focus in the sBelle era.

 One such analysis is the time-dependent $CP$ asymmetry measurement
 in $\PBz\to\phi\PKz$ and related decay modes to search for the effect
 of new physics.
 Since this analysis needs good $\Delta t$ resolution, the optimal
 beam energy asymmetry is similar to the case of
 $\PBz\to J/\psi\PKzS$.
 Analyses that include neutrinos in the final state such as $B\to\tau\nu$,
 $B\to K\nu\nu$, and $B\to D^\ast\tau\nu$ are also important. Since detector
 inefficiency 
 is the source of the peaking background in these modes that drastically
 degrades the sensitivity, better hermeticity is desired,
 while better $\Delta t$ resolution is not required.
 
 In the subsections~\ref{sec:boostphiks} and \ref{sec:boosttaunu},
 physics sensitivities for
 $\PBz\to\phi\PKzS$ and $B\to\tau\nu$ are examined for various energy
 asymmetry cases. A summary is given in subsection~\ref{sec:boostsummary}.


 \subsection{$\PBz\to\phi\PKzS$ and the beam energy asymmetry}
 \label{sec:boostphiks}
For the time-dependent $CP$ asymmetry measurement, the most important
variable is the quantity $\Delta t$.
Better $\Delta t$ resolution is also important for extracting
$\PBz\to\phi\PKzS$ signals from the large $e^+e^-\to q\bar{q}$
continuum background. This is because the signal events decay
with a finite
lifetime of 1.5\,ps, while the $q\bar{q}$ background events decay promptly.
So far, we have achieved a $\Delta t$ resolution of 1.4\,ps with the
current Belle detector.
This corresponds to a $\Delta z$ resolution of about 180\,$\mu$m with a
$\beta\gamma$ of 0.425 in asymmetric collisions of 
3.5\,GeV positrons (LER) on 8.0\,GeV electrons (HER).

We study the sensitivity of the $CP$ asymmetry measurement in $\PBz\to\phi\PKzS$ and
$J/\psi\PKzS$ decays for various energy asymmetries with toy MC experiments.
We evaluate the cases of $\beta\gamma = 0.56$, $0.425$,
$0.38$ and $0.34$, corresponding to LER$\times$HER energies of $3.1
\times 9.0$,
$3.5\times 8.0$, $3.65\times 7.67$ and $3.8\times 7.36$.
The $\Delta z$ resolutions and the S/N ratios are fixed to the values
obtained from
the Belle analysis using 492\,fb$^{-1}$ of data\cite{bib:b2s492fb-1}.
The change in the detector acceptance due to the
change of the boost factor is also considered.

The sensitivity is shown as the solid red line for $B\to\phi K^0_S$ 
and the blue dashed line for $B\to J\psi K^0_S$ in
Fig.~\ref{fig:boostsummary}.
The vertical axis shows the ratio of the luminosity to obtain the
same sensitivity with the current Belle boost. The smaller
value corresponds to the better sensitivity.

\subsection{$B\to\tau\nu$ and the beam energy asymmetry}
\label{sec:boosttaunu}
\begin{figure}
 \begin{center}
  \includegraphics[width=0.85\textwidth]{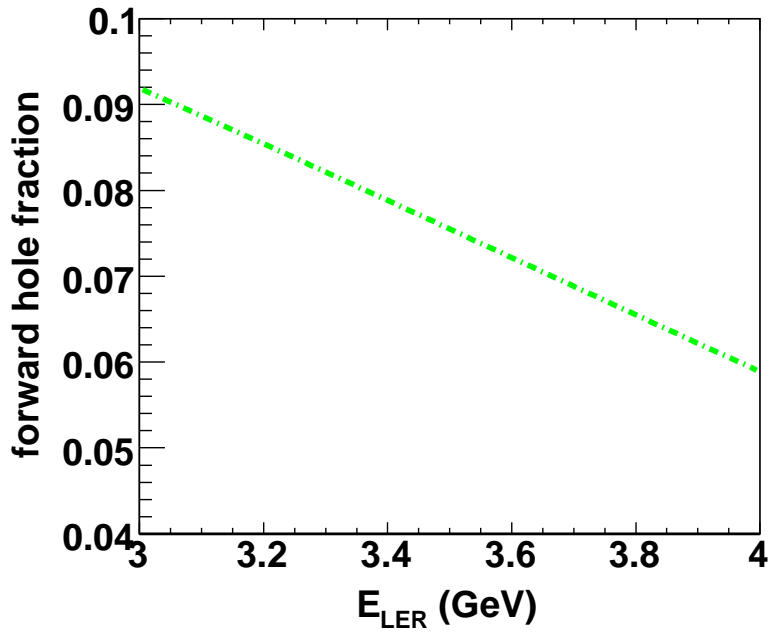}
  \caption{
  \label{fig:boost-fwdhole}
  The dependence of the size of the forward hole in the detector acceptance 
  on the boost factor in terms of LER energy.
  }
 \end{center}
\end{figure}

For the $B\to\tau\nu$ analysis,
detection of all particles from the two $B$ decays
except for neutrinos is required; the absence of additional energy
clusters in the ECL is the identifying feature of the signal.
The most significant effect of changing the
boost factor is the change of the detector acceptance.
Figure~\ref{fig:boost-fwdhole} shows the size of the hole in the
acceptance in the forward side of the Belle detector depending on the
LER energy. At an LER energy of 3\,GeV that is better for time-dependent
$CP$ asymmetry measurement, the hole
becomes about 20\% larger compared to the case of the present
Belle detector, which is not optimal for the $B\to\tau\nu$ analysis.

\begin{figure}
 \begin{center}
  \includegraphics[width=0.85\textwidth]{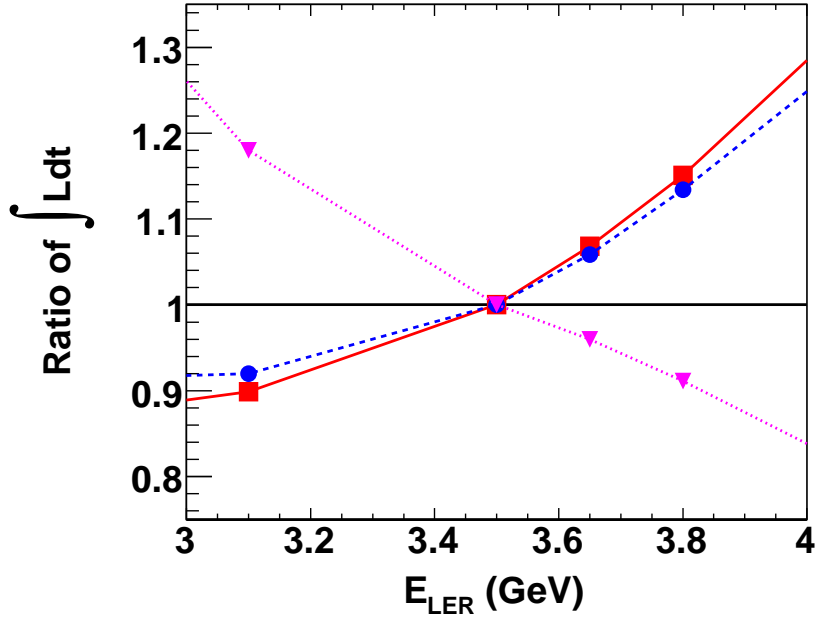}
  \caption{
  \label{fig:boostsummary}
  The sensitivity dependence on the boost factor corresponding to
  the LER energy.
  The purple dotted line is the sensitivity for $B\to\tau\nu$,
  the blue dashed line is for $B\to J\psi K^0_S$ and 
  the red line is for $B\to\phi K^0_S$.
  In the vertical axis is the luminosity ratio to achieve the same
  sensitivity (smaller is the better).
  }
 \end{center}
\end{figure}

There are two major effects from the change of the detector acceptance.
One is the detection efficiency of the tagging-side $B$ that is fully
reconstructed in hadronic $B$ decays. Another is the peaking background
fraction in the remaining ECL energy distribution from the loss of ECL 
energy cluster reconstruction.
We study the change in the full-reconstruction efficiency of the
tagging-side $B$ based on the geometrical acceptance. We assume the
reconstruction efficiency of the tagging $B$ is uniform in 
the detector acceptance and only examined the distribution at the
generator level.
The effect of decreasing the detector 
acceptance on the peaking background fraction is studied using a gsim MC event sample of generic $B$ decays.
For a variety of boost factors, we examine each ECL cluster if it is inside the detector acceptance 
and estimate the change in the number of events remaining in the
ECL energy signal region.
The effect on the peaking background fraction from an increase of the detector acceptance is
also studied using a toy MC and using the gsim event sample in the ECL signal region of 
generic $B$ decays.
If there are $\gamma$'s in the MC generator information inside of the detector 
acceptance for a given boost factor, we add the $\gamma$ energy to the ECL energy based
on the reconstruction efficiency obtained from gsim. We then count the number
of events remaining in the ECL signal region.
Table~\ref{tab:taunu-effpeak} shows the ratios of the full-reconstruction tagging
efficiencies and the peaking backgrounds to the current Belle boost.
The sensitivity in terms of the requrired luminosity ratio is shown as
the purple dotted line in Fig.~\ref{fig:boostsummary}.

\begin{table}
 \begin{center}
    \caption{
    \label{tab:taunu-effpeak}
    The ratio of the full-reconstrucion tagging efficiency and 
    the peaking BG for $B\to\tau\bar{\nu}$ depending on the 
    LER energy.
    }
    \begin{tabular}{lcccccc}
     \hline\hline
     Full-rec. eff. & $1.09\pm0.01$ & $1.05\pm0.01$ & $1.02\pm0.01$ & 1   &
     $0.92\pm0.01$ & $0.86\pm0.01$ \\
     Peaking BG     & $0.90\pm0.04$ & $0.96\pm0.04$ & $0.98\pm0.04$ & 1   & $0.86\pm0.04$ & $1.04\pm0.04$ \\
     E$_{\text LER}$      & 4.1  & 3.8  & 3.65 & 3.5 & 3.1  & 2.9 \\
     \hline\hline
     \end{tabular}
 \end{center}
\end{table}

\subsection{Summary of the beam energy asymmetry studies}
\label{sec:boostsummary}
Figure~\ref{fig:boostsummary} shows the improvement for one mode 
results in the loss of the sensitivity for the other mode by a similar amount.
If both modes are equally important, 
we conclude that the current beam energy asymmetry is the best point.




%% file: closing.tex
\section{Closing Remarks}
Extensive simulation studies have been performed to evaluate the impact
of various detector upgrade options and related parameters on several
important physics analyses. Studies are based on the current best
knowledge of the expected beam background, which is extrapolated from
ten years of Belle operation.

This report concentrates on the baseline design that
we had when we started this activity in the spring of 2007.
There are, in fact, more studies done, but not included in this report.
As an example, the detector hermeticity will be extremely important for
the analyses that include neutrinos in the final states. The impact of
a silicon-based tracker in the very low angle regions in the forward
and the backward direction has been studied; these show that an
ILC-FCAL-like detector could be very important.
Another missing item in this report is a study of a pixel detector.
Since we had no feasible option for day one, we did not include the
pixel detector in the baseline design.
The missing topics as well as updates especially in the background
simulation that are consistent with the latest design of the IR will be
included in a separate report, or possibly in the technical design
report (TDR).

Most of the studies are performed with tentative versions of
the sBelle simulator such as an integrated toy simulator and several
modified versions of the Belle's full simulator. An integrated full
simulator of the sBelle is being prepared, which will be used
for studies toward the TDR. Studies in this
report will be updated in the TDR; nevertheless, this report provides
a clear direction for the pre-TDR design of the sBelle detector.
We hope that this report will be helpful for each detector subsystem.